\documentclass[prl,aps,twocolumn,showpacs,preprintnumbers,amsmath,amssymb]{revtex4}
\usepackage{graphicx}
\usepackage{amssymb,amsmath,amsfonts}
\usepackage[hypertex]{hyperref}
\usepackage{epsfig}

\begin{document}


\def \be  {\begin{equation}}
\def \ee  {\end{equation}}
\def \ba  {\begin{eqnarray}}
\def \ea  {\end{eqnarray}}

\def\sel{ \tilde{e}}
\def\smu{\tilde{\mu}}
\def\stau{\tilde{\tau}}
\def\LSP{\tilde{\chi}^0_1}
\def\Evis {E_{\mbox{\tiny vis}}}
\def\chap{\tilde{\chi}_1^+}
\def\cham{\tilde{\chi}_1^-}
\def\sneumu{\tilde{\nu}_\mu}
\def\neu2{\tilde \chi_2^0}
\def\neu1{\tilde \chi_1^0}
\def\Emiss{E_{\mbox{\tiny miss}}}
\def\etc{{\it etc}}
\def\etal{{\it et al.}}
\def\ie{{\it i.e.}}
\def\eg{{\it e.g.}}
\def\etal{{\it et al.}}
\def\ibid{{\it ibid}.}
\def\tev{\,{\rm TeV}}
\def\gev{\,{\rm GeV}}
\def\inpb{\ifmmode {\rm pb}^{-1}\else ${\rm pb}^{-1}$\fi}
\def\infb{\ifmmode {\rm fb}^{-1}\else ${\rm fb}^{-1}$\fi}
\def\epem{\ifmmode e^+e^-\else $e^+e^-$\fi}
\newskip\zatskip \zatskip=0pt plus0pt minus0pt
\def\matth{\mathsurround=0pt}
\def\lsim{\mathrel{\mathpalette\atversim<}}
\def\gsim{\mathrel{\mathpalette\atversim>}}
\def\atversim#1#2{\lower0.7ex\vbox{\baselineskip\zatskip\lineskip\zatskip
  \lineskiplimit 0pt\ialign{$\matth#1\hfil##\hfil$\crcr#2\crcr\sim\crcr}}}


\pagestyle{plain}

\preprint{SLAC--PUB--12797, MIT-CTP-3882, ANL-HEP-PR-07-89}
\

\vskip 1cm
\title{General Features of Supersymmetric Signals at the ILC: Solving the LHC Inverse Problem}

\author{{C.F. Berger$^{1,2}$}\footnote{This work is
supported in part by funds provided by the U.S. Department of
Energy (D.O.E.) under cooperative research agreement
DE-FC02-94ER40818.}$^\ddagger$}
\email[email: ]{cfberger@mit.edu}
\author{{J.S. Gainer$^2$}\footnote{Work supported in part
by the Department of Energy, Contract DE-AC02-76SF00515.}}
\email[email: ]{jgainer@slac.stanford.edu}
\author{{J.L. Hewett$^2$}$^\ddagger$}
\email[email: ]{hewett@slac.stanford.edu}
\author{{B. Lillie$^{3,4}$}\footnote {Research
 supported in part by the US Department of Energy under contract
 DE--AC02--06CH11357}}
\email[email: ]{lillieb@uchicago.edu}
\author{{T.G. Rizzo$^2$}$^\ddagger$}
\email[email: ]{rizzo@slac.stanford.edu}

\affiliation{1) Center for Theoretical Physics,
Massachusetts Institute of Technology, Cambridge, MA 02139, USA}
\affiliation{2) Stanford Linear Accelerator Center, 2575 Sand Hill Rd.,
Menlo Park, CA  94025, USA}
\affiliation{3) High Energy Physics Division, Argonne National Laboratory,
Argonne, IL 60439, USA}
\affiliation{4) Enrico Fermi Institute, University of Chicago, 5640 South Ellis Avenue,Chicago, IL 60637, USA}


\begin{abstract}

We present the first detailed, large-scale study of the Minimal Supersymmetric Standard
Model (MSSM) at a $\sqrt s=500$ GeV International
Linear Collider, including full Standard Model backgrounds and
detector simulation. We investigate 242 points in the MSSM parameter space, which we
term models, that have been shown by Arkani-Hamed \etal\ to
be difficult to study at the LHC. In fact, these points in MSSM
parameter space correspond to 162 pairs of models which
give  indistinguishable signatures at the
LHC, giving rise to the so-called LHC Inverse Problem.
We first determine whether the production of the various
SUSY particles is visible above the Standard Model background for
each of these parameter space points, and then make a detailed
comparison of their various signatures. Assuming an integrated
luminosity of 500 fb$^{-1}$, we find that only 82 out of 242
models lead to visible signatures of some kind with a significance
$\geq 5$ and that only 57(63) out of the 162 model pairs are
distinguishable at $5(3)\sigma$. Our analysis includes PYTHIA and
CompHEP SUSY signal generation, full matrix element SM backgrounds
for all $2\to 2\,, 2\to 4$, and $2\to 6$ processes, ISR and
beamstrahlung generated via WHIZARD/GuineaPig, and employs the
fast SiD detector simulation org.lcsim.

\end{abstract}

\maketitle


\section{Introduction}

The LHC is scheduled to begin operations within a year and is expected
to change the landscape of particle physics. While the
Standard Model (SM) does an excellent job describing all
strong interaction and electroweak
data to date~{\cite{LEPEWWG,Yao:2006px}}, there are many
reasons to be dissatisfied with the SM. Chief among them are
issues related to electroweak symmetry breaking. As is by now
well-known, the SM with a single Higgs doublet that is responsible for
generating the masses of both the electroweak
gauge bosons and fermions encounters difficulties associated
with stability, fine-tuning, and naturalness. Addressing these issues
necessitates the existence of new physics at the Terascale. To this end,
numerous creative candidate theories that go beyond the SM have been proposed
and many yield characteristic signatures at the LHC.
When the ATLAS and CMS detectors start taking data at the LHC, they will
explore this new territory. They will then begin the process
of identifying the nature of physics at the Terascale and of determining how it
fits into a broader theoretical structure.

Of the several proposed extensions of the SM that resolve the issues
mentioned above,
the most celebrated is Supersymmetry (SUSY)~{\cite{SUSY}}. Our working
hypothesis in this paper is that SUSY has been discovered at the LHC, \ie,
that new particles have been observed and it has been determined that they
arise from Supersymmetry. Identifying new physics as Supersymmetry is in
itself a daunting task, and we would be lucky to be in such a situation!
However, even in this optimistic scenario, much work would be left to be done
as SUSY is a very broad framework. We would want
to know which version of SUSY nature has realized and for this we would need
to map the LHC observables to the fundamental parameters in the weak
scale SUSY Lagrangian. A question that would arise is whether this
Lagrangian can be uniquely reconstructed given the full set of LHC
measurements. This issue has been recently quantified in some detail by the
important work of Arkani-Hamed, Kane, Thaler and Wang
(AKTW)~{\cite{Arkani-Hamed:2005px}},
which demonstrates what has come to be known as the LHC Inverse Problem.
AKTW found that even in the simplest realization of Supersymmetry, the
Minimal Supersymmetric Standard Model (MSSM), such a unique mapping does
not take place given the LHC observables alone and that many points in
the MSSM parameter space
cannot be distinguished from each other.
Here, we extend their study and examine whether data from the proposed
International Linear Collider (ILC) can uniquely perform this inverse
mapping and resolve the model degeneracies found by AKTW.

In brief, AKTW considered a restricted Lagrangian parameter
subspace of the MSSM. They forced all
SUSY partner masses to lie below 1 TeV (in order to obtain a large
statistical sample at the LHC), fixed the third
generation $A$-terms to 800 GeV and set the
pseudoscalar Higgs mass to be 850 GeV. Points in the MSSM
parameter space, hereafter referred to as models for brevity,
were generated at random with the conditions that
$\tan \beta$ lies in the range 2-50,
squark and gluino Lagrangian mass terms lie
above 600 GeV, and Lagrangian mass terms for the non-strongly interacting
particles be greater than 100 GeV. 43,026 models were generated
in this 15-dimensional parameter space under the assumption
that all parameter ranges were uniformly distributed,
\ie, flat priors were employed.
No further constraints, such as the LEP lower bound
on the Higgs mass {\cite {LEPEWWG}} or consistency with the relic density
of the universe were applied. For each model,
PYTHIA~{\cite{Sjostrand:2006za}}
was used to calculate the resulting physical
SUSY spectrum and to generate 10 fb$^{-1}$ of SUSY `data' at the LHC,
including all decays and hadron showering effects.
This `data' was then piped through the PGS fast
detector simulation {\cite {PGS}} to mimic the effects
of the ATLAS or CMS detectors. From this `data,'
AKTW constructed a very large number of
observables associated with the
production and decay of the SUSY partners. No SM backgrounds
were included in their `data' sample. AKTW then observed that
a given set of values for these observables along with their
associated errors, \ie, a fixed region in LHC
signature space, corresponded to several distinct
regions in the 15-dimensional MSSM parameter space.
This implies that the mapping from experimental data to the underlying theory
is far from unique at the LHC.  Normally phenomenologists determine the values
of observables for a given point in the parameter space of some model.
Thus, the task of experimentalists, to determine parameters of the underlying
theory from the data, can be viewed as an inverse mapping.  The analysis of
AKTW found that this mapping is many-to-one, that is, the underlying parameters
cannot be uniquely determined.  Thus there is an ``LHC Inverse Problem.''
Clearly, if one incorporates the existing SM backgrounds as well as
systematic effects into this kind of study, the number of possible
models that share indistinguishable signatures will only increase,
potentially significantly. The LHC Inverse Problem is thus a very serious one.
It is also important to be reminded that AKTW was a general MSSM study
in the sense that no assumptions were made about the SUSY breaking
mechanism; almost all LHC studies do make such assumptions, and therefore
study potential signatures only within a few restricted scenarios, which
may or may not be realized in Nature.  A general MSSM parameter scan gives
us a much better feeling for the range of possibilities in SUSY scenarios.

However, the fact that an LHC Inverse Problem exists is not overly
surprising and the real issue we face is how to resolve it.
In this paper we will begin to address the question of whether the
models that AKTW found to be indistinguishable at the LHC can be resolved by a
high luminosity $e^+e^-$ collider operating at
500 GeV in the center of mass with a polarized initial electron
beam, \ie, the ILC. Traditional ILC lore indicates this is the case, as
studies have shown \cite{ILClit,Weiglein:2004hn},
\eg, the mass
and couplings of any kinematically accessible weakly
interacting state should be measured at the $1\%$ level or
better at the ILC. Such precise determinations imply
that decay signatures and distributions produced by new
particles such as the SUSY partners will be
observed with relative cleanliness and be well measured.
The LHC Inverse Problem provides us with a unique
opportunity to test this lore over a very wide range of the
MSSM parameter space by comparing the signatures
of hundreds of models. We will show that, as believed, the ILC can
generally distinguish models, at least in the case of this restricted
scenario of the MSSM,
and we will explore the reasons why it fails when it does.
We will
find that some SUSY measurements are more difficult to
obtain than previously thought, and we will identify some
problematic areas of the MSSM parameter space which
require further study.
We note that although AKTW did not impose all of the
many possible experimental constraints on the MSSM parameter space, the resulting large model
sample should be representative of the range of somewhat more difficult possibilities
that one might face at the ILC.

On our way to addressing the Inverse Problem at the ILC, we face the more
immediate, and perhaps even more important, issue of the visibility of the various SUSY particles in
the AKTW models. We find that this is surprisingly non-trivial and is
perhaps a more important task as one cannot differentiate between models
which have
no visible SUSY signatures. In our analysis below, we will perform a detailed
study of the visibility of the various SUSY particles in all of the models.
We will employ an extensive menu of search techniques and examine when they
succeed and how they fail. Our philosophy will be to apply a general
search strategy that performs uniformly well over the full MSSM parameter
region, rather than make use of targeted searches for particular
parameter points.
We believe this mirrors the reality of an experimental search for new physics
and reflects the fact that not all of the SUSY particles in these models
will have been observed at the LHC (recall that the models we have
inherited from AKTW are difficult cases at the LHC).
It is important to point out that given the large set of models we examine,
not linked to a particular SUSY breaking mechanism, provides a wide window on what
studying generic SUSY may be like at the ILC. This is the first analysis of this kind and
the results of this study were found to be unexpected and quite surprising,
at least to the present authors.

The possibility of measuring specific SUSY particle
properties at the ILC for particular special points in the MSSM
parameter space has a long history {\cite {ILClit}}.
Our approach here, however, provides several
aspects which have not been simultaneously featured in earlier
analyses: ($i$) We examine several hundred,
essentially random, points in MSSM parameter space, providing
a far wider than usual sampling of
models to explore and compare. This gives a much better indication
of how an arbitrary MSSM parameter point behaves and what experimental
techniques are necessary to adequately cover the full parameter space.
($ii$) We include all effects
arising from initial state radiation (ISR), \ie, bremsstrahlung,
as well as the specific ILC beamstrahlung spectrum for the superconducting
RF design, including finite beam energy
spread corrections. The beam spectrum is
generated by GuineaPig {\cite {Guinea,TimB}}.
($iii$) We incorporate
all $2\to 2,~2\to 4$ and $2\to 6$ SM background processes,
including those resulting from initial state
photons (\ie, from the corresponding
$\gamma \gamma$ and $\gamma e^{\pm}$ interactions).
These are generated with full matrix elements
via WHIZARD/O'MEGA {\cite {WHIZARD}} for arbitrary beam polarization
configurations and are fragmented using PYTHIA. There
are well over 1000 of these processes {\cite {TimB}}.
($iv$) We include ILC detector effects by making use of the
java-based SiD {\cite {SiD}} detector fast simulation
package org.lcsim {\cite {lcsim,SiDDOD}}. All in all, we believe that
we have performed our analysis in as realistic a manner as possible.

Given the present cost and timing controversies in the world community surrounding the
ILC (or, generally, any $e^+e^-$
collider), this is a critical time to ascertain in as realistic way as possible the capabilities
of such machines to explore new physics that may be discovered at the LHC. Thus, as will be described
below, we present a very detailed study of the MSSM at the ILC using the SiD detector, an important
milestone on the road to achieving this goal.

The outline of this paper is as follows: Section 2 contains
a discussion of the various kinematical
features of the AKTW models under consideration, while Section 3
provides an overview of our analysis procedure
as well as a general discussion of the SM backgrounds. In the
next Sections we separately consider the
individual SUSY particle analyses for sleptons (Section 4),
charginos (Section 5) and neutralinos (Section 6).
This leads to an overall set of model observation and comparisons in Section 7 where we discuss
the ability of the ILC to distinguish the AKTW models and resolve
the LHC Inverse Problem.
This is followed by a discussion of our results and conclusions.

\section{Spectrum and Kinematical Features of the Models}

Before beginning our analysis, we first examine the kinematical traits
and features of the SUSY models that AKTW found to be indistinguishable
at the LHC.\footnote{We thank AKTW \cite{Arkani-Hamed:2005px} for
giving us the weak scale parameters for these models.}
This consists of a set of 383 models (\ie, 383 points in a
15-dimensional MSSM parameter space; we hereafter refer to distinct points
in the MSSM parameter space as models). In their study,
AKTW compared models pairwise, so
that these 383 models correspond to 283 pairs of models which gave
indistinguishable signatures at the LHC. In some cases, models were found
to give degenerate signatures multiple times.
While this may naively seem to be a relatively
small number of inseparable models, one needs to recall that AKTW
performed a small sampling of a large parameter space (due to computational
limitations). Based on the number of models AKTW generated, the number
of degeneracies they found led AKTW to estimate
that a more complete statistical
sampling of the available parameter space volume would yield a degeneracy
of each model with ${\cal O}(10-100)$ other points.

\begin{figure*}[hpb]
\parbox{7cm}{
\centerline{\includegraphics[height=7.5cm,angle=270,clip=]{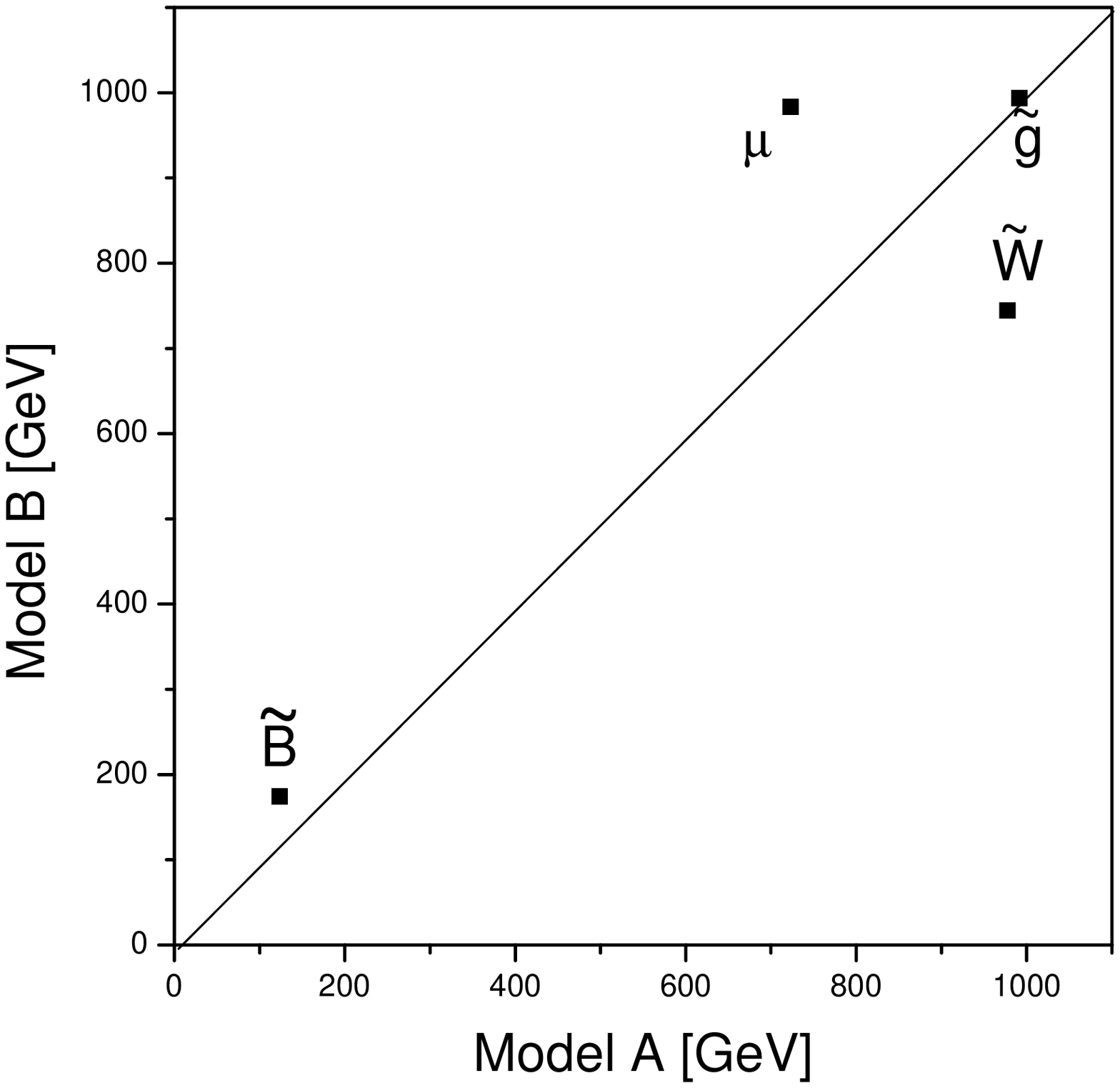}}
\vspace*{2.2cm}}\hspace*{5mm}\parbox{8cm}{
\centerline{\includegraphics[height=9cm,clip=]{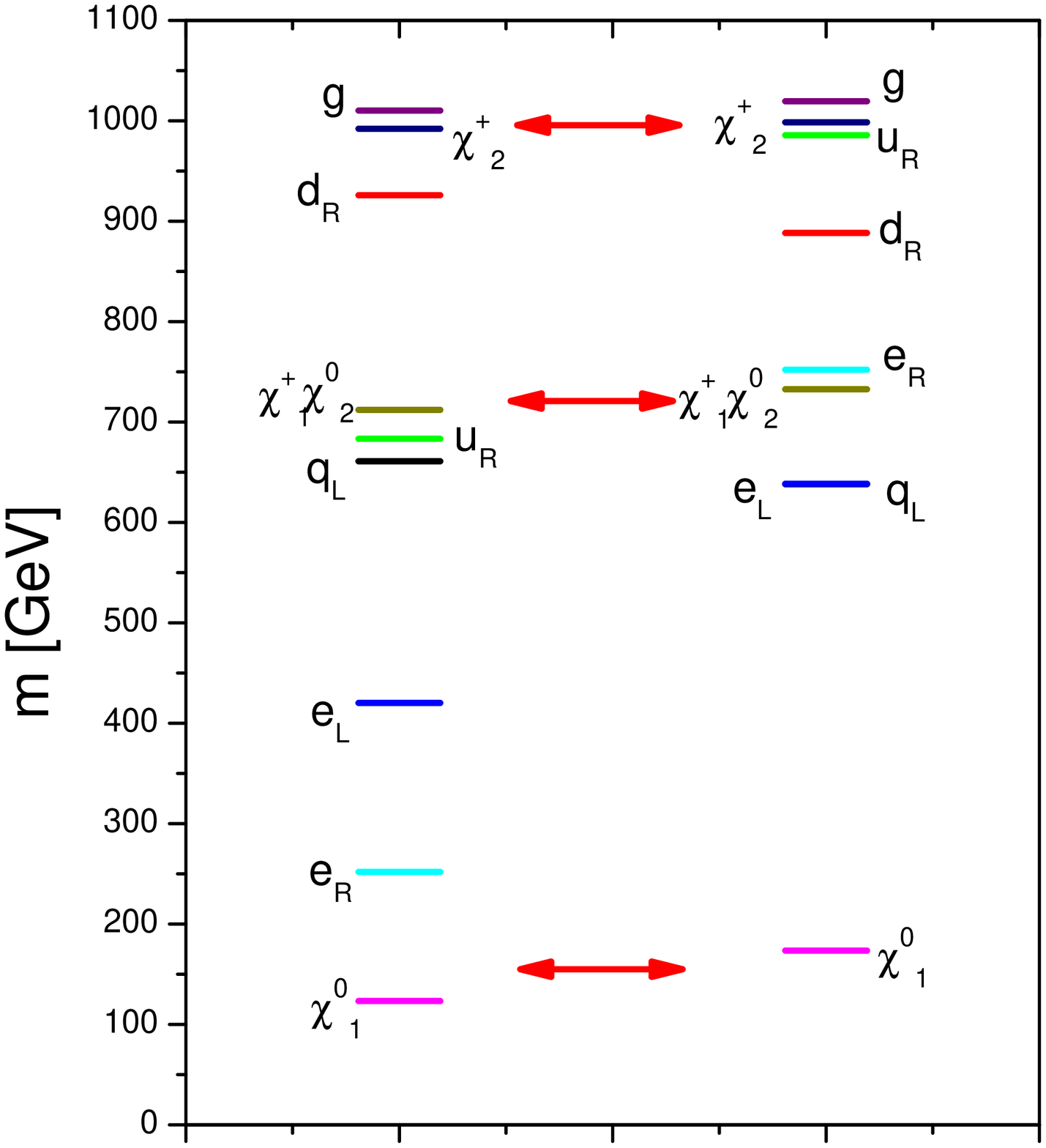}}}
\vspace*{0.1cm}
\caption{Illustration of the Flipper ambiguity in the MSSM spectrum.
The left panel displays a typical mass value for the Bino, Wino, and Higgsino
mixing parameter in the weak eigenstate basis for two models A and B.
The right panel shows the corresponding sparticle spectrum in two of the
AKTW models, with the red arrows indicating the Flipper effect.}
\label{flipper}
\end{figure*}
\begin{figure*}[hptb]
\parbox{7cm}{
\centerline{\includegraphics[height=6.5cm,angle=0,clip=]{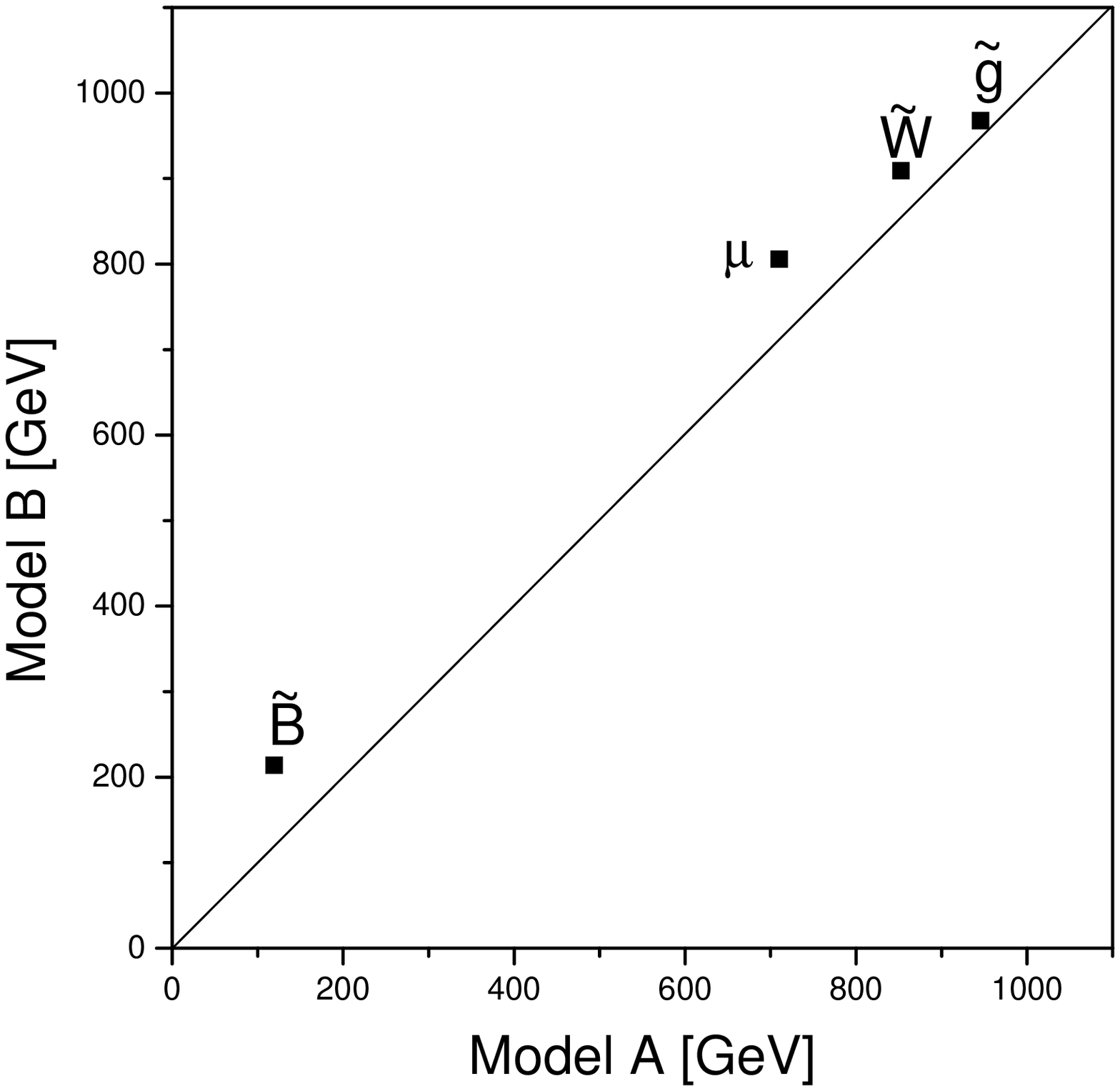}}
\vspace*{2.2cm}}\hspace*{4mm}\parbox{8cm}{
\centerline{\includegraphics[height=9cm,clip=]{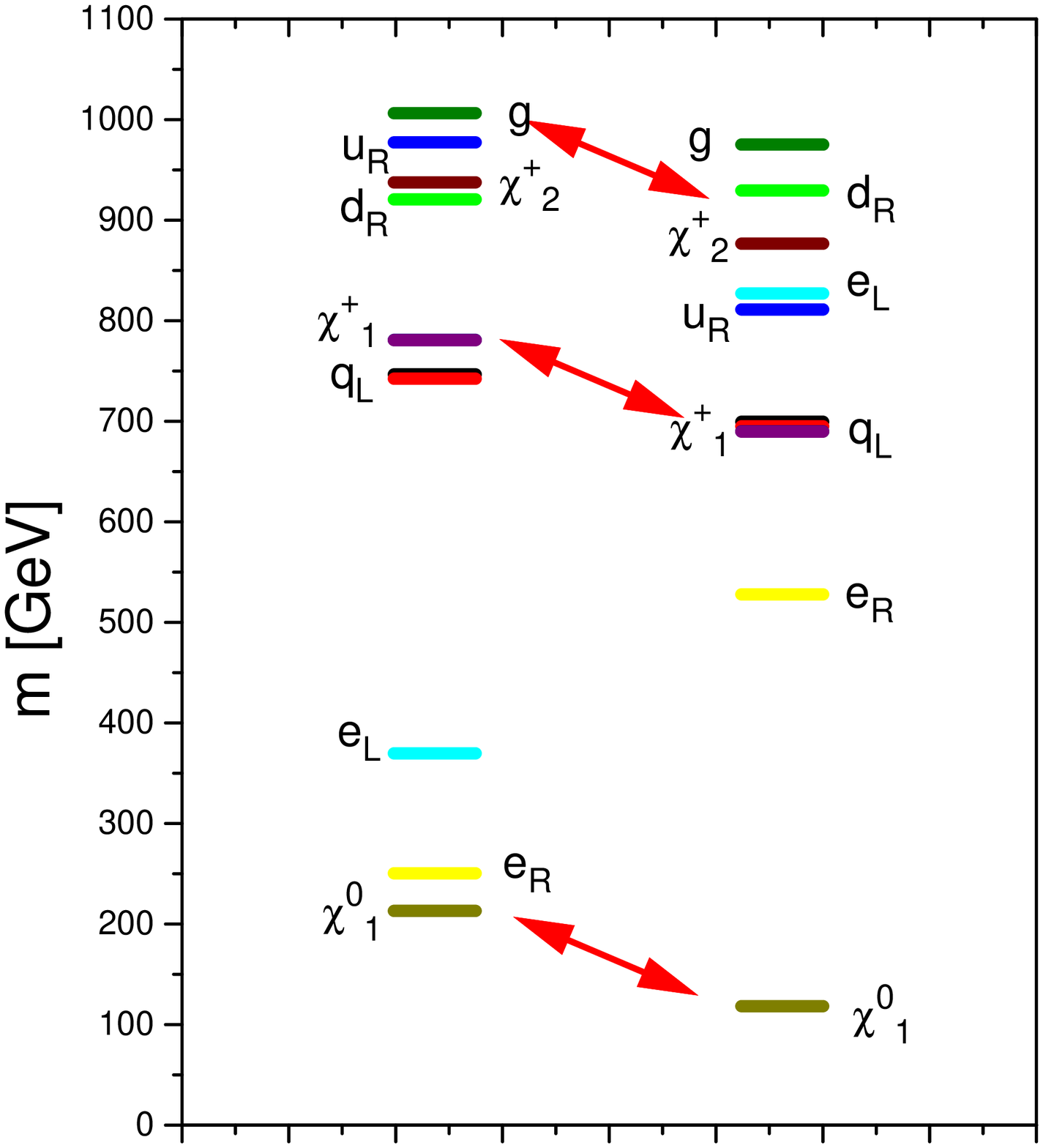}}}
\vspace*{0.1cm}
\caption{Illustration of the Slider ambiguity in the MSSM spectrum.
The left panel displays a typical mass value for the Bino, Wino, and Higgsino
mixing parameter in the weak eigenstate basis for two models A and B.
The right panel shows the corresponding sparticle spectrum in two of the
AKTW models, with the red arrows indicating the Slider effect.}
\label{slider}
\end{figure*}
\begin{figure*}[hptb]
\parbox{7cm}{
\centerline{\includegraphics[height=7.5cm,angle=270,clip=]{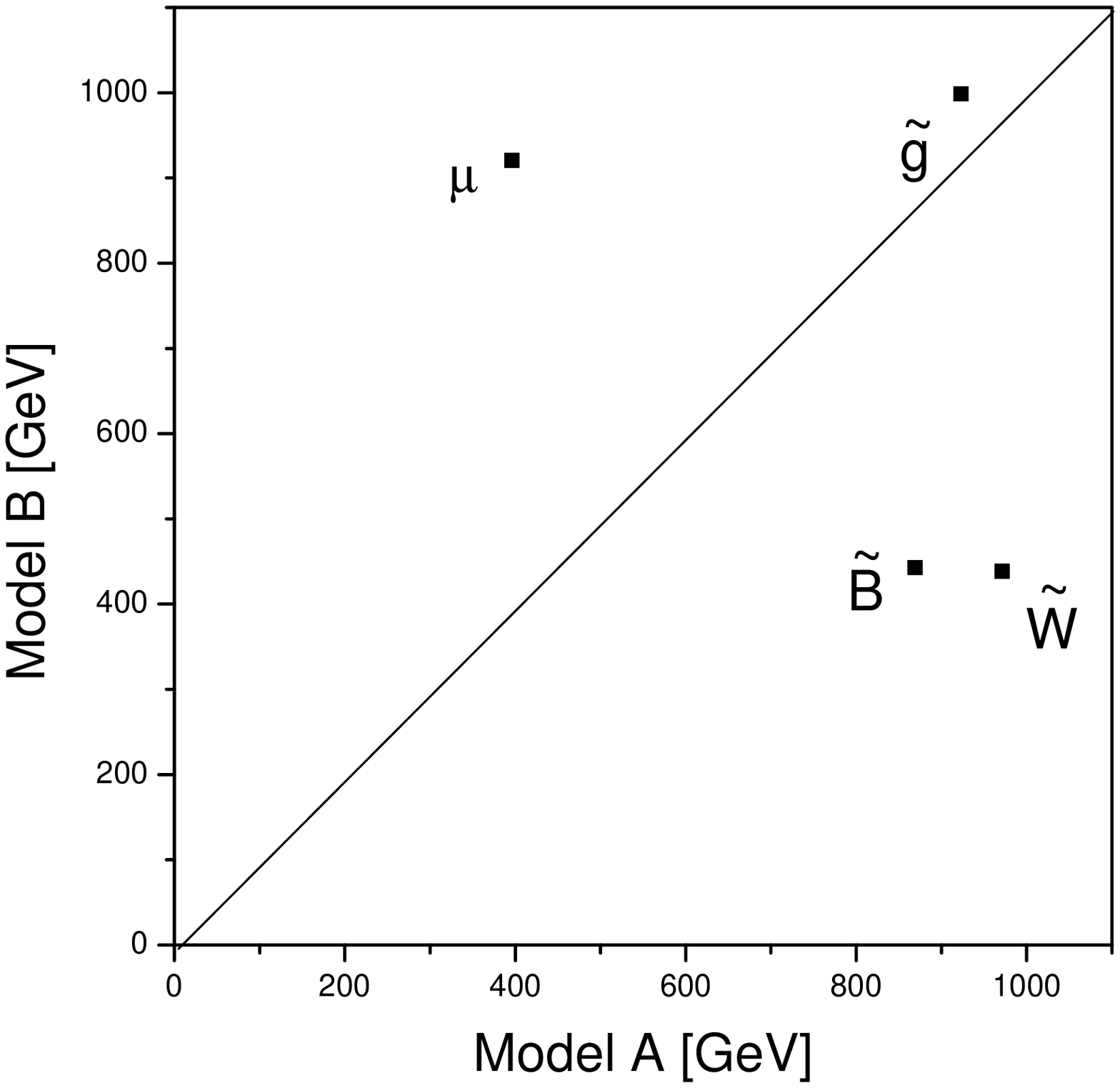}}
\vspace*{1.9cm}}\hspace*{4mm}\parbox{8cm}{
\centerline{\includegraphics[height=9cm,clip=]{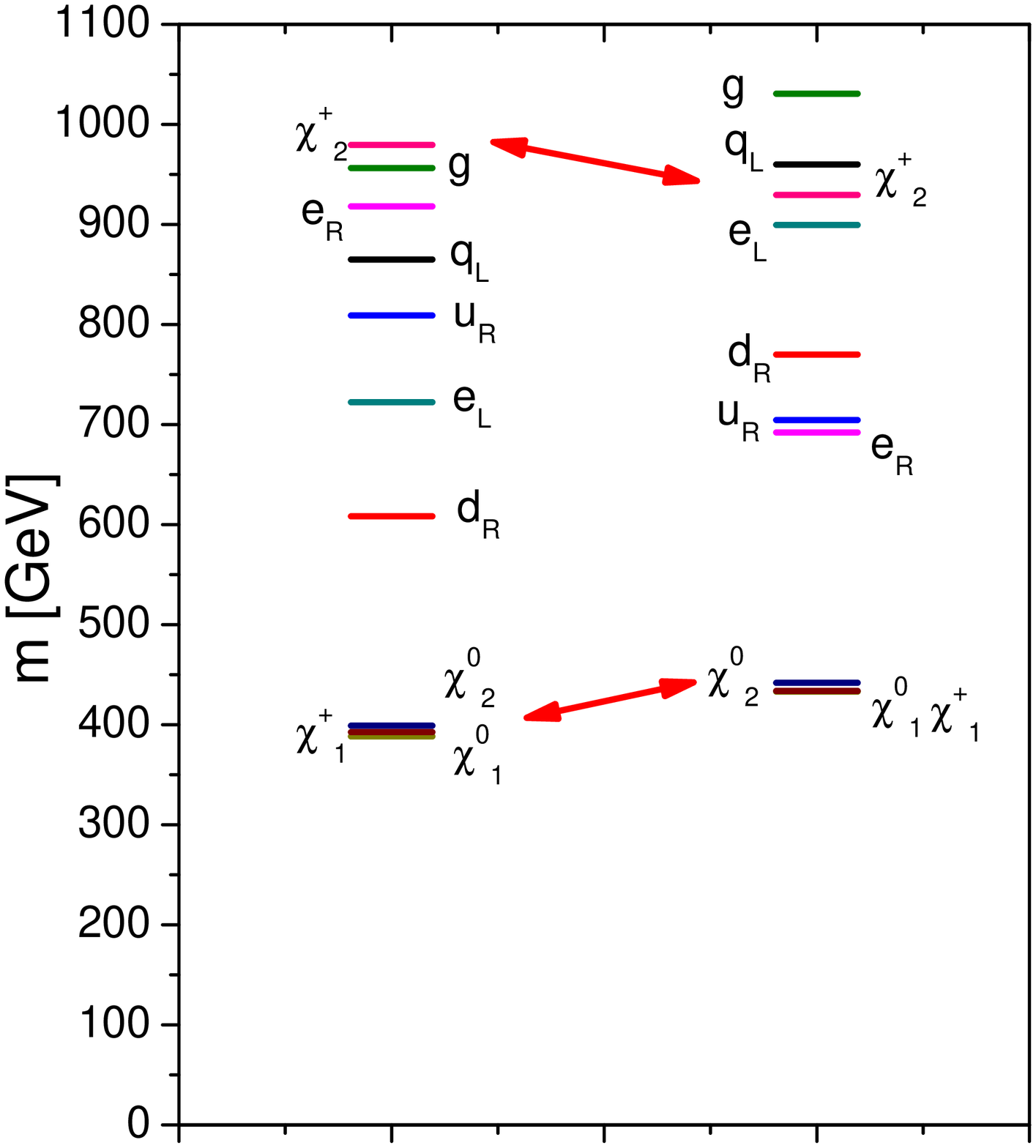}}}
\vspace*{0.1cm}
\vspace*{0.1cm}
\caption{Illustration of the Squeezer ambiguity in the MSSM spectrum.
The left panel displays a typical mass value for the Bino, Wino, and Higgsino
mixing parameter in the weak eigenstate basis for two models A and B.
The right panel shows the corresponding sparticle spectrum in two of the
AKTW models, with the red arrows indicating the Squeezer effect.}
\label{squeezer}
\end{figure*}

One may wonder if there are any common features of these models
that give rise to their indistinguishability at the LHC.
AKTW demonstrated that these degeneracies are essentially
the result of three possible characteristics that involve the relative
composition of the physical electroweak gaugino sector in terms of the
higgsino, wino, and bino weak eigenstates. These mechanisms
are referred to as `Flippers', `Sliders' and `Squeezers'
and are schematically shown in
Figs.~\ref{flipper}, \ref{slider}, and \ref{squeezer}. The ambiguities
that arise from these model characteristics
originate directly from the manner in which SUSY is
produced and observed at the LHC; several of these mechanisms
can be simultaneously present. As is
well-known, the (by far) dominant production mechanism for
R-parity conserving SUSY at the LHC is via the strong
interactions, \ie, the production of squarks and
gluinos. These particles then decay through
a long cascade chain via the generally lighter
electroweak gaugino/higgsino partner states. This
eventually leaves only the SM fields in the final state
together with the stable Lightest SUSY Particle
(LSP), which is
commonly the lightest neutralino, appearing as additional missing energy.
The decays of the SM fields
produce additional jets, leptons, and missing transverse energy
from neutrinos. Unfortunately sleptons do not
always play a major role in these cascades
(due to phase space considerations
in the sparticle spectrum, see, \eg, the models in Fig.~\ref{flipper})
so that much
valuable information associated with their
properties is generally lost.
When comparing the possible
decay chains which result from the
produced squarks and gluinos, similar final states can occur
if the identities of the higgsino, wino
and bino weak
states in the spectrum are interchanged while their masses are
held approximately fixed. This is an example of the
so-called `Flipper' ambiguity (Fig.~\ref{flipper})
where two spectra with interchanged electroweak quantum numbers
can produce very similar
final state signatures. A second possible source of degeneracy
can arise from the fact that
absolute masses, and in particular the mass of the LSP, are
not well measured at the LHC in contrast to the mass
differences between states~{\cite{Weiglein:2004hn,Aguilar-Saavedra:2005pw}}.
Thus models can have similar spectra but
be somewhat off-set from each other in their absolute mass scale and
hence be difficult to distinguish; this represents the
`Slider' degeneracy (Fig.~\ref{slider}). Lastly, pairs of
states in the spectra with relatively small mass differences
compared to the overall SUSY scale lead to
relatively soft decay products in the cascade chain. Such
a possibility can cause significant loss of information
as well as general confusion in parameter
extractions and are termed `Squeezers' (Fig.~\ref{squeezer}).
Of course in all
these cases some shifts are needed in the
strongly interacting part of the SUSY spectrum
to keep the various production rates and decay
distributions comparable between potentially indistinguishable
models. It goes without saying that some degeneracies can
also arise when more than one of these mechanisms are active simultaneously.

\begin{figure*}[hptb]
\centerline{
\psfig{figure=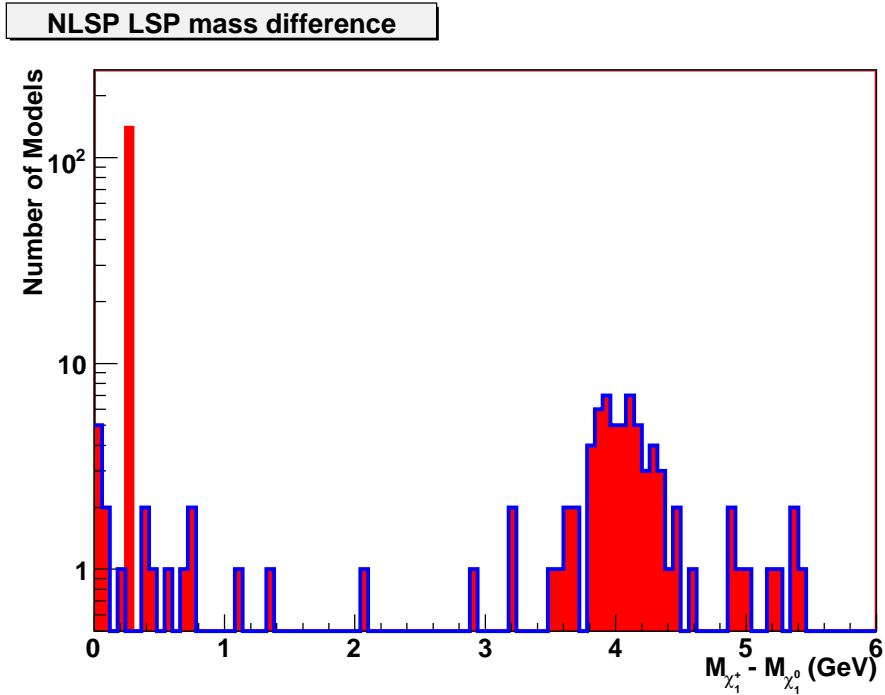,angle=0,width=13cm,clip=}}
\vspace*{0.1cm}
\caption{Low mass end of the lightest chargino-LSP mass difference
which displays the
PYTHIA feature in the sparticle spectrum generator discussed in the
text. }
\label{fig:pythfeature}
\end{figure*}

We now examine the physical particle spectra
in the 383 models found by AKTW to be indistinguishable at the LHC.
First, we note that since AKTW have required
squarks and gluinos to have Lagrangian masses greater than 600 GeV
in their parameter scans, the only states potentially accessible to
the ILC will be the sleptons and the sparticles associated with the
electroweak gaugino/higgsino sector.
Of particular phenomenological interest is the mass splitting between
the Next-to-LSP(NLSP) and LSP
(see Fig.~\ref{fig:pythfeature}). Here, this is usually that between the
lightest chargino, $\tilde \chi_1^{\pm}$, and lightest neutralino
state, $\tilde \chi_1^0$. Generally this distribution for our set
of AKTW models appears rather flat {\it except} for a huge and puzzling
feature near $\sim 270$ MeV. It would seem that
almost $40\%$, \ie, $141$ of these models, experience
this exact mass splitting between these two states.

\begin{table}
\centering
\begin{tabular}{|c|c|c|} \hline\hline
Final State & 500 GeV & 1 TeV \\ \hline
$\tilde e_L^+ \tilde e_L^-$ & 9 & 82 \\
$\tilde e_R^+ \tilde e_R^-$ & 15 & 86 \\
$\tilde e_L^\pm \tilde e_R^\mp$ & 2 & 61 \\
$\tilde \mu_L^+ \tilde \mu_L^-$ & 9 & 82 \\
$\tilde \mu_R^+ \tilde \mu_R^-$ & 15 & 86 \\
Any selectron or smuon & 22 & 137 \\
$\tilde \tau_1^+ \tilde \tau_1^-$ & 28 & 145 \\
$\tilde \tau_2^+ \tilde \tau_2^-$ & 1 & 23 \\
$\tilde \tau_1^\pm \tilde \tau_2^\mp$ & 4 & 61 \\
$\tilde \nu_{e\mu}\tilde \nu_{e\mu}^*$ & 11 & 83 \\
$\tilde \nu_\tau \tilde \nu_\tau^* $ & 18 & 83 \\
$\tilde \chi_1^+ \tilde \chi_1^-$ & 53 & 92 \\
Any charged sparticle & 85 & 224 \\
$\tilde \chi_1^\pm \tilde \chi_2^\mp$ & 7 & 33 \\
$\tilde \chi_1^0 \tilde \chi_1^0$ & 180& 236 \\
$\tilde \chi_1^0 \tilde \chi_1^0$ only & 91 & 0 \\
$\tilde \chi_1^0 + \tilde \nu$ only & 5 & 0 \\
$\tilde \chi_1^0 \tilde \chi_2^0 $ & 46 & 178 \\
$\tilde \chi_1^0 \tilde \chi_3^0 $ & 10 & 83 \\
$\tilde \chi_2^0 \tilde \chi_2^0 $ & 38 & 91 \\
$\tilde \chi_2^0 \tilde \chi_3^0 $ & 4 & 41 \\
$\tilde \chi_3^0 \tilde \chi_3^0 $ & 2 & 23 \\
Nothing & 61 & 3 \\ \hline\hline
\end{tabular}
\caption{Number of models at $\sqrt s=500$ GeV and 1 TeV
which have a given final
state kinematically accessible. Note that for the 500 GeV case,
96/242 models have
only LSP or neutral pairs accessible while 61/242 models have
no SUSY particles accessible.}
\label{finalstates}
\end{table}

An investigation shows that this result is an artifact of
the manner in which PYTHIA6.324
generates the physical SUSY particle spectrum at tree-level
from the Lagrangian parameters. Recall that
AKTW randomly generated points in a
15-dimensional weak scale MSSM parameter space, described in the
previous Section, from which the physical SUSY particle
masses are then calculated at tree-level via PYTHIA6.324.
With this procedure,
it is possible that sometimes the mass of the lightest chargino
$\tilde \chi_1^{\pm}$
turns out to be {\it less} than that of the
$\tilde \chi_1^0$ once the mass eigenstates are computed;
this is usually considered to be `unphysical'
as it would imply charged Dark Matter in the standard cosmological
picture. PYTHIA6.324 handles this situation by
artificially resetting the chargino mass to be greater than that of the LSP
by $m_{\tilde \chi_1^{\pm}}=m_{\tilde \chi_1^0}+2m_{\pi}$
without an associated warning message.
This apparently happens
frequently and causes the large peak in the distribution
shown in Fig.~\ref{fig:pythfeature}. This feature is mentioned in the
PYTHIA manual (where it is noted that the tree-level SUSY spectrum
calculator is not for publication quality), and has been further clarified
in later versions of PYTHIA {\cite {peter}}. However, here we
need to follow the analysis of AKTW as closely as
possible to reproduce their sparticle spectra and specific model
characteristics. Due to this and additional reasons discussed
below in the text, in our
analysis we use a slightly modified version of PYTHIA6.324.
In the strictest sense, these models are only
`unphysical' at the tree-level since loop corrections
restore the correct mass hierarchy. We have checked that all 383 of the
AKTW models have an appropriate mass spectrum
when the SuSpect2.34
routine \cite{suspectfolks}, which includes the higher order corrections,
is employed to generate the physical spectrum. However, in the present
work, the 141 models cannot be artificially saved simply by employing this
mass re-assignment or by using SuSpect
as their collider production and signature
properties would be modified as compared to the AKTW study. We thus drop them
completely from further consideration. This leaves us with
a sample of 242 models which consist of 162 degenerate model
pairs to examine.{\footnote {Note again that some models are members
of degenerate triplets or quartets which
influences this counting.}}

\begin{figure*}[hptb]
\centerline{
\includegraphics[width=10.0cm,height=13.0cm,angle=-90]{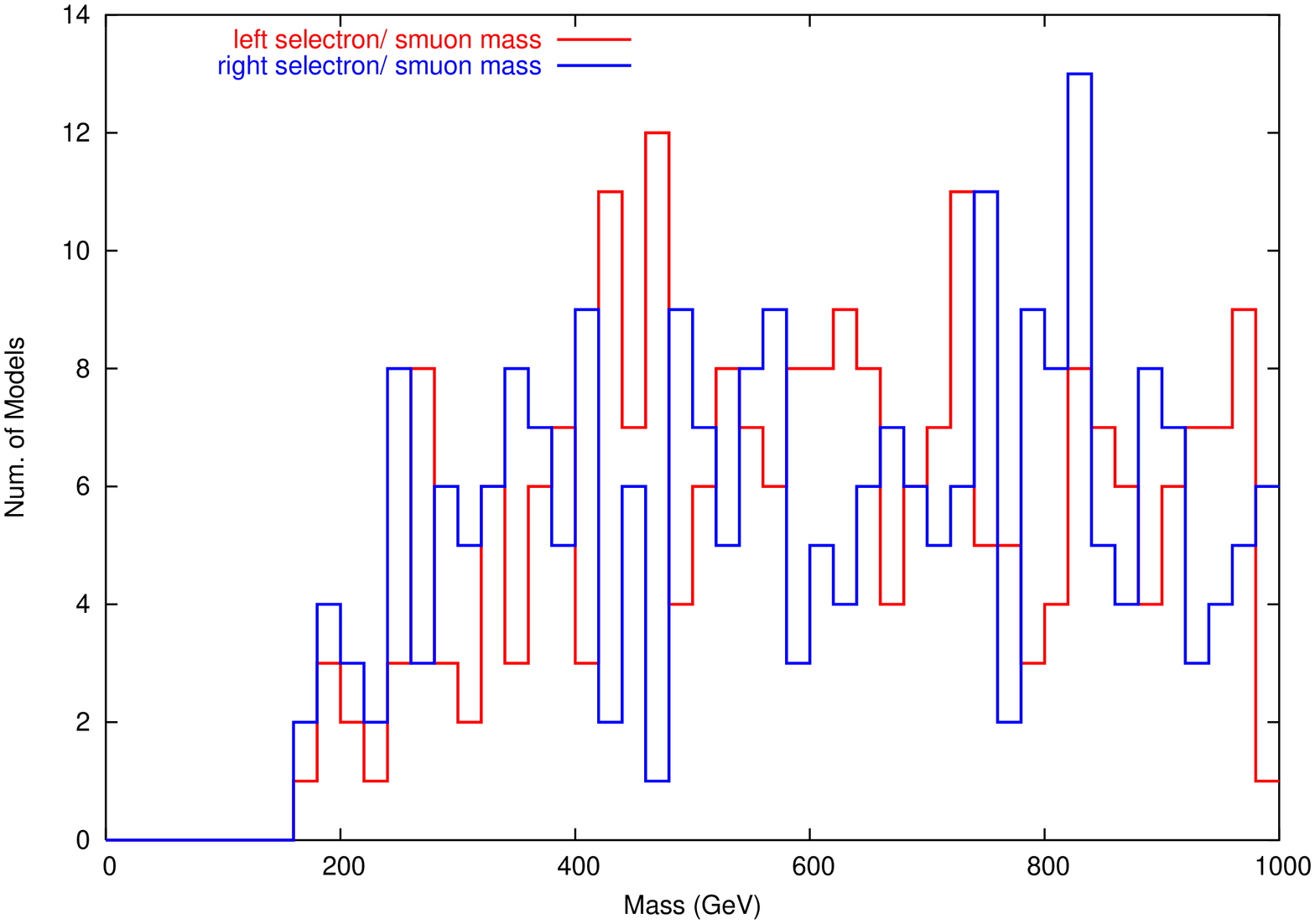}}
\vspace*{0.1cm}
\centerline{
\includegraphics[width=10.0cm,height=13.0cm,angle=-90]{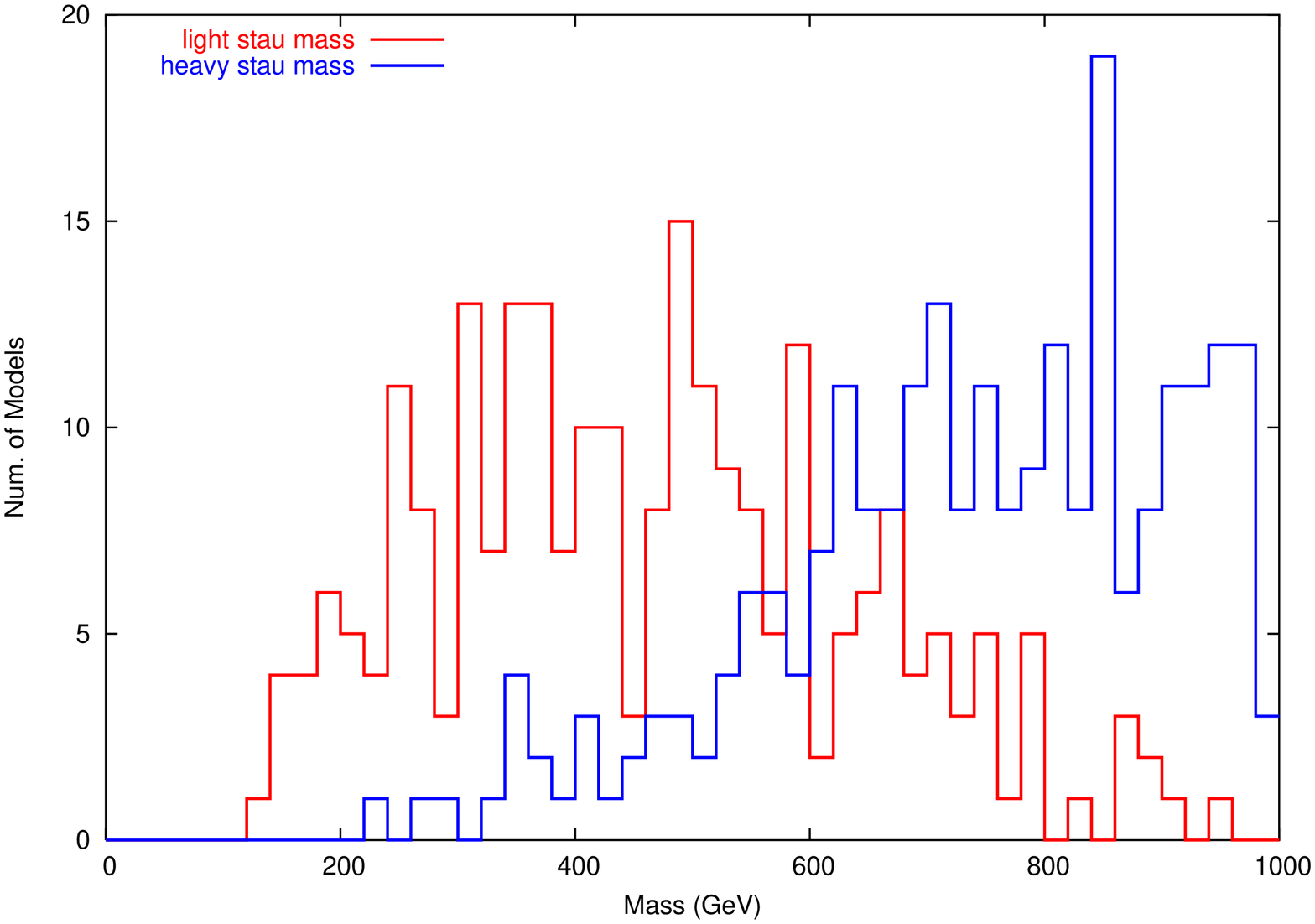}}
\vspace*{-0.1cm}
\caption{Selectron/smuon (top) and stau (bottom) mass spectra in the 242 models
under study.}
\label{specfig1}
\end{figure*}
\begin{figure*}[hptb]
\centerline{
\includegraphics[width=10.0cm,height=13.0cm,angle=-90]{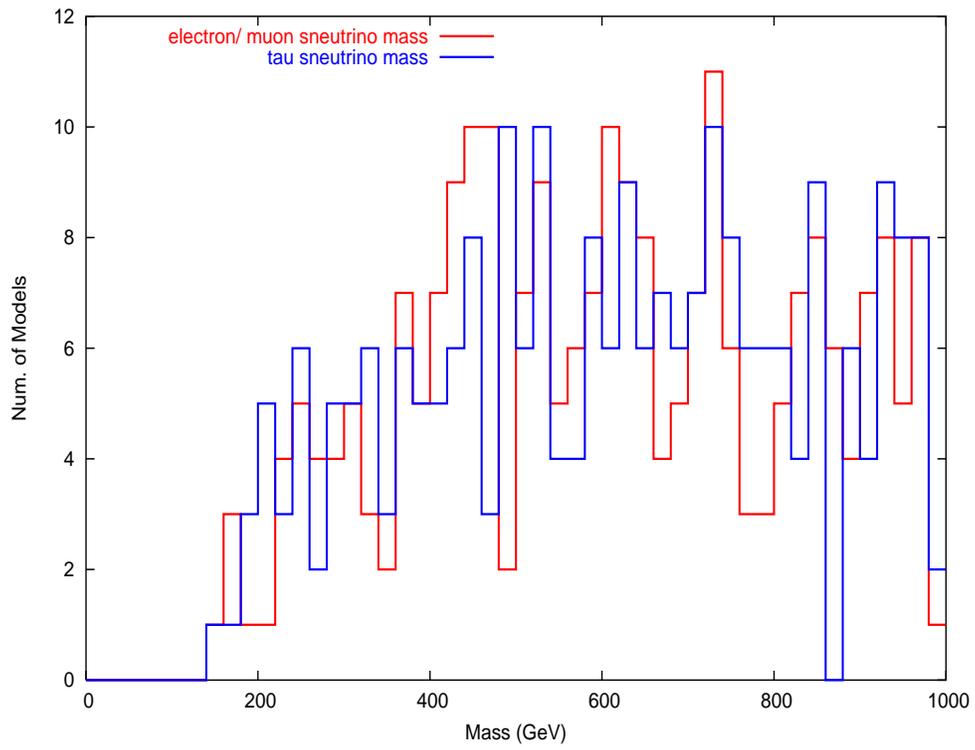}}
\vspace*{0.1cm}
\caption{Sneutrino mass spectra in the 242 models under study.}
\label{specfig2}
\end{figure*}
\begin{figure*}[hptb]
\centerline{
\includegraphics[width=10.0cm,height=13.0cm,angle=-90]{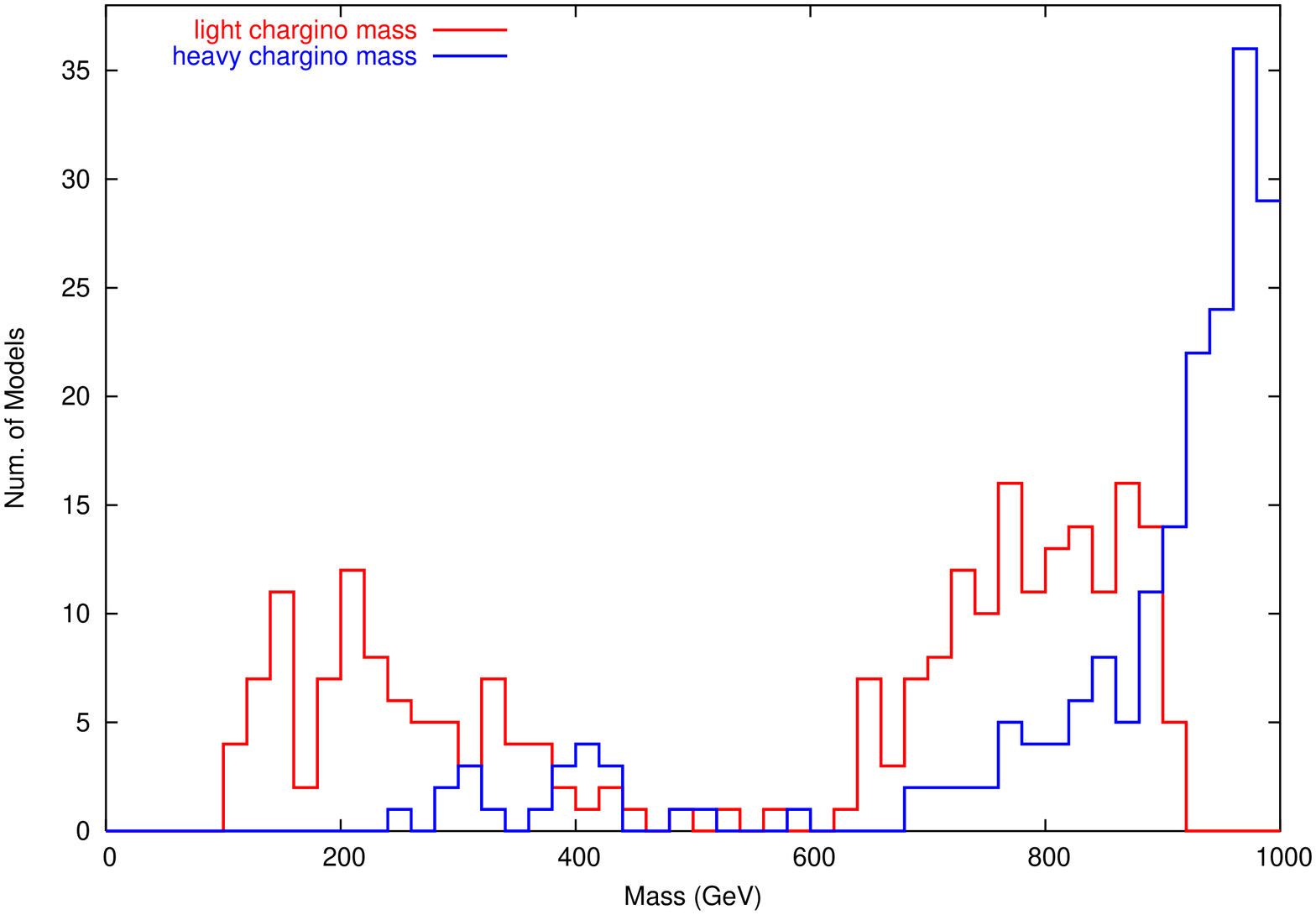}}
\vspace*{0.1cm}
\centerline{
\includegraphics[width=10.0cm,height=13.0cm,angle=-90]{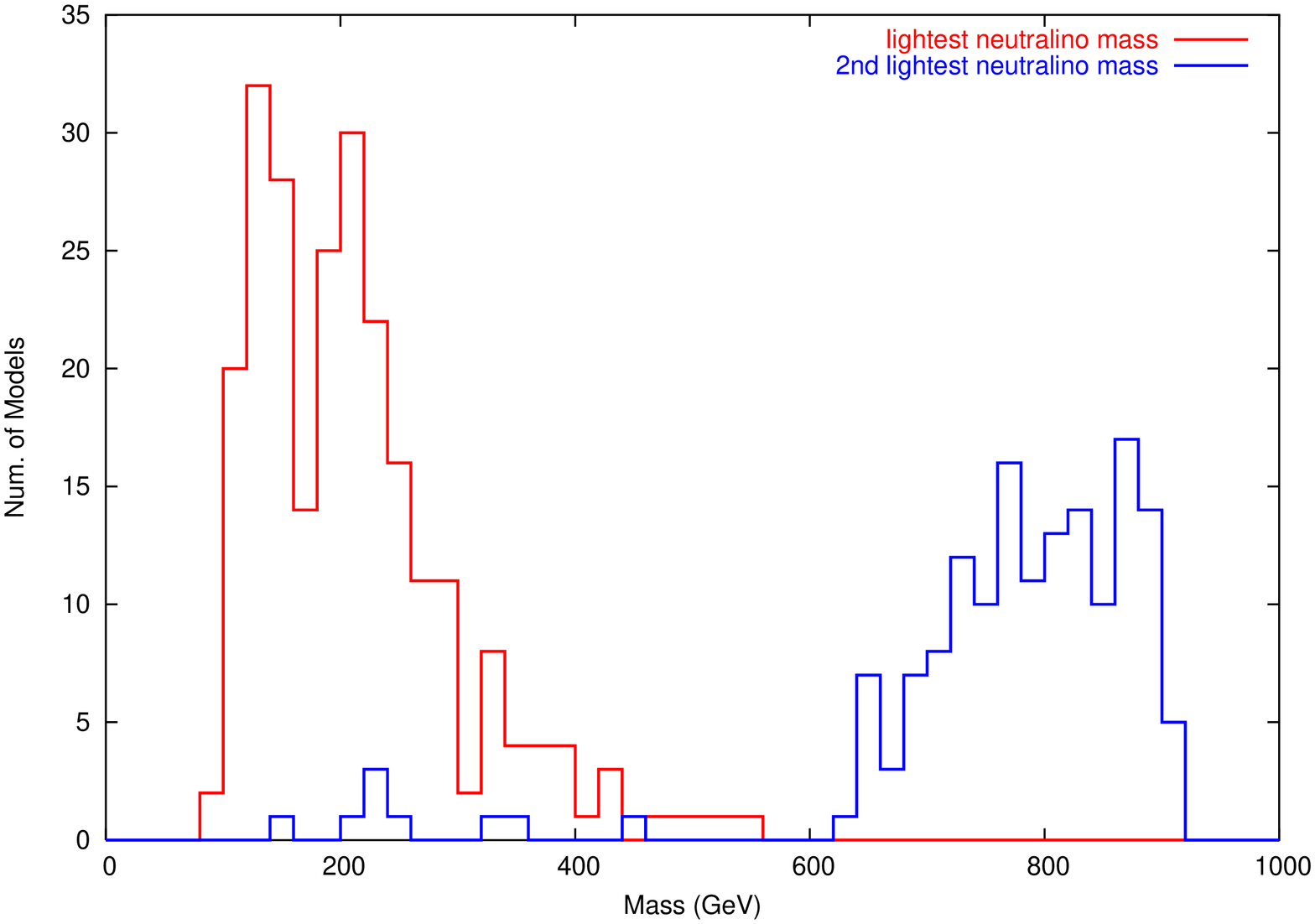}}
\vspace*{-0.1cm}
\caption{Chargino (top) and neutralino (bottom) mass spectra in the
242 models under study.}
\label{specfig3}
\end{figure*}
\begin{figure*}[hptb]
\centerline{\epsfig{figure=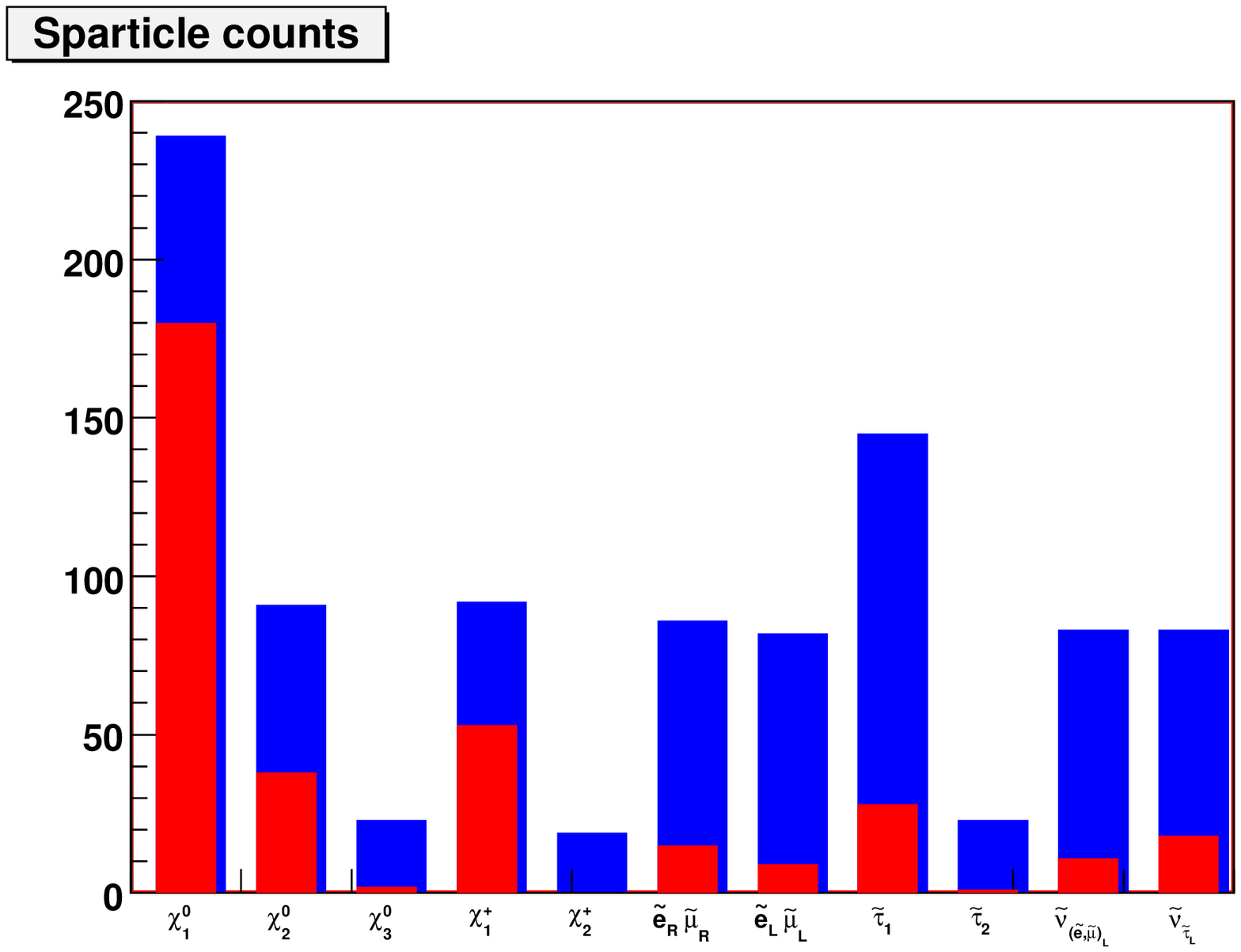,width=13cm,angle=0,clip=}}
\vspace*{0.1cm}
\caption{Number of models with various kinematically accessible
MSSM states at a 500 (red) or 1000 (blue) GeV
$e^+e^-$ collider via pair production.}
\label{benj2}
\end{figure*}

Given these 242 models, we next address the question of what fraction
of their SUSY
spectra are kinematically accessible at a 500 or 1000 GeV ILC.
The results are shown
in Figs.~\ref{specfig1},~\ref{specfig2}
and ~\ref{specfig3}, which display the individual
mass spectra for the weakly interacting sectors of the various SUSY
models under consideration. The full accessible sparticle count for
$\sqrt s=500$ and 1000 GeV is presented in Fig.~\ref{benj2}.
There are many things to observe by examining these Figures.
First, we recall from
the discussion above that in all cases the squarks are too massive
to be pair produced at the ILC so that we are restricted to the
slepton and electroweak gaugino sectors. Here in Table \ref{finalstates}
and in the Figures we see that for a
500(1000) GeV collider, there are only 22(137)/242, {\it i.e.}, 22(137)
out of 242, models with kinematically accessible (which here means via
pair production) selectrons and smuons at $\sqrt s=500(1000)$ GeV; note
that these two sparticles are
degenerate in the MSSM. Similarly, 28(145)/242 of the models have
accessible light staus,
6(55) of which also have kinematically accessible
selectrons/smuons. 53(92)/242 models have kinematically accessible
light charginos, 4(12) of which
also have accessible selectrons/smuons and 6(12) of which also have
accessible staus. At $\sqrt s=1$ TeV, 19 of these 92 models with accessible
light charginos also have the
second chargino accessible by pair production. Very importantly, at
$\sqrt s=500$ GeV,
in 96/242 models the {\it only} kinematically accessible
sparticles are neutral, \eg, $\tilde\chi_1^0$ or $\tilde\nu$,
while 61/242 other models have
{\it no} SUSY particles accessible whatsoever. At $\sqrt s=1$
TeV, these numbers drop
to only 0/242 and 3/242 in each of these latter categories,
respectively. Recalling
that we are looking at essentially random points in the MSSM parameter space,
we see from this simple counting exercise that $\sim 60\%$ of the
models will have no `traditional' SUSY signatures at a 500 GeV ILC,
whereas a 1 TeV machine essentially covers almost all the cases. This
is a strong argument for having the capability of upgrading to 1 TeV at the
ILC as quickly as possible. However, in the
analysis that follows we will consider only the case of a 500 GeV
ILC with the 1 TeV case to be considered separately in the future.
Table~\ref{finalstates} summarizes the kinematic accessibility of all the
relevant MSSM final states for
$\sqrt s=500$ GeV in our study as well as the corresponding
results for 1 TeV.

Given that so many models have such a sparse SUSY spectrum at
$\sqrt s=500$ GeV, it is not uncommon for one of
the two models in the pair we are comparing to have no
kinematically accessible sparticles. In such a case,
breaking the model degeneracy at the LHC
might seem to be rather straightforward,
as for one model we might observe SUSY signals
above the SM background but not for the other in the pair. Of course,
at the other end of the spectrum of difficulty,
one can imagine cases where both models being compared are
Squeezers, in which case model differentiation will
be far more difficult and having an excellent ILC detector
will play a much more important role.

\clearpage

\section{Analysis Procedure and General Discussion of Background}
\label{Sec:analysis}

To determine whether or not the ILC resolves the LHC
inverse problem, we compare the ILC experimental signatures for the
pairs of SUSY models that AKTW found to be degenerate, and
see whether these signatures can be distinguished. We examine
numerous production channels and signatures for supersymmetric
particle production in \epem\ collisions. Before the model comparisons
can be carried out, we must first ascertain
if the production of the kinematically accessible
SUSY particles is visible above the SM background.
Our analysis procedure is described in this Section.

\subsection{Event Generation of Signal and Background}

We generate 250 fb$^{-1}$ of SUSY events at $\sqrt s = 500$ GeV
for each of the AKTW models for both 80\% left- and 80\% right-handed
electron beam polarization with unpolarized positron beams, providing a
total of 500 fb$^{-1}$ of integrated luminosity. To generate the signal
events, we use \linebreak
PYTHIA6.324 \cite{Sjostrand:2006za} in order to retain
consistency with the AKTW analysis. However, as will be described in
detail below, we find that PYTHIA underestimates the production cross
section in two of our analysis channels, and in these two cases we
employ CompHEP \cite {Boos:2004kh}.
We also analyze two statistically independent 250 fb$^{-1}$ sets
of Standard Model background events for each of the two electron beam
polarizations. We then study numerous different analysis channels.
When we determine if a signal is observable over the SM background
in a particular channel, we statistically compare the combined
distribution for the signal plus the background from our first
background sample with the distribution from our second, independent
background sample. When we perform the model comparisons,
we add each set of SUSY events to a distinct Standard Model
background sample generated for the same beam polarization. We
then compare observables for the many different analysis channels
for these two samples of signal and background (\ie, model A $+$
background sample 1 is compared to model B $+$ background sample 2).
It is important to note that we take into account the
full Standard Model background in all analysis channels rather than
only considering the processes that are {\it thought} to be the dominant
background to a particular channel; surprisingly, sometimes many
small contributions can add up to a significant background.

\begin{figure*}[hpt]
\centerline{
e\psfig{figure=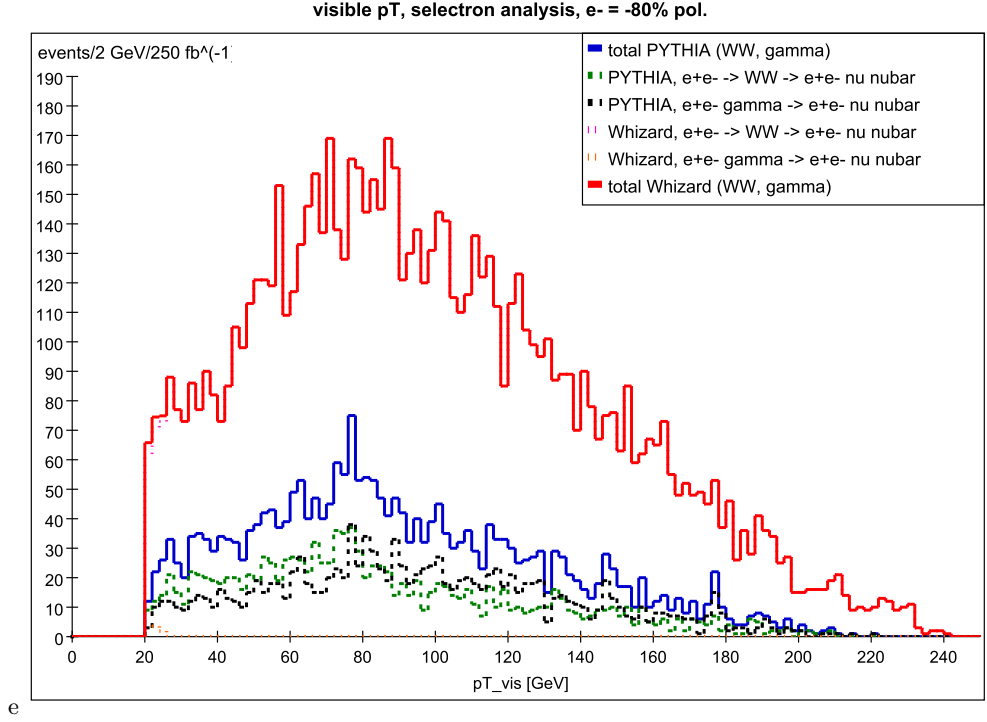,angle=0,width=13cm,clip=}}
\vspace*{0.1cm}
\caption{Transverse momentum distribution in $\epem\to\epem\nu_e\bar\nu_e$
as generated via PYTHIA and WHIZARD,
for 250 fb$^{-1}$ of integrated luminosity with $80\%$ left-handed
electron beam polarization at $\sqrt s = 500 $ GeV. Our selectron selection
cuts (discussed in the text below) have been applied.}
\label{fig:ptdist}
\end{figure*}

Our background contains all SM
$2\to2$, $2\to4$, and $2\to6$ processes with the initial states
$e^+e^-$, $e^\pm\gamma$, or $\gamma\gamma$; in total there are 1016
different background channels.
These events were generated by T. Barklow~\cite{TimB} with
O'MEGA as implemented in WHIZARD~\cite{WHIZARD}, which uses
full tree-level matrix elements and incorporates a realistic beam treatment
via the program GuineaPig~\cite{Guinea}.
The use of full matrix elements leads to qualitatively different
background characteristics in terms
of both total cross section and kinematic distributions compared to those
from a simulation that uses only the production and decay of on-shell
resonances, e.g., the procedure generally employed in PYTHIA.
WHIZARD
models the flux of photons in $e^\pm \gamma$ and $\gamma \gamma$ initiated
processes via the equivalent photon approximation. However, in the
standard code, the electrons and positrons which emit the photon(s) that
undergo hard scattering do not receive a corresponding kick
in $p_T$, in contrast to the electrons or positrons that undergo
initial state radiation. The version of WHIZARD used here to generate the
background events was thus amended to correct
this slight inconsistency in the treatment of transverse momenta.
An illustration of the resulting effects from employing exact
matrix elements and modelling the transverse
momentum distributions in a realistic fashion is presented in
Fig.~\ref{fig:ptdist}. This Figure
compares the transverse momentum distribution for the process
$e^+e^- \rightarrow e^+e^-\nu_e \bar{\nu}_e$ in the SM
after our selectron selection cuts (see Section 4.1) have been applied,
as generated with PYTHIA versus the modified version of WHIZARD,
using the same beam spectrum in both codes. We see that in this
case, the $p_T$ distribution generated by PYTHIA is smaller and has a
shorter tail.

We now discuss our treatment of the beam spectrum in further detail.
The backgrounds were generated using a
realistic beam treatment, employing the program GuineaPig~\cite{Guinea}
to model beam-beam interactions.
Finite beam energy spread was taken into account and combined
with a beamstrahlung spectrum specific to a cold technology linear
collider, \ie, the ILC.
The effect of beamstrahlung is displayed in
Figure~\ref{fig:beammu}, which
shows the invariant mass of muon pairs formed by $e^+e^-$ collisions
with the beam spectrum we employ. The resulting spectrum is somewhat
different qualitatively
from a commonly used purely analytic approximate approach
\cite{Yokoya:1989jb,Chen:1991wd,Peskin:1999pk}.

{
\begin{figure*}[hpt]
\centerline{
\epsfig{figure=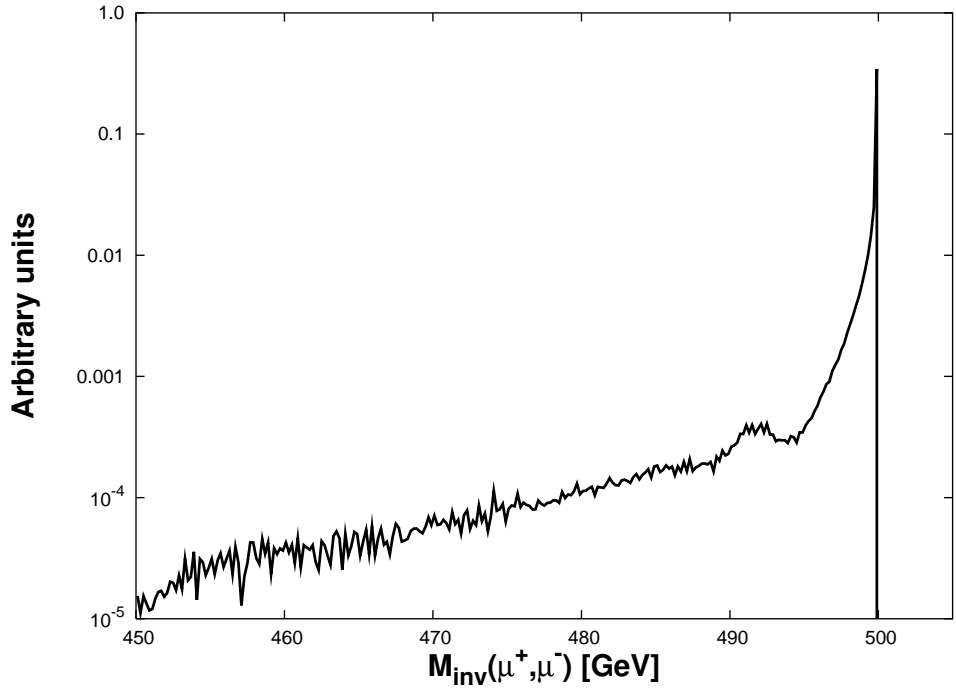,angle=270,width=13cm,clip=}}
\vspace*{0.1cm}
\caption{Invariant mass of the muon pairs in $e^+e^- \rightarrow \mu^+ \mu^-$
at $\sqrt{s} = 500$ GeV, using the beamspectrum described in the text.}
\label{fig:beammu}
\end{figure*}

While our backgrounds contain 500 fb$^{-1}$ of total integrated
luminosity for processes with initial $e^+e^-$ or $e^\pm\gamma$ states,
some $\gamma\gamma$ initiated processes yield very high cross sections,
and thus a smaller number of events had to be generated
and then rescaled due to limited storage space.
In total, our background sample uses approximately 1.7 TB of disk space.
This rescaling of some $\gamma \gamma$ processes introduces
artificially large fluctuations in the corresponding analysis distributions.
In order to remedy this, we employ the
following procedure: we combine
the two independent background samples for the affected reactions,
and then randomly reallocate each entry on a bin by bin basis to one
of the two background sets.
Thus, on average,
each histogram contains an equal amount of entries bin per
bin, while remaining
statistically independent.
Of course, this procedure does not completely eliminate the fluctuations.
However, due to the random reallocation
of entries, the contribution of these fluctuations to the statistical
analyses in our comparison of models performed in Section 7 is greatly
reduced.

\subsection{Analysis Procedure}
\label{Sec:detectorandanalysis}

For each SUSY production process, we perform a cuts-based analysis,
and histogram the distributions of various
kinematic observables that we will describe in
detail in the following Sections. We apply a general analysis strategy
that performs uniformly well over the full MSSM parameter region.
Each analysis is thus applied to every
model in exactly the same fashion; there are no free parameters, and
we do not make use of any potential information from the LHC;
in particular we assume that the LSP mass is not known. Recall that
the AKTW models that we have inherited are difficult
cases at the LHC, and thus in general we cannot make any assumptions
about what measurements, if any, will have been performed by the LHC
detectors. We also note that AKTW did not impose any additional
constraints from flavor physics, cosmological
observations, etc. Such a global analysis is clearly desirable
but is beyond the
scope of the present study and is postponed to a future publication.

Our background and signal events described above are piped through a
fast detector simulation using the org.lcsim
detector analysis package~\cite{lcsim}, which is currently
specific to the SiD detector
design~\cite{SiD}. org.lcsim is part of the Java Analysis
Studio (jas3)~\cite{jas3},
a general purpose java-based data analysis tool.
The org.lcsim fast detector simulation incorporates the specific
SiD detector geometry, finite energy resolution, acceptances,
as well as other detector specific processes and
effects. Unfortunately, the identification of
displaced vertices and a measurement of dE/dx are not yet
implemented in the standard, fully tested version of the simulation
package employed here, although preliminary versions of these
functions are under development. The present study represents the first
large end user application of the lcsim software package and hence
we prefer to use the standard, tested, version of the software
without these additional features.
The output of our lcsim-based analysis code is given in terms of
AIDA histograms, where AIDA refers to Abstract
Interfaces for Data Analysis \cite{Serbo:2003sm}, a
standard set of java and C++ interfaces
for creating and manipulating histograms, which is incorporated into
the jas3 framework.

org.lcsim allows for the study of various different detectors,
whereby an xml description of
the specific detector is loaded
into the software in a modular fashion. Currently, xml
descriptions for various slightly different versions of the SiD
detector geometry are publicly available.
We use the SiD detector version studied extensively at
Snowmass 2005 (\texttt{sidaug05}) \cite{snowmass}.
In addition, two files, called \texttt{ClusterParameters.properties} and
\texttt{TrackingParameters.properties} allow the user to adjust
various tracking and
energy resolution parameters. We set the parameters such that we
closely follow the SiD
detector outline document (DOD)~\cite{SiDDOD}. In particular, we employ
the following configuration in our study:
\begin{itemize}
\item The minimum transverse momentum of registered tracks is given
by $p_T > 0.2$ GeV.
\item There is no tracking capability below 142 mrad,
which corresponds to
$\left|\cos \theta\right| < 0.99$.
\item Between 142 mrad and 5 mrad, electromagnetically charged
particles appear as neutral clusters.
\item There is no detector coverage below 5 mrad.
\item The jet energy resolution is set to $30\%/\sqrt{E}$.
\item The electromagnetic energy resolution is set to $18\%/\sqrt{E}$.
\item The hadronic energy resolution is set to $50\%/\sqrt{E}$.
\item The hadronic degradation fraction is $r = 1.0$.
\item The electromagnetic jet energy fraction is $w_\gamma = 0.28$.
\item The hadronic jet energy fraction is $w_h = 0.1$.
\end{itemize}
For a detailed explanation of these parameters we refer to the SiD DOD,
specifically Section IV.B
regarding the energy resolution parameters.

However, we note that the lack of tracking capability below 142 mrad
causes highly energetic
forward muons to not be reconstructed. They are too energetic to
deposit energy into clusters and are thus undetected and appear as
missing energy. This effect produces a substantial Standard Model
background to, \eg, our stau analysis (see Section \ref{Sec:stau}),
where we allow one tau to decay
hadronically and the other leptonically. In this case, we keep
events with one electron and one muon of opposite charge,
which can be mimicked by
$\gamma \gamma \rightarrow \mu^+ \mu^-$ events where one of the beam electrons
is kicked out sufficiently to be detected, but one of the final state muons is
too energetic and too close to the beam axis to be reconstructed.
We find that this background is substantial and,
given these detector parameters,
can only be eliminated by discarding all tau
events with electrons/positrons in the final state.

The default jet finding algorithm of org.lcsim is the JADE jet
algorithm~\cite{Bethke:1988zc} in
the E scheme with $y_{\mbox{\tiny cut}} = 0.005$ employed as the
default setting. The JADE jet algorithm in
the E scheme is defined as follows:
\begin{eqnarray}
\min \left(p_i + p_j\right)^2 \!& = &\!
\min 2 E_i E_j (1- \cos \theta_{ij}) > y_{\mbox{\tiny cut}} s
\\
p_{ij} = p_i + p_j & \!\!& \mbox{ for the recombination scheme }
\nonumber
\end{eqnarray}
The default $y_{\mbox{\tiny cut}}$, however, is too small, and causes
soft gluons to produce far too many jets. We therefore set the value
of $y_{\mbox{\tiny cut}}$ to $y_{\mbox{\tiny cut}} = 0.05$ within
the JADE jet algorithm. In addition, one must take care
when using the default org.lcsim jet finder, as every parton,
including leptons and photons,
is in principle identified as a jet. More sophisticated jet finders
are in the development
stage. We thus use the default jet finder, but with additional checks
on the jet particle
content to discard non-hadronic ``jets''.

We perform searches for slepton, chargino, and neutralino production
and in many cases design
analyses for several different decay channels of these sparticles. Each of our
analyses is designed to optimize a particular
signature, and we apply each analysis to
every AKTW model. A particular model may or may not produce a visible
signature in a specific channel. We will describe our cuts in detail for each
analysis channel in the Sections below.

As a starting point, we incorporate sets of kinematic
cuts that were developed in various previous supersymmetric
studies in the literature
(the specific references are given in the following Sections).
However, in many
cases we find that some of these cuts are optimized for specific
Supersymmetry benchmark points, \eg, the
Snowmass Points and Slopes (SPS)~\cite{Aguilar-Saavedra:2005pw},
and are too stringent for the general
class of models we study here. In other cases,
we find that the cuts employed in the literature are not
stringent enough to sufficiently reduce the Standard
Model background in order
to obtain a good signal to background ratio. Through a seemingly endless
series of iterations, we have thus designed sets of cuts
(which are described in the following Sections) that
optimize the signal to
background ratio for an arbitrary point in MSSM parameter space.
This scenario
corresponds to a first sweep of ILC data in search of a SUSY signal,
and is therefore a reasonable course of action. We also remind the
reader that these AKTW models are difficult at the LHC and hence the
slepton, chargino, and neutralino states will not necessarily be observable
at the LHC.

The signatures that we have developed analyses for are summarized in
Table~\ref{Tab:signature}, which lists the signature, dominant background
source, and the observable kinematic distribution
for each SUSY production process.
We note that in some cases, the same signature can arise from
different sources of sparticle production, \eg, $\mu^+\mu^-+$ missing energy
can occur from both smuon and chargino production.
Indeed, it is well known
that sometimes SUSY is its own background and we will note this in
the following Sections. Our cuts, however, are chosen such to
minimize this effect.

\begin{table*}
\centering
\begin{tabular}{|l|l|l|l|}
\hline

Sparticles Produced & Signature & Main Background & Observable \\ \hline
$\tilde{e}^+ \tilde{e}^-$ & $e^+ e^-$ + missing E & $e^+ e^- \rightarrow
e^- \bar{\nu}_l \nu_{l} {e}^{+},\,$ & energy of $e^+,\,e^-$ \\
$\tilde{\mu}^+ \tilde{\mu}^-$ & $\mu^+ \mu^-$ + missing E &
$e^+ e^- \rightarrow \mu^- \bar{\nu}_l \nu_{l} {\mu}^{+},\,$ & energy of
$\mu^+,\,\mu^-$ \\
$\tilde{\tau}^+ \tilde{\tau}^-$ & $\tau^+ \tau^-$ + missing E & $e^\pm \gamma
\rightarrow e^\pm \nu \bar{\nu}$ & energy of tau jets \\
& & $e^\pm \gamma \rightarrow e^\pm l^+ l^-$ & \\
$\tilde{\nu} \tilde{\nu}^*$ & $jjjjl^+ l^-$ + missing E & $\gamma
\gamma \rightarrow
c \bar{c}, b \bar{b}$
& missing energy \\
$\tilde{\nu} \tilde{\nu}^*$ & $jjjjjj$ + missing E & none & missing energy \\
$\tilde{\chi}^+_1 \tilde{\chi}^-_1$ & $\mu^+ \mu^-$ + missing E &
$e^+e^- \rightarrow l^- \bar{\nu}_l \nu_{l'} {l'}^+$ & energy of
$\mu^+,\,\mu^-$ \\
$\tilde{\chi}^+_1 \tilde{\chi}^-_1$ & $jj\mu^\pm$ + missing E &
$e^+e^- \rightarrow q
\bar{q}' l
\bar{\nu}_l$ & energy and invariant \\
&& $\gamma \gamma \rightarrow \tau^+ \tau^-$ & mass of dijet pair \\
&& $\gamma \gamma \rightarrow q \bar{q}$ & \\
$\tilde{\chi}^+_1 \tilde{\chi}^-_1$ & $jjjj$ + missing E &
$\gamma \gamma \rightarrow
q \bar{q}$ & energy and invariant \\
&&& mass of dijet pairs, \\
& & & missing energy \\
$\tilde{\chi}^+_1 \tilde{\chi}^-_1$ & $\gamma$ + charged tracks &
$e^+e^- \rightarrow
l^- \bar{\nu}_l \nu_{l'} {l'}^+$ & recoil mass \\
& & $e^\pm \gamma \rightarrow e^\pm l^+ l^-$ & \\
$\tilde{\chi}^+_1 \tilde{\chi}^-_1$ or
$\tilde{\tau}^+
\tilde{\tau}^-$ & 2 stable charged tracks & $e^+e^-
\rightarrow e^+e^- + \mbox{ISR,BS}$ & $p/E$ \\
$\tilde{\chi}^0_2 \tilde{\chi}^0_1$ & $l^+l^-$ + missing E & $\gamma \gamma
\rightarrow l^+ l^-$ & invariant mass \\
&&& of lepton pair \\
$\tilde{\chi}^0_2 \tilde{\chi}^0_1$ & $jj$ + missing E & $e^\pm
\gamma \rightarrow
\nu_e q \bar{q}$ &
invariant mass \\
&&& of jet-pair \\
$\tilde{\chi}^0_1 \tilde{\chi}^0_1$ & $\gamma$ + nothing &
$e^+e^- \rightarrow \gamma
\nu \bar{\nu}$ & photon energy \\ \hline
\end{tabular}
\caption{Summary of signatures and observables in all analysis
channels that we study and sources of the main
standard model background. $l = e,\mu,\tau$}
\label{Tab:signature}
\end{table*}

As discussed in the introduction, the first step in our analysis is to
determine whether or not a given SUSY particle is visible above the
SM background. Specifically, for a kinematic
distribution resulting from our analysis of a given observable, we ask
whether or not there is sufficient evidence to claim a `discovery' for a
SUSY particle within a particular model. There are many ways to do this,
but we
follow the Likelihood Ratio method, which we base on Poisson statistics.
(See, \eg, Ref. \cite{Ball:2007zz}).
In this method, we introduce the general Likelihood distribution:
\begin{equation}
L(n,\mu)=\prod_{i}^{bins} {{\mu_i^{n_i} e^{\mu_i}}\over {n_i!}}\,,
\end{equation}
where $n_i(\mu_i)$ are the number of observed (expected) events in each bin
$i$ and we take the product over all the relevant bins in the histogram. As
discussed above, we have generated two complete and
statistically independent background samples, which we will refer to as
$B1$ and $B2$. Combining the
pure signal events, $S$, which we generate for any given model with one of
these backgrounds, we form the Likelihood Ratio
\begin{equation}
R=L(S+B1,B2)/L(B1,B2)\,.
\end{equation}
The criterion for a signal to be observed above background is
that the significance, $\cal S$, satisfy
\begin{equation}
{\cal S}=\sqrt{2 \log R}>5\,.
\end{equation}
This corresponds to the one-sided Gaussian probability that a fluctuation in
the background mimics a signal of $\simeq 2.9\cdot 10^{-7}$, which is the
usual $5\sigma$ discovery criterion. When employing this method, we sometimes
encounter bins within a given histogram for which there is no background due
to low statistics but where a signal is observed. In this case, the function
$L$ is not well-defined. When this occurs we enter a single event into the
empty background histogram in that bin.

Given that our full SM background samples are only available at
fixed energies, our toolbox does not include the
ability to do threshold scans. As is well known,
this is a very powerful technique that can be used to obtain
precision mass determinations for charged SUSY (or any other new)
particles that are kinematically accessible. Such
measurements would certainly aid in the discrimination between
models, especially in difficult cases where the measurements we employ
do not suffice. In addition, especially for sparticles which decay
inside the detector volume, input from the excellent SiD vertex
detector could prove extremely useful. In the analysis presented
below, the vertex detector is used only for track matching and not
as a search tool for long-lived states.

\section{Slepton Production}
\label{Sec:sleptons}

\subsection{Charged Slepton Pair Production}

For detecting the production of charged sleptons, we focus on
the decay channel
\begin{equation}
\tilde{l}^+ \tilde{l}^- \rightarrow l^+ l^- \LSP \LSP \, ,
\end{equation}
that is, the signature is a lepton pair plus missing energy. In
the cases of selectrons and smuons these signatures are fairly straightforward
to study; the stau case is slightly more complicated due to the more involved
tau identification.

As is well known, the main Standard Model background for all
of these cases arises from the
production of $W$ pairs followed by their subsequent decay
into lepton-neutrino pairs and from $Z$-boson pair production,
where one $Z$ decays into
a charged lepton pair and the other into a neutrino pair. A
significant background also arises from gamma-induced processes through
beam- and bremsstrahlung.

The $W$ pair background
produces leptons that are predominantly along the beam axis towards
$|\cos \theta| \approx 1$, where $\theta$ takes on the conventional
definition. This is because
the decaying $W$ bosons are produced either through
$s$-channel $Z$- or $\gamma$-exchange, for
which the differential cross section is proportional to
$(1+ \cos^2 \theta)$, or through $t$-channel
neutrino-exchange, which behaves as $1/\sin^4 (\theta/2)$.
The photon-induced
background also yields electrons that are peaked along the beam axis
because they are mainly produced at low $p_T$ from beam-
and bremsstrahlung, although our more realistic beamspectrum
has a larger $p_T$ tail than the PYTHIA-generated backgrounds
studied conventionally
(cf. the discussion in Sec. \ref{Sec:analysis}). As we will
illustrate below in Section~4.1.3, having
the best possible forward detector coverage in terms of tracking and
particle identification (ID) is therefore of utmost importance to reduce the
Standard Model background.

To reduce the SM background, we employ a series of cuts that have been
adapted and expanded from previous studies
\cite{Goodman,Martyn:2004jc,Bambade:2004tq}. Our cuts
are fairly similar
for all slepton analyses. We will discuss them in detail
in our selectron analysis presented
below, and then will list the cuts with only brief comments in our
discussion of smuon and stau production.

\subsubsection{Selectrons}
\label{Sec:selectron}

As discussed above, in the case of selectron production
we study the clean decay channel
\be
\sel^+ \sel^- \rightarrow e^+ e^- \LSP \LSP \, ,
\ee
that is, the signature is an electron pair plus missing energy.
The main backgrounds arising from the SM originate
in $W$ and $Z$ pair production, followed by their leptonic decays,
along with several processes originating
from both $\gamma \gamma$ and $\gamma e $ interactions. To reduce
these backgrounds
we employ the following cuts, which are
expanded from those in \cite{Goodman}:

\begin{enumerate}
\item We require exactly two leptons, identified as an electron and a positron,
in the event and that there be no other charged particles.
This removes SM backgrounds where, for example, both $Z$ bosons decay into
charged leptons.

\item $\Evis < 1 $ GeV for $ | \cos \theta | \geq 0.9$, where $\Evis$
corresponds to the visible energy in the event.
This helps to reduce large SM background from forward $W$ production,
as well as beam-/bremsstrahlung reactions
that yield leptons predominantly along the direction of the beam axis.

\item $\Evis < 0.4 \sqrt{s}$ in the forward hemisphere.
Here, the forward hemisphere is defined as the hemisphere around the
thrust axis
which has the greatest visible energy. Since
we only have 2 visible particles in the final state, this amounts to
defining the forward hemisphere about the particle with the
highest energy.
The SUSY signal has missing energy in both hemispheres,
whereas the SM reaction $\epem\to ZZ\to \epem \nu\bar\nu$
has missing
energy in only one of the hemispheres since the decay $Z\to\nu\bar\nu$ occurs
in the hemisphere opposite of the $Z$ decay to charged leptons.

\item The angle between the reconstructed electron-positron pair is
restricted to have $\cos \theta > -0.96$.
Since SUSY has a large amount of missing energy, the selectron
pair will not be back-to-back, in contrast to the SM background events.

\item We demand that the visible transverse momentum, or
equivalently, the transverse momentum
of the electron-positron pair, $p_{T \mbox{\tiny vis}} =
p_T^{\,e^+ e^-} > 0.04 \sqrt{s}$.
This cut significantly reduces both the $\gamma \gamma$ and $e^\pm \gamma$
backgrounds which are mostly at low $p_T$.

\item The acoplanarity angle must satisfy $\Delta \phi^{e^+e^-} > 40$ degrees.
Since we demand only an electron and positron pair, the
acoplanarity angle
is equivalent to $\pi$ minus the angle between the transverse momentum of
the electron and positron,
$\Delta \phi^{e^+e^-}
= \pi - \theta_T$. This requirement translates to a
restriction on the transverse angle of
$\cos \theta_T > 0.94$.
This cut further reduces contributions from both the $W$-pair and
$\gamma\gamma$ backgrounds where the \epem\ pair tends to be more back-to-back.

\item $M_{e^+e^-} < M_Z - 5 \mbox{ GeV}$ or $M_{e^+e^-} > M_Z +
5 \mbox{ GeV}$, where $M_{\epem}$ is the invariant mass of the lepton pair.
This cut is to further
remove events from $Z$ boson pair production with
subsequent decays into \epem\ pairs.

\end{enumerate}

As already mentioned above, we note that below 142 mrad
($\left|\cos \theta\right| > 0.99$), the SiD detector does
not have particle tracking information according to the
current detector design~\cite{SiDDOD},
and any charged particle in this region
appears only as a neutral electromagnetic cluster.
However the first cut listed above,
where we demand exactly an \epem\ pair in the final state, substitutes for
potentially more detailed cuts that assume tracking capabilities down
to much smaller angles.

\begin{figure*}[hpb]
\centerline{
\epsfig{figure=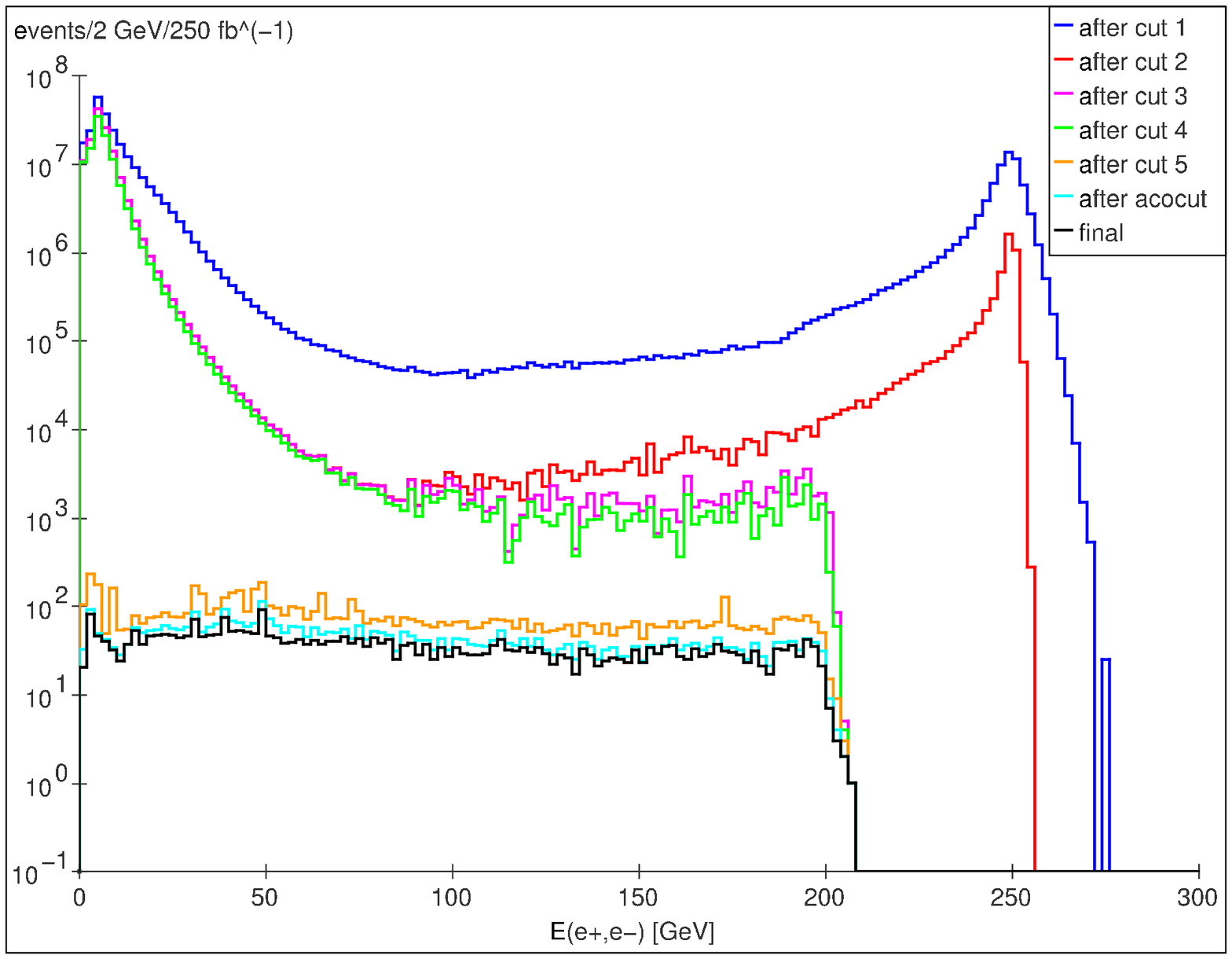,width=13cm,clip=}}
\caption{Standard Model background after each of the cuts listed in the text
is successively imposed for an incoming right-handed electron beam with
80\% polarization
and $250 \mbox{ fb}^{-1}$ of integrated luminosity.}
\label{fig:epemcuts}
\end{figure*}

The standard selectron search analysis is based on the energy distribution
of the final state electron or positron. Since the selectron decays into
a clean 2-body final state, the $e^-/e^+$ energy distribution has a
box-shaped ``shelf'' in a high statistics, background-free, perfect
detector environment in the absence of radiative effects.
Kinematics determines the minimum and maximum electron energies which
are related to the two unknown masses of the selectron and LSP by
\ba
E_{\max} & = & \frac{\sqrt{s}}{4}
\left( 1 + \sqrt{1-\frac{4 m_{\sel}^2}{s}} \right)
\left( 1 - \frac{m_{\LSP}^2}{m_{\sel}^2} \right) \, , \label{Emax} \\
E_{\min} & = & \frac{\sqrt{s}}{4} \left( 1 -
\sqrt{1-\frac{4 m_{\sel}^2}{s}} \right)
\left( 1 - \frac{m_{\LSP}^2}{m_{\sel}^2} \right) \, . \label{Emin}
\ea
The sharp edges of this box-shaped energy distribution allow for a precise
determination of the selectron and LSP masses. However,
due to beamstrahlung, the effective $\sqrt{s}$ above will
vary, and once detector effects are also included the edges of this
distribution will tend to be slightly
washed out. Since our goal here is to simply detect superpartners
and then distinguish
models with different sets of underlying parameters, we do not need to
perform a precise mass determination
in the present analysis. We also consider additional kinematic
observables, such as the distribution of $p_T^{vis}$ and the $e^+e^-$
invariant
mass $M_{ee}$, as they will be useful at separating different SUSY
signal sources.

The successive effect of each of the above cuts on the SM background is
illustrated in Fig.~\ref{fig:epemcuts}. Here, we show the
electron and positron energy distribution for 250 $\mbox{fb}^{-1}$
of simulated Standard Model background for RH electron
beam polarization
at $\sqrt{s} = 500\,$ GeV. The $y$-axis corresponds to the number of events
per 2 GeV bin. We note that cuts number 1-5 essentially eliminate any
potential background arising from the large Bhabha scattering and
$\gamma \gamma \rightarrow e^+ e^-$ cross sections. The main contribution to
the background remaining after these cuts
arises from processes involving $W$ and $Z$ pair production from
electron positron initial states, \ie, $\epem\to
\ell^- \bar{\nu}_\ell \nu_{\ell'} {\ell'}^{+}$, with $\ell^{(')}=e,\tau$,
as shown in Fig.~\ref{fig:selectronbg}. We find that most of the photon
initiated background has been removed by our cuts.
Note that applying these cuts
in a different order would necessarily show a different
level of effectiveness.

\begin{figure*}[hpbt]
\centerline{
\epsfig{figure=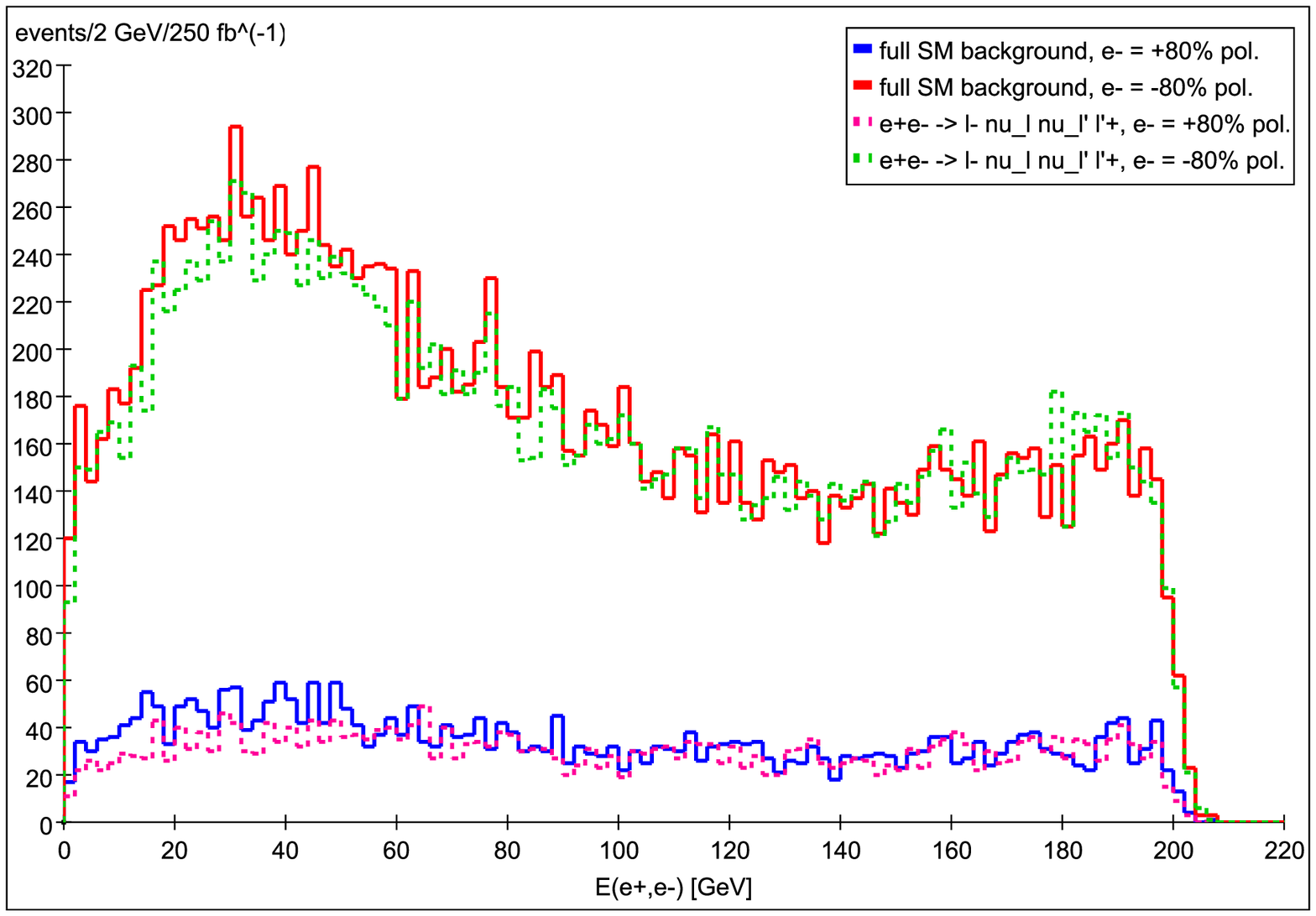,width=13cm,clip=}}
\caption{Remaining SM background after the full set of selectron
selection cuts listed in the text are imposed. This is generated
from $250 \mbox{ fb}^{-1}$ of SM events with
80\% right-handed (solid blue line) and 80\% left-handed (solid red line)
electron beam polarization, and unpolarized positron beam
at $\sqrt{s} = 500$ GeV. The dotted lines show the main process contributing
to the
background, $e^+ e^- \rightarrow
\ell^- \bar{\nu}_\ell \nu_{\ell'} {\ell'}^{+}$, with $\ell^{(')}=e,\tau$,
for 80\% right-handed (dotted pink line) and 80\% left-handed (dotted green
line) electron polarization. Note that here and in other Figures below the
spikes in the full background
and the main contributions are because of rescaling issues and the
thus necessary mixing of the two independent background samples (cf. the
discussion in Section 3.1). This
mixing is a random procedure which explains why the spikes are not all in
the same bins.}
\label{fig:selectronbg}
\end{figure*}

A further comment on cut number 5 is in order. One finds that increasing
this cut from $p_T^{\,e^+ e^-} >20$ GeV to $p_T^{\,e^+ e^-}> 30$ GeV at
$\sqrt{s} = 500$ GeV to further reduce the photon initiated background,
introduces a dip in the center of the ``shelves'' for MSSM models that
have edges near $E_e \simeq$
30-40 GeV. This apparently occurs when both visible leptons have
approximately the same amount of energy, so that their
transverse momenta partially cancel, leaving insufficient
visible transverse momentum to
pass the cut. When one of the leptons is more energetic than the other,
the visible $p_T$ is generally above the cut.
Thus increasing the cut on visible $p_T$ in order to reduce the background
further also affects
the signal in a perhaps somewhat unexpected way. The same observation also
applies in the smuon analysis below.

Figure~\ref{fig:selectroncuts} shows how a typical signal from a model
with kinematically
accessible selectrons responds to the same cuts
that were applied to the backgrounds above. Note that while the cuts
reduce the backgrounds by many orders of magnitude, the signal is
reduced only by a factor of $2-3$.

\begin{figure*}[hptb]
\centerline{
\epsfig{figure=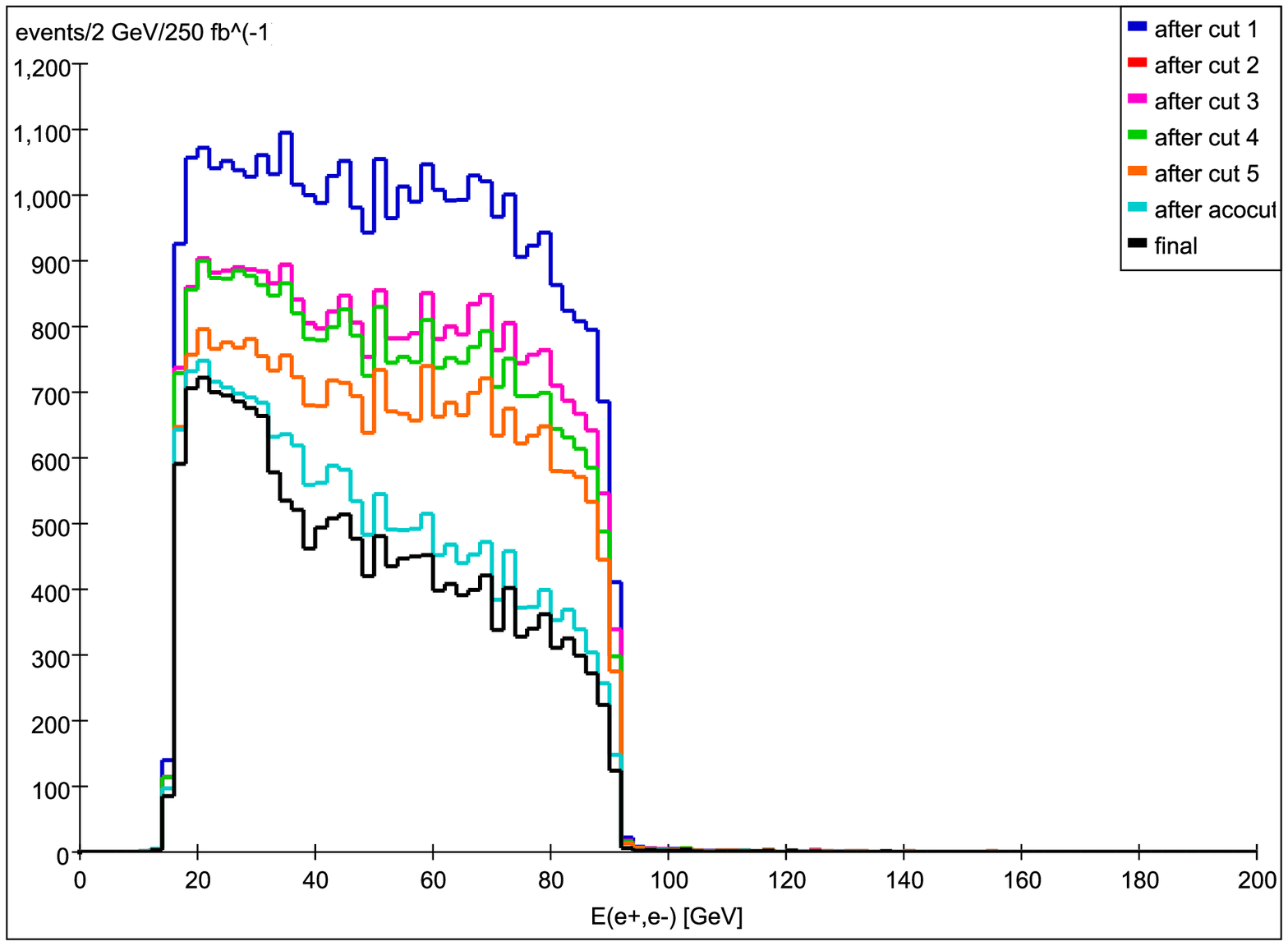,angle=0,width=13cm,clip=}}
\vspace*{0.1cm}
\caption{The electron energy distribution for selectron production after
successively
imposing each of the cuts listed in the text for the case of a right-handed
incoming electron beam with 80\% polarization and
250 fb$^{-1}$ of integrated luminosity.}
\label{fig:selectroncuts}
\end{figure*}

We now examine selectron pair production for the AKTW models.
In this case, there are 22/242 (22 out of 242)
models with kinematically accessible selectrons at
$\sqrt s=500$ GeV. The $\tilde e_{L(R)}$ is accessible in 9(15)
of these 22 models; for 2 models we find that both states are
potentially visible.
Note that fewer than $10\%$ of our models have this relatively
clean channel accessible. Selectrons are pair produced via $s$-channel
$\gamma$ and $Z$ exchange as well as $t$-channel neutralino exchange.
For the case of the well-studied $\tilde e \to e \tilde \chi_1^0$
decay mode, which we examine here,
selectrons are usually searched for by examining the detailed structure
of the resulting individual $e^\pm$ energy spectra
and looking for any excesses above the expected SM background. As is by
now well-known and briefly discussed above, in
the absence of such backgrounds, with high statistics and neglecting any
radiative effects, the 2-body
decay of the selectrons lead to flat, horizontal shelves in this
distribution. In a more
realistic situation where all such
effects are included and only finite statistics are available, the general
form of the shelf structure remains but they are now jagged, tilted
downwards (towards higher $e^\pm$ energies), and have somewhat smeared edges.
These effects are illustrated in Fig.~\ref{selectronfig} which shows
examples of the $e^\pm$ spectra (adding signal and background) for some
representative AKTW models containing kinematically accessible selectrons
with either beam polarization configuration. There are
several important features to note in these
Figures. The detailed nature of the $\tilde e$ signal in the $e^\pm$
energy spectrum shows significant variation
over a wide range of both magnitude and width depending upon the
$\tilde e_{L,R}$
and $\tilde \chi_1^0$ masses and the resulting production cross sections.
Recall that $t$-channel sneutrino exchange can be
important here and dramatically affects the size of this cross section.
The ratio of signal to background is not always as large as
in most cases discussed in the literature. In addition, we note
that the range of possible signal shapes relative to the SM background can be
varied; not all of our signals appear to be truly shelf-like. In some models,
the background overwhelms the signal.
Note that RH polarization leads to far smaller backgrounds than does LH
polarization as would be expected, this being
due to the diminished contribution from $W$-pairs which prefer LH coupling.

\begin{figure*}[hptb]
\centerline{
\includegraphics[width=13.0cm,height=10.0cm,angle=0]{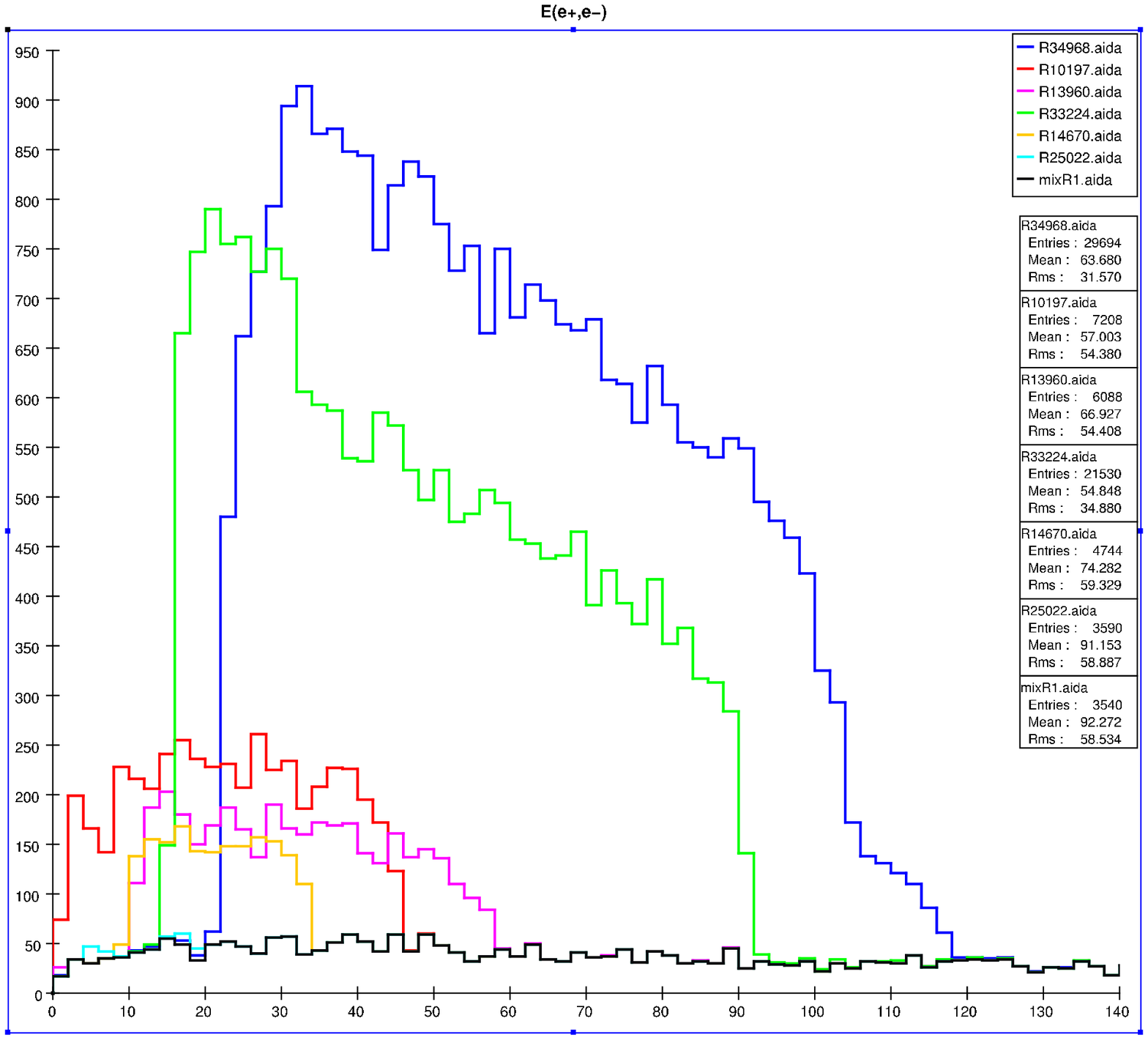}}
\vspace*{0.1cm}
\centerline{
\includegraphics[width=13.0cm,height=10.0cm,angle=0]{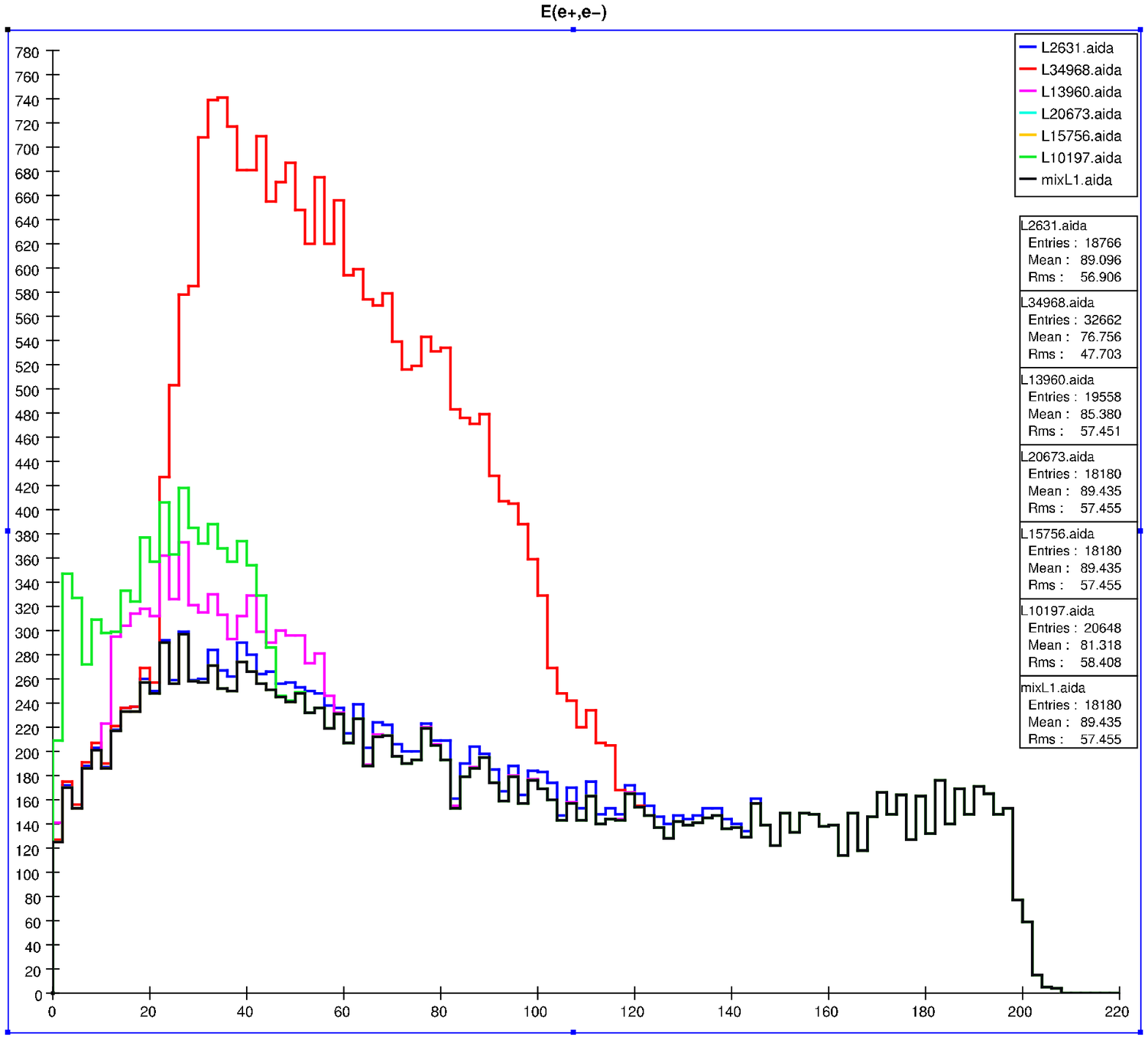}}
\vspace*{-0.1cm}
\caption{Electron energy distribution: the number of events/2 GeV bin
after imposing the full set of cuts discussed in the text for several
representative models. RH(LH) beam polarization is employed in the top(bottom)
panel, assuming $80\%$ electron beam polarization
an integrated luminosity of 250 fb$^{-1}$
for either polarization. The SM background is shown as the black histogram.}
\label{selectronfig}
\end{figure*}

Of the 22 kinematically accessible models, 18(15) lead to signals
with a visibility significance over background of ${\cal S} >5$
at these integrated luminosities assuming RH(LH)
beam polarization.
Combining the LH and RH polarizations channels we
find that 18/22 models with selectrons lead to signals with significance
$>5$. Furthermore, $8/9$ models with kinematically accessible
$\tilde e_L$ are observable while
$12/15$ models with $\tilde e_R$ are visible.
Note that 4 models have selectrons with
masses that are in excess of 241 GeV. This leads to a strong
kinematic suppression
in their cross sections and, hence, very small
signal rates, so they are missed by the present analysis.
Some of the models
in both the RH and LH polarization
channels have a rather small $S/B$ and are not easily visible
at this level of integrated luminosity;
typical examples of `difficult' models are presented in
Figs.~\ref{selectronfig2} and ~\ref{selectronfig3}.

\begin{figure*}[hptb]
\centerline{
\includegraphics[width=13.0cm,height=10.0cm,angle=0]
{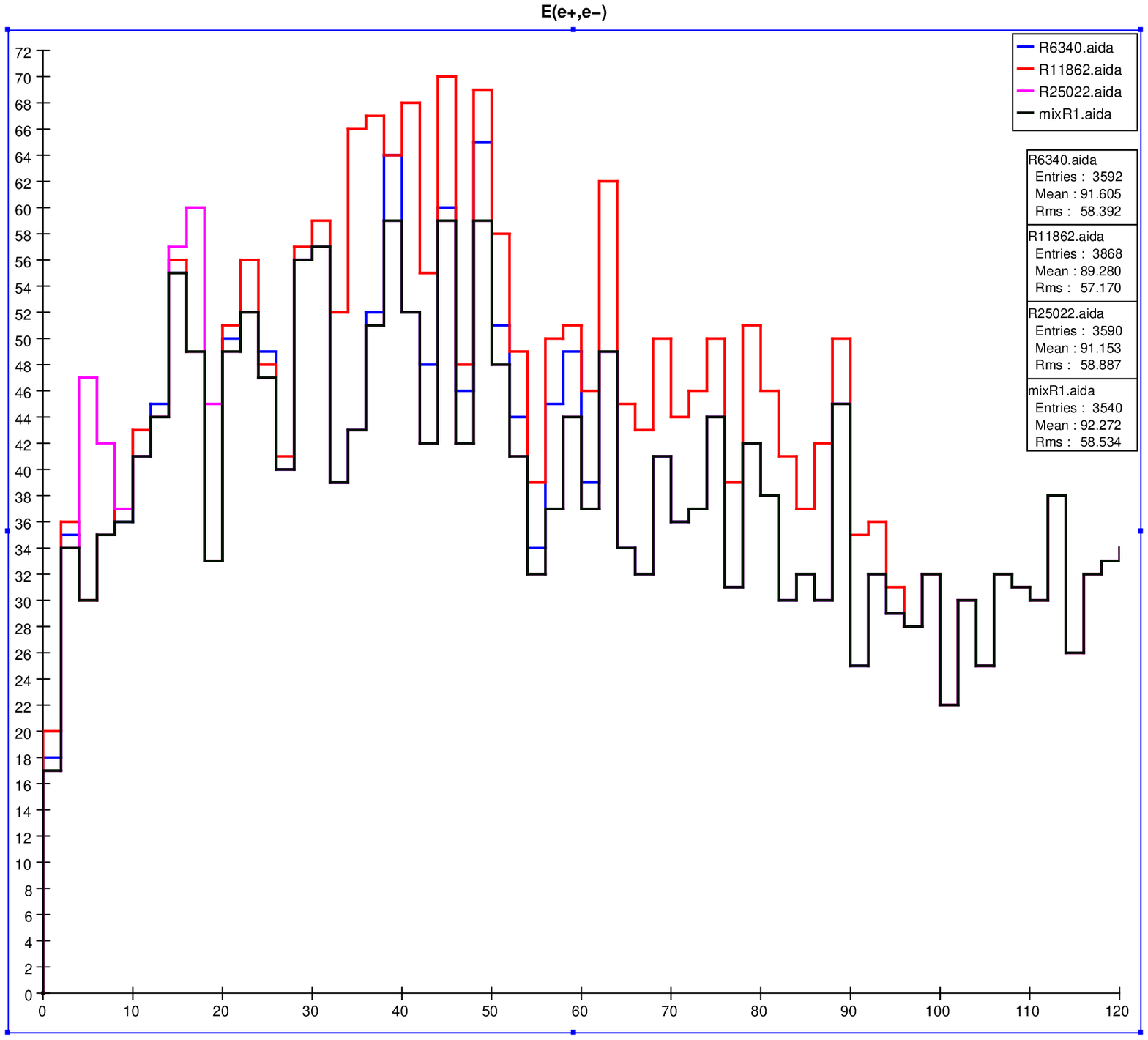}}
\vspace*{-0.1cm}
\caption{Same as the previous Figure but now for three
models which are difficult to observe due to small cross sections in the RH
polarization channel.}
\label{selectronfig2}
\end{figure*}
\begin{figure*}[hptb]
\centerline{
\includegraphics[width=13.0cm,height=10.0cm,angle=0]
{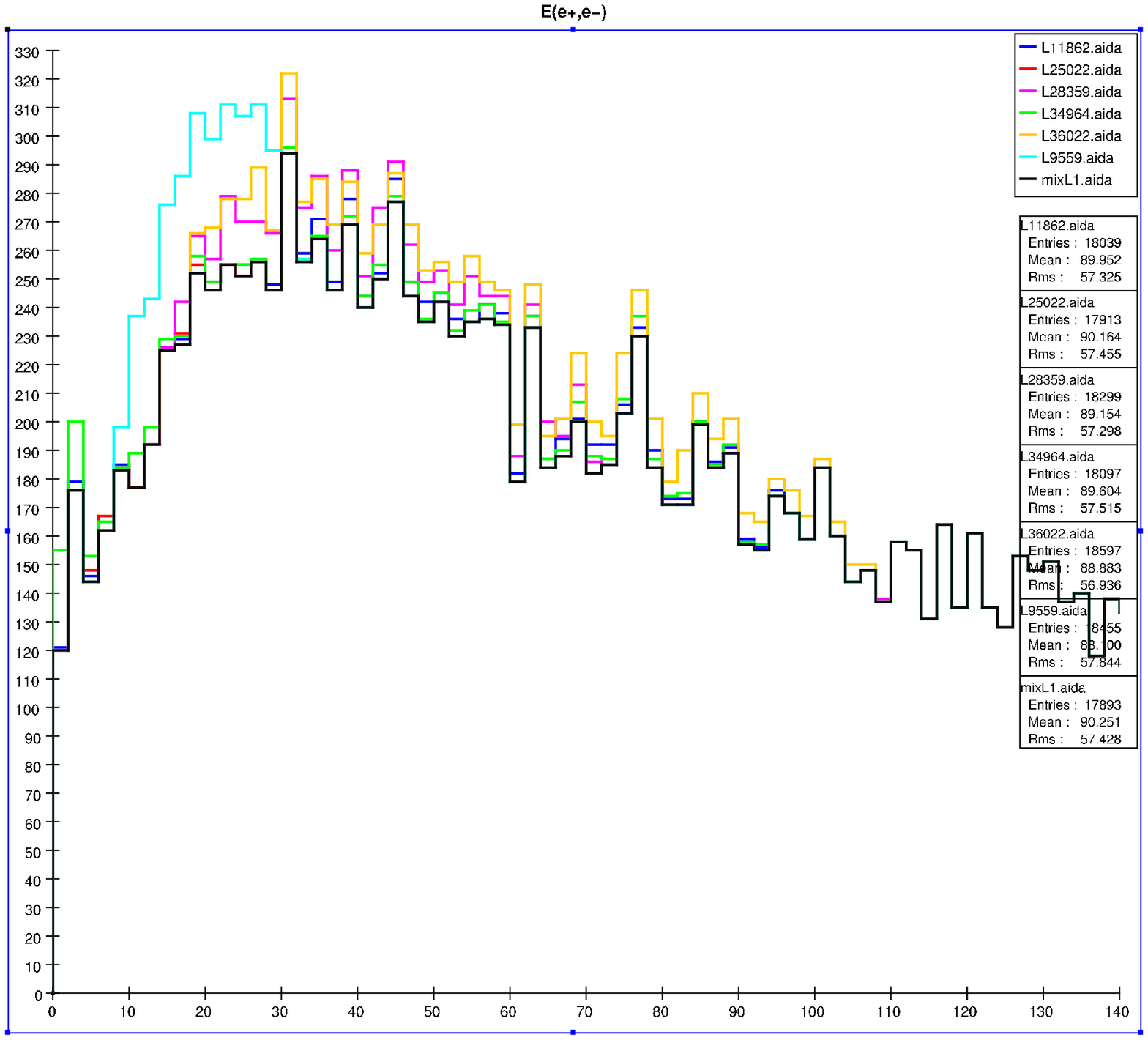}}
\vspace*{0.1cm}
\centerline{
\includegraphics[width=13.0cm,height=10.0cm,angle=0]
{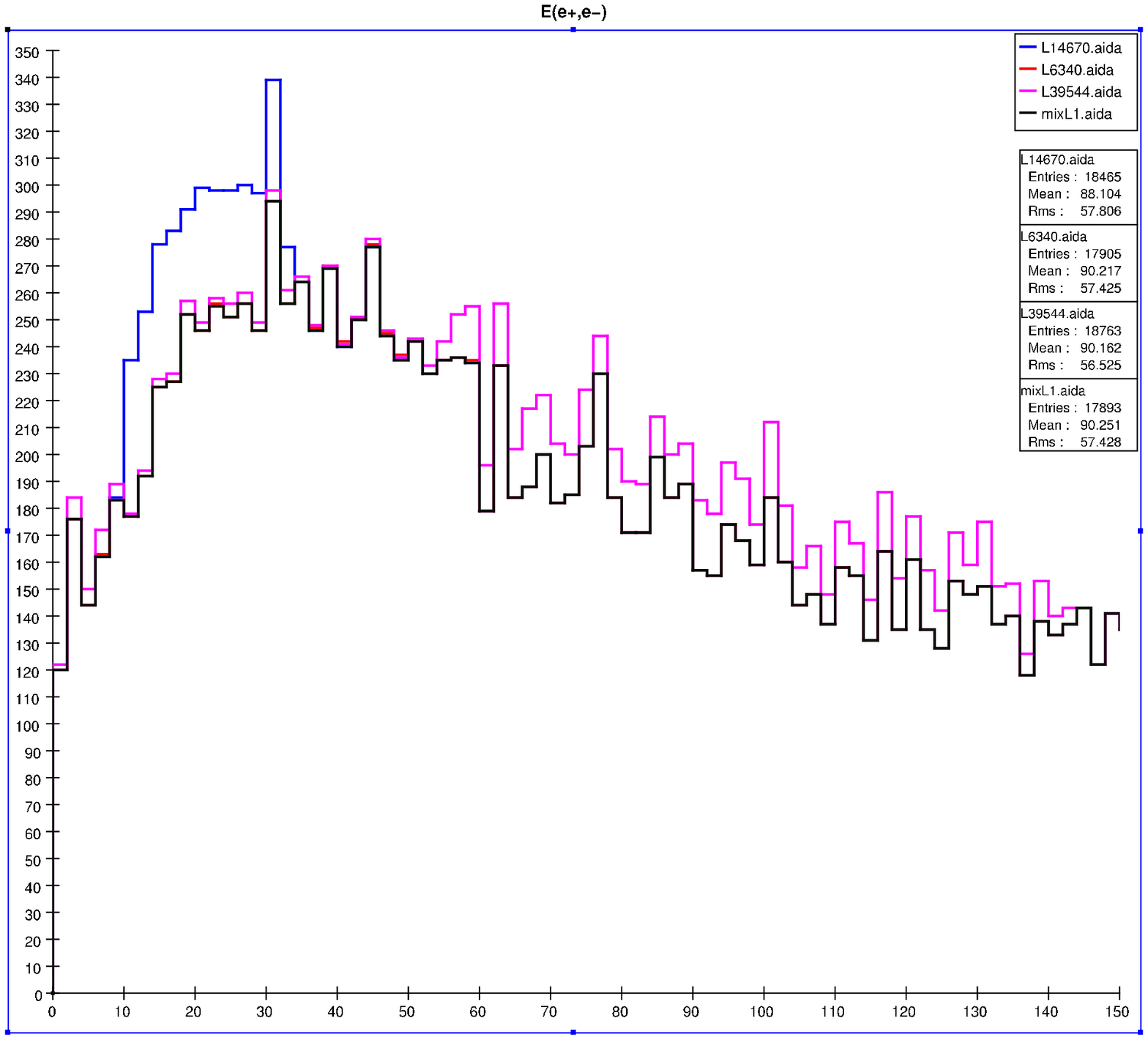}}
\vspace*{-0.1cm}
\caption{Same as the previous Figure but now for the LH polarization channel
for two sets of models which are a bit more difficult to observe
due to small cross sections.}
\label{selectronfig3}
\end{figure*}

It is interesting to compare these results to what we obtain in the case of the
well-studied SPS1a' benchmark model \cite{Aguilar-Saavedra:2005pw};
Fig.~\ref{selectrons_sps1a} shows the electron
energy distribution for this model for both beam polarization choices in
an analogous manner to that shown in the previous Figures for the AKTW models.
Due to the large production cross section, detecting
the signal in this case is rather trivial as we would expect
from the detailed studies made in
the literature. The most important thing to notice from this Figure is
that SPS1a'
leads to substantially larger signal rates than in {\it any} of
the models we are investigating in the present analysis. In fact, the
SPS1a' signal
rates can be almost two orders of magnitude larger than some of the
models we are
examining here. We also observe the obvious
presence of two shelves, especially in the case of LH beam polarization,
clearly
indicating that both the $\tilde e_L$ and $\tilde e_R$ states are
kinematically accessible and
are being simultaneously produced.

\begin{figure*}[htpb]
\centerline{
\includegraphics[height=10cm,width=13cm,angle=0]{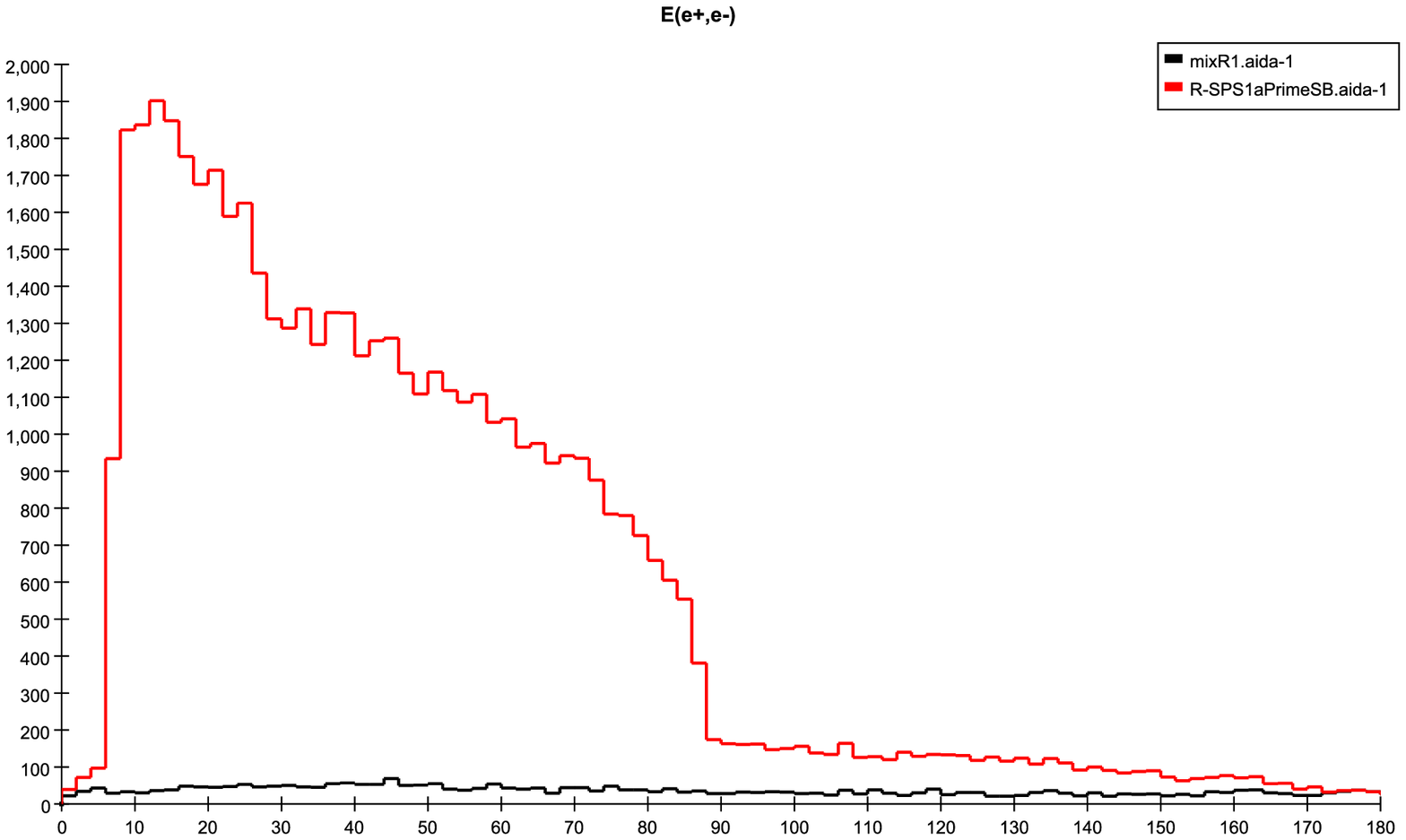}}
\vspace*{0.1cm}
\centerline{
\includegraphics[height=10cm,width=13cm,angle=00]{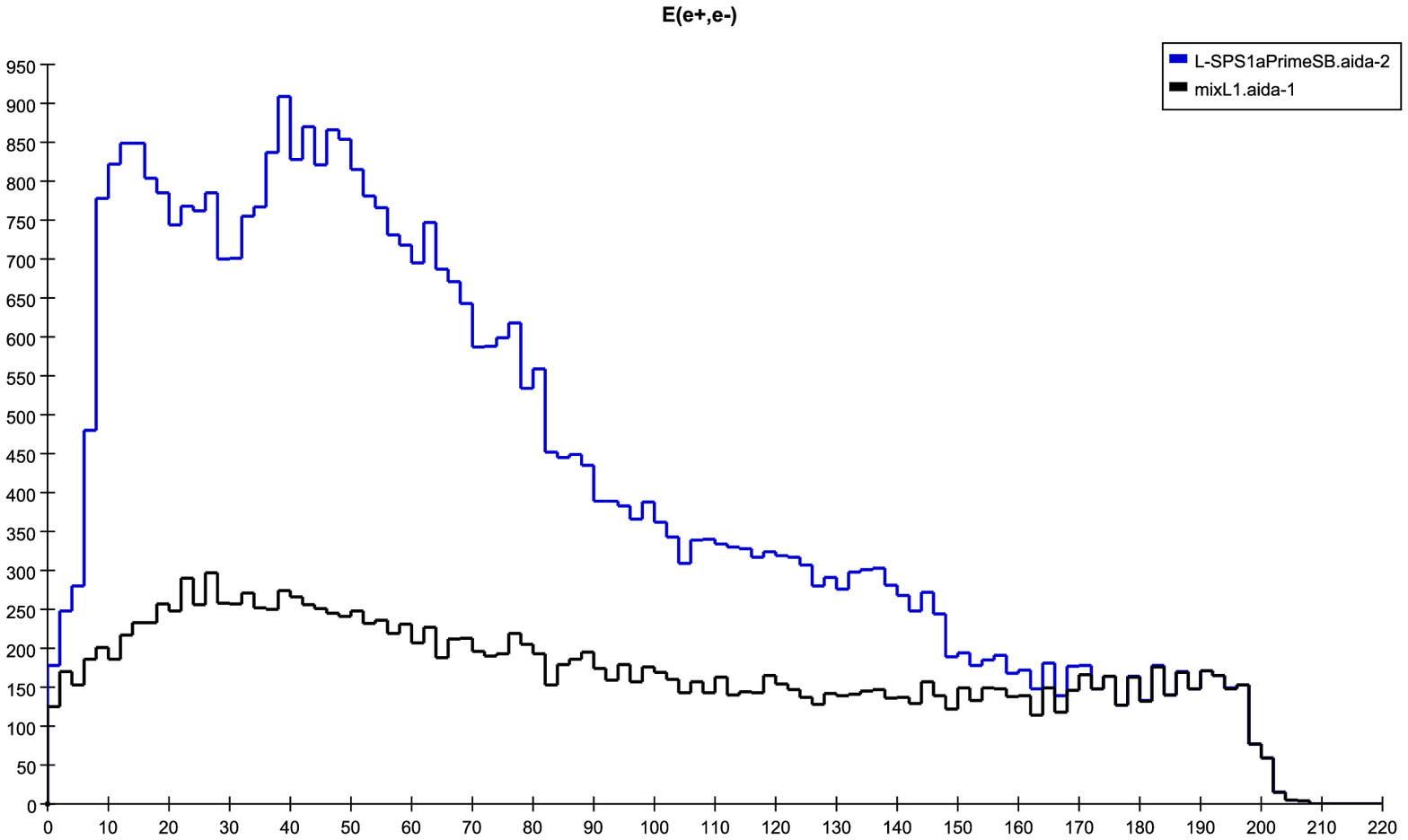}}
\vspace*{0.1cm}
\caption{The electron energy distribution
after imposing the full set of cuts discussed in the text for the benchmark
model SPS1a'. RH(LH) beam polarization is employed in the top(bottom)
panel, assuming
an integrated luminosity of 250 fb$^{-1}$
for either polarization. The SM background is shown as the black histograms.}
\label{selectrons_sps1a}
\end{figure*}

Interestingly, one finds that there are a number of models, particularly
in the case of RH polarization, which {\it do not} have
kinematically accessible selectrons but which have visible signatures in
the $\tilde e$-pair analysis. This is an example of SUSY
being a background to SUSY. There are, of course, other SUSY particles
which can decay into $e^\pm$ and missing
energy, \eg, chargino pair production followed by the decay
$\tilde \chi_1^+\to W^*\tilde\chi_1^0$ with $W^*\to e \nu$, or associated
$\tilde\chi_2^0 \tilde\chi_1^0$ production followed by the decay
$\tilde\chi_2^0\to e^+e^-\tilde\chi_1^0$. Both of these processes result
in the same observable final state.
Figure~\ref{selectronfig4} shows some of these `fake' models that appear
in our selectron analysis in the case of RH polarization. Fake
models, by which we mean models where other SUSY particle production
leads to a visible signature in the selectron analysis, also appear
for LH polarization and, in fact, we find 14 counterfeit models for either
polarization. Note that the shapes of these fake model
signatures are somewhat different than those in a typical model
with actual selectrons present; there are no truly shelf-like
structures and the $e^\pm$ energies are all peaked at
relatively low values. This occurs because in these examples the
final state electrons are the result of a 3-(or more) body
decay channel when the $W$ boson is off-shell and because the
$\tilde \chi_1^\pm-\tilde \chi_1^0$ mass splitting is relatively small.
Both of these conditions are present
in most of our models.

\begin{figure*}[hptb]
\centerline{
\includegraphics[width=13.0cm,height=10.0cm,angle=0]{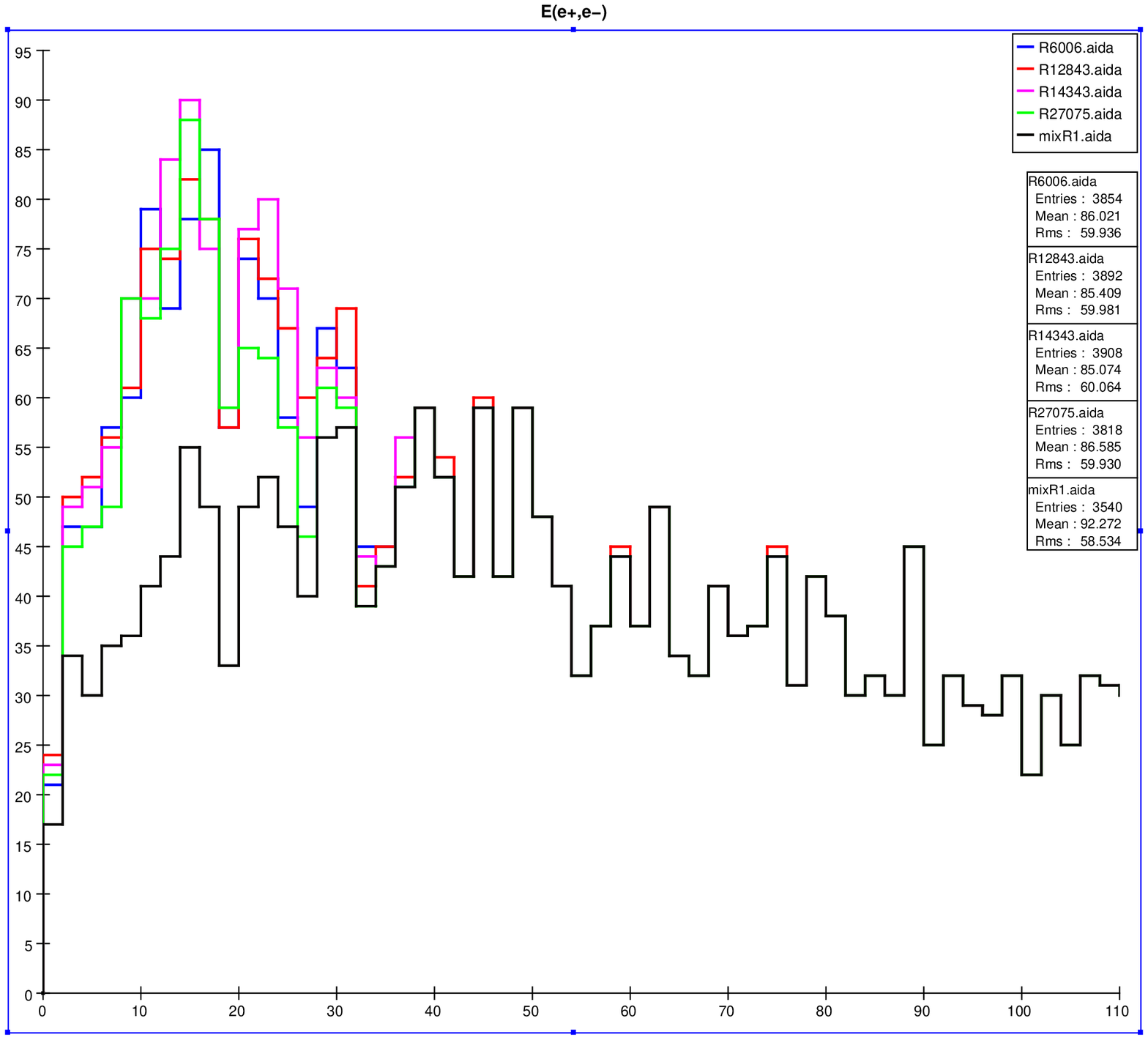}}
\vspace*{-0.1cm}
\caption{The electron energy distribution
after imposing the full set of cuts discussed in the text for a subset of
fake selectron models in the RH polarization channel.
The SM background is shown as the black histogram as usual.}
\label{selectronfig4}
\end{figure*}

If further differentiation from the fake signatures is required, we need to
examine a different kinematic distribution, \eg, the invariant mass
of the $e^+e^-$ pair, $M_{ee}$. One would expect the counterfeit signals to
populate small values of $M_{ee}$ while the models with
actual selectrons will have a higher number of events with larger values of
$M_{ee}$. This is indeed the case as can be clearly seen in
Fig.~\ref{selectronfig5}. As we will see below, fake signals
occur quite commonly in almost all of our analyses. Though
specific analyses are designed to search for a particular SUSY partner
it is quite easy for other SUSY states to also contribute to a given final
state and be observed instead, \eg, a similar signature can be generated
using the visible $p_T$ of the electron.

It would be interesting to return to this issue with a wider set of
models that lead to
larger mass splittings in the electroweak gaugino sector to see how well
selectrons and charginos
can be differentiated under those circumstances.

\begin{figure*}[htpb]
\centerline{
\includegraphics[width=13.0cm,height=10.0cm,angle=0]{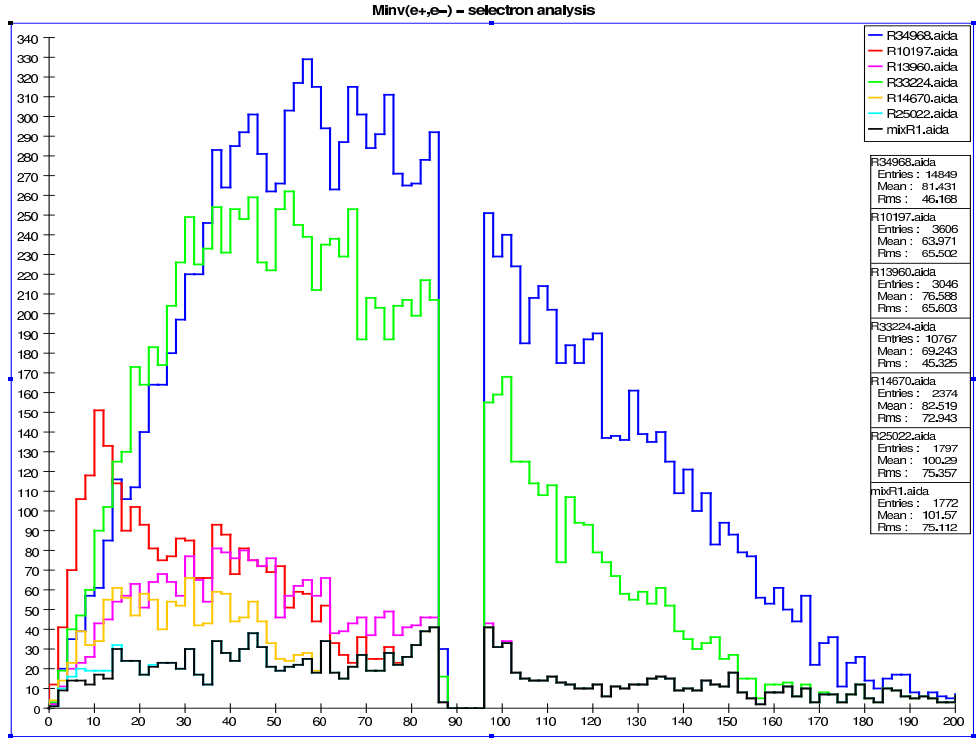}}
\vspace*{0.1cm}
\centerline{
\includegraphics[width=13.0cm,height=10.0cm,angle=0]{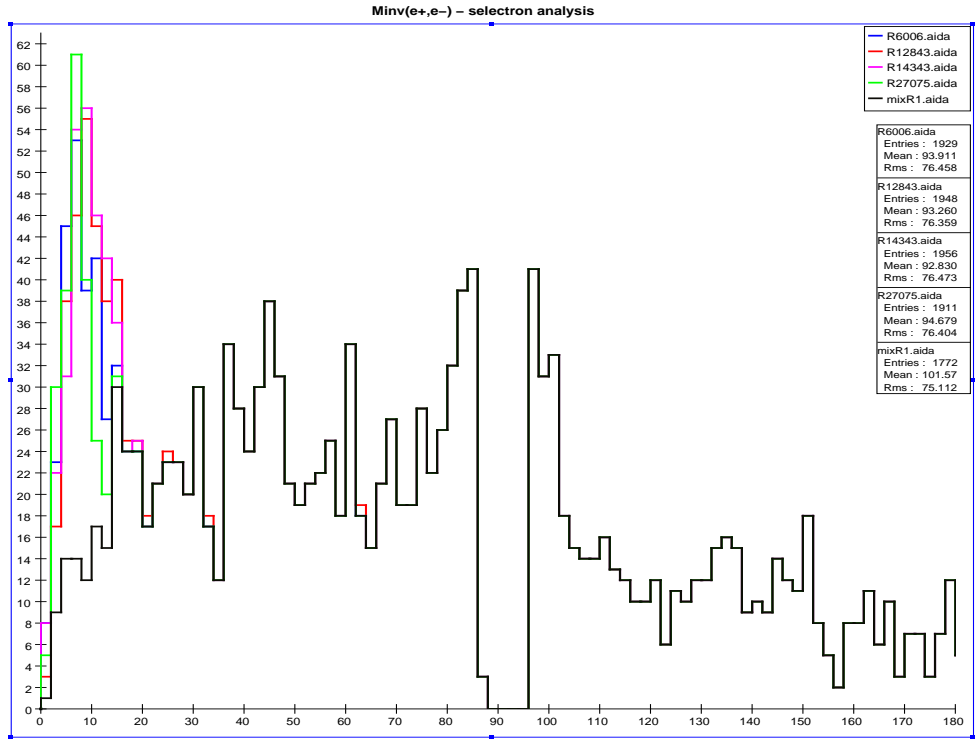}}
\vspace*{-0.1cm}
\caption{The dielectron invariant mass spectrum in the RH polarization channel
for true selectron models (top) and those caused by other SUSY particles (bottom).}
\label{selectronfig5}
\end{figure*}

\clearpage

\subsubsection{Smuons}
\label{Sec:smuon}

For $\smu$ pair production the standard search/analysis channel is
\be
\smu^+ \smu^- \rightarrow \mu^+ \mu^- \LSP \LSP \, ,
\ee
\ie, the signature is a muon pair plus missing energy. Smuon production
occurs via $s$-channel $\gamma$ and $Z$ exchange; there is no corresponding
$t$-channel contribution as in the case of selectrons. As in the
selectron analysis, the dominant background
arises from leptonic decays of $W$-pair and $Z$-pair production, as well
as the ubiquitous (though somewhat less important in this case)
$\gamma\gamma$ background. Since the background and signal are similar
to those for selectron production, our choice of cuts here will follow
those employed in the selectron analysis above and are adapted from those
proposed by
Martyn~\cite{Martyn:2004jc} (see also \cite{Bambade:2004tq}):

\begin{enumerate}
\item No electromagnetic energy (or clusters) $> 0.01 \sqrt{s}$
in the region $|\cos\theta| > 0.995$.

\item Exactly two muons are in the event with no other charged particles
and they are weighted by their charge within the polar angle
$-0.9 < Q_\mu\cos \theta_\mu < 0.75$ with no other visible particles.
This removes a substantial part of the $W$-pair background.

\item The acoplanarity angle satisfy $\Delta \phi^{\mu\mu} > 40$ degrees.
This reduces both the $W$-pair and $\gamma\gamma$ backgrounds.

\item $|\cos \theta_{p_{\mbox\tiny missing}}| < 0.9$.

\item The muon energy is constrained to be $E_\mu > 0.004 \sqrt{s}$.

\item The transverse momentum of the dimuon system, or equivalently,
visible transverse momentum (since only the muon pair is visible),
satisfy $p_{T \mbox{\tiny vis}} = p_T^{\mu \mu} > 0.04 \sqrt{s}$.
This removes a significant portion of the remaining $\gamma\gamma$ and
$e^\pm \gamma$ backgrounds.

\end{enumerate}

The remaining SM background after these cuts have been imposed are
displayed in Fig.~\ref{fig:smuonbg} for both polarization configurations.
The main background to $\smu$-pair production, $e^+ e^-
\rightarrow
l^- \bar{\nu}_l \nu_{l'} {l'}^{+},\, l = \mu, \tau$, is also shown in
the Figure, and we see that it essentially comprises the full background
sample. The background is somewhat smaller here than in the case for
selectron production, as beam remnants from $\gamma$-induced reactions
are not confused with the signal for smuon production.

\begin{figure*}[hpb]
\centerline{
\epsfig{figure=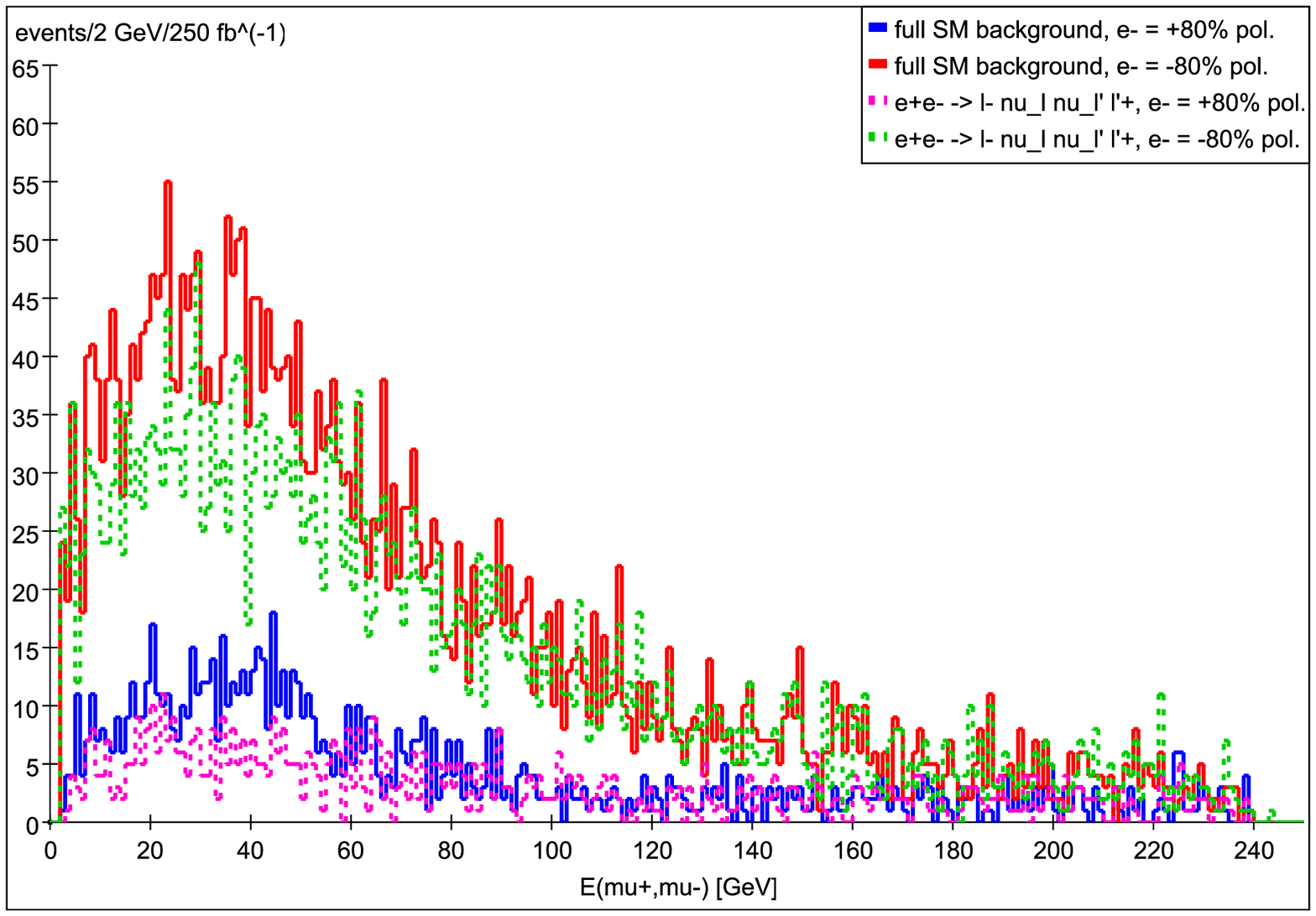,width=13cm,clip=}}
\vspace*{0.1cm}
\caption{The muon energy distribution of the
SM background after the smuon selection cuts described in the text
have been imposed. The red(blue) solid curves correspond to the full
background sample with $80\%$ LH(RH) electron beam polarization, and the
green(pink) dotted lines represent the contribution from the process
$e^+ e^- \rightarrow
l^- \bar{\nu}_l \nu_{l'} {l'}^{+},\, l = \mu, \tau$, with LH(RH) beam
polarization, respectively. 250 fb$^{-1}$ of integrated luminosity for
each beam polarization has been assumed.}
\label{fig:smuonbg}
\end{figure*}

As in the case of selectron production, there are only 22 out of 242 models
which have kinematically accessible smuons at $\sqrt s=500$ GeV. The
$\tilde \mu_{L(R)}$ is found to be kinematically accessible in
9(15) of these cases, there again being 2 models
where both smuon states can be simultaneously produced. In the $\tilde
\mu \to \mu \chi_1^0$ decay channel, smuons are observed by detecting
a structure above the SM background in the muon energy distribution,
similar to the search for selectrons. Here too, as is
well-known, in the absence of such backgrounds, with high statistics
and neglecting radiative effects, the 2-body
decay of the $\tilde \mu$'s leads to horizontal shelf-like
structures. In the more realistic situation where all such effects are
included, the shelves remain but are now tilted
downward (towards higher muon energies), as in the selectron case
and have somewhat rounded edges. Examples of the muon energy spectra for
some representative AKTW models displaying these effects are shown in
Fig.~\ref{smuonfig1} for either beam polarization
configuration. Several things are to be noted in these Figures.
The $\tilde \mu$ signals in the muon energy distribution vary over a
wide range
of both height and width depending upon the values of the $\tilde \mu$
and $\chi_1^0$ masses and the production cross sections.
The range of possible signal shapes relative to the background is varied,
but generally the signal is separable from the background in most models.
We note again that the background is somewhat reduced compared to selectron
production since there are
fewer issues with beam remnants here. Again, we see that
RH electron beam polarization leads to far smaller backgrounds
than does LH beam polarization, as expected
due to the diminished contribution from $W$-pair production.

Of the 22 kinematically accessible models, 19(17) lead to
signals with significance $>5$ at these integrated luminosities with
RH(LH) electron beam polarization.
Combining the LH and RH polarization channels, we
find that 19/22 models with accessible smuons lead to signals
that meet our visibility criteria. We display a representative set
of these models in Fig.~\ref{smuonfig1} for both beam polarizations.
The three models that do not pass our discovery criteria have smuons
with masses in excess of 241 GeV; this leads to a strong
kinematic suppression in their cross sections, and hence, very small
signal rates. The $S/B$ ratio is somewhat small for some of the models
in the LH polarization channel, as can be seen in
Fig.~\ref{smuonfig2}, and are not so easily visible at this
integrated luminosity. However, they nonetheless pass our significance
tests for discovery.

\begin{figure*}[htbp]
\centerline{
\includegraphics[height=10.0cm,width=13.0cm,angle=0]{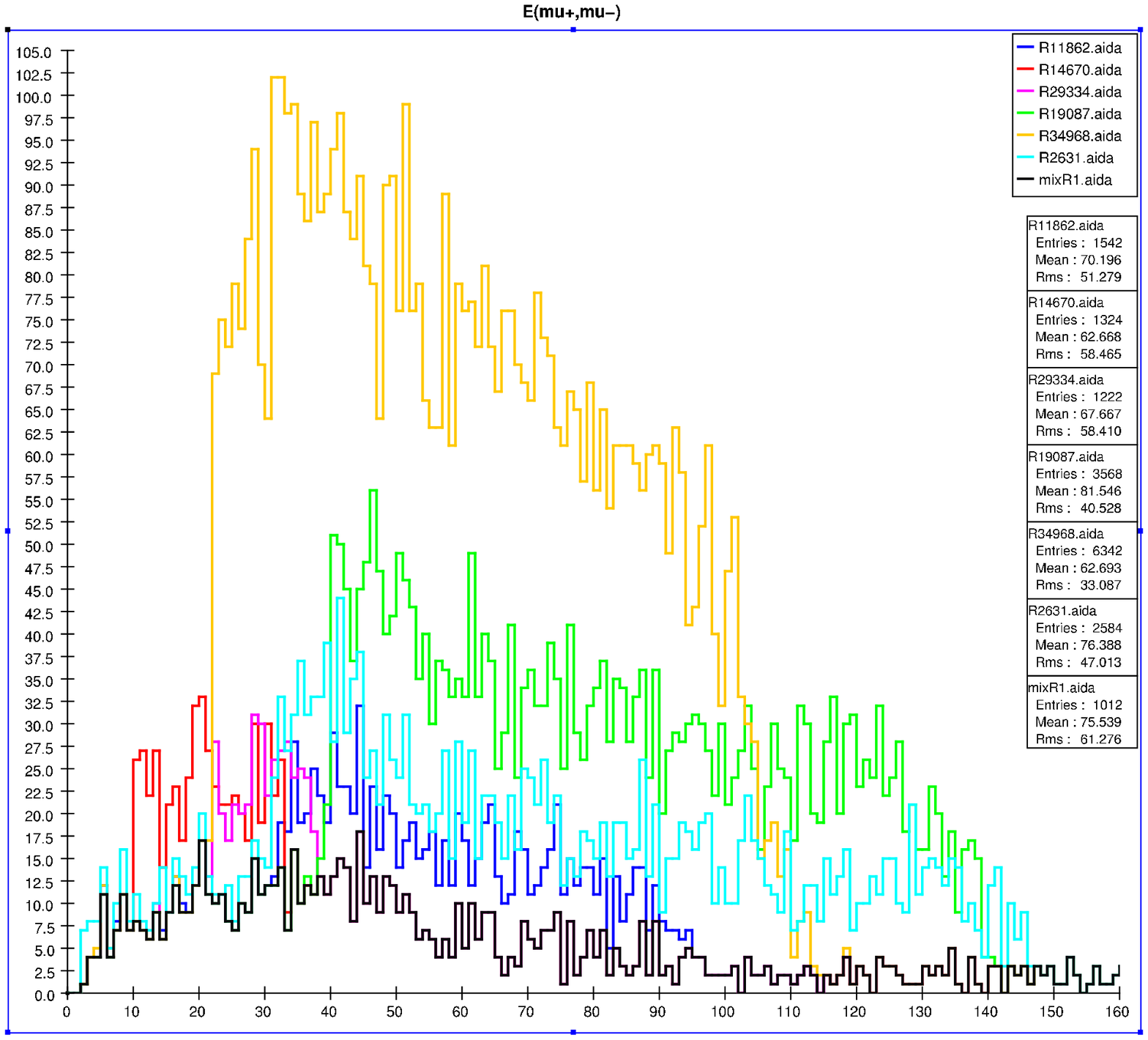}}
\vspace*{0.1cm}
\centerline{
\includegraphics[height=10.0cm,width=13.0cm,angle=0]{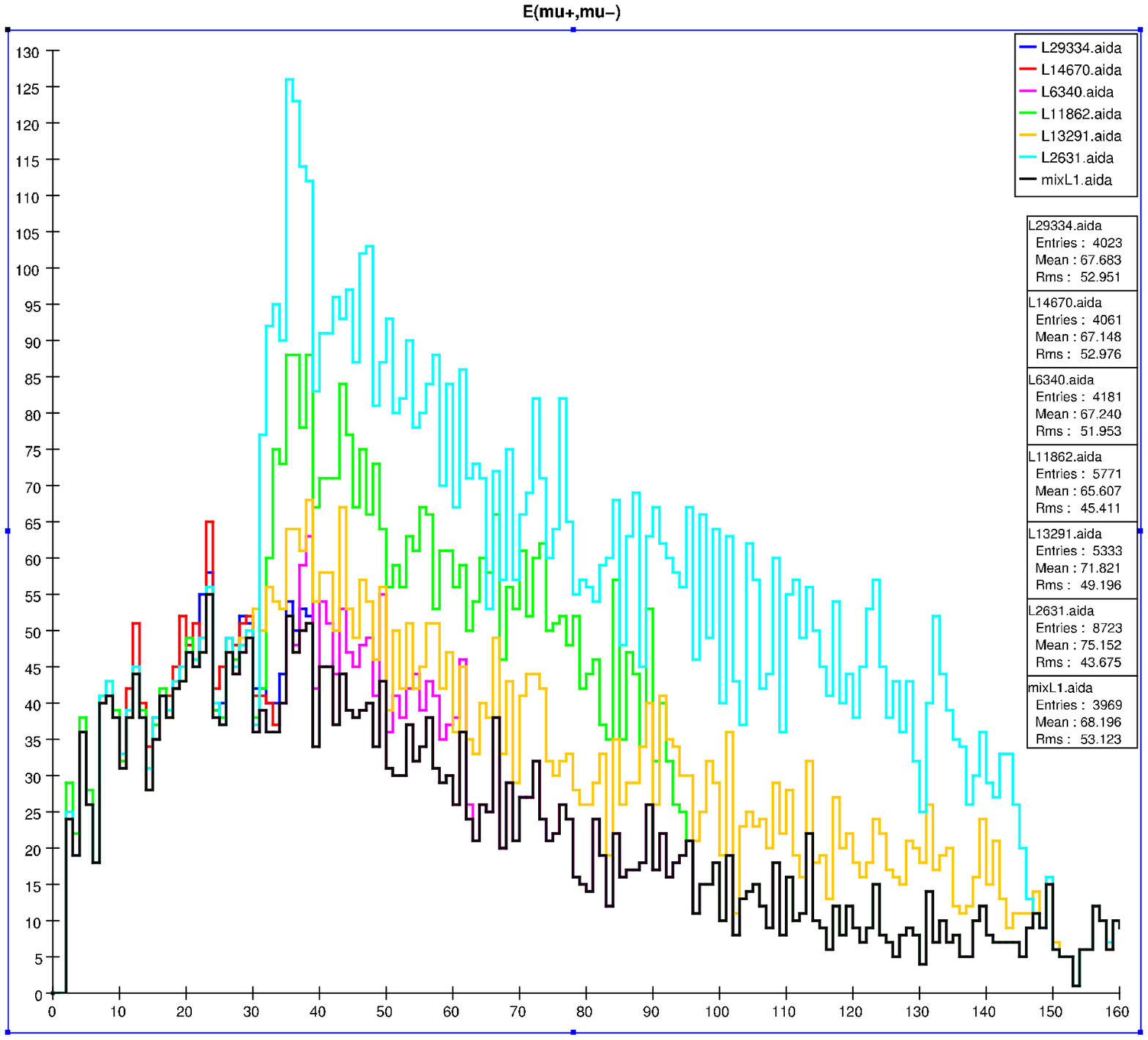}}
\vspace*{0.1cm}
\caption{Muon energy distribution: the number of events/2 GeV bin
(combined signal and background) after
imposing the cuts described in the text for several representative
models, with RH(LH) beam polarization in the top(bottom) panel, assuming
an integrated luminosity of 250 fb$^{-1}$
for either polarization. The SM background is represented by
the black histogram.}
\label{smuonfig1}
\end{figure*}
\begin{figure*}[htbp]
\centerline{
\includegraphics[height=10cm,width=13cm,angle=0]{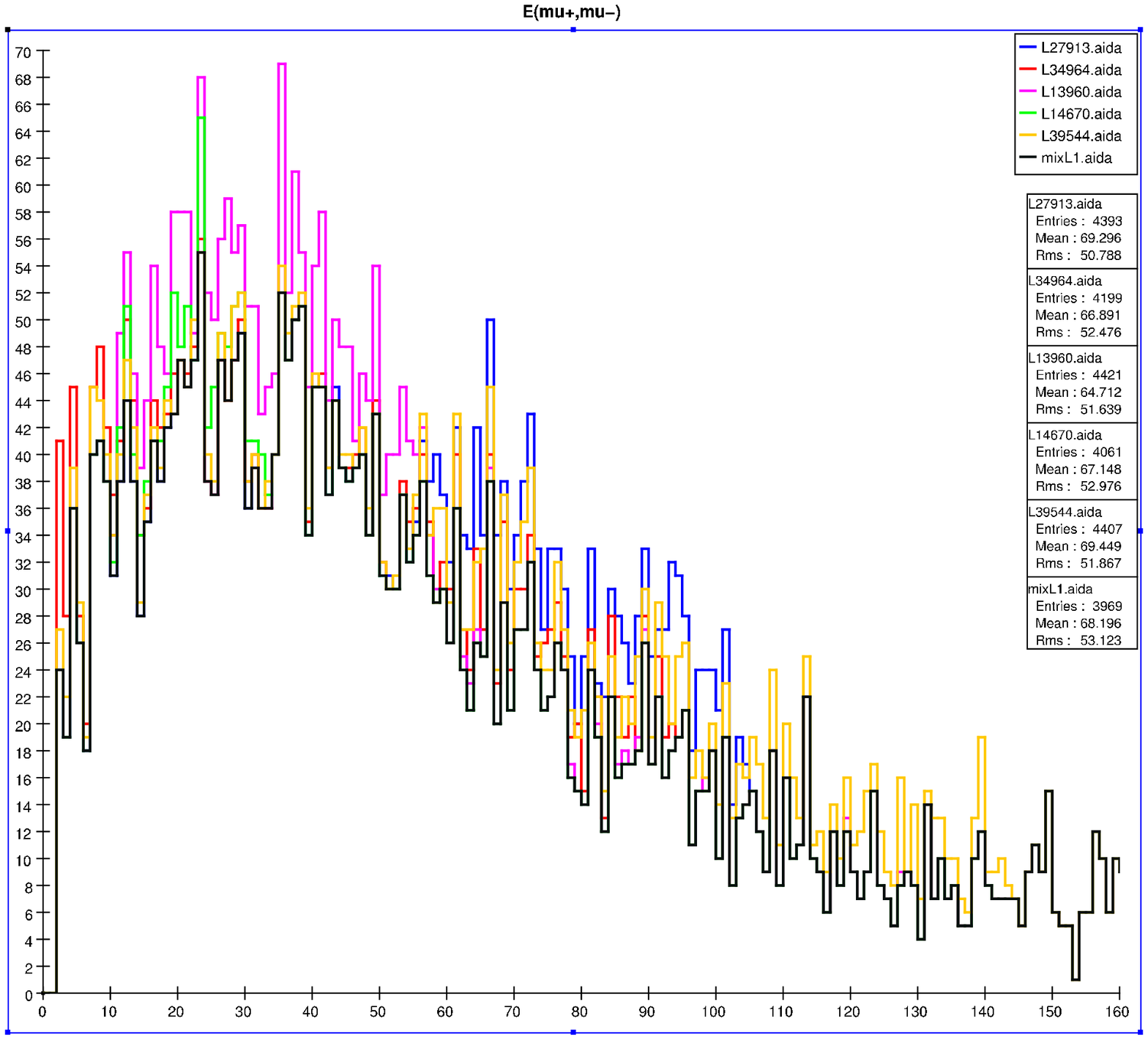}}
\vspace*{0.1cm}
\caption{Same as the previous Figure but now showing AKTW models that
result in a small
$S/B$ ratio in the LH polarization channel in the
$\smu$-pair analysis.}
\label{smuonfig2}
\end{figure*}

It is again interesting to compare the AKTW models that have visible
smuons at the ILC with the
well studied case of SPS1a'. Fig.~\ref{smuons_sps1a} shows the
muon energy spectrum we obtain after imposing our kinematic cuts
in the case of SPS1a' for both beam polarizations.
As in our selectron analysis, we observe that the event yield for
SPS1a' is far larger than all
the AKTW models we study here, in some cases by as much as a
factor of order 50.
Also, as in the previous analysis, two distinct
shelves are observed since both
$\tilde \mu_{L,R}$ are being simultaneously produced. The muon energy
distribution is slightly different from that obtained for electrons
in this model, not only because of the small differences in our cuts,
but also due to the fact that
the mixed final state $\tilde \mu_L \tilde \mu_R$ is not produced
due to the absence of the $t$-channel contribution. Clearly, in
comparison to the bulk of our models, it is rather trivial to
discover and make precise determinations of the smuon properties in
SPS1a'.

\begin{figure*}[htpb]
\centerline{
\includegraphics[height=10cm,width=13cm,angle=0]{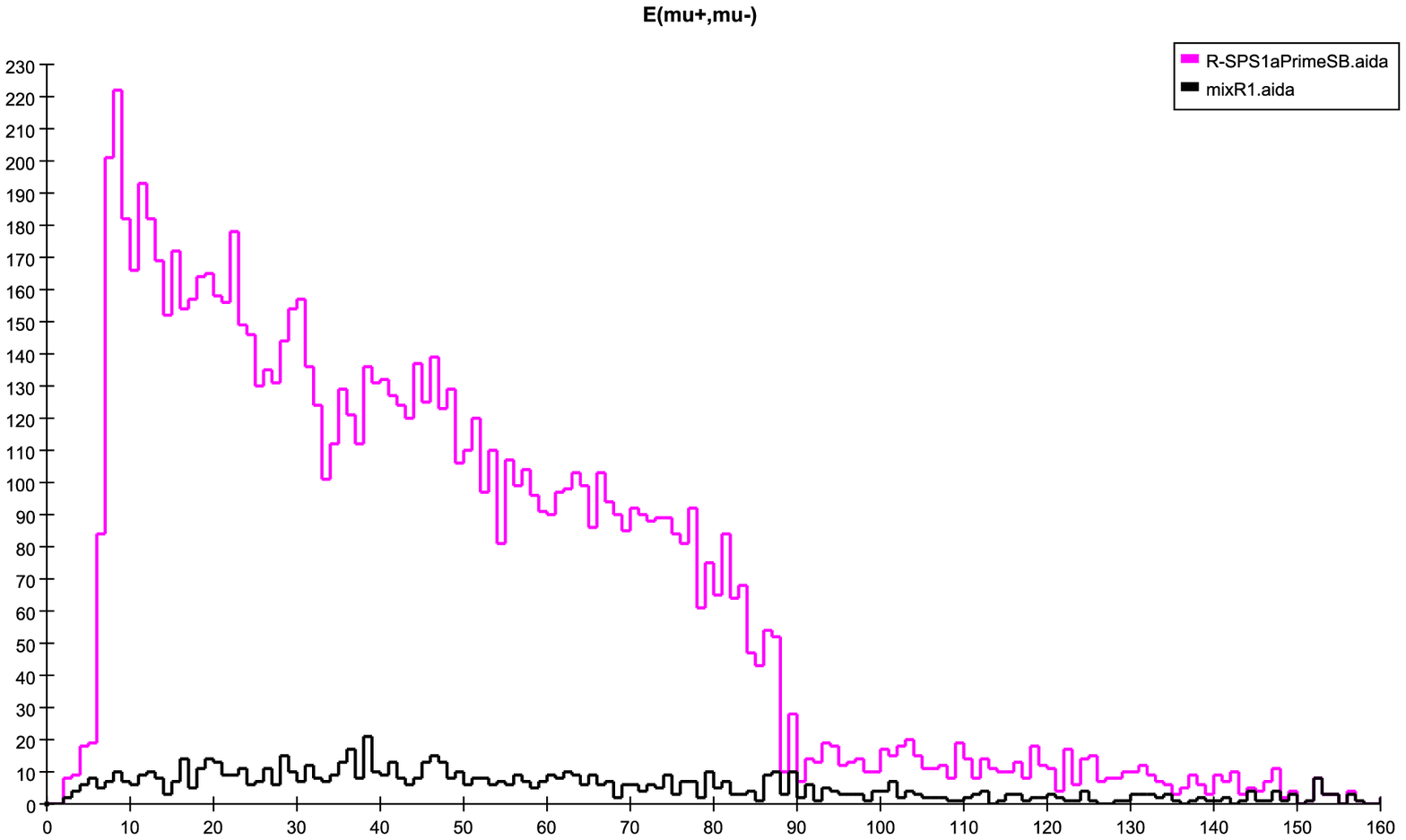}}
\vspace*{0.1cm}
\centerline{
\includegraphics[height=10cm,width=13cm,angle=00]{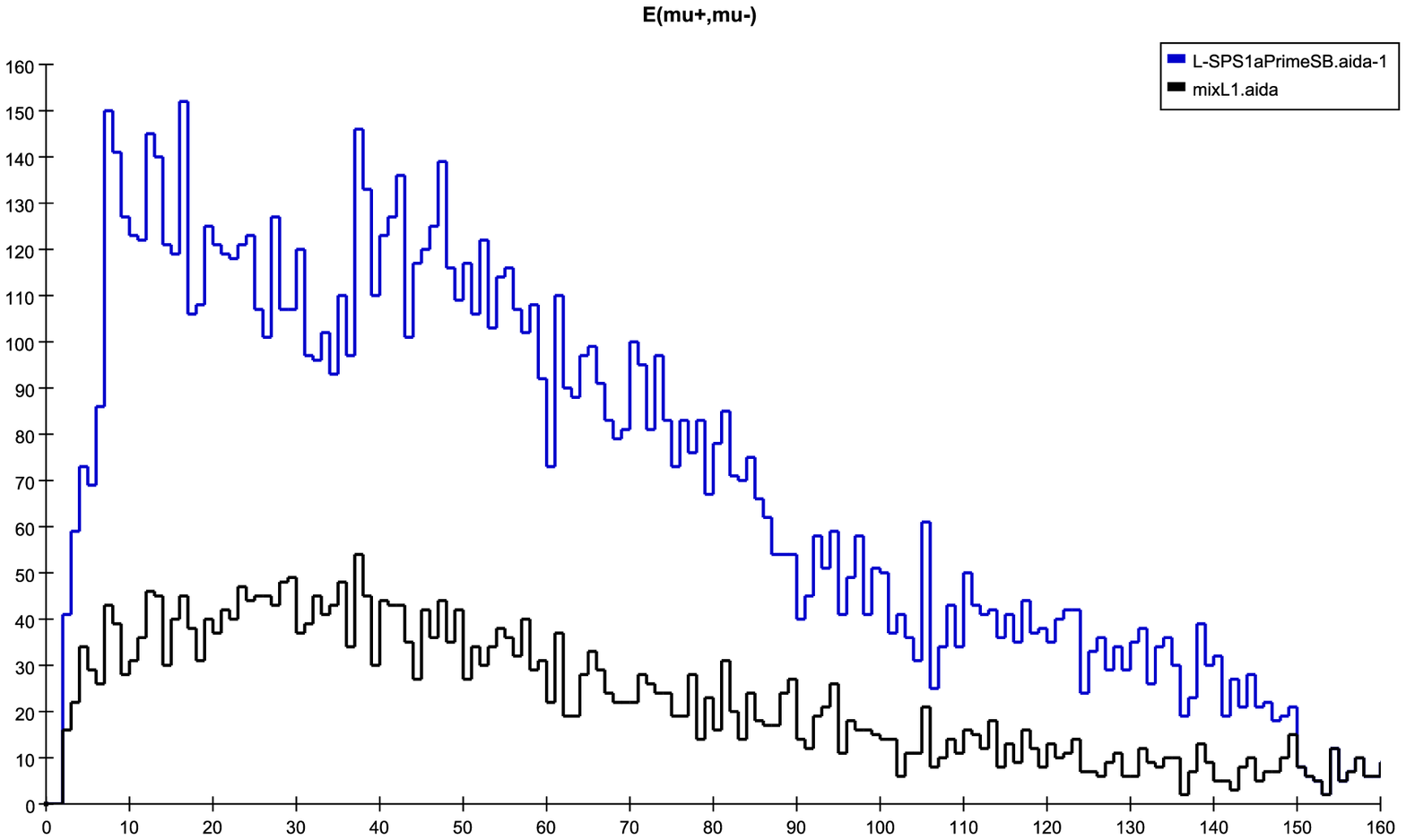}}
\vspace*{0.1cm}
\caption{Muon energy distribution: the number of events/2 GeV bin
(combined signal and background) after
imposing the cuts described in the text for the benchmark model SPS1a',
with RH(LH) beam polarization in the top(bottom) panel, assuming
an integrated luminosity of 250 fb$^{-1}$
for either polarization. The SM background is represented by
the black histogram.}
\label{smuons_sps1a}
\end{figure*}

Note that 4 of the models with kinematically accessible smuons
{\it also} have kinematically accessible lightest
chargino states. However, since all of these charginos are rather
close in mass to the LSP, \ie,
within 5 GeV, the existence of the charginos does not constitute
a large additional source of background and does not significantly
affect the qualitative structure of the muon energy spectra.
They could, however, modify the
extracted values of the particle masses obtained from an analysis of the
endpoints of the muon energy spectra and this possibility should be
studied further.

Interestingly, as in the selectron case above, a number of models
which {\it do not} have kinematically accessible smuons give rise to
visible signals in the $\smu$-pair analysis. This is just another
example of the well-known phenomenon where SUSY
is a background to itself. We find that there are 20(15) models which
yield fake signals in the case of RH(LH) polarization.
As in the previous analysis, decays of other SUSY
particles into muons, \eg, $\tilde\chi_1^+\to W^*\chi_1^0$ with $W^*\to
\mu \nu$,
can lead to the same observable final
state in both polarization channels. We present examples of such
misleading signals in Fig.~\ref{smuonfig3}. Note that in the case of
RH beam polarization, the fake signature looks quite different than in a
typical model which really has smuons present; there are
no shelf-like structures and the muon energies are relatively low.
This is to be expected when the final state muons are
the result of a 3-(or more) body decay channel, and when
chargino-neutralino mass splittings are small. However, for LH polarization,
two representative fake models (labelled 8324 and 39331 in the Figures)
appear to mimic the smuon shelf-like feature. This is due
to the fact that in these particular models, as will be
discussed further below in Section 5.1,
the $W$ boson in the $\tilde \chi_1^\pm$ decay
process is
on-shell so that the final state muons
are the result of true 2-body decays.

\begin{figure*}[htbp]
\centerline{
\includegraphics[width=13.0cm,height=10.0cm,angle=0]{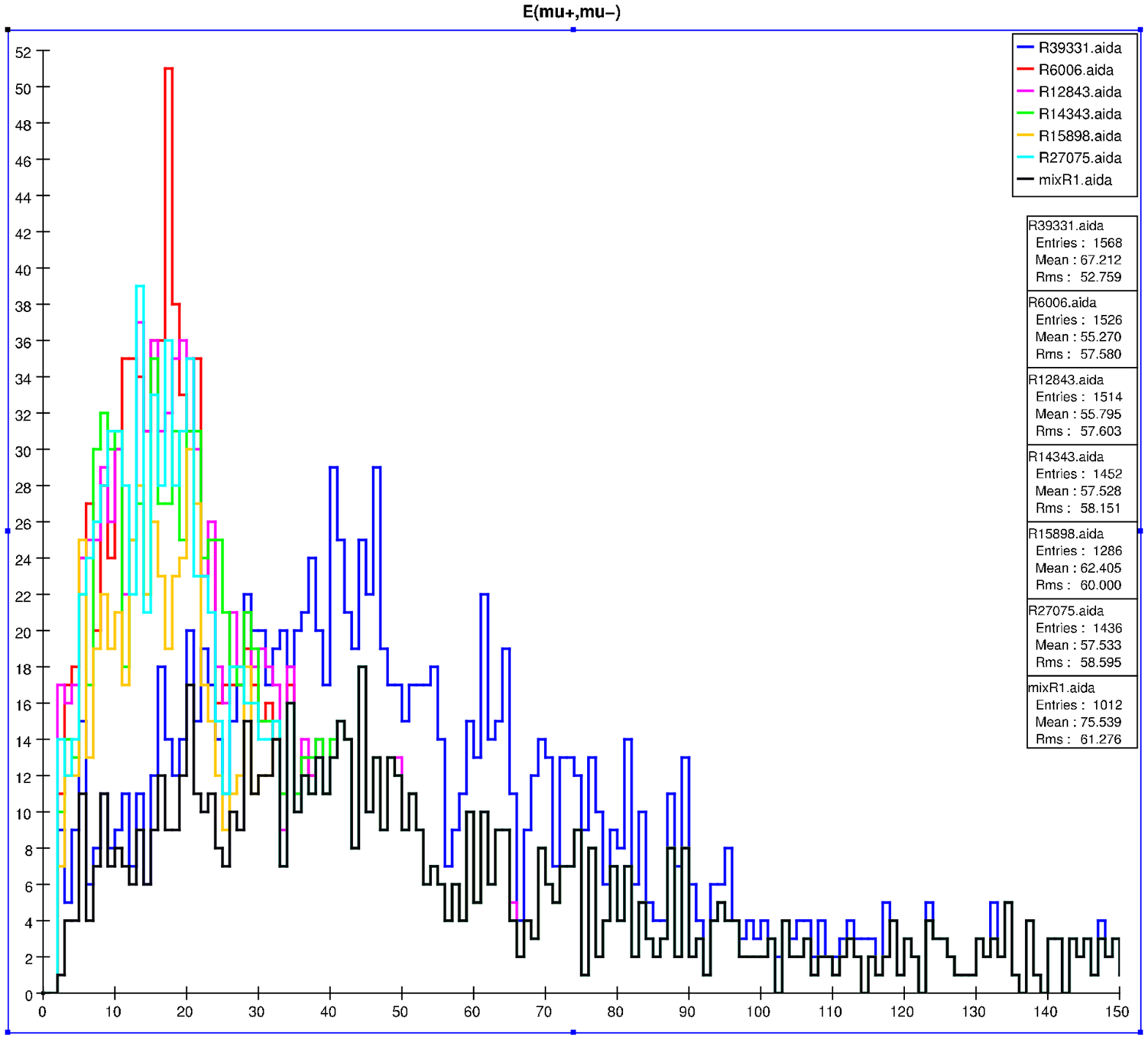}}
\vspace*{0.1cm}
\centerline{
\includegraphics[width=13.0cm,height=10.0cm,angle=0]{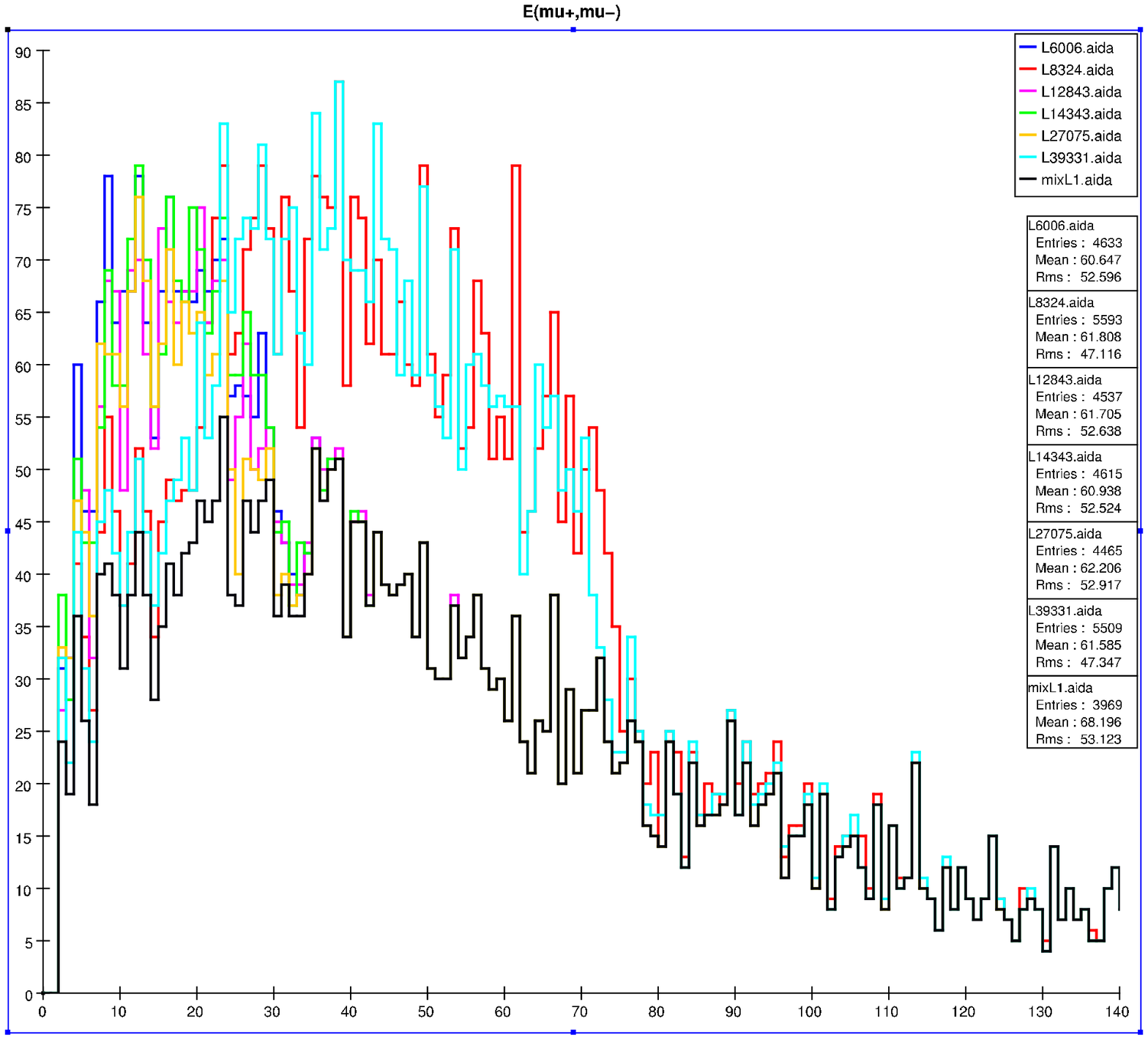}}
\vspace*{0.1cm}
\caption{Muon energy distribution: the number of events/2 GeV bin
(combined signal and background) after
imposing the cuts described in the text for representative models
which lead to fake smuon signatures from chargino and neutralino
decays. Here, we show
RH(LH) beam polarization in the top(bottom) panel, assuming
an integrated luminosity of 250 fb$^{-1}$
for either polarization. The SM background is represented by
the black histogram.}
\label{smuonfig3}
\end{figure*}

In order to assist in the differentiation of models with real smuons from
ones that do not, it is necessary to
examine other kinematic distributions. Figures \ref{smuonfig4}
and ~\ref{smuonfig5} show the $p_T^{vis}$ distributions
for the real smuon and fake models, respectively. Here we see that the
models with real smuons generally lead to
harder muons in the final state than do the counterfeit cases;
this holds
to some extent in the fake models where the
charginos decay to on-shell $W$ bosons.

\begin{figure*}[htbp]
\centerline{
\includegraphics[width=13.0cm,height=10.0cm,angle=0]{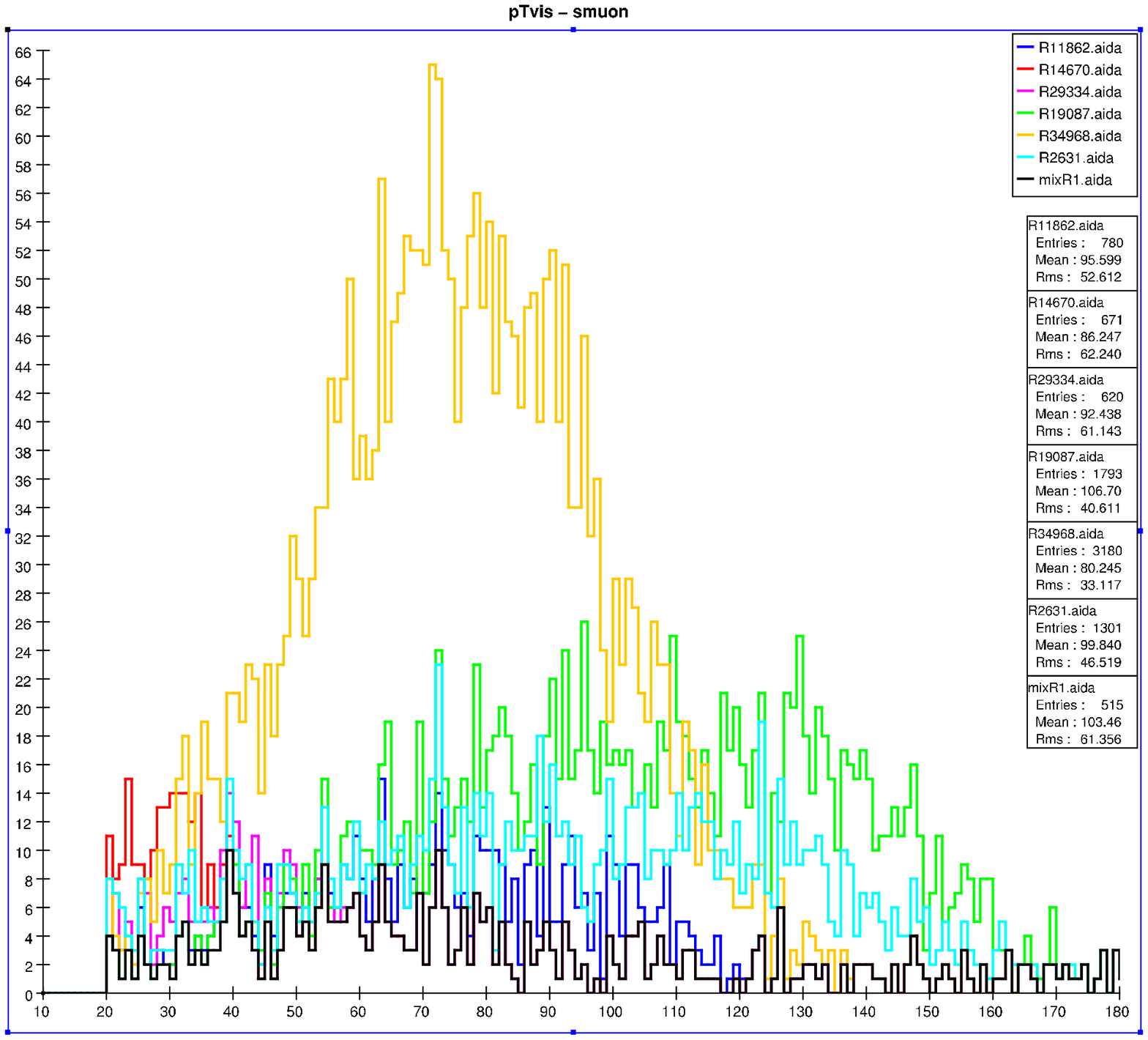}}
\vspace*{0.1cm}
\centerline{
\includegraphics[width=13.0cm,height=10.0cm,angle=0]{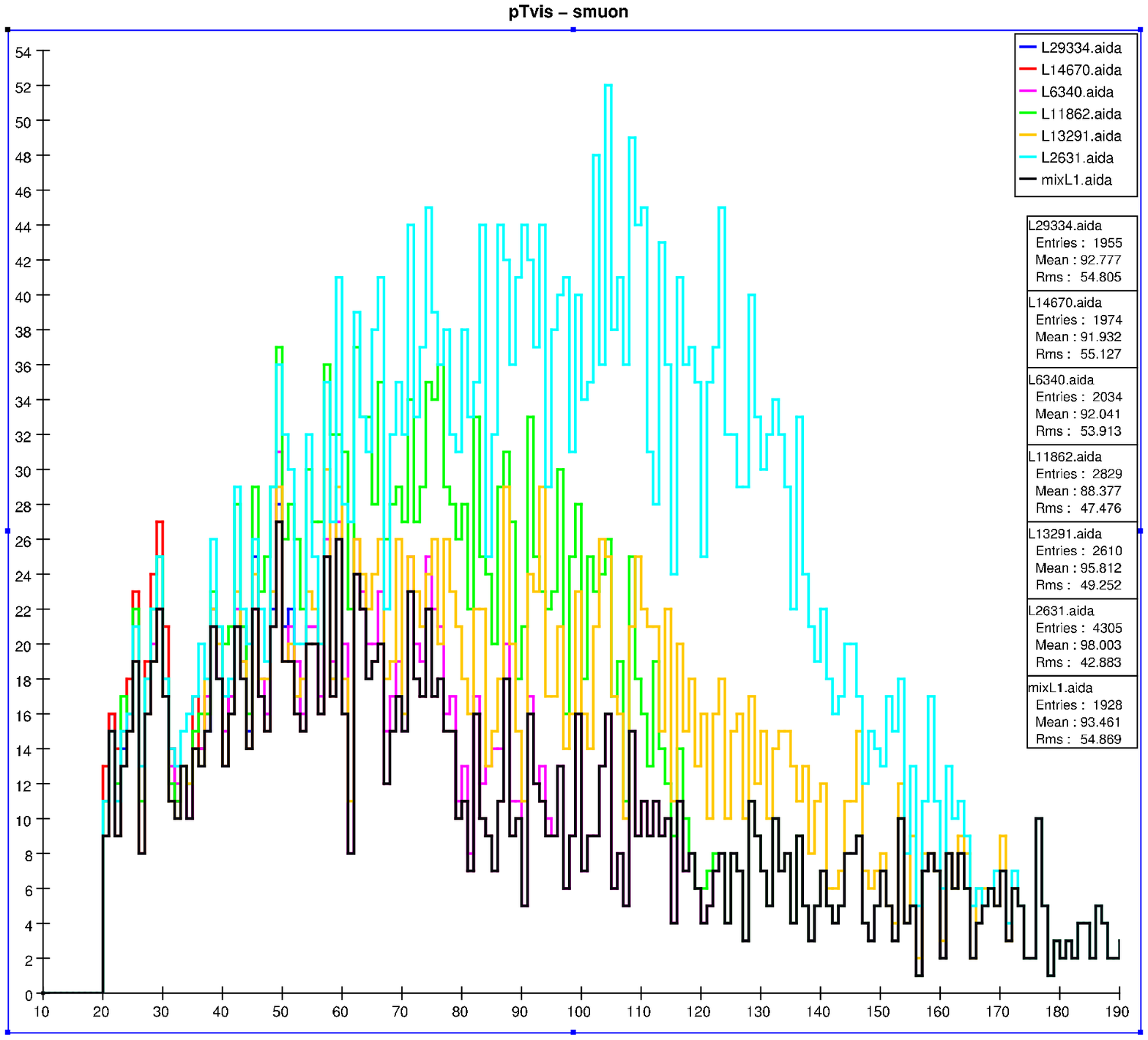}}
\vspace*{0.1cm}
\caption{The $p_T^{vis}$
distribution for the same models shown in Fig.~\ref{smuonfig1} with real
smuons. Here, we show
RH(LH) beam polarization in the top(bottom) panel, assuming
an integrated luminosity of 250 fb$^{-1}$
for either polarization. The SM background is represented by
the black histogram.}
\label{smuonfig4}
\end{figure*}
\begin{figure*}[htbp]
\centerline{
\includegraphics[width=13.0cm,height=10.0cm,angle=0]{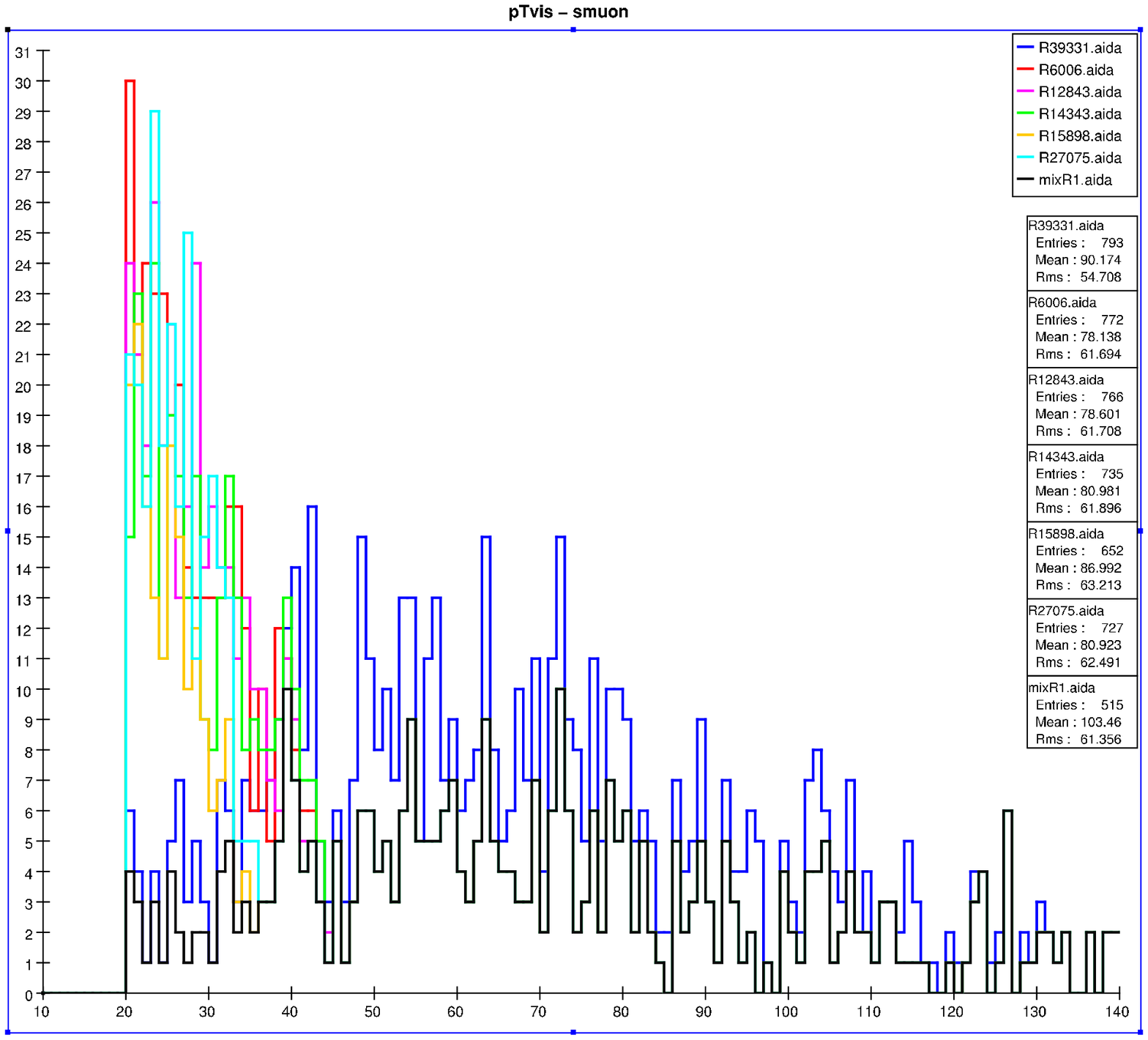}}
\vspace*{0.1cm}
\centerline{
\includegraphics[width=13.0cm,height=10.0cm,angle=0]{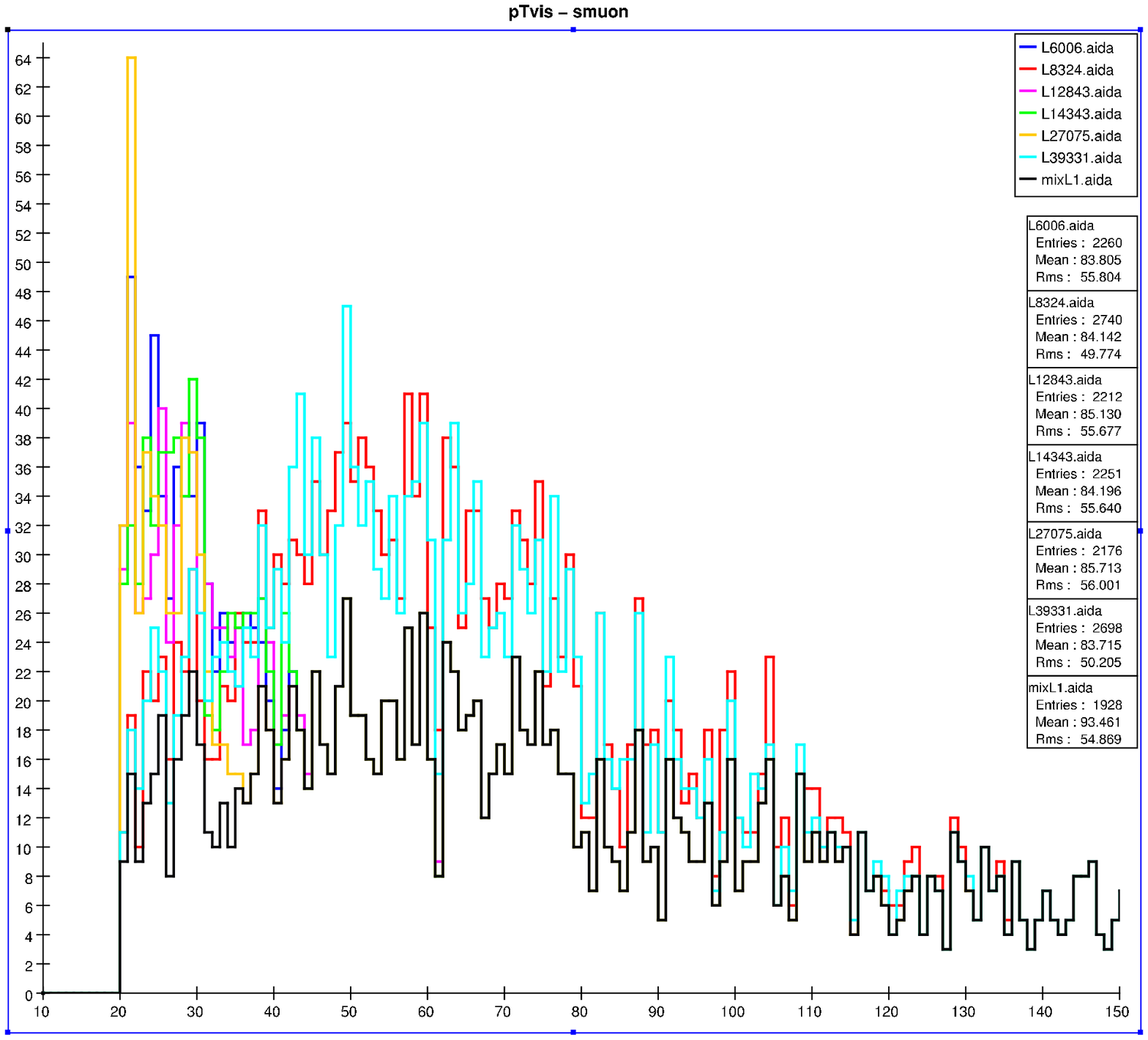}}
\vspace*{0.1cm}
\caption{The $p_T^{vis}$
distribution for the same models shown in Fig.~\ref{smuonfig3} with fake
smuons. Here, we show
RH(LH) beam polarization in the top(bottom) panel, assuming
an integrated luminosity of 250 fb$^{-1}$
for either polarization. The SM background is represented by
the black histogram.}
\label{smuonfig5}
\end{figure*}

\clearpage

\subsubsection{Staus}
\label{Sec:stau}

This analysis is similar to the other charged slepton analyses
discussed above. We analyze the channel
\be
\stau^+ \stau^- \rightarrow \tau^+ \tau^- \LSP \LSP \, ,
\ee
that is, the signature is a tau pair plus missing energy. Staus
are pair produced via $s$-channel $\gamma$ and $Z$ exchange and receive
no $t$-channel contribution. The left- and right-handed staus mix
to form two mass eigenstates, which have mixing-dependent couplings to
the $Z$ boson.
In contrast to the other charged slepton analyses above, the identification
of the final state tau leptons is nontrivial,
because the tau decays in the detector, predominantly into
hadrons.

We focus on the hadronic decays of taus into pions,
$\tau \rightarrow \pi \nu_\tau; \,
\tau \rightarrow \rho \nu_\tau \rightarrow \pi^{\pm} \pi^0 \nu_\tau; \,
\tau \rightarrow 3 \pi \nu_\tau$, the latter being a 3-prong jet, but
also include the leptonic decays of the $\tau$.
In the hadronic decay channel,
taus are identified as jets with a charged multiplicity of 1 or 3, and
with invariant mass less than some maximum value. Our tau selection
criteria are as follows \cite{TimNorm,Martyn:2004jc} (note that we
employ the notation tau-jet to describe the visible $\tau$ decay
products):
\begin{enumerate}

\item We require 2 jets in the event, each with charged multiplicity of 1
(where the tau decays into a lepton, $\rho$, $\pi$,
or $3 \pi$-decay with $2 \pi^0$s)
or 3 (where the tau decays into 3 charged pions).

\item The invariant mass of tau-jet, \ie, the visible tau decay
products, must be $<$ 1.8 GeV.

\item If the tau-jet is 3-prong (charged multiplicity of 3), then
none of the charged particles should be an electron or muon.

\item If both tau-jets in the event are 1-prong, then we reject events
where both jets are same flavor leptons, that is,
an electron-positron or a muon pair. However we keep pairs of
tau-jets that are, for example,
an electron and a muon, or an electron and a pion,
whereby a pion is defined as a charged track that is not identified
as an electron or a muon.
\end{enumerate}

As an \emph{alternative} analysis, we follow the above criteria and
allow leptonic tau decays into muons, but reject
taus that decay into electrons. This reduces contamination
from photon-induced backgrounds.

As mentioned above in our description of the SiD detector in
Section~\ref{Sec:detectorandanalysis}, the current detector design does
not allow for tracking, and hence does not have the capability for
particle ID, below $142$ mrad. Thus
muons at low angles are completely missed if they are too energetic to
deposit energy into clusters. As we will see,
certain $\gamma$-induced processes
constitute a significant background to stau production, particularly
in the case where such energetic muons are produced but not reconstructed
and the beam electron (or positron) receives a sufficient transverse
kick to be detected. In this case, the final visible state is an electron
and a muon, which would pass the standard tau ID preselection described
above. The alternative tau
preselection criteria, which rejects the electron decay channel,
eliminates this background at the price
of reducing the signal correspondingly by roughly 30\%.

After these tau identification criteria are imposed, we employ cuts to
reduce the SM background. Following~\cite{Martyn:2004jc}, we demand:
\begin{enumerate}

\item No electromagnetic energy (or clusters) in the
region $|\cos\theta| > 0.995$.

\item Two tau candidates as identified above,
are weighted by their charge within the polar angle $-0.75 < Q_\tau
\cos \theta_\tau < 0.75$.
This reduces the $W$-pair background.

\item The acoplanarity angle must satisfy $\Delta \phi^{\tau\tau} > 40$ degrees.
Here, since we demand two tau candidates, the acoplanarity angle
is equivalent to $\pi$ minus
the angle between the $p_T$ of the taus, $\Delta \phi^{\tau \tau} = \pi -
\theta_T$. The above requirement then
translates to $\cos \theta_T > 0.94$.
This cut reduces the $W$-pair and $\gamma\gamma$-induced background.

\item $|\cos \theta_{p_{\mbox\tiny missing}}| < 0.8$.

\item The transverse momentum of the ditau system be in the range
$0.008 \sqrt{s} <
p_T^{\tau \tau} < 0.05 \sqrt{s}$.
This decreases the $\gamma\gamma$-induced background.

\item The transverse momentum of each of the tau candidates be $p_T >
0.001 \sqrt{s}$.
This cut is crucial to reduce the $\gamma\gamma$ and $e \gamma$ background.

\item The combined cut on $\sum p_{\perp, \vec{T}}^\tau$ and
$\Delta \phi^{\tau \tau}$,
\begin{eqnarray}
\sum p_{\perp, \vec{T}}^\tau & < & 0.00125 \sqrt{s}
\left(1 + 5\, \sin \Delta \phi^{\tau \tau} \right)\, \nonumber \\
& & = 0.00125 \sqrt{s} \left(1 + 5\, \sqrt{ 1 - \cos^2
\theta_T^{\tau\tau}} \right)
\end{eqnarray}
is imposed. Here, $\sum p_{\perp, \vec{T}}^\tau$ is the sum of the tau
momenta projected onto the
transverse thrust axis $\vec{T}_\perp$, where the
transverse thrust axis is given
by the $xy$-components of the thrust axis.
This last cut is necessary in the tau decay channel to further
decrease the $\gamma\gamma$ background.
\end{enumerate}

As in the other slepton analyses, we histogram the resulting $\tau^\pm$
energy spectrum as well as $p_{T \mbox{\tiny vis}}$ in this case. We show the remaining
SM backgrounds after these cuts are imposed
in Fig.~\ref{fig:staubg}; the dominant background left after the cuts
stems from $\gamma$-induced lepton-pair production processes.

In Fig.~\ref{fig:staubg},
we also display the effect of the alternative tau ID criteria discussed
above. This alternative technique
nearly completely eliminates the background since events where
a beam electron/positron is falsely identified as a tau decay
product are rejected.
Of course, as mentioned above this technique also reduces the signal by
approximately 30\%. Augmenting the detector with muon ID capabilities at
lower angles could reduce the $\gamma$-induced background without having to
pay the
price of introducing a restricted tau identification. As we will see below,
a significant portion of the AKTW models have
very low stau signal rates, and an improved tracking
capability could be crucial if in fact this portion of the
SUSY parameter space is realized in nature.

\begin{figure*}[hptb]
\centerline{
\epsfig{figure=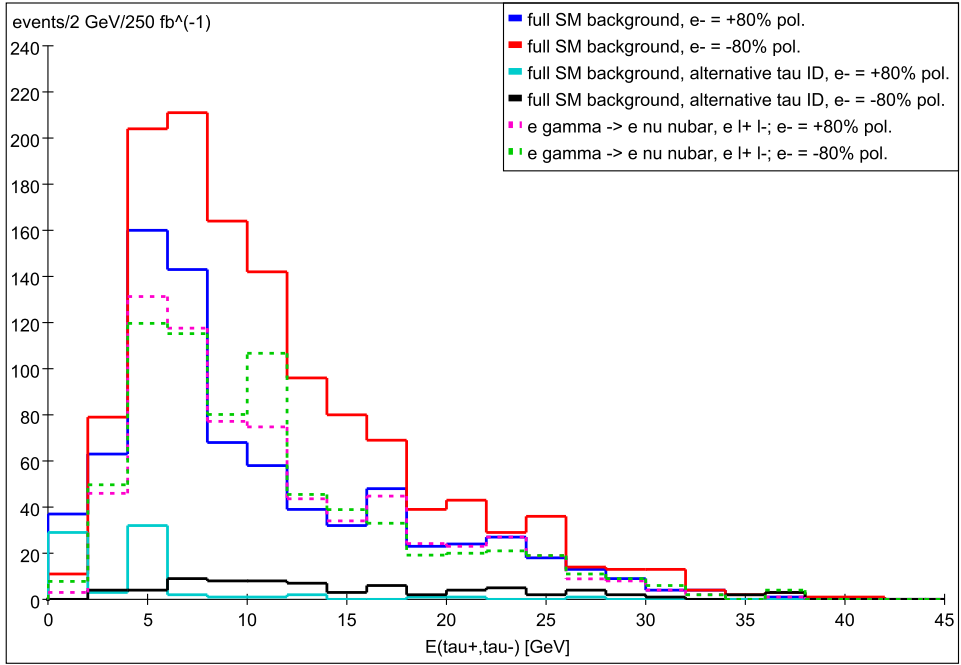,width=13cm,clip=}}
\vspace*{0.1cm}
\caption{The $\tau^\pm$ energy spectrum of the SM background after the
tau identification and stau selection cuts have been imposed.
The solid blue and
red lines are the full SM background for 80\% right- (blue) and left-handed
(red) electron polarization, the
dashed lines (pink, right-handed, green left-handed) represent
the dominant background source, $e^\pm \gamma
\rightarrow e^\pm \nu \bar{\nu}, e^\pm l^+ l^-,\, l = e,\mu,\tau$.
Furthermore, the effect of eliminating the
misidentification of beam electrons as tau decay products
via the alternative tau ID described in the text is represented by the
solid cyan line (80\% right-handed beam polarization) and solid black line
(80\% left-handed polarization).}
\label{fig:staubg}
\end{figure*}

In 28 of the 242 AKTW models, the lightest stau is kinematically
accessible for pair
production at the 500 GeV ILC, and in one of these models the heavier
stau partner can also be produced. The signal for stau production is somewhat
different than in the case for selectrons and smuons as the final
state tau decays in the detector. In this case, we no longer have the
distinctive shelf-like feature in the resulting energy spectrum of
the reconstructed tau. The shape of these spectra is highly dependent
on the mass difference between the stau and the LSP as shown in
Fig.~\ref{staudist}, which displays the pure signal in 3 models before
our selection cuts have been imposed.
Here we see that small mass differences result
in a sharply peaked distribution at low tau energies, while a larger
mass difference yields a flatter distribution, albeit at a lower event rate.
In principle, the search strategy, and hence the set of selection cuts,
could be tailored to maximize the signal to background ratio once
the stau$-$LSP mass difference is known.
However, until the stau is discovered a general search strategy, such
as that presented here, that applies for all mass regions must be
employed.

\begin{figure*}[hptb]
\centerline{
\psfig{figure=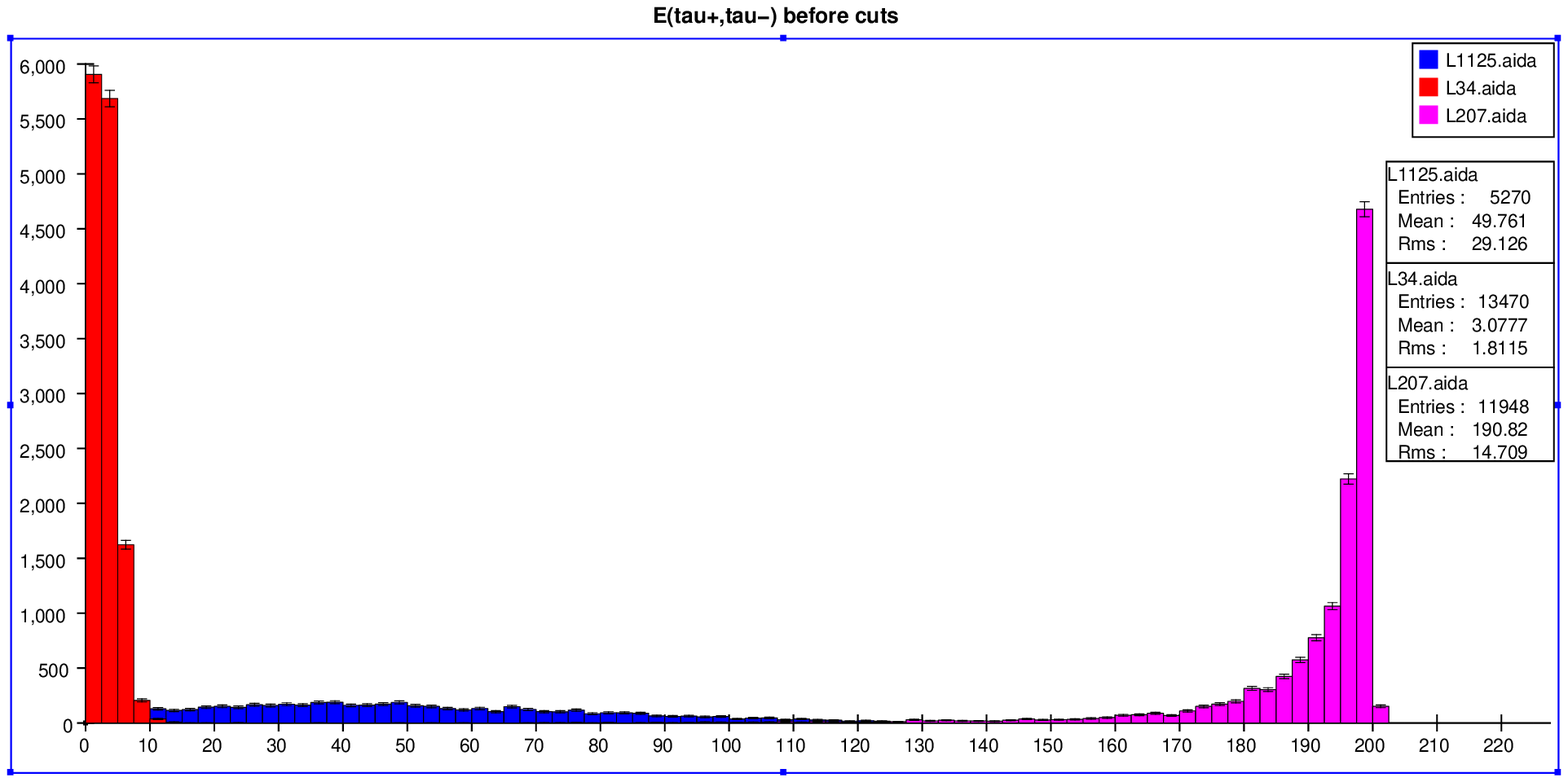,angle=0,width=13cm,clip=}}
\vspace*{0.1cm}
\caption{Tau energy distribution stemming from stau pair production
before our set of kinematic cuts have been imposed
in three AKTW models. The red, blue, magenta histograms correspond to a
stau-$\tilde \chi_1^0$ mass difference of order 25, 100, and 0.5 GeV,
respectively.}
\label{staudist}
\end{figure*}

Of these 28 models, we find that 18 lead to
signals which can be observed at the significance level ${\cal S}>5$.
We also find that the heavier stau with $m_{\tilde\tau_2}=240$ GeV is
not produced with large enough event rates to be visible above the SM
background.
In addition, in 3 of these 28 models the mass difference between the
lightest stau and the LSP is small enough such that the stau decays
outside the detector, and
it can be observed in our stable charged particle search (described
below in Section 5.3). Of the 18 detectable stau models, 9(10) are
visible via our
standard search criteria in the LH(RH) beam polarization configuration.
The total (combining signal and background) $\tau^\pm$ energy distribution
is shown for a representative set of these models
for both beam polarizations in Fig.~\ref{stausearch1}. Here, we see that
some of these models cleanly rise above the SM background, while others
are just barely visible.
The number of detectable stau signals is greatly increased when we
apply our alternative set of preselection criteria discussed above. One finds that
17(12) models are observable with LH(RH) electron beam polarization; the
tau energy spectrum for a sampling of these models is displayed in
Fig.~\ref{stausearch2}.
In this case, we see that the signal is cleanly visible above the
background for all models.

In all of the 18 models with observable staus, we find that the number of
stau events that pass our cuts is dramatically reduced compared to
the event rate in Fig.~\ref{staudist} before the cuts were applied.
In addition, due to our cuts, the tau energy
distribution is always peaked at low values, regardless of the
stau-LSP mass difference, which ranges from $7-94$ GeV in these 18 models.
In the 7 models where the signal is not observable, 2 are phase-space
suppressed with $m_{\tilde\tau_1}>240$ GeV. The remaining 5 models
all have reasonable $\tilde\tau_1$ masses and the $\tilde\tau_1 -
\tilde\chi_1^0$ mass difference ranges from $42-108$ GeV, but
nonetheless have a small production cross section due to stau
mixing.

\begin{figure*}[htpb]
\centerline{
\includegraphics[height=10cm,width=13cm,angle=0]{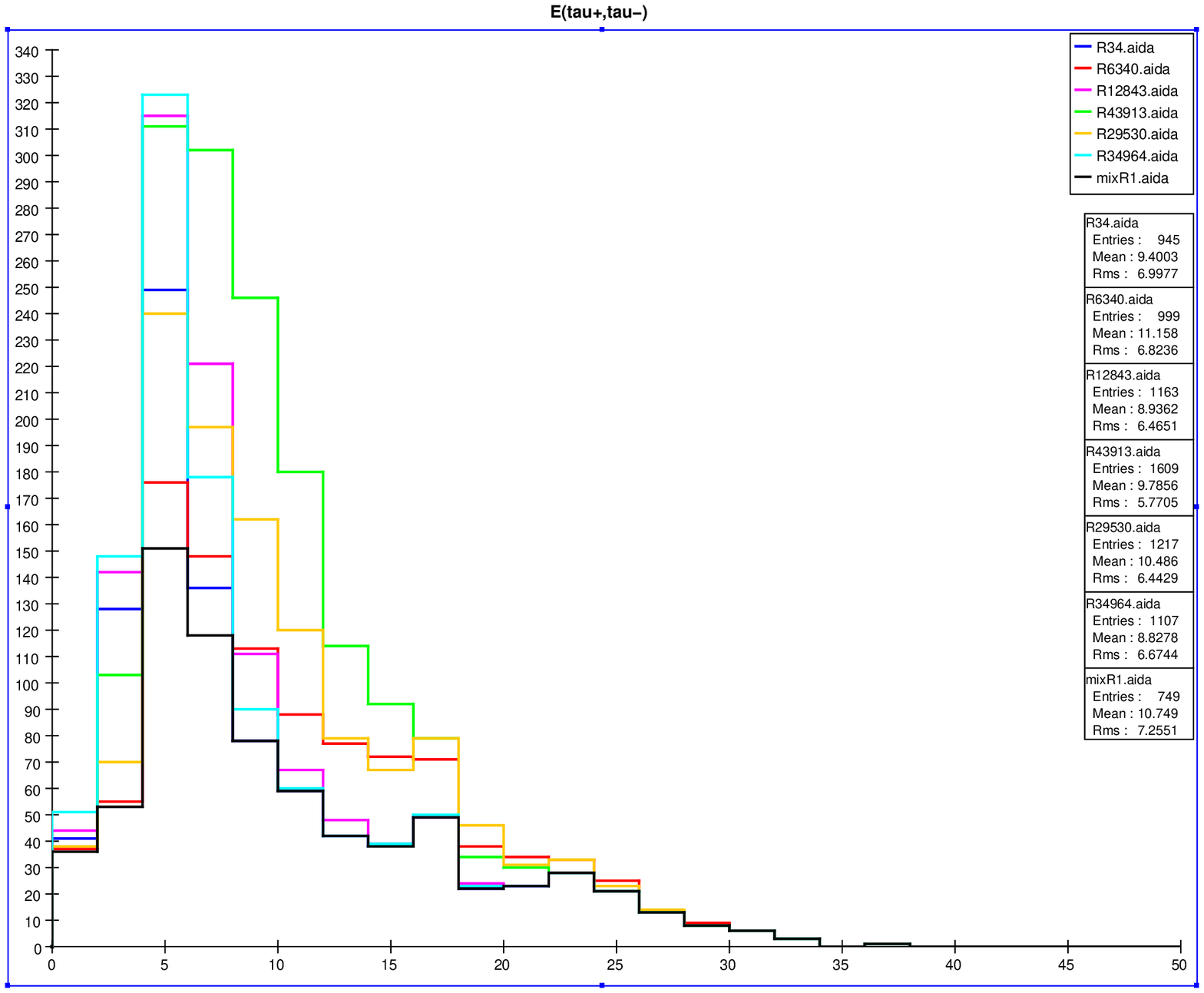}}
\vspace*{0.1cm}
\centerline{
\includegraphics[height=10cm,width=13cm,angle=0]{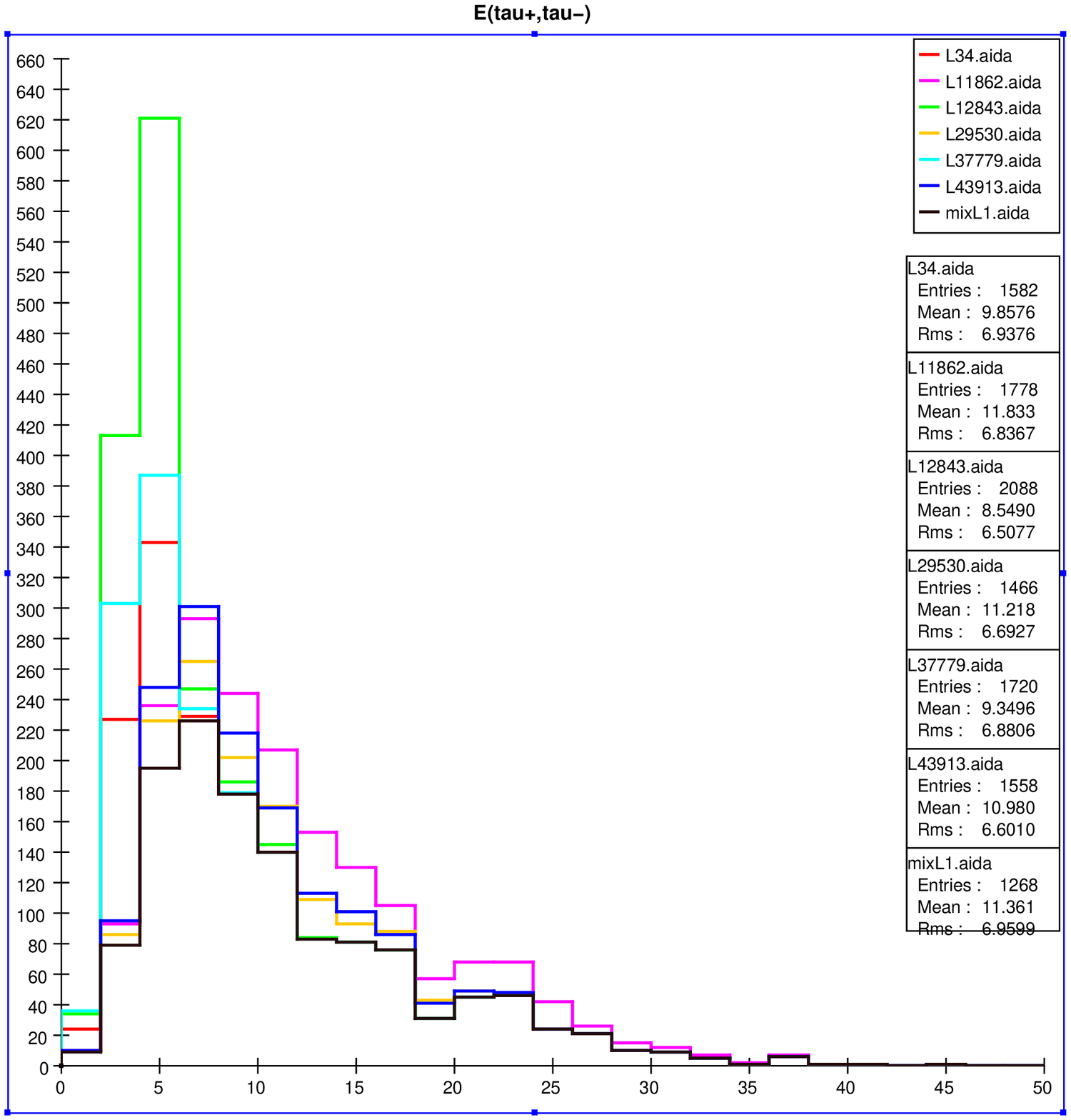}}
\vspace*{0.1cm}
\caption{Tau energy distribution: the number of events/2 GeV bin after
imposing the standard set of tau ID and $\tilde\tau$ selection criteria
described in the text
for several representative AKTW models (signal and background combined).
RH(LH) beam polarization is
employed in the top(bottom) panel, assuming an integrated luminosity of
250 fb$^{-1}$ for each polarization configuration. The SM background
corresponds to the black histogram.}
\label{stausearch1}
\end{figure*}

\begin{figure*}[htpb]
\centerline{
\includegraphics[height=10cm,width=13cm,angle=0]{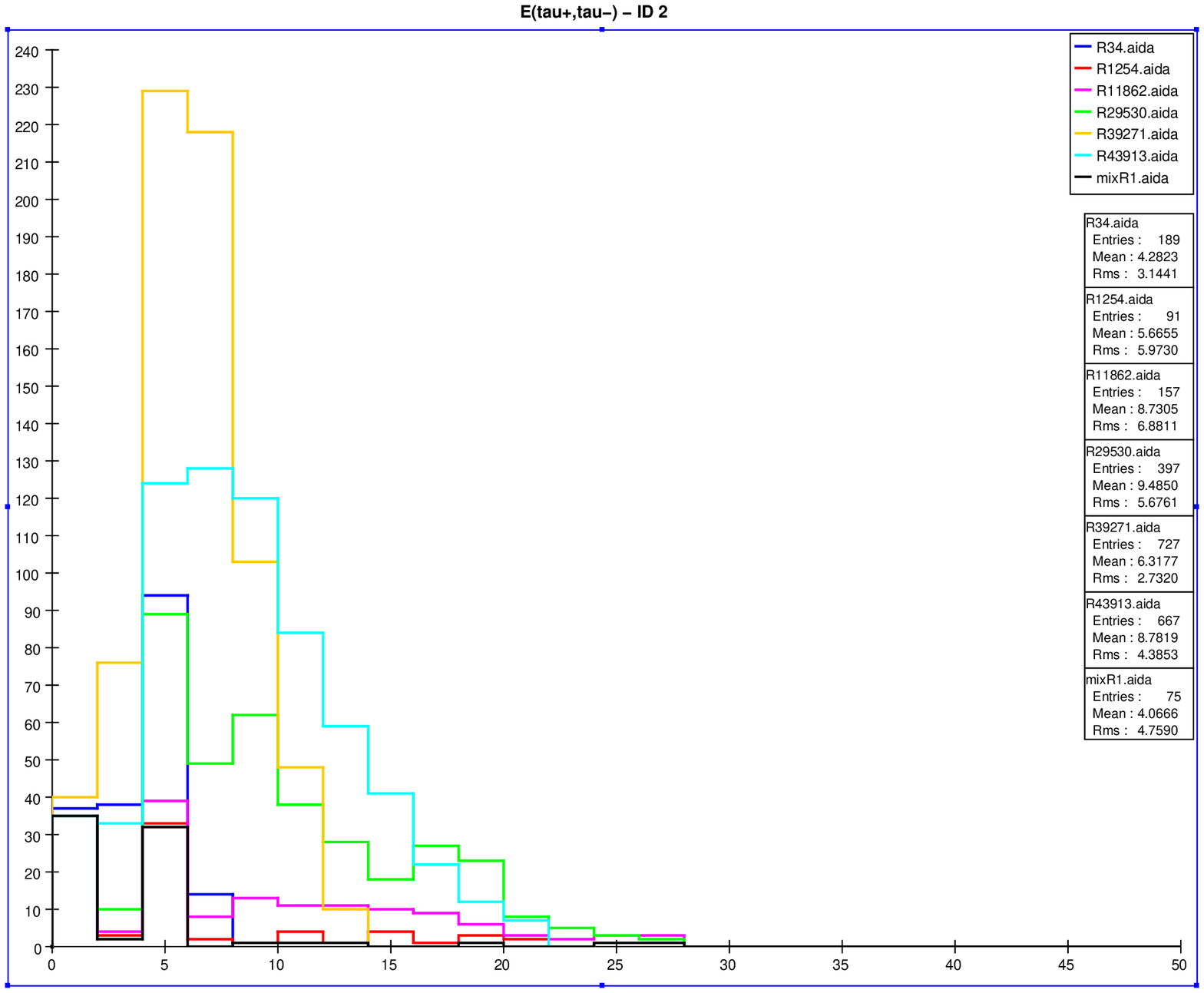}}
\vspace*{0.1cm}
\centerline{
\includegraphics[height=10cm,width=13cm,angle=0]{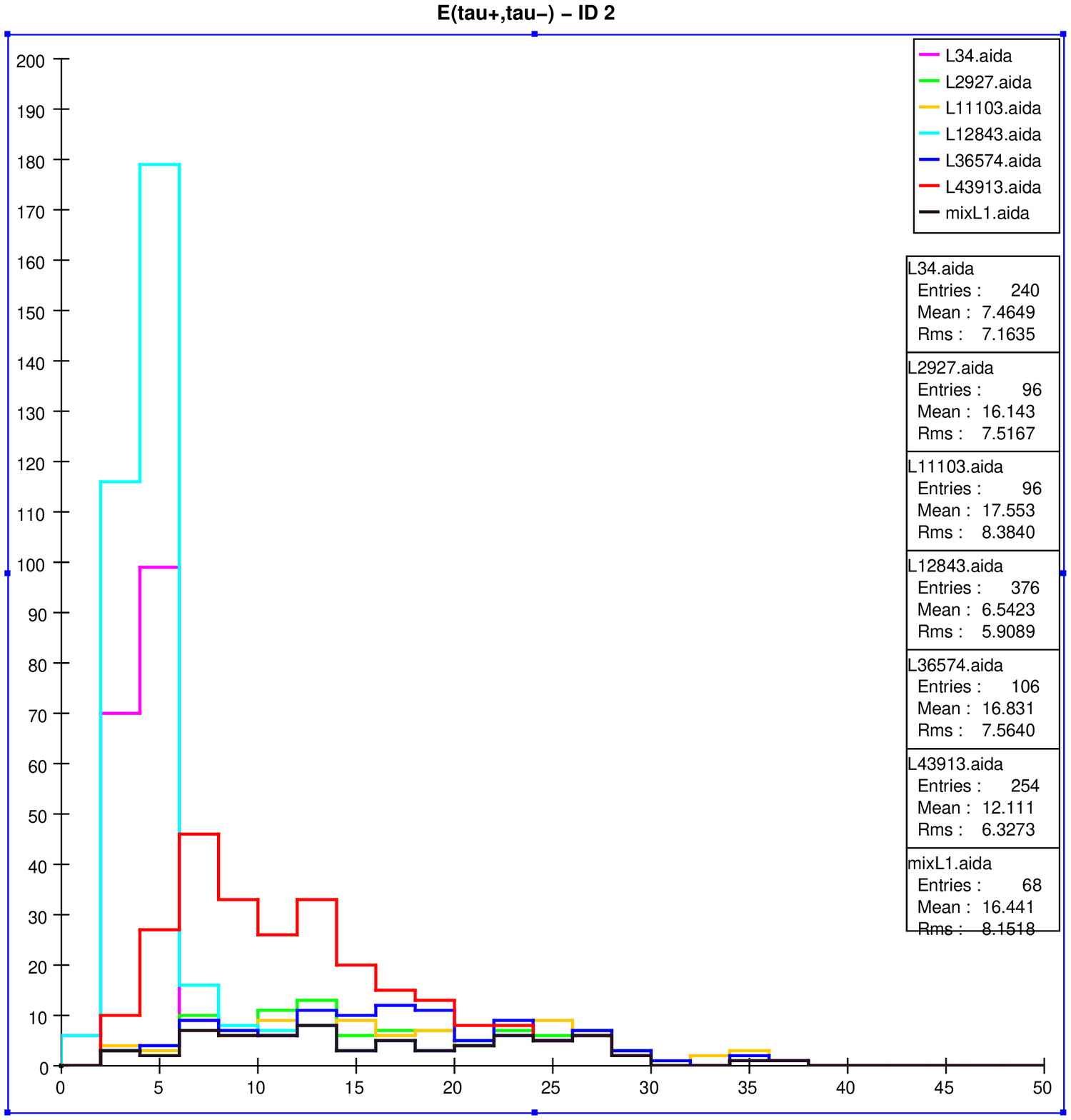}}
\vspace*{0.1cm}
\caption{Tau energy distribution: the number of events/2 GeV bin after
imposing the alternative set of tau ID and $\tilde\tau$ selection criteria
described in the text
for several representative AKTW models (signal and background combined).
RH(LH) beam polarization is
employed in the top(bottom) panel, assuming an integrated luminosity of
250 fb$^{-1}$ for each polarization configuration. The SM background
corresponds to the black histogram.}
\label{stausearch2}
\end{figure*}

Unlike many of the AKTW models we are examining, stau production at the
ILC is straightforwardly observable in the
case of the benchmark model SPS1a'. This holds in either of the
analysis channels as can
be observed in Figs.~\ref{staus_sps1a} and
~\ref{staus_ID2_sps1a}. Here we see that the stau signal is
quite substantial and can be cleanly
observed over the SM background for both choices of the
electron beam polarization.

\begin{figure*}[htpb]
\centerline{
\includegraphics[height=10cm,width=13cm,angle=0]{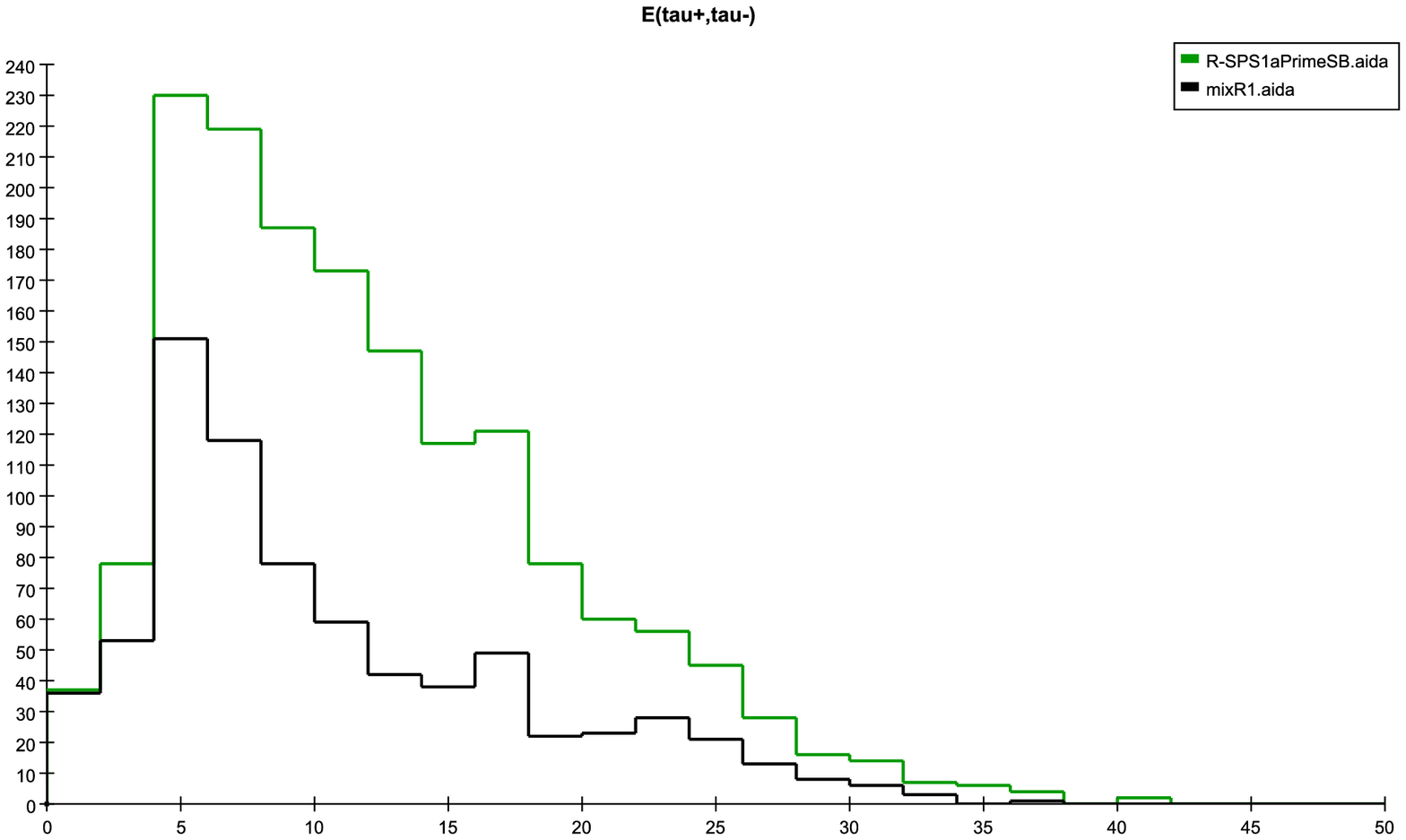}}
\vspace*{0.1cm}
\centerline{
\includegraphics[height=10cm,width=13cm,angle=0]{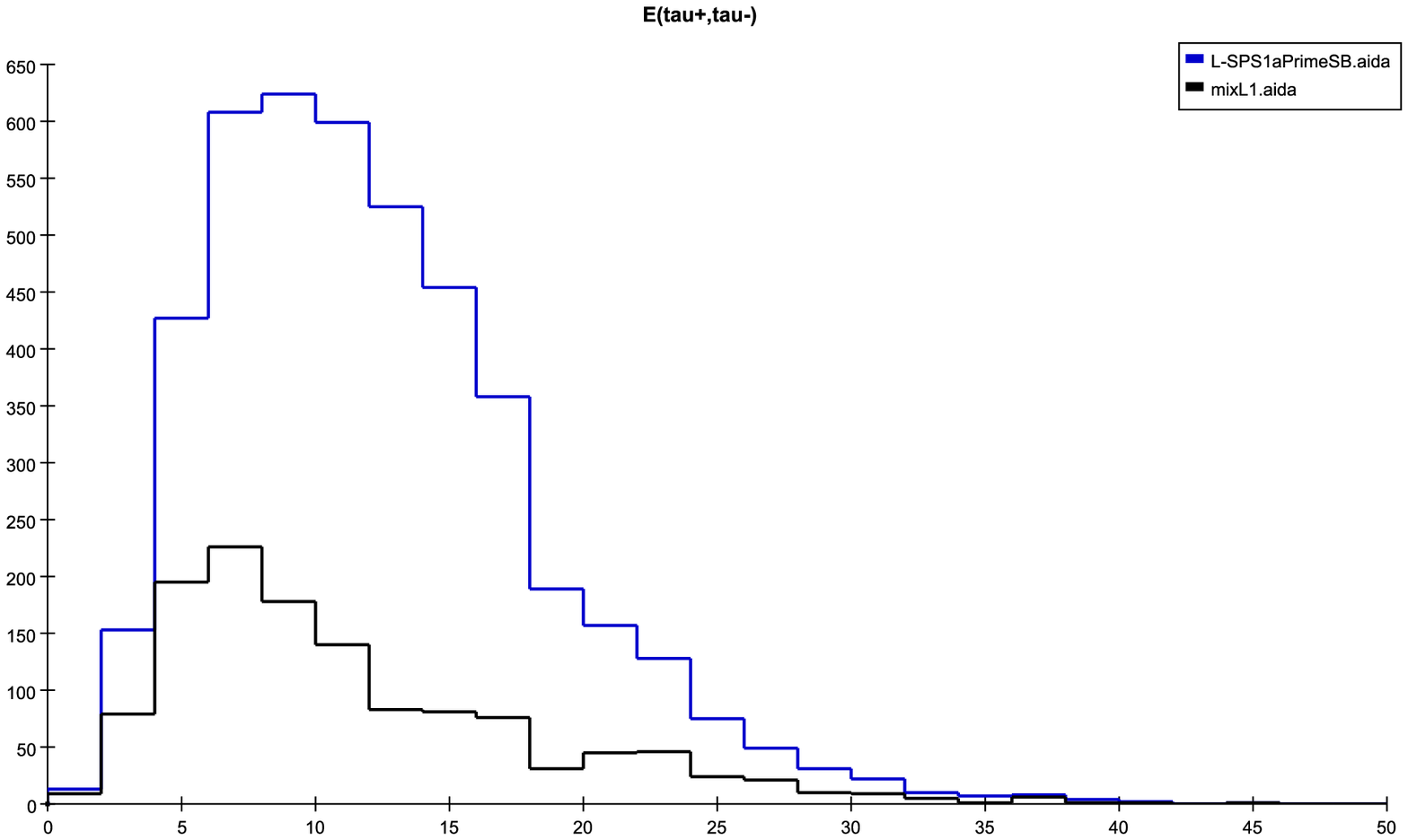}}
\vspace*{0.1cm}
\caption{Same as in Fig.~\ref{stausearch1} except now for SPS1a'.}
\label{staus_sps1a}
\end{figure*}

\begin{figure*}[htpb]
\centerline{
\includegraphics[height=10cm,width=13cm,angle=0]{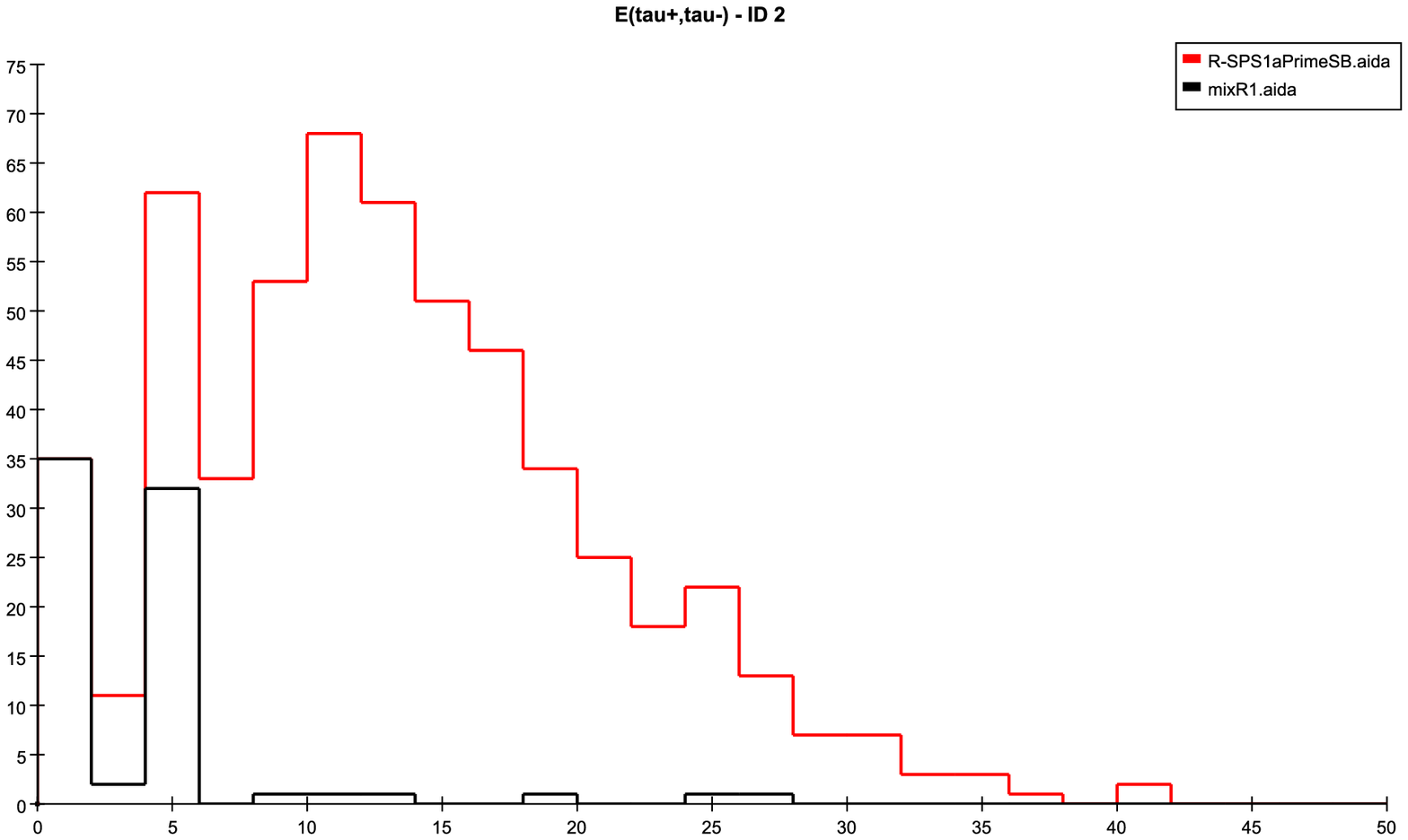}}
\vspace*{0.1cm}
\centerline{
\includegraphics[height=10cm,width=13cm,angle=0]{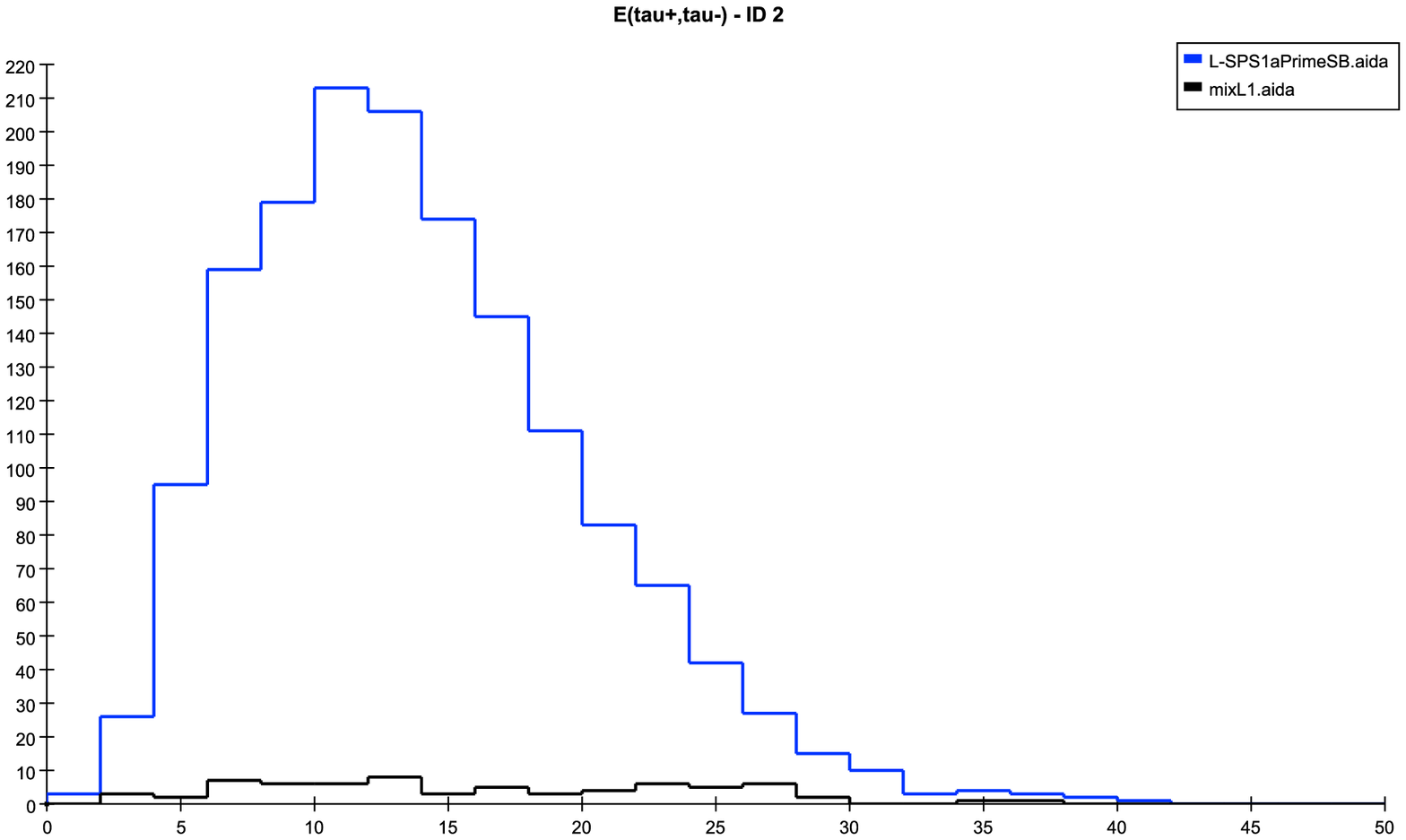}}
\vspace*{0.1cm}
\caption{Same as in Fig.~\ref{stausearch2} except now for SPS1a'.}
\label{staus_ID2_sps1a}
\end{figure*}

We find that many AKTW models which do not contain kinematically
accessible stau states nonetheless give rise to visible signatures
with significance $>5$
in this analysis, providing yet another example of SUSY being a
background to itself. The tau energy distribution for a representative
sampling of these SUSY background models is presented in
Fig.~\ref{staufake}, using the alternative set of kinematic cuts.
We find that there are 29(28) models which
yield fake signals with LH(RH) electron beam polarization in our
standard set of kinematic cuts. For our
alternative analysis which rejects electrons in the final state,
there are 30(28) models with false signatures for the LH(RH) polarization
configuration. This analysis clearly has a very large number of
false signals. We note that in every one of these "fake" models,
$\tilde\chi_1^\pm\tilde\chi_1^\mp,\, \tilde\chi_2^0\tilde\chi_1^0$,
and $\tilde\chi_2^0\tilde\chi_2^0$ production is kinematically
possible, and in one case selectron and smuon production is also
viable. There are thus several sources of SUSY background which
can lead to the same final state as that for stau
pair production.

\begin{figure*}[htpb]
\centerline{
\includegraphics[height=10cm,width=13cm,angle=0]{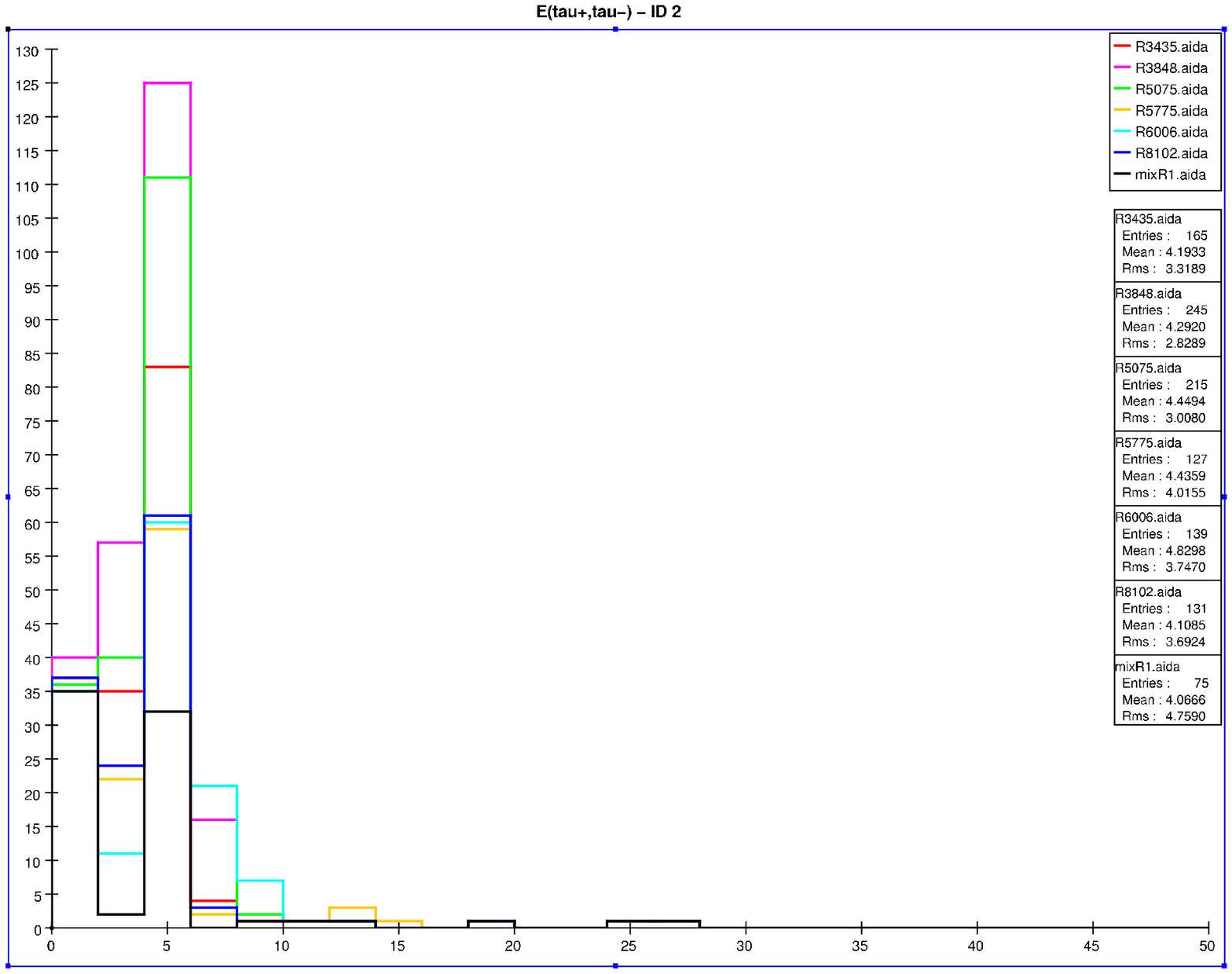}}
\vspace*{0.1cm}
\centerline{
\includegraphics[height=10cm,width=13cm,angle=0]{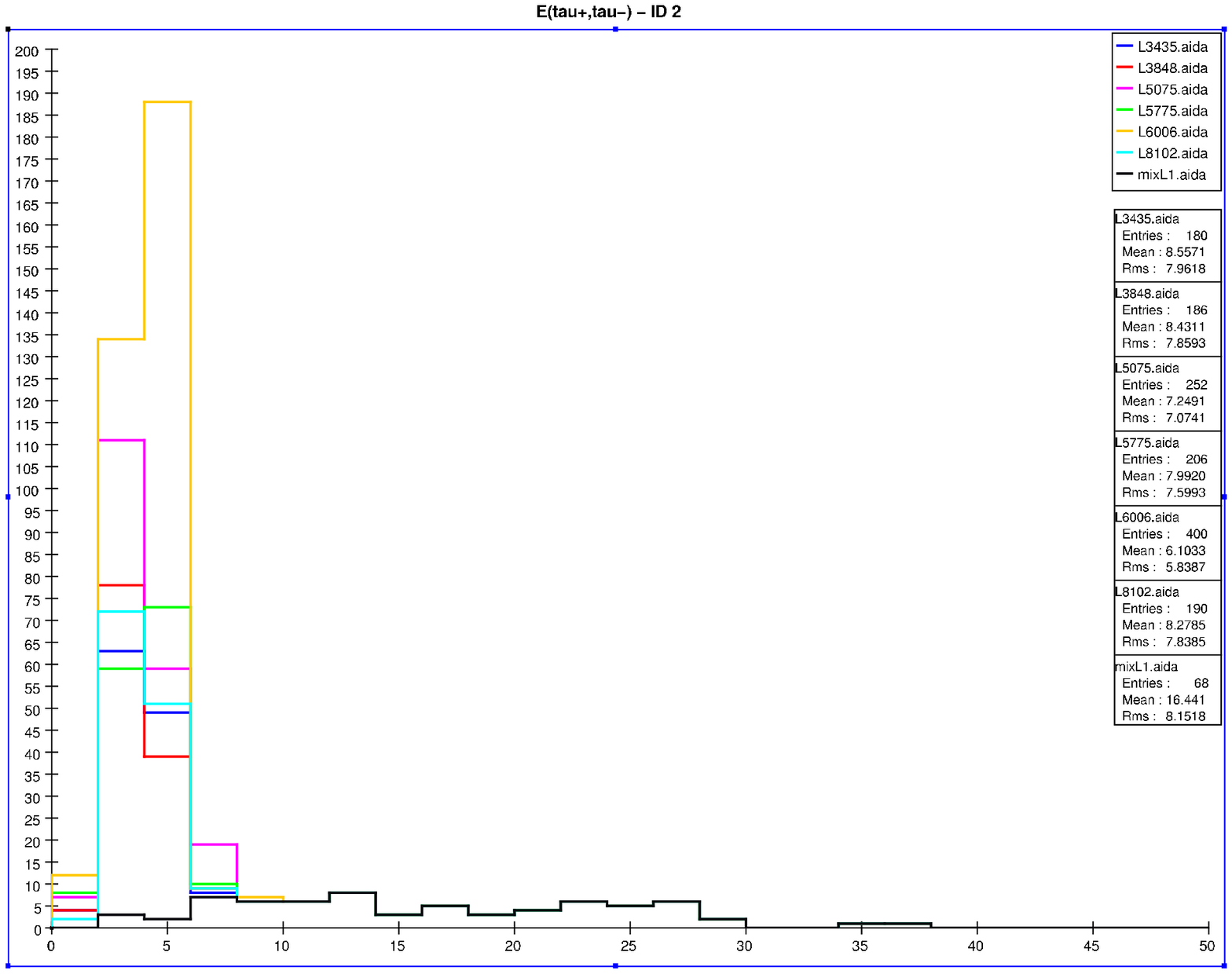}}
\vspace*{0.1cm}
\caption{
Tau energy distribution: the number of events/2 GeV bin after
imposing the alternative set of tau ID and $\tilde\tau$ selection criteria
described in the text
for several representative AKTW models that give a fake signal in this
channel. RH(LH) beam polarization is
employed in the top(bottom) panel, assuming an integrated luminosity of
250 fb$^{-1}$ for each polarization configuration. The SM background
corresponds to the black histogram.}
\label{staufake}
\end{figure*}

In order to distinguish between stau production
and these SUSY background sources, we investigate the variable
\be
M_{\mbox{eff}}=E_T^{\mbox{miss}}+\sum_{i=1,2} |E_T^{\tau_i}|\,.
\end{equation}
This variable is presented in Fig.~\ref{Meffstau} for both real and
fake stau production. Here, we see that the false signals are slightly
more peaked at lower values of $M_{\mbox{eff}}$ than does actual stau
production. However, the distinction is not as clear as in the
identification of selectron and smuon fake signatures discussed above,
which makes use of the observable $p_T^{vis}$. This is because the
full $\tau$ energy is not carried by its visible decay products.
We note that the $p_T^{vis}$ observable
is not effective in distinguishing stau states
from SUSY background sources, as in this case both the staus
and the background sparticles have multi-body decays.

\begin{figure*}[htpb]
\centerline{
\includegraphics[height=10cm,width=13cm,angle=0]{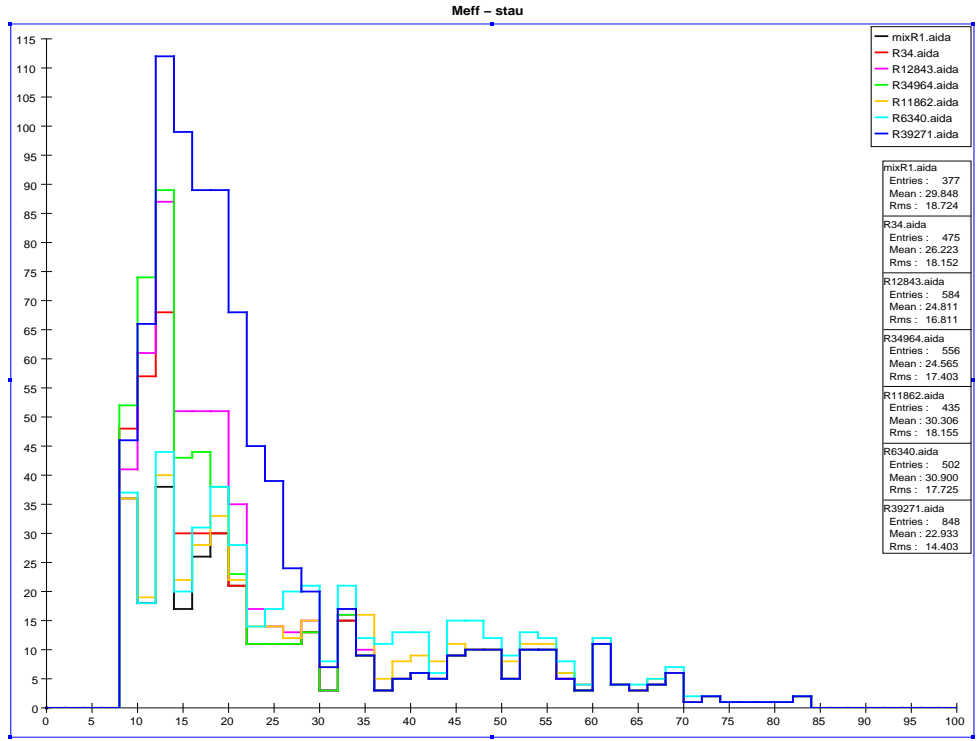}}
\vspace*{0.1cm}
\centerline{
\includegraphics[height=10cm,width=13cm,angle=0]{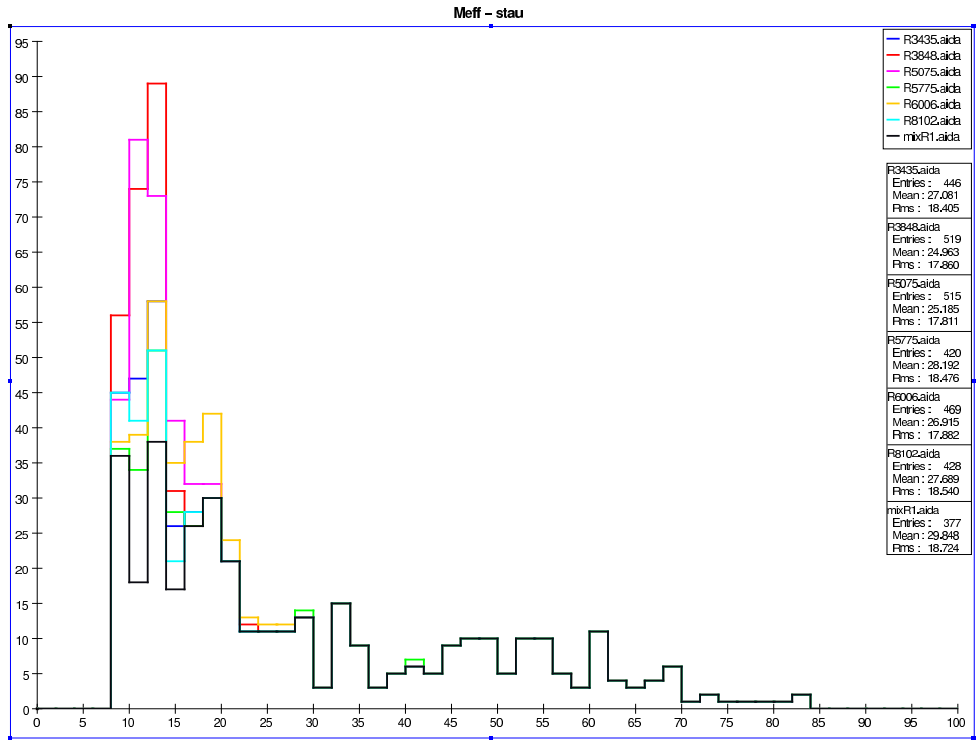}}
\vspace*{0.1cm}
\caption{Effective mass (as employed at the LHC) of the products from the decays of objects
identified as tau pairs
produced from stau production (top panel) and from other, non-stau
SUSY sparticles (bottom panel). Here we present
the number of events/2 GeV bin assuming RH polarization and 500 fb$^{-1}$
of integrated luminosity for several models shown in comparison to the
SM background represented as the black histogram.}
\label{Meffstau}
\end{figure*}

\clearpage

\subsection{Sneutrino Pair Production}
\label{Sec:sneu}

We now examine the neutral slepton sector, \ie, sneutrinos, which
provides another potential handle for distinguishing between models.
For all three sneutrino families there is the usual
$Z$ boson exchange contribution to the production amplitude
in the $s$-channel, while for
electron sneutrinos there is an additional
$t$-channel graph due to $\tilde\chi_{1,2}^\pm$ exchange. If the
charginos are heavy,
then the $Z$-exchange graph dominates for all three
generations and the resulting production
cross section is determined solely by the amount of available phase
space.
11/242 of our AKTW models have electron or muon
sneutrino pairs which are kinematically accessible at $\sqrt s=500$ GeV, while
18/242 models contain accessible tau sneutrinos. In one of the
models the sneutrino is the LSP.

Sneutrinos, being neutral and weakly-interacting, are essentially only visible
through their decay modes, of which there
are several possible channels to consider:
($i$)~$\tilde \nu_\ell \to \nu_\ell \tilde \chi_1^0$ largely dominates
in most cases,
but leads to an invisible final state which, by itself, is clearly
useless for either discovery or model comparison.
($ii$)~$\tilde \nu_\ell \to W \tilde \ell$ is kinematically
forbidden as an on-shell mode when $\ell=e,\mu$ in all of our AKTW models
and thus the corresponding 3-body branching fraction mediated by
off-shell $W$ bosons
is very small. However, due to large $\stau$ mixing, this 2-body
mode is allowed for 6 of our models in the case of $\ell=\tau$. When both
the $W$ and $\tau$ decay hadronically, we can search
for final states with multiple jets plus missing energy in
this case. ($iii$)~$\tilde \nu_\ell \to \nu_\ell \tilde \chi_2^0$ can
also occur, with the subsequent decay $\tilde \chi_2^0 \to
\tilde \chi^0_1 jj$ via a $Z$ or Higgs boson. This occurs in 1(3) models
in our sample when $\ell=e/\mu(\tau)$. However, in
this case the resulting jets are likely to be relatively
soft, due to a smaller $\tilde\chi_2^0-\tilde\chi_1^0$ mass difference,
making this mode difficult to observe above background.
$(iv)$ $\tilde \nu_\ell \to \tilde \chi_1^+\ell^-$ is accessible in 1(6)
of these models when $\ell=e/\mu(\tau)$, and
leads to a final state of multiple jets plus two charged leptons
plus missing energy.
As before, it is probable that these jets will be soft
due to a smaller mass splitting between the chargino and the
LSP and will most likely be difficult to observe
depending upon the details of the rest of the spectrum.

We first study the final state $jjjjl^+ l^-$ + missing energy. This
final state results from the decays
\ba
\tilde{\nu} & \rightarrow & W \tilde{l} \rightarrow jj l \LSP \,,
\nonumber\\
\tilde{\nu} & \rightarrow & l \tilde{\chi}_1^\pm \rightarrow ljj \LSP \, .
\ea
For our signal selection, we require:
\begin{enumerate}
\item Precisely one opposite sign lepton pair and two jet-pairs and no
other charged particles in the event.

\item No particles/clusters below the angular region of 100 mrad.

\item Missing energy to be $> 2 \, m_{\LSP,\mbox{\tiny min}}$.
We take $m_{\LSP,\mbox{\tiny min}} = 46$ GeV, which is the current
(weak, yet model-dependent) bound on the mass of the
lightest neutralino~\cite{Yao:2006px}.
This bound arises from the invisible decay width of the $Z$ boson
and holds unless the $\tilde\chi_1^0$ is very fine-tuned to be a
pure Bino state and thus has no couplings to the $Z$ \cite{herbie2}.
However, in order to estimate the effect on the background if this
bound is increased, we perform a second analysis with
$m_{\LSP,\mbox{\tiny min}} = 100$ GeV.

\item In order to eliminate background that originates from very
soft leptons or jets, we demand $E_{\mbox{\tiny jet}, l} > 0.01 \sqrt{s}$.
\end{enumerate}

These cuts effectively remove most of the SM background as can be
seen in Fig.~\ref{sneubg}, which displays the missing energy distribution
for the background sample. Here, we see that at large values of
missing energy, the dominant background
remaining after the cuts arises from the process
$\gamma\gamma\to c\bar c,b\bar b$. Unfortunately, the
sneutrino signal rates are also small.
Fig.~\ref{sneutrinofig} presents the results of our analysis:
none of the sneutrino models rise above the background, but several
of the `fake'
models, where the signals arise from other SUSY particles, do appear.
4(3) fake models yield visible signals in the case of
RH(LH) beam polarization. The counterfeit
signals here are due to chargino and neutralino production and decay.
We find that increasing the minimum LSP mass to 100 GeV does not improve
these results.

\begin{figure*}[hptb]
\centerline{
\epsfig{figure=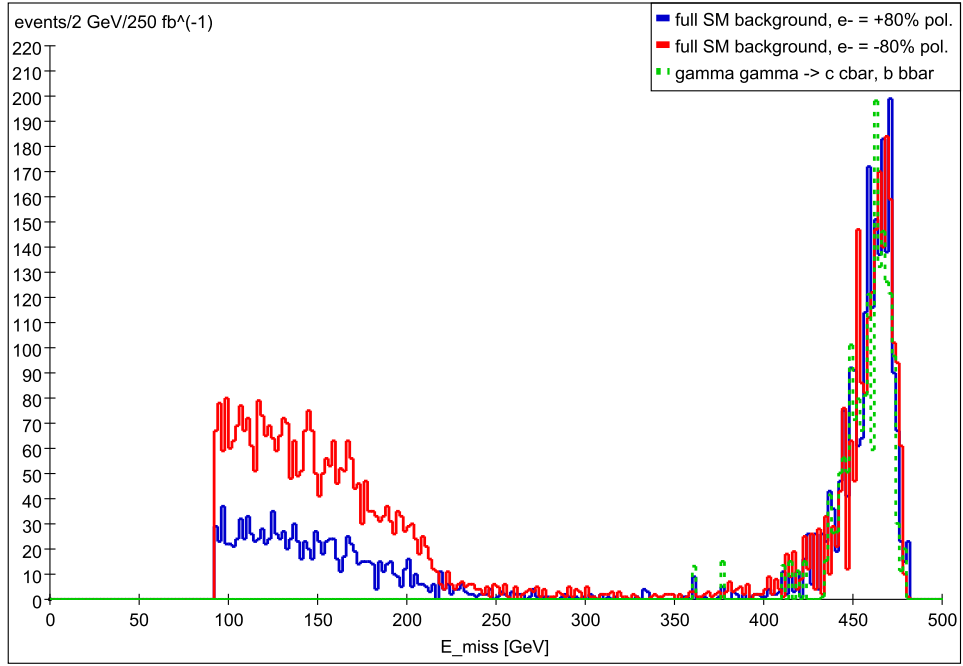,width=13cm,clip=}}
\vspace*{0.1cm}
\caption{Missing energy distribution in 2 GeV bins for the SM background
after the sneutrino selection cuts for the 4jet $+2\ell$ channel have
been imposed. The blue(red) histogram corresponds to $80\%$ RH(LH)
electron beam polarization. The green dotted curve corresponds to the
dominant background source, $\gamma\gamma\to c\bar c,b\bar b$. 250
fb$^{-1}$ of integrated luminosity was assumed for each polarization
channel at 500 GeV.}
\label{sneubg}
\end{figure*}

\begin{figure*}[hptb]
\centerline{
\includegraphics[width=13.0cm,height=10.0cm,angle=0]{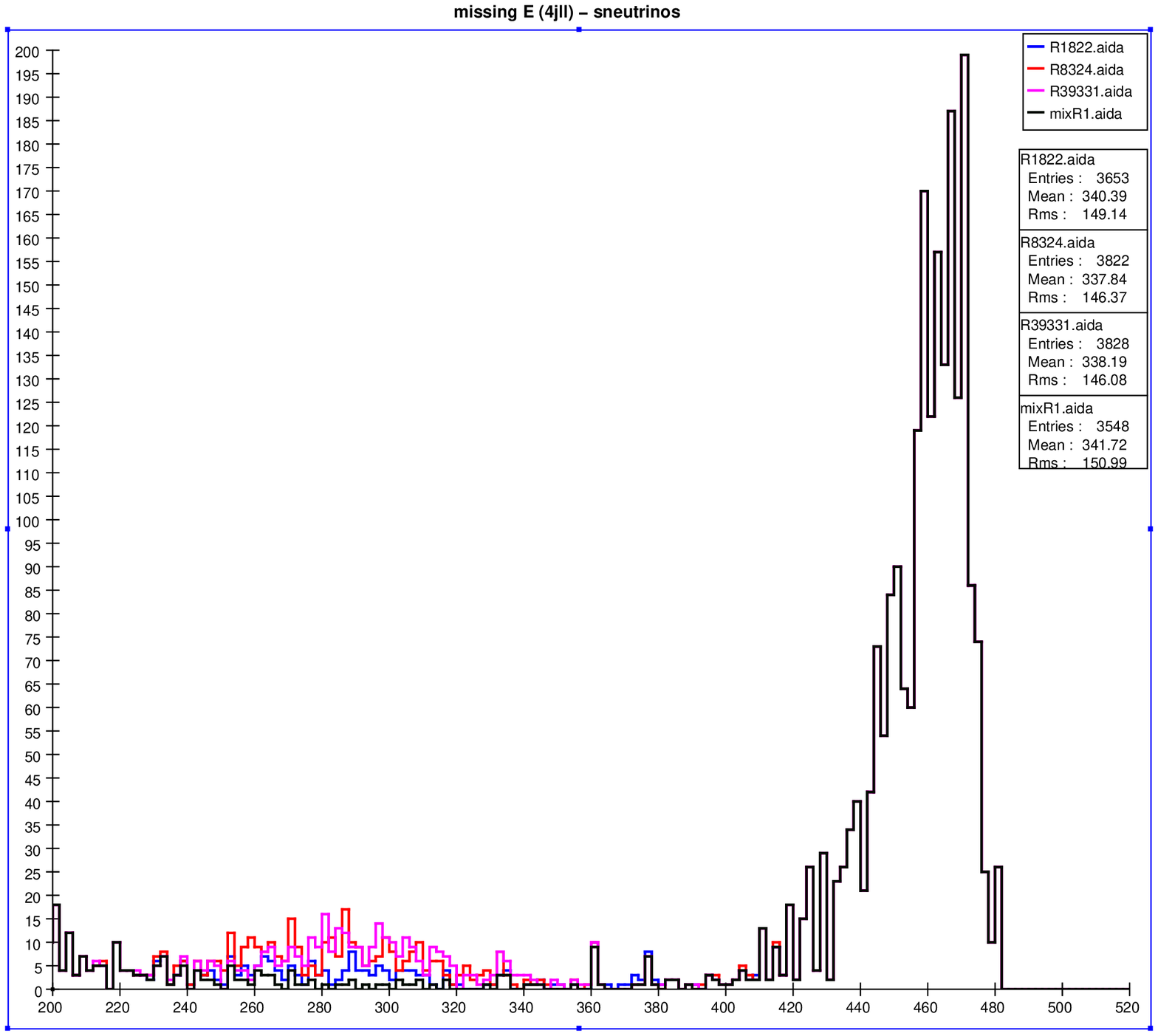}}
\vspace*{0.1cm}
\centerline{
\includegraphics[width=13.0cm,height=10.0cm,angle=0]{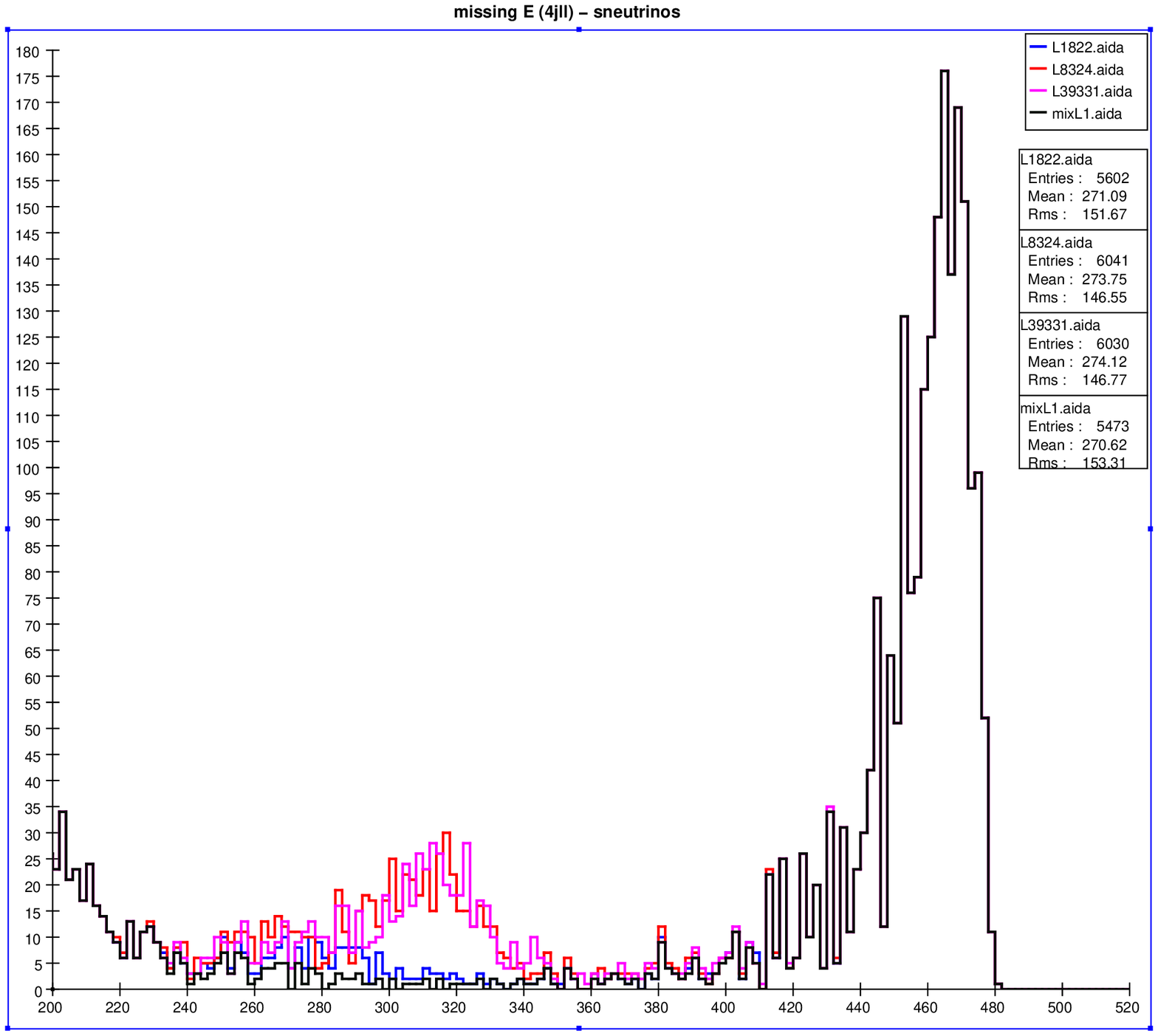}}
\vspace*{-0.1cm}
\caption{Missing energy distribution for the sneutrino $4j+$lepton pair
analysis: the number of events/2
GeV bin for several the fake models, with RH(LH) beam polarization
in the top(bottom) panel, assuming
an integrated luminosity of 250 fb$^{-1}$ for either polarization.
The SM background is shown as
the black histogram.}
\label{sneutrinofig}
\end{figure*}

We also study a second channel, with 6 jets and missing
energy in the final state. This is produced from the decay
$\tilde\nu\to W\tilde\tau \to jjj\tilde\chi^0_1$.
The cuts and observables are similar to those of the $4j2\ell+$ missing
energy analysis, with the obvious substitution that we demand
precisely 6 jets and no other charged particles to appear in the
event. We find that there is little to no SM background in this
channel. However, we also find that none of our models are
observable in this channel.

An additional possible way to observe sneutrinos is via the
radiative process
$e^+e^- \to \tilde \nu_\ell \tilde \nu_\ell +\gamma$. This may be
particularly
useful in the case where the decay channel $\tilde \nu_\ell \to \nu_\ell +
\tilde \chi_1^0$
dominates. This reaction leads to a final state with a photon and missing
energy and is thus similar to
radiative LSP pair production, which we will discuss in detail
below in Section 6.2. We find that radiative sneutrino production
generally occurs with a smaller cross section than its LSP counterpart.
As we will see below, the SM background from $\epem\to\nu\bar\nu\gamma$
are generally too large
to see this radiative process.

In the case of the SPS1a' benchmark model, the sneutrino pair production
mode is invisible as the sneutrinos dominantly
decay into the $\nu \chi_1^0$ final state, as in most of our models here.

Taking these results together for these various sneutrino analyses,
we conclude that the direct observation of
$\tilde \nu_\ell$ production is very difficult, if not essentially hopeless
for the set of AKTW models.

\clearpage

\section{Chargino Production}

The chargino sector is simpler than that of
the neutralinos as in the CP-conserving MSSM; it depends on only three real
Lagrangian parameters at tree-level: $M_2, \mu$, and $\tan \beta$.
The resulting mass eigenstates, $\tilde\chi_{1,2}^\pm$, are thus general
admixtures of the charged Wino and Higgsino weak states.
Figure~\ref{chargmix} displays the Wino/Higgsino
content of the lightest chargino, $\tilde \chi_1^\pm$, in the 53
AKTW models that contain kinematically accessible charginos at $\sqrt s=500$
GeV. We see that the lightest chargino tends to be an almost pure Wino
or Higgsino state in most of our models. In $e^+e^-$ collisions, such
particles can be produced via two mechanisms; $s$-channel
$\gamma,Z$ exchange produces either pure Wino or Higgsino pairs
but no mixed Wino-Higgsino final states as the analogous $WZH^\pm$
coupling is absent. In addition, the $t-$channel sneutrino
exchange amplitude produces pairs of charged winos only. Clearly,
the cross section for chargino pair production not only depends on the
eigenstate masses but also on the various mixing angles present in
the chargino sector. At $\sqrt s=500$ GeV,
$\tilde\chi_2^\pm$ pairs are typically too heavy to pair produce, so we
will consider only $\tilde\chi_1^\pm$ pair production in our analysis
below.{\footnote {Associated $\tilde\chi_1^\pm \chi_2^\mp$ production
is also possible in 7 of our models,
but has not been considered in this analysis.
However, the analysis of associated $\tilde\chi_2^0 \tilde\chi^0_1$
production, presented below in Section 6.1,
also picks up some contributions from $\tilde\chi_1^\pm \chi_2^\mp$.}}

\begin{figure*}[hptb]
\centerline{
\includegraphics[width=10.0cm,height=13.0cm,angle=-90]{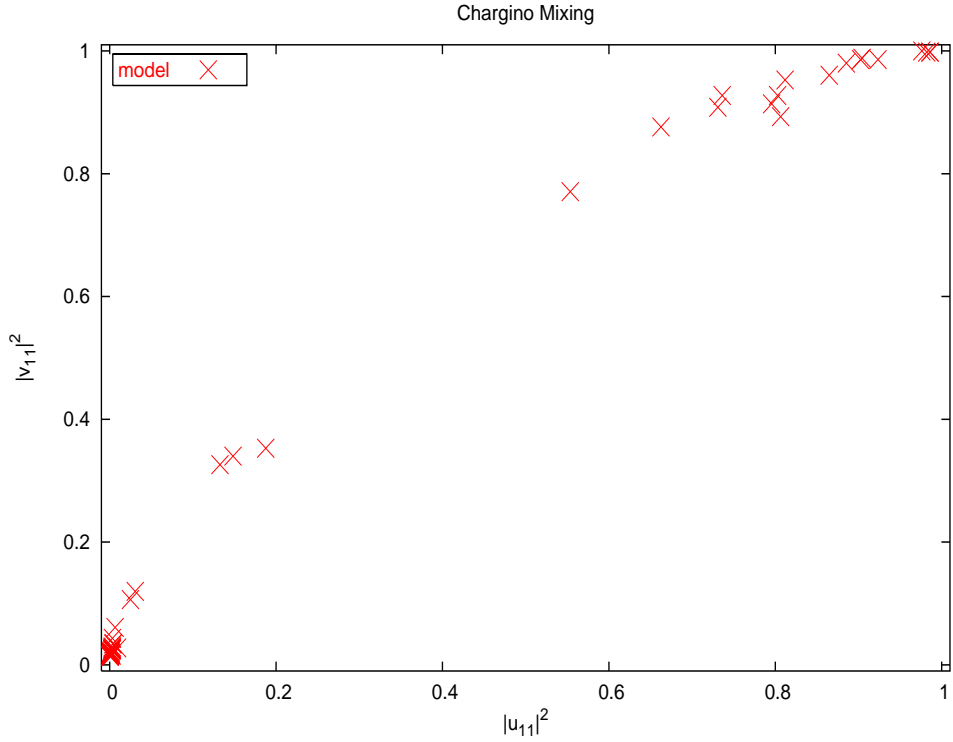}}
\vspace*{0.1cm}
\caption{Wino/Higgsino content of the $\tilde\chi_1^\pm$
for the 53 models that have kinematically
accessible charginos at $\sqrt s=500$ GeV.
$U(V)$ is the left(right) diagonalizing matrix. In
the lower left (upper right) corner of the Figure, the physical chargino is
dominantly Higgsino (Wino).}
\label{chargmix}
\end{figure*}

Once produced, the detailed nature of the
$\tilde\chi_1^\pm$ decays critically depends upon the mass difference
$\Delta m_{\tilde \chi}=m_{\tilde\chi_1^\pm}-m_{\tilde\chi_1^0}$, with
the latter being
the LSP. In all cases, $\chi_1^\pm$ decay can proceed
via either a $W$, $\tilde\chi_1^\pm\to W\tilde\chi_1^0$, or through an
intermediate slepton, \eg, $\tilde\chi_1^\pm\to \ell \tilde \nu,
\tilde \nu \to \nu\tilde\chi_1^0$. As seen above in Fig.~\ref{fig:pythfeature},
and in Figs.~\ref{highdm} and ~\ref{lowdm}, the mass spectrum
of the models that have chargino states kinematically accessible at
$\sqrt s=500$ GeV is such that the
distribution of values for $\Delta m_{\tilde \chi}$ are concentrated
in the region $<5$ GeV.
The variation in $\Delta m_{\tilde\chi}$ yields several
distinct signatures for $\tilde\chi_1^\pm$ production and thus
all possible values of $\Delta m_{\tilde \chi}$
must be considered when performing our analysis. For
example, if $\Delta m_{\tilde \chi}>M_W$ then on-shell $W$ bosons
can be produced and may be identified either through their
leptonic or hadronic decay modes. We thus consider channels such
as $\tilde\chi_1^\pm \to jj\,\Emiss$, with the dijets
reconstructing
to the $W$ mass, or $\tilde\chi_1^\pm \to \mu \Emiss$, with the latter
mode also covering decays through the slepton
channels: $\tilde\chi_1^\pm \to \tilde \mu^\pm \nu,\mu^\pm \tilde
\nu \to \mu \Emiss$. Hence, in this region, we search for the final states
$\tilde\chi_1^+ \tilde\chi_1^- \to 4j \, \Emiss$,
$2j \,\mu^\pm\, \Emiss$ or
$\mu^+ \mu^-\, \Emiss$. For smaller values of $\Delta m$, but still
greater than a few GeV, we search for the same
final states although the dijets will no longer reconstruct
to $M_W$. A more difficult region is reached when
$\Delta m_{\tilde \chi}$ is only a few GeV or less, as then the
visible part of the $\tilde\chi^\pm_1$ decays
are very $\tau$-like.
Fig.~\ref{fig:pythfeature} shows this region of chargino-LSP mass splitting
for the range $\Delta m_{\tilde \chi} <6$ GeV as a
function of the chargino mass; we see that this region comprises
the bulk of our models with accessible $\tilde\chi_1^\pm$ states at
$\sqrt s=500$ GeV.

At the other end of the spectrum, we must consider the case of
small values of $\Delta m_{\tilde \chi} \lsim 1$ GeV. The branching
fractions for the $\tilde\chi_1^\pm$ decay channels in this case
are presented in Fig.~\ref{fig:charginoBF}.
If $\Delta m_{\tilde \chi}$ is very small, $\lsim 100$ MeV, then the
chargino is long lived, and will travel many meters before
decaying into an electron and missing energy. In such a case, we
perform a massive stable charged particle search,
determining the velocity of the $\tilde\chi_1^\pm$
via momentum and energy measurements.
As $\Delta m_{\tilde \chi}$ increases from this tiny value and the
thresholds for $\tilde\chi_1^\pm$ decay into
the $\mu$ and pion(s) are passed, the chargino lifetime decreases
substantially and the chargino now decays to soft charged
particles. In this mass
range there are two possible search techniques: one can either look for
decays in the detector from (semi-)long-lived $\tilde\chi_1^\pm$
states, or tag these soft decays via photon emission
off of the initial and final state particles. The latter corresponds
to the radiative
process $e^+e^- \to \tilde\chi_1^+\tilde\chi_1^-\gamma$
{\cite {Gunion:2001fu,Riles:1989hd}}, and is the approach we will pursue below.

\begin{figure*}[hptb]
\centerline{
\psfig{figure=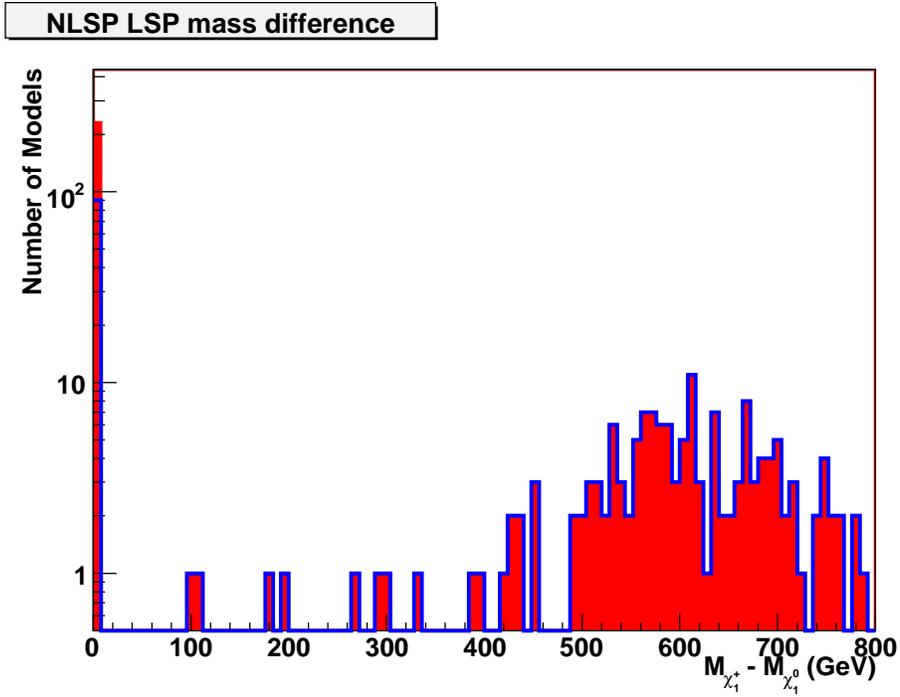,angle=0,width=13cm,clip=}}
\vspace*{0.1cm}
\caption{Lightest chargino-LSP mass difference for the
region $0 < \Delta m_{\tilde \chi} <
800$ GeV. Note that the chargino states in the
models where $\Delta m{\tilde\chi}>150$ GeV are not kinematically
accessible at $sqrt s = 500$ GeV.}
\label{highdm}
\end{figure*}

\begin{figure*}[hptb]
\centerline{
\epsfig{figure=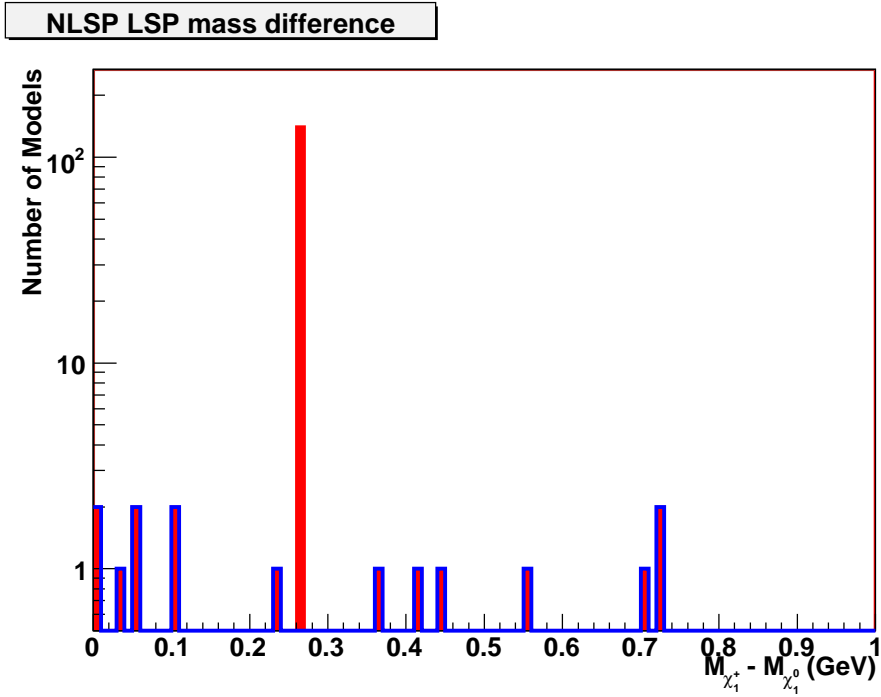,width=13cm,clip=}}
\vspace*{0.1cm}
\caption{Lightest chargino-LSP mass difference for the close
mass region $0 < \Delta m_{\tilde{\chi}} < 1$ GeV.}
\label{lowdm}
\end{figure*}

\begin{figure*}[hpb]
\centerline{
\epsfig{figure=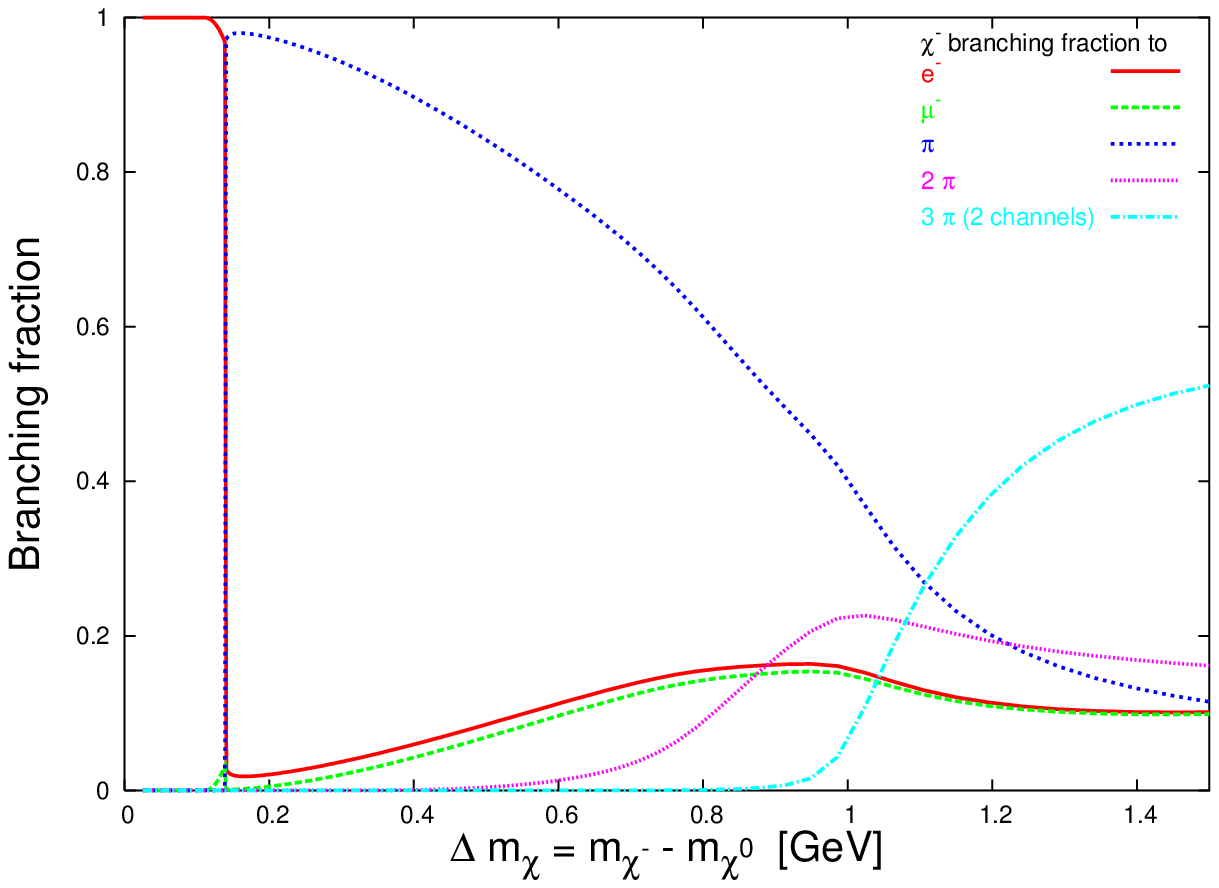,width=13cm,clip=}
}
\caption{Branching fraction of $\tilde \chi_1^\pm$ as a function
of $\Delta m_{\tilde{\chi}}$.}
\label{fig:charginoBF}
\end{figure*}

At $\sqrt s=500$ GeV, we find that 53/242 of the AKTW models have
kinematically accessible charginos. Figure~\ref{highdm}
shows that in all but two cases (which we label here as models 8324 and
39331)
the models populate the region $\Delta m_{\tilde \chi}
\leq 5.5$ GeV; for the
two exceptions we see that $\Delta m_{\tilde \chi}> 100$ GeV.
Interestingly, we note that models with small
$\Delta m \leq 1$ GeV tend to have a large Wino content, while those in
the range $\Delta m_{\tilde \chi} \simeq
4\sim 5$ GeV are found to have a large Higgsino content as can be seen
in Fig.~\ref{chargmix}.

We now discuss our analyses for each region of $\Delta m_{\tilde\chi}$.

\subsection{Non-Close Mass Case}
\label{Sec:charginogen}

We first examine the case where
\be
\Delta
m_{\tilde{\chi}} \equiv m_{\tilde{\chi}^{\pm}} - m_{\LSP} > 1
\mbox{ GeV}.
\ee
As already mentioned above, there are several possible final states that
can be studied in this mass region. Analysis techniques have been
developed for the cases where the chargino decay proceeds through
on- or off-shell $W$ bosons
(with $W\rightarrow jj$ or $W\rightarrow l \bar{\nu}_l$)
or via sleptons (\eg, $\tilde{\mu}\rightarrow \mu \LSP$). We discuss
each of these in turn.

\subsubsection{Chargino Decays via On-Shell $W$ bosons}

This analysis applies to the case $\Delta
m_{\tilde{\chi}}>M_W$. Here,
we examine four-jet final states, stemming from the decays of
the chargino pair into $W$ bosons with subsequent decays into quark pairs,
\be
\chap \cham \rightarrow W^+ W^- \LSP \LSP \, ,
\ee
with
\be
W^\pm \rightarrow q \bar{q} \, .
\ee
Taking the hadronic decay mode of both $W$ bosons yields the final
state with the largest statistical sample.

As always, the SM background is significant. In order to reduce the
background, we demand, as expanded and adapted
from~\cite{Williams,Abbiendi:2003sc}:
\begin{enumerate}

\item There be precisely 4 jets in the final state and no other
charged particles.
As mentioned in section~\ref{Sec:detectorandanalysis},
we employ the JADE jet algorithm
in the E scheme, with $y_{\mbox{\tiny cut}} = 0.05$.
This choice of $y_{cut}$ avoids the situation where soft
gluons produce too many jets.

\item $\Evis > 0$ in the backward direction. Here, the
backward direction is defined with respect to the thrust axis, and
corresponds to the hemisphere with the lesser amount of energy.
This cut is designed to reduce SM background from $Z$ pair production,
where one $Z$ boson decays into neutrinos,
and the other $Z$ decays into quarks, which then radiate hard gluons.

\item In the forward direction, the visible energy is constrained to
be $\Evis < \frac{1}{2} \sqrt{s} - m_{\LSP,\mbox{\tiny min}}$. As in
the case of stau production,
we take $m_{\LSP,\mbox{\tiny min}} = 46$ GeV, which is the current
bound on the mass of the lightest
neutralino~\cite{Yao:2006px} in the case where the $\tilde\chi_1^0$
is not a pure Bino eigenstate.

\item The visible energy is constrained to be
$\Evis < 1$ GeV in the region $ 0.9 \leq | \cos \theta | \leq 0.99$.
This is to decrease the $W$ pair background which is strongly peaked
in the forward direction.

\item The acoplanarity angle satisfy
$\Delta \phi^{\mbox{\tiny jetpair}\, \mbox{ \tiny
jetpair}} > 30$ degrees.
Since we demand that two jet pairs recombine into
$W$ bosons, the acoplanarity angle
is equivalent to $\pi$ minus
the angle between the $p_T$ of the $W$ bosons, \ie,
$\Delta \phi^{\mbox{\tiny jetpair}\,\mbox{\tiny jetpair}}
= \pi - \theta_T$.
This significantly reduces the $W$ pair and $\gamma\gamma$ background,
since the $W$ bosons from the chargino decays are accompanied by
missing energy from the LSP.
\end{enumerate}

The standard search analysis for this final state is based on the
energy distribution of the jet-pairs which reconstruct into a $W$ boson.
As in the case of the selectron analysis, this distribution should have
a box-like shape with a shelf and sharp endpoints in the presence of a high
statistics, background-free, perfect detector environment with the
absence of radiative effects. Here, the 2-body decay is taken to be
$\tilde\chi_1^\pm\to W+\tilde\chi_1^0$ and the expressions in
Eqs.~(\ref{Emax})-(\ref{Emin}) need to be adapted to include the massive
$W$ boson. In this case, one can solve the equations for the
chargino and LSP masses and finds
\ba
m_{\tilde{\chi}_1^\pm}^2 & = & \frac{
A \pm B}
{2( E_{\max} + E_{\min})^2/s}\, , \\
m_{\LSP}^2 & = & m_{\tilde{\chi}_1^\pm}^2 + m_W^2
\left(1 - 2 \frac{( E_{\max} + E_{\min})}{\sqrt{s}}
\right) \, ,
\ea
where
\ba
A = m_W^2 + E_{\max} E_{\min}, \\
B =  \sqrt{(m_W^2 - E_{\max}^2)(m_W^2 - E_{\min}^2)},
\ea
and where $E_{\max}$, $ E_{\min}$ are determined experimentally.

We refrain from presenting the remaining SM background after these cuts
are imposed, as only a handful of events pass the cuts.
As mentioned above, only 2 models in our sample lie in the kinematic
region $\Delta m_{\tilde\chi}>M_W$. The jet-pair energy distribution
for these two cases is displayed in Fig.~\ref{ejetpairs_charg} for
left- and right-handed electron beam polarization. Here, we see that
the overall event rate that survives the kinematic cuts is not large,
but the signal cleanly towers above the even smaller background.
We can see the effects of the cuts and the detector environment in these
cases, as the shape of the spectrum does not display the shelf-like
behavior discussed above.
Note that an additional model (labelled 1822), which has
$\Delta m_{\tilde\chi}\sim 1$ GeV, also passes the kinematic
cuts for this analysis. This model yields a smaller event rate than
the cases with on-shell $W$ bosons,
but populates a different region of the spectrum. However, this model
contains a light $\tilde\chi_{1,2}^\pm\,, \tilde\chi_{1,2,3}^0$ sector
and the signal that passes our cuts here is due to the production of
these heavier gaugino states and not from the $\tilde\chi_1^+$ and is
thus a fake. Three out of 53 AKTW
models with accessible charginos are thus visible with a significance
${\cal S}>5$ in this channel. We note that fake signals from the
production of other SUSY particles do not satisfy our visibility
criteria in this channel, except for model 1822. It would thus
seem that this analysis is relatively free from Supersymmetric
backgrounds, at least in the case of the AKTW models.

In summary, if $\Delta
m_{\tilde{\chi}}$ is large enough to produce on-shell $W$ bosons, this
is clearly a very clean channel.

\begin{figure*}[hptb]
\centerline{
\includegraphics[width=13.0cm,height=10.0cm,angle=0]{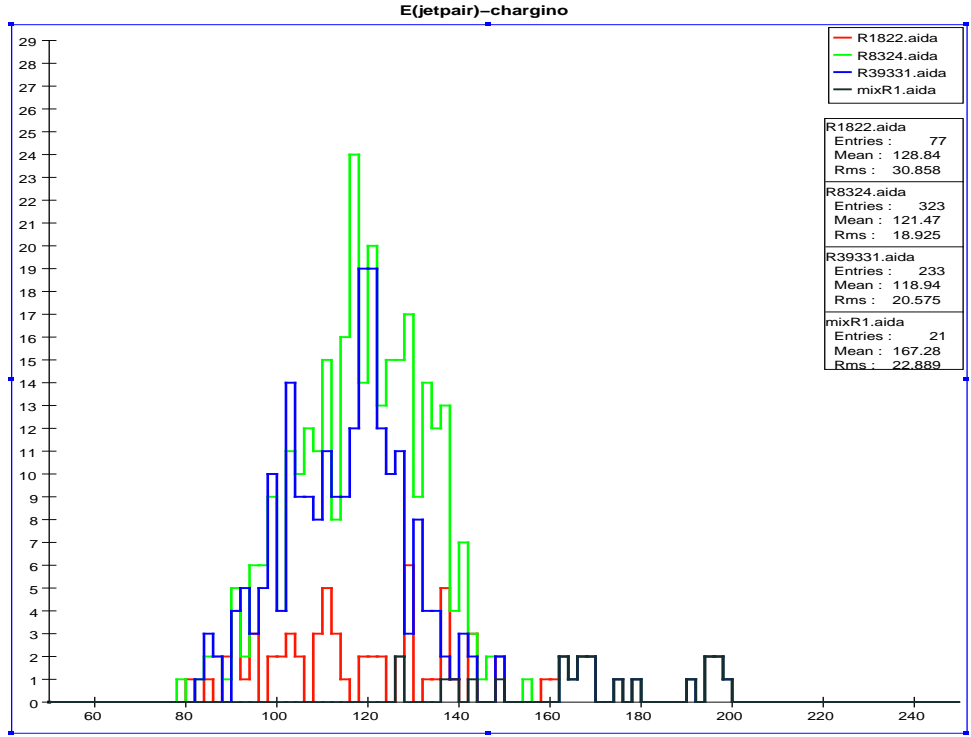}}
\vspace*{0.1cm}
\centerline{
\includegraphics[width=13.0cm,height=10.0cm,angle=0]{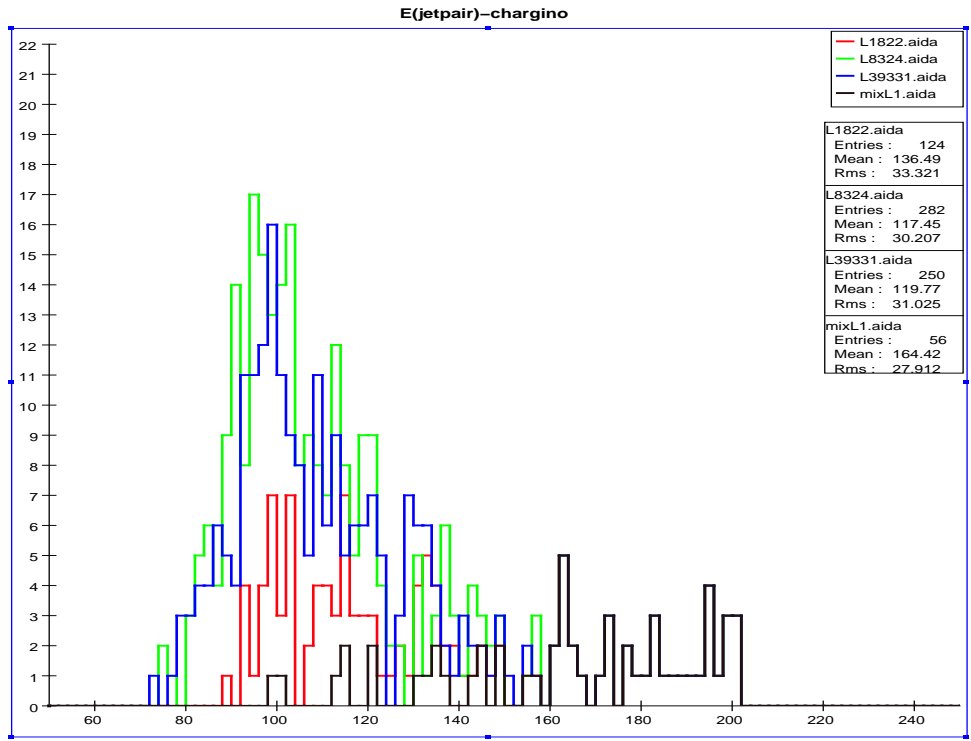}}
\vspace*{-0.1cm}
\caption{Jet-pair energy distribution: the number of events/2 GeV bin
after imposing the full set of cuts discussed in the text for
chargino production and on-shell decays to a $W$ boson for the three
models which are visible in this channel.
RH(LH) beam polarization is employed in the top(bottom)
panel, assuming
an integrated luminosity of 250 fb$^{-1}$
for either polarization. The SM background is shown as the black histogram.}
\label{ejetpairs_charg}
\end{figure*}

\subsubsection{Chargino Decays via Off-Shell $W$s and/or Sleptons}

We search for three final states in the kinematic region $M_W > \Delta
m_{\tilde{\chi}}> 1$ GeV where the $W$ boson is produced off-shell in the
$\tilde\chi_1^\pm$ decay:
four jet events plus missing energy, two jets and a lepton plus missing energy
or two leptons plus missing energy. In order to avoid the large SM
background from the beam remnants in $\gamma e^\pm$, $\gamma\gamma$
reactions, we require that the final state leptons be muons.

In the fully leptonic decay channel, the kinematic
cuts we employ \cite{barron} to distinguish signal
from background are very similar to the slepton searches described in
Section~\ref{Sec:sleptons} above. As in the case of smuon production,
one searches for a structure above the SM background in the final state
muon energy distribution. Here, however, the signal distribution is not
expected to display the by-now familiar shelf-like behavior, due to the
3-body nature of the chargino decay. Examples of the muon energy
spectra for some representative AKTW models are shown in Fig.~\ref{emumucharg}
for both beam polarization states. Here, we see that the $E_\mu$
spectrum for the signal varies greatly in size and shape between the various
models depending on the value of $\Delta m_{\tilde\chi}$ and the
production cross section, but is nonetheless
clearly separable from the SM background for these cases. Comparing
with Fig.~\ref{smuonfig1} we note that the SM background has a similar
shape, but is slightly larger throughout the spectrum in this analysis.
As usual, RH electron beam polarization leads to a smaller SM background
since the $t$-channel contribution to $W$ pair production is suppressed
in this case.

Of the 53 models with kinematically accessible $\tilde\chi_1^\pm$ states,
we find that 12(11) lead to visible signals over the background at a
significance of ${\cal S}>5$ for RH(LH) electron beam polarization.
Combining the two polarization channels, a total of 14/53 models meet
our visibility criteria. From Fig.~\ref{emumucharg} we note, however,
that $S/B$ can be small enough in some cases to render a detailed
study of the chargino properties difficult.

SUSY can be a substantial background to itself in this channel,
with smuon production being a particularly large background source.
We find that 14(12) models yield fake signals that pass our
visibility criteria for a RH(LH) polarization configuration.
We note that in all cases, the counterfeit signal indeed
arises from the production of smuons.
The muon energy distribution for some representative
examples of such misleading signals
are presented in Fig.~\ref{emumuchargfakes}. From the Figure, we see
that the signal tends to have a shelf-like behavior, as would be expected
for smuon production, and thus looks quite different than the case
of chargino production.

\begin{figure*}[hptb]
\centerline{
\includegraphics[width=13.0cm,height=10.0cm,angle=0]{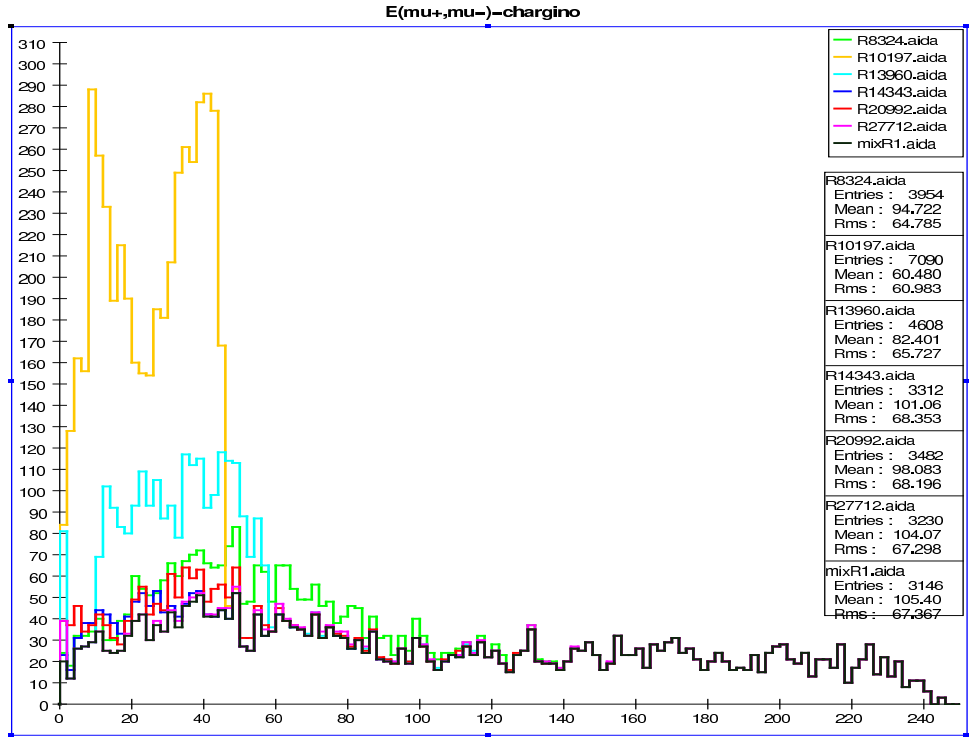}}
\vspace*{0.1cm}
\centerline{
\includegraphics[width=13.0cm,height=10.0cm,angle=0]{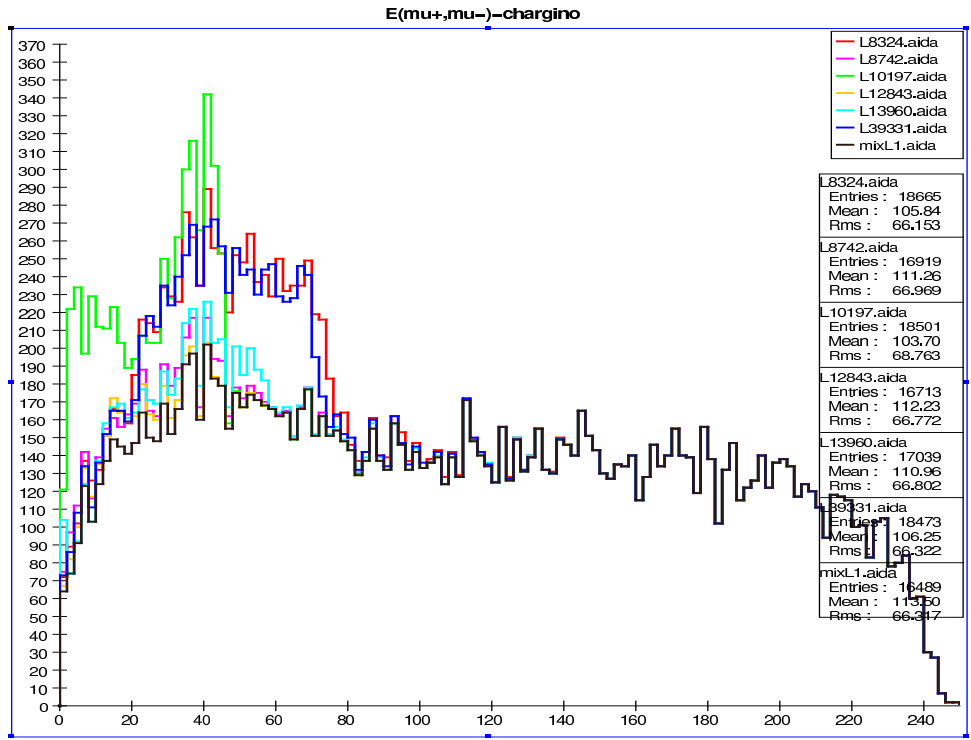}}
\vspace*{-0.1cm}
\caption{Muon energy distribution: the number of events/2 GeV bin
after imposing the full set of cuts discussed in the text for
chargino production for representative
models which are visible in this channel.
RH(LH) beam polarization is employed in the top(bottom)
panel, assuming
an integrated luminosity of 250 fb$^{-1}$
for either polarization. The SM background is shown as the black histogram.}
\label{emumucharg}
\end{figure*}

\begin{figure*}[hptb]
\centerline{
\includegraphics[width=13.0cm,height=10.0cm,angle=0]{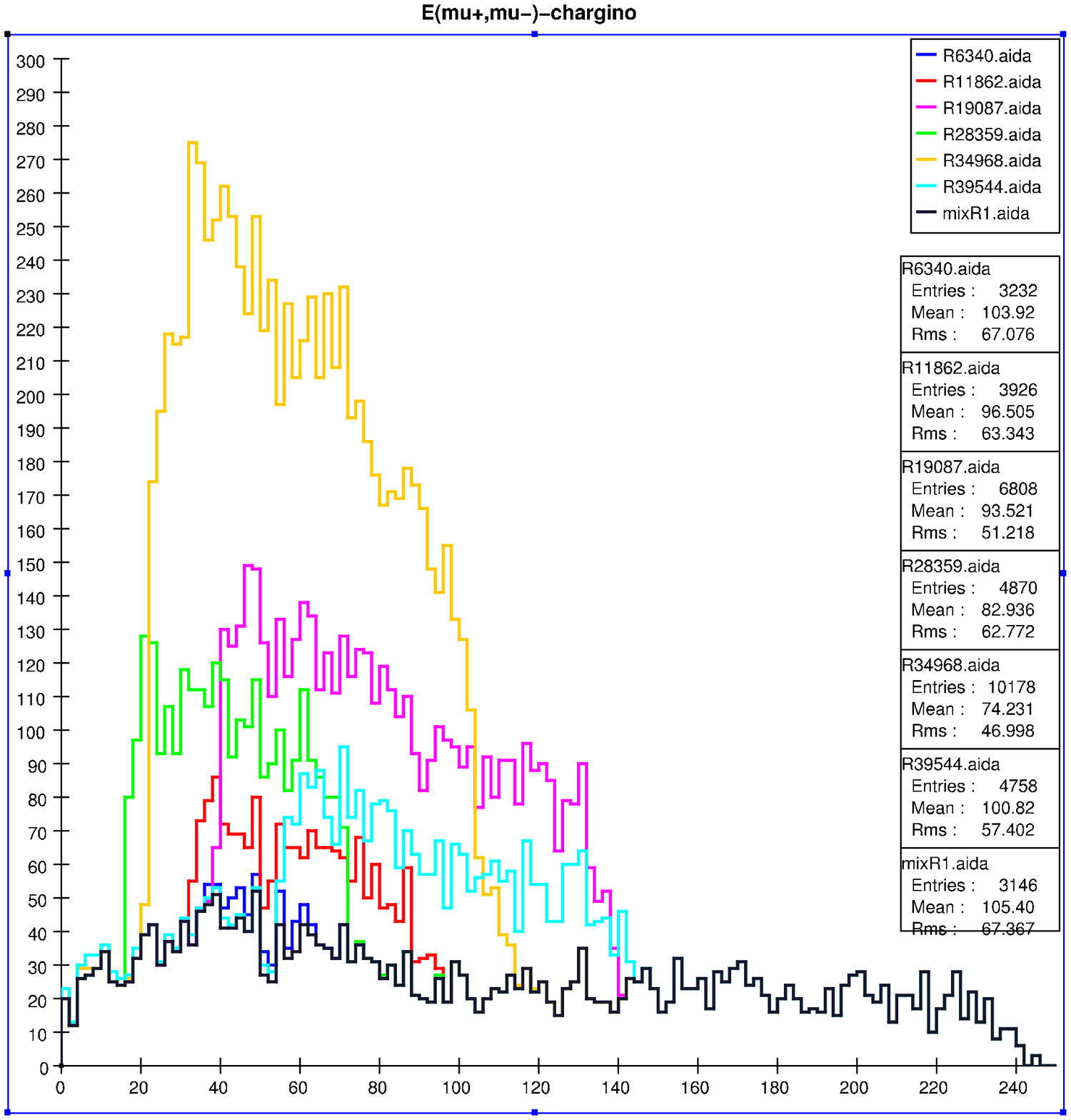}}
\vspace*{0.1cm}
\centerline{
\includegraphics[width=13.0cm,height=10.0cm,angle=0]{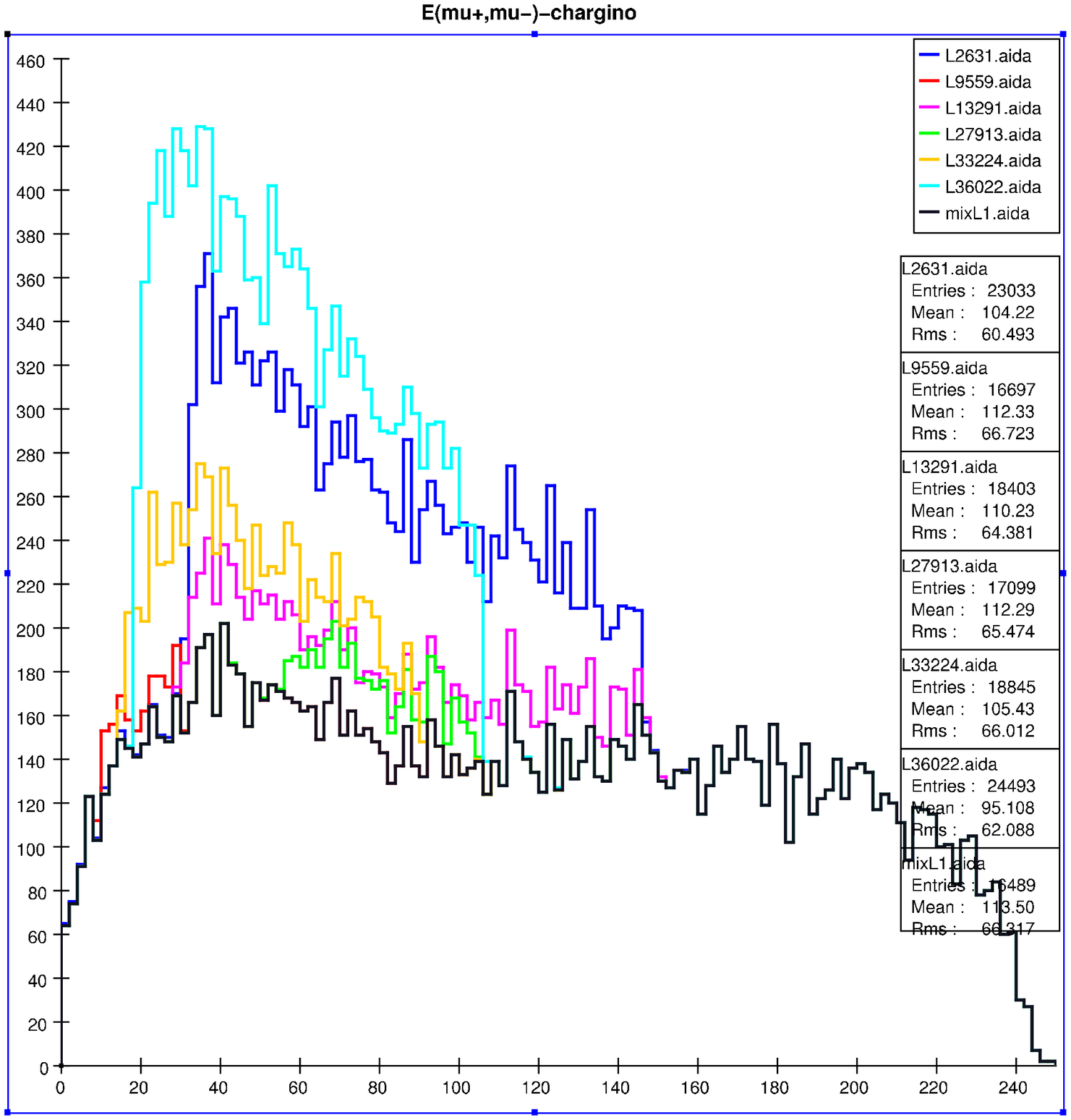}}
\vspace*{-0.1cm}
\caption{Muon energy distribution: the number of events/2 GeV bin
after imposing the full set of cuts discussed in the text for representative
models which fake a chargino signal in this channel.
RH(LH) beam polarization is employed in the top(bottom)
panel, assuming
an integrated luminosity of 250 fb$^{-1}$
for either polarization. The SM background is shown as the black histogram.}
\label{emumuchargfakes}
\end{figure*}

We now turn to the fully hadronic channel, where the final state is 4 jets
plus missing energy. We employ the following kinematic cuts
(based on, \eg, \cite{Abbiendi:2003sc}):
\begin{enumerate}

\item There be precisely 4 jets in the final state and no other charged
particles.

\item No tracks (or clusters) be present below an angle of 100 mrad.
This reduces photon-initiated backgrounds.

\item Missing energy is constrained to be $> 0.5 \sqrt{s}$. This favors
the signal, which contains a large amount of missing energy compared to
many background sources.

\item In the forward direction, the visible energy is constrained to be
$\Evis < \frac{1}{2} \sqrt{s} - m_{\LSP,\mbox{\tiny min}}$.
We, again, take $m_{\LSP,\mbox{\tiny min}} = 46$ GeV.
However, in order to estimate the effect on the background
if this bound is increased, for example by future studies at the LHC,
we perform a second analysis with $m_{\LSP,\mbox{\tiny min}} = 100$ GeV.

\item We require precisely two jets in each hemisphere as determined
by the thrust axis.
This cut eliminates jets stemming from $\tau$ decays arising from
tau pair production, where
one $\tau$ has a one-prong decay,
and the other is 3-prong.

\item We reconstruct the
off-shell $W$ bosons by coalescing the 4 jets into 2, one for each
$W$ boson. We force this by adjusting the
$y_{\mbox{\tiny cut}}$ parameter of the JADE jet finding algorithm
until two jets are found (see Section~\ref{Sec:analysis} for details
of the algorithm), and then require
the resulting dijet invariant masses to be $> 2$ GeV.

\end{enumerate}

The first observable we consider for this channel is the missing energy
distribution, where we expect a peak from the signal at large values of
$E_{\rm miss}$. This spectrum is presented in Fig.~\ref{me4jets}
where the black histogram corresponds to the SM background, as usual.
We find that a particularly troublesome background arises from the
process $\gamma \gamma \rightarrow q\bar q$.
It is clear from the Figure that this background reaction stubbornly
yields a significant event rate even after the above cuts are imposed.
The missing energy spectra for representative AKTW
models are also shown in the Figure
for both polarization states and by eye seem barely visible above the
background. However, the statistical sample is large, and
nonetheless, chargino production in several AKTW models are observable
in this channel. We find that 9(31)
models are observable with significance $>5$ in the RH(LH) polarization
state. All models that are visible with RH beam polarization are also
observed with LH beam polarization. If we increase the minimal value
of the bound on the LSP mass to 100 GeV as mentioned in the kinematic
cuts above,
we find that 4 additional models satisfy our visibility criteria. We
note that the production of other Supersymmetric particles do not
pass our cuts in this channel and hence there are no `fake' signals.

\begin{figure*}[hptb]
\centerline{
\includegraphics[width=13.0cm,height=10.0cm,angle=0]{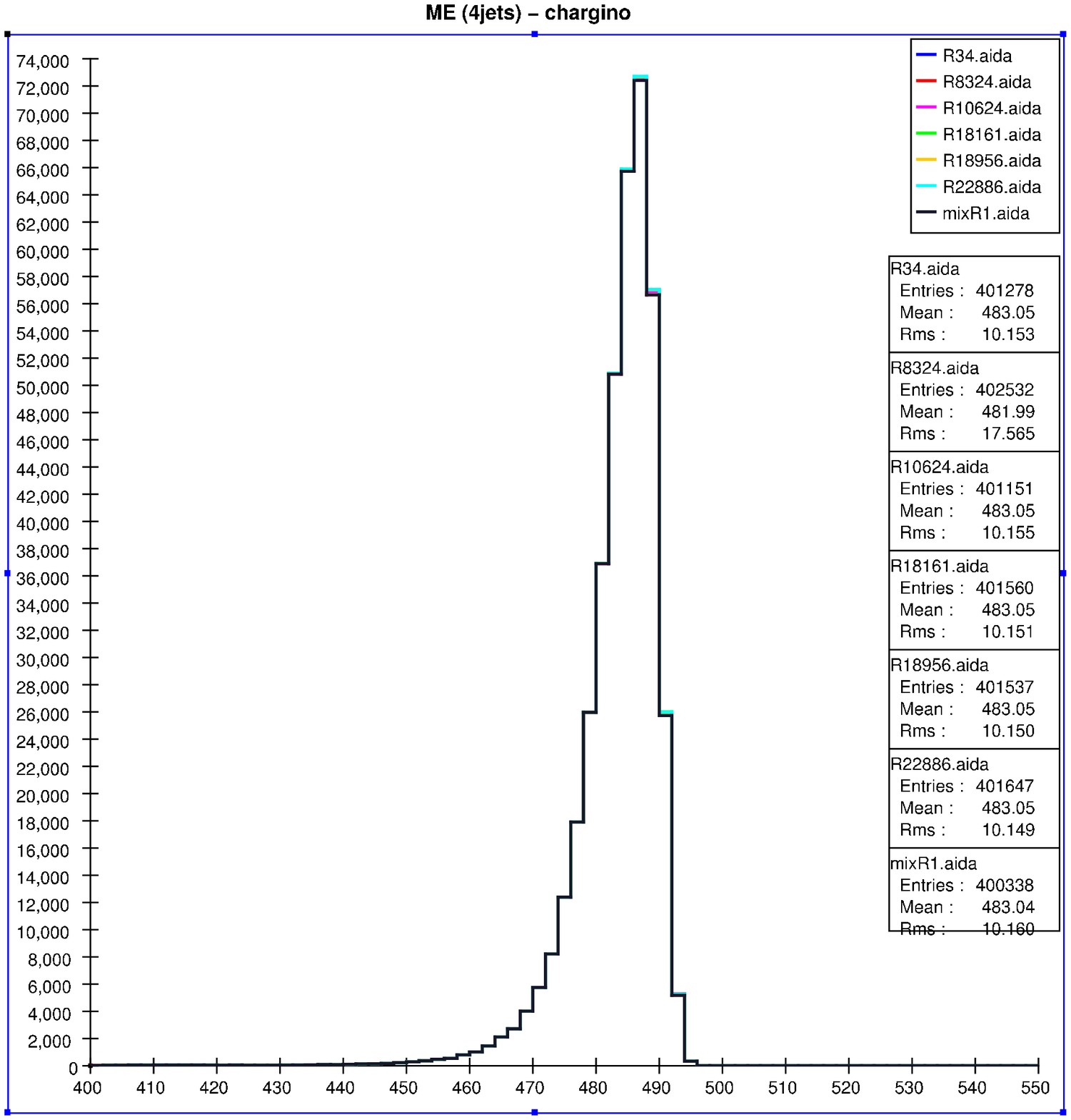}}
\vspace*{0.1cm}
\centerline{
\includegraphics[width=13.0cm,height=10.0cm,angle=0]{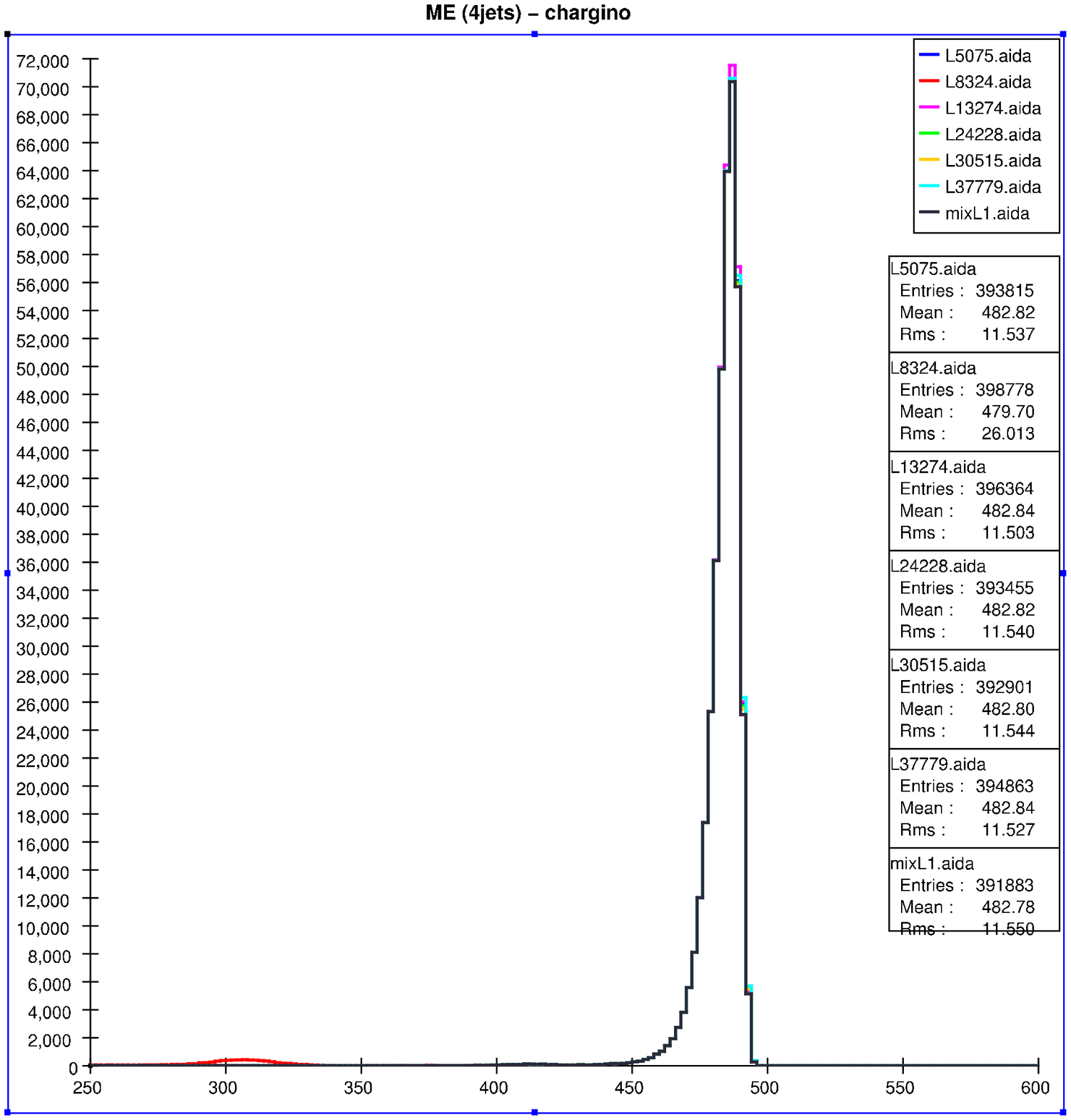}}
\vspace*{-0.1cm}
\caption{Missing energy distribution: the number of events/2 GeV bin
after imposing the full set of cuts discussed in the text for the
fully hadronic chargino decay channel for representative
models which are visible in this channel.
RH(LH) beam polarization is employed in the top(bottom)
panel, assuming
an integrated luminosity of 250 fb$^{-1}$
for either polarization. The SM background is shown as the black histogram.}
\label{me4jets}
\end{figure*}

We next examine the distribution of the two dijet invariant masses
to see if the photon-induced background is less problematic for this
observable.
Figure~\ref{fig:4jetbg} displays the SM background for this case for
both electron beam polarizations. Here, we see that the background from
two-photon initiated processes is still significant, with the dominant
channel passing the cuts being $\gamma\gamma\to q \bar q$. The results
for the AKTW models are presented in Fig.~\ref{mjetpairs} for
both electron beam polarizations. We find that with RH polarization
none of the AKTW models with chargino decays into off-shell $W$ bosons
are visible over the background in this observable.
In the case of LH electron beam polarization, we find that
2 such models (labelled as 12843 and 14343)
are detectable with a significance $>5$. These models
yield an excess of events in the first two bins of the distribution;
this excess is not visible by eye, but is statistically significant due
to the large sample size. The two
models with chargino decays into on-shell $W$ bosons (labelled as 8324 and
39331) display a clear signal for both beam polarizations as shown in
the Figure. We see
that the invariant $M_{jj}$ spectrum is broader for these two models
and yields a high event rate at large invariant masses.
If the minimum value of the LSP mass is raised to 100 GeV in our
kinematic cuts, we find that
only the on-shell $W$ boson decays are visible above the background.
One additional
model (labelled as 1822) is also distinguishable from the background.
In this case, however, it is due to the pair production of $\tilde\chi_2^0$
states, with their subsequent decays into $Z+\tilde\chi_1^0\to
2 {\rm jets} + \tilde\chi_1^0$, that passes our
cuts and provides a fake signal.

\begin{figure*}[hpb]
\centerline{
\epsfig{figure=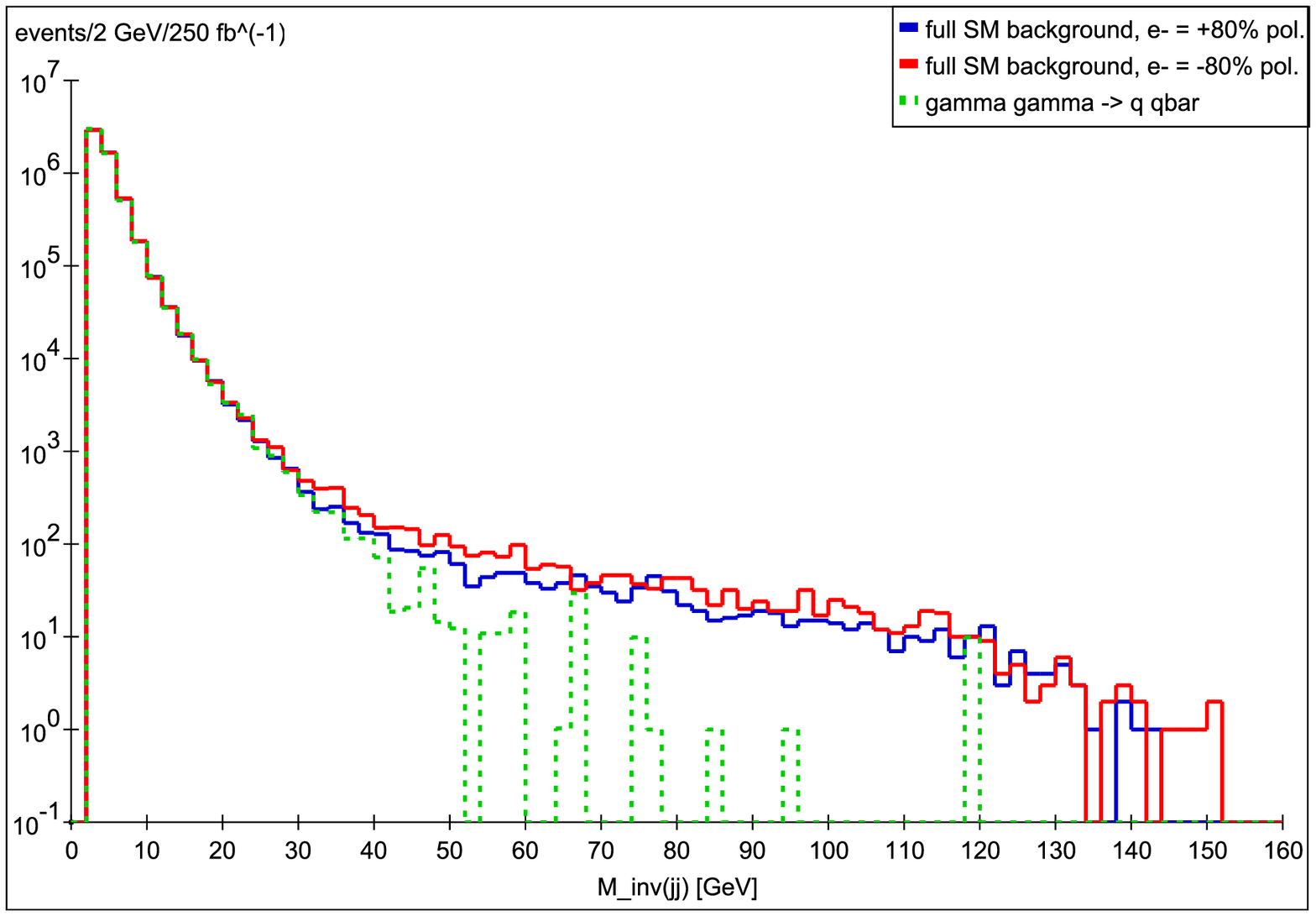,width=13cm,clip=}}
\vspace*{0.1cm}
\caption{Distribution of dijet invariant masses
from the remaining SM background after the chargino 4-jet
selection cuts listed in the text have been imposed. This is generated for
$250 \mbox{ fb}^{-1}$ of SM events with
80\% right-handed (solid blue line) and 80\% left-handed (solid red line)
electron beam polarization, and unpolarized positron beam
at $\sqrt{s} = 500$ GeV. The dashed green line shows the
main processes contributing to the
background, $\gamma \gamma \rightarrow
q \bar{q}$, which is independent of the beam polarization.}
\label{fig:4jetbg}
\end{figure*}

\begin{figure*}[hptb]
\centerline{
\includegraphics[width=13.0cm,height=10.0cm,angle=0]{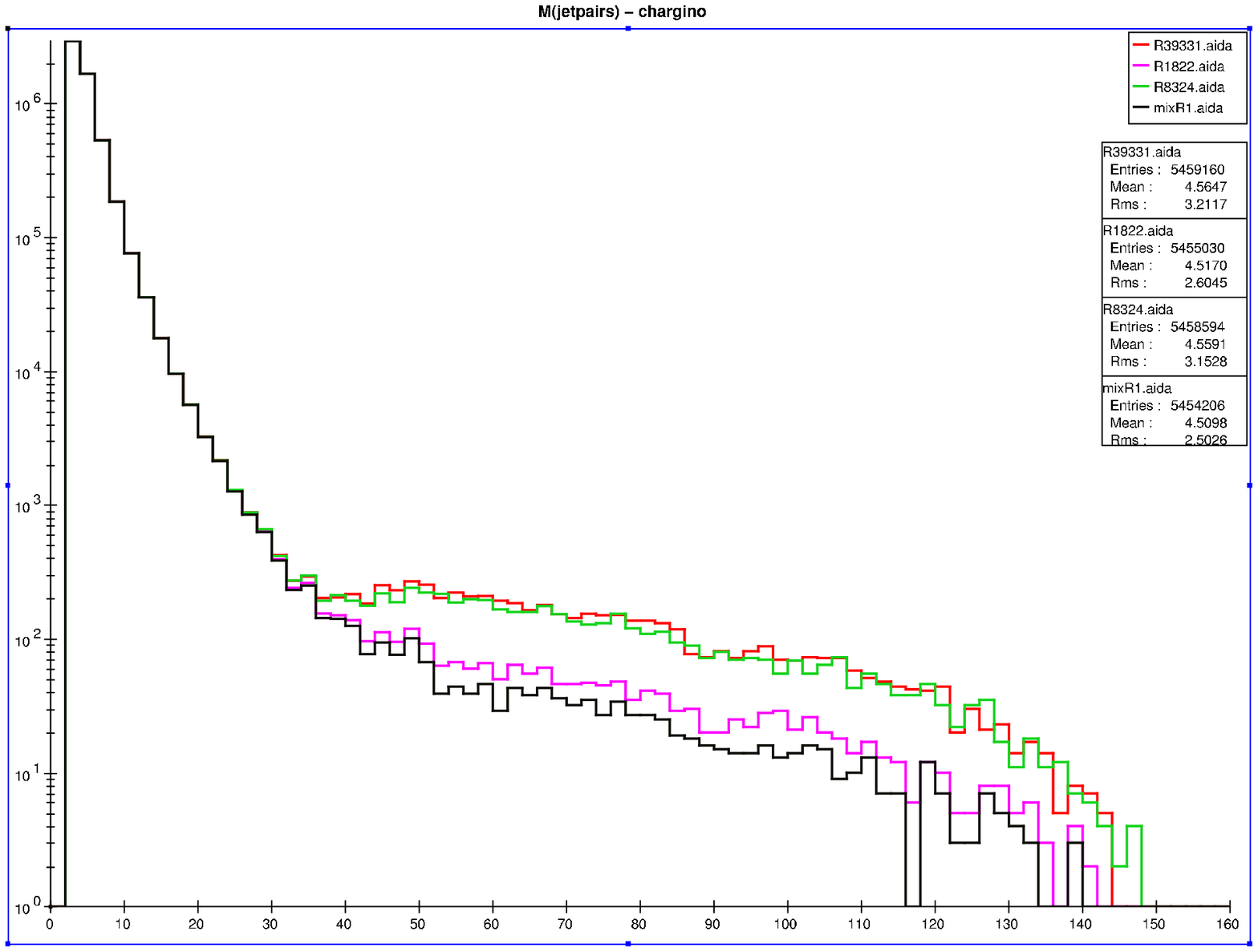}}
\vspace*{0.1cm}
\centerline{
\includegraphics[width=13.0cm,height=10.0cm,angle=0]{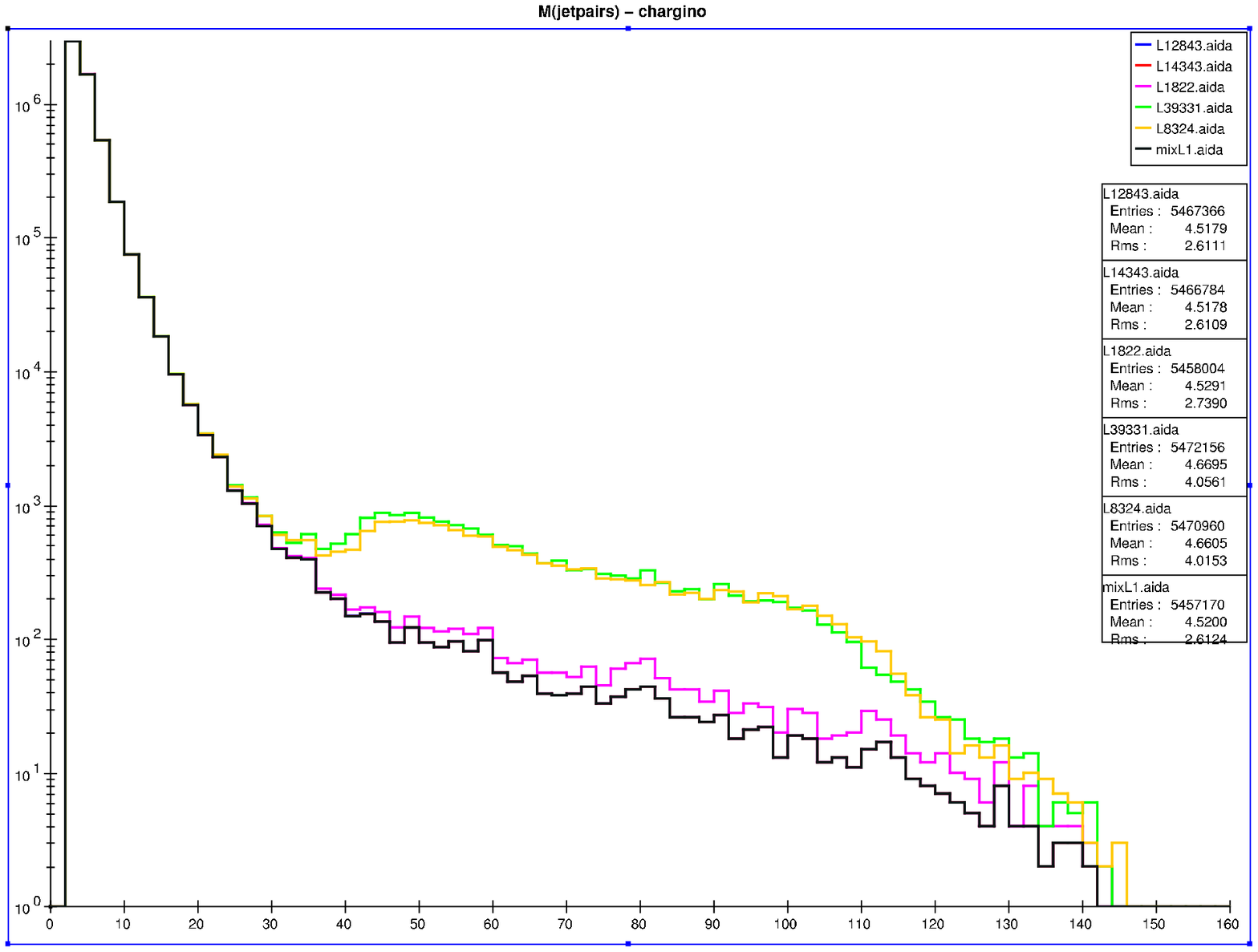}}
\vspace*{-0.1cm}
\caption{Jet-pair invariant mass distribution: the number of events/2 GeV bin
after imposing the full set of cuts discussed in the text for the
fully hadronic decays of the chargino for representative
models which are visible in this channel.
RH(LH) beam polarization is employed in the top(bottom)
panel, assuming
an integrated luminosity of 250 fb$^{-1}$
for either polarization. The SM background is shown as the black histogram.}
\label{mjetpairs}
\end{figure*}

Further attempts to decrease the SM background in the 4-jet channel
prove to be difficult and tend to
remove the signal as well as the background. This is because the
characteristics of the
signal and remaining background are similar for variables such as the
missing energy spectrum, acoplanarity, and
$p_{T \mbox{\tiny vis}}$ (see Fig.~\ref{fig:pTchargino}).
In particular, both signal and background distributions peak at low values of
$p_{T \mbox{\tiny vis}}$ and acoplanarity and at
high values of missing energy as discussed above. Previous searches for
charginos in the literature (\eg, \cite{Abbiendi:2003sc})
have employed additional cuts on $p_{T \mbox{\tiny vis}}$
and/or acoplanarity. In particular, we find that an additional restriction
on the transverse momentum,
\begin{enumerate}
\setcounter{enumi}{6}
\item $p_{T \mbox{\tiny vis}} > 0.06 \sqrt{s}$
\end{enumerate}
effectively reduces the background as shown in Fig.~\ref{fig:4jetpTcutbg},
but also removes the signal for all of the AKTW models with chargino decays
into off-shell $W$ bosons. We show the effect of this additional cut
on transverse momentum in the missing energy and jetpair invariant mass
distributions in Figs.~\ref{ME_ptcut} and \ref{Mjj_ptcut} for both electron
beam polarizations. Here, we display all of the AKTW models which yield
a visible signal with significance $>5$.
We see that the signal for the models where
the chargino decays into on-shell $W$ bosons (labelled as 8324 and 39331)
towers above the background for both observables with both beam
polarizations. As we saw above, the visible signal for model 1822
is due to the production of $\tilde\chi_2^0$ states and is thus fake.
We emphasize that none of the AKTW models with chargino decays into
off-shell $W$ bosons are observable once this additional $p_T$ cut
is applied.

\begin{figure*}[hptb]
\centerline{
\epsfig{figure=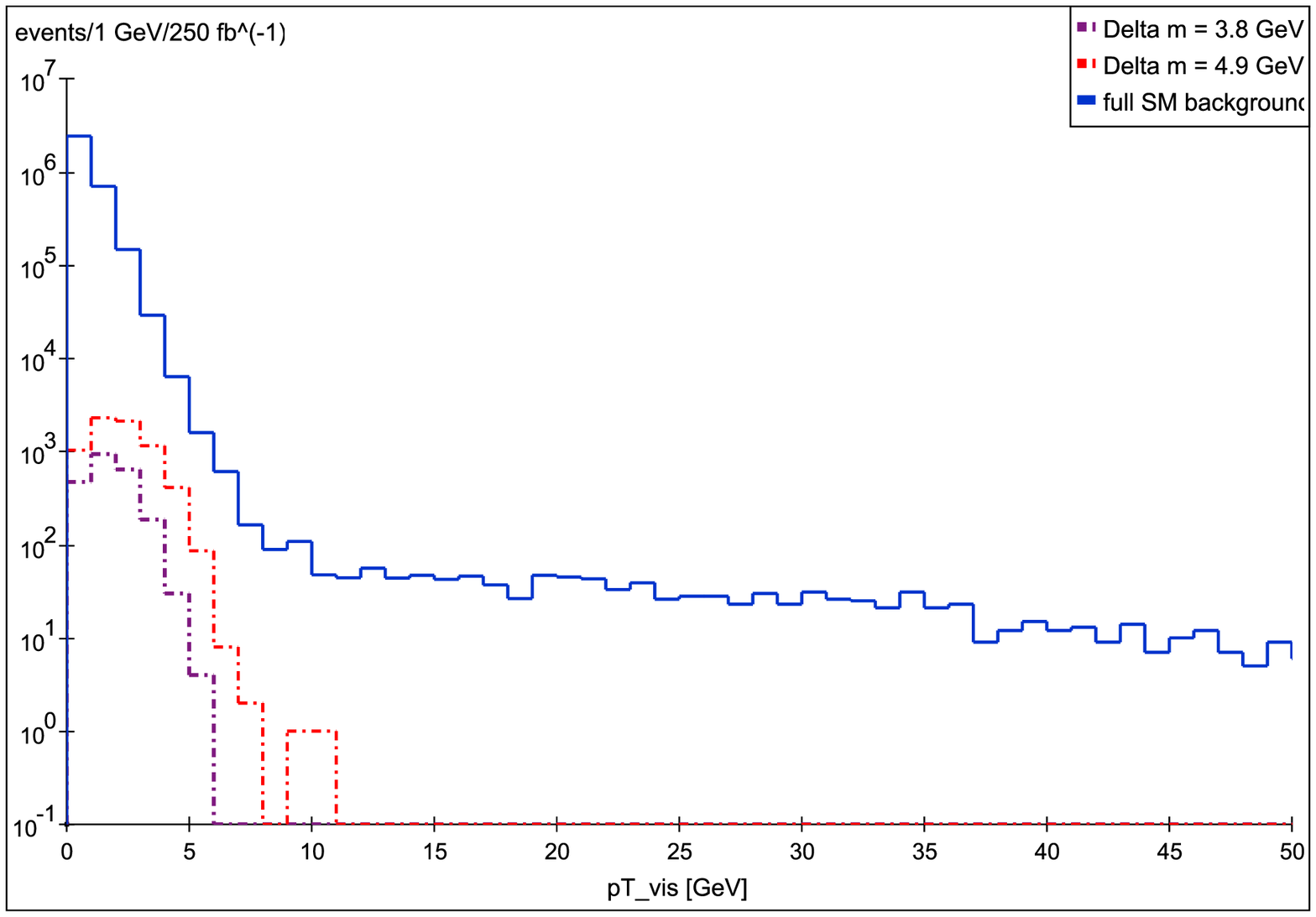,width=13cm,clip=}}
\vspace*{0.1cm}
\caption{Transverse momentum distribution of remaining
SM background after the chargino 4-jet
selection cuts listed in the text have been imposed. This is generated
from $250 \mbox{ fb}^{-1}$ of SM events with
80\% left-handed
electron beam polarization, and unpolarized positron beam
at $\sqrt{s} = 500$ GeV (solid blue line).
The dashed purple and red lines show
signal events produced in two AKTW models that are representative
for the class of models with $\Delta
m_{\tilde{\chi}}$ of the order of a few GeV.}
\label{fig:pTchargino}
\end{figure*}

\begin{figure*}[hptb]
\centerline{
\epsfig{figure=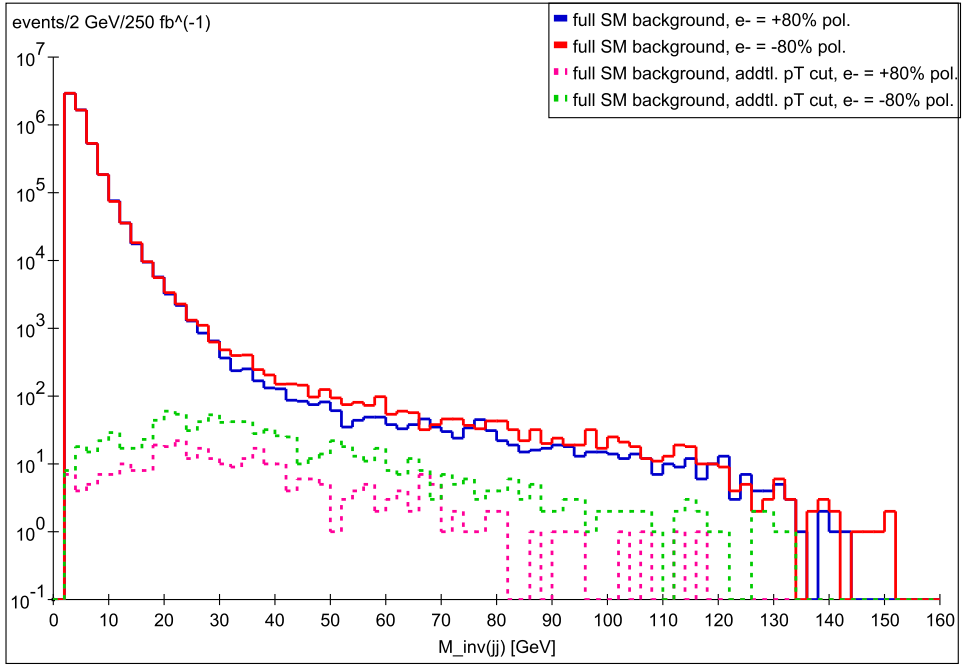,width=13cm,clip=}}
\vspace*{0.1cm}
\caption{Distribution of dijet invariant masses
from the remaining SM background after the chargino 4-jet
selection cuts listed in the text have been imposed with the additional cut on
transverse momentum. This is generated from
$250 \mbox{ fb}^{-1}$ of SM events with
80\% right-handed (dashed pink line) and 80\% left-handed (dashed green line)
electron beam polarization, and unpolarized positron beam
at $\sqrt{s} = 500$ GeV. The solid lines are as
in fig.~\ref{fig:4jetbg}.}
\label{fig:4jetpTcutbg}
\end{figure*}

\begin{figure*}[hptb]
\centerline{
\includegraphics[width=13.0cm,height=10.0cm,angle=0]{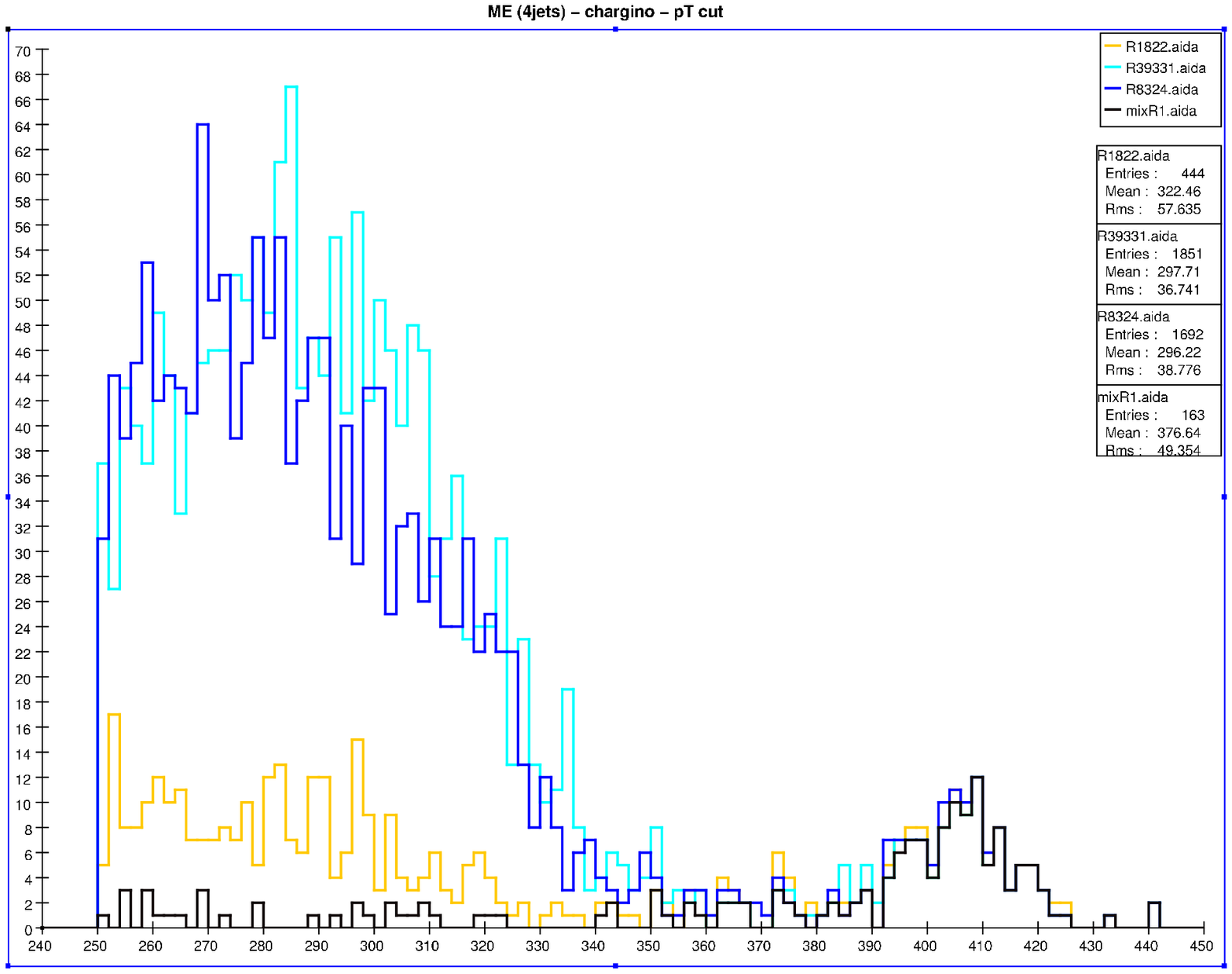}}
\vspace*{0.1cm}
\centerline{
\includegraphics[width=13.0cm,height=10.0cm,angle=0]{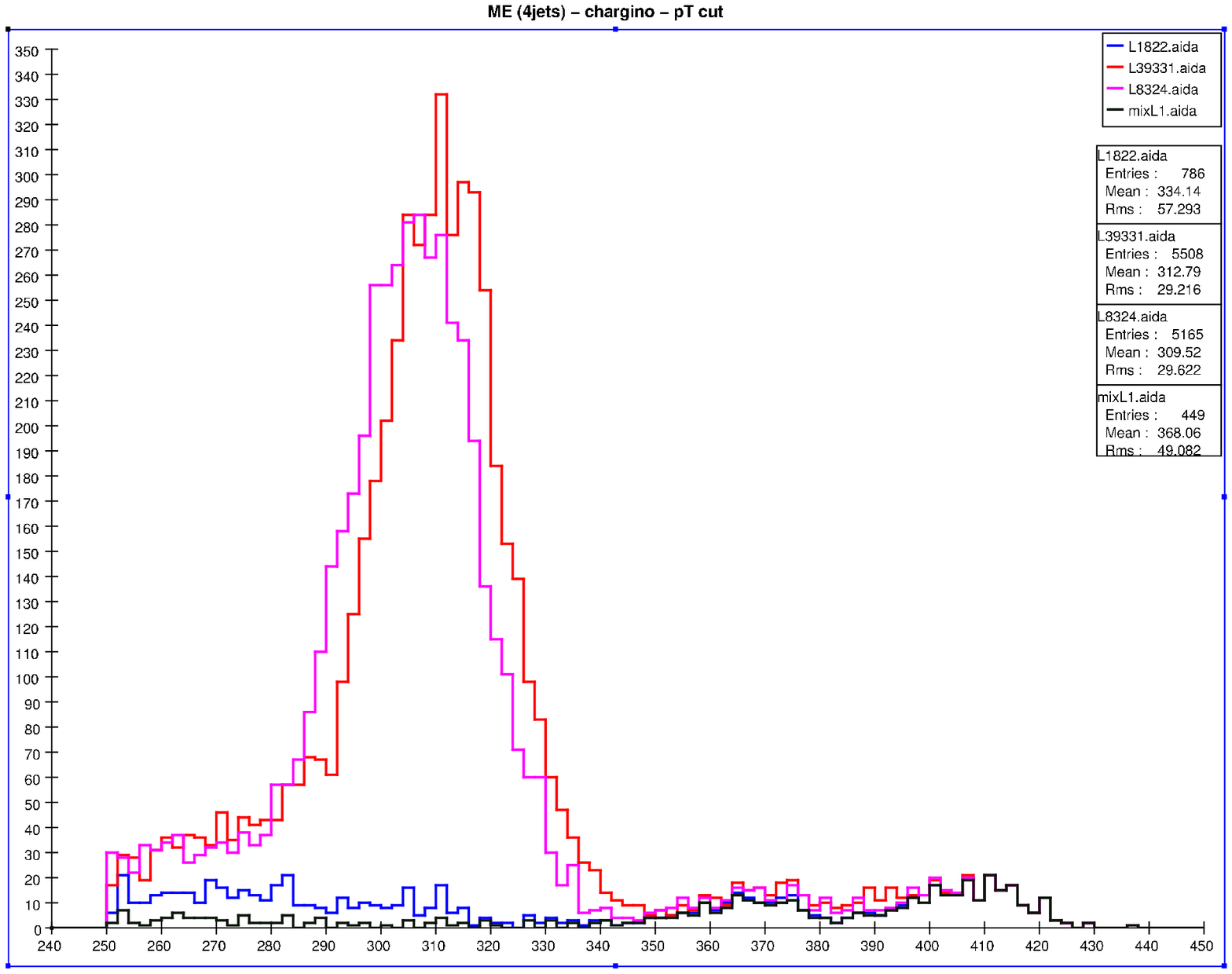}}
\vspace*{-0.1cm}
\caption{Missing energy distribution: the number of events/2 GeV bin
after imposing the full set of cuts discussed in the text for the fully
hadronic chargino decay channel, including an additional cut on transverse
momentum, for representative
models which are visible in this channel.
RH(LH) beam polarization is employed in the top(bottom)
panel, assuming
an integrated luminosity of 250 fb$^{-1}$
for either polarization. The SM background is shown as the black histogram.}
\label{ME_ptcut}
\end{figure*}

\begin{figure*}[hptb]
\centerline{
\includegraphics[width=13.0cm,height=10.0cm,angle=0]{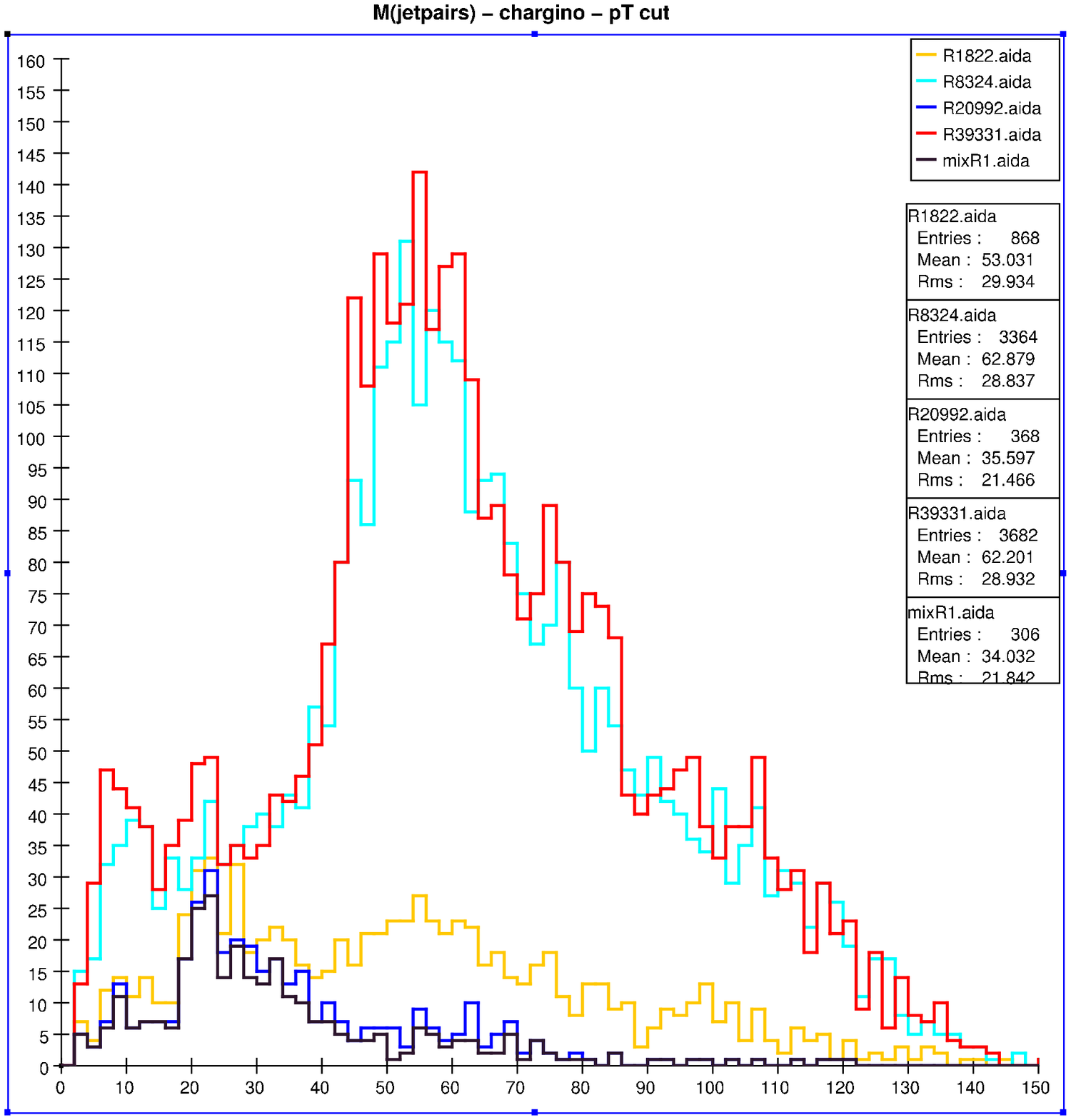}}
\vspace*{0.1cm}
\centerline{
\includegraphics[width=13.0cm,height=10.0cm,angle=0]{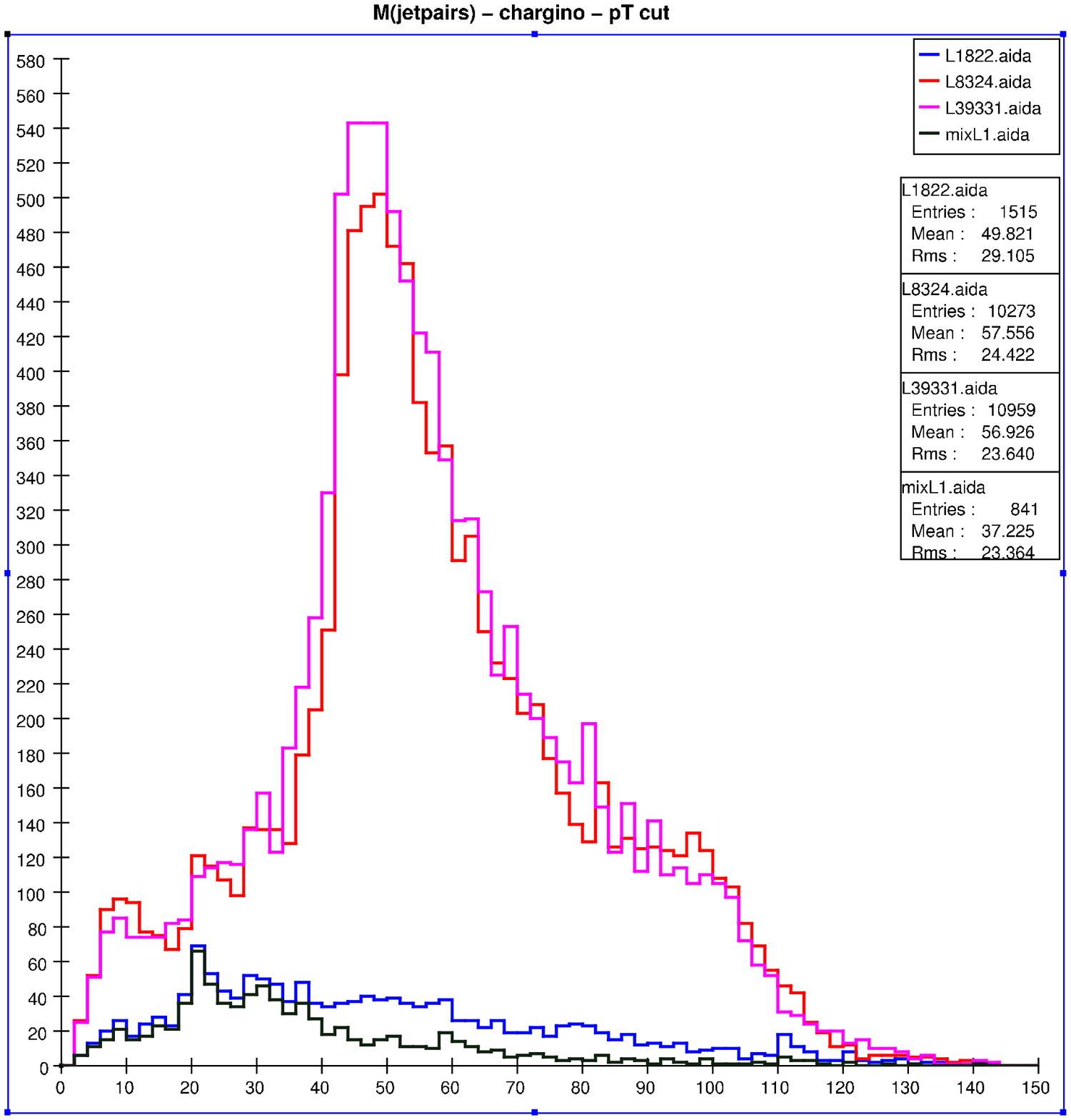}}
\vspace*{-0.1cm}
\caption{Jet-pair invariant mass distribution: the number of events/2 GeV bin
after imposing the full set of cuts discussed in the text for the fully
hadronic chargino decay channel, including an additional cut on transverse
momentum, for representative
models which are visible in this channel.
RH(LH) beam polarization is employed in the top(bottom)
panel, assuming
an integrated luminosity of 250 fb$^{-1}$
for either polarization. The SM background is shown as the black histogram.}
\label{Mjj_ptcut}
\end{figure*}

Lastly, we examine the mixed decay channel,
\be
\chap \cham \rightarrow q\bar q\tilde\chi_1^0 +\mu\bar\nu_\mu\tilde\chi_1^0
\, ,
\ee
which can proceed via (if kinematics allow)
\be
\chap\cham\to W^{*\,\pm}\LSP+ \mu^{\mp} \sneumu ,
W^{*\,\pm} \LSP+\tilde \mu^{\mp} \nu_\mu
\ee
with
\be
W^{*\,\pm} \rightarrow q \bar{q} \, ,
\ee
or via
\be
\chap \cham \rightarrow W^{*\,+} W^{*\,-} \, ,
\ee
where one of the virtual $W$ bosons decays hadronically while the other
decays leptonically into a muon.

For this channel, we employ the cuts
\begin{enumerate}
\item There be 2 jets (with no muonic component) plus one muon with no
other visible particles in the final state.

\item There be no tracks or clusters of energy within 100 mrad of the
beampipe as the signal is peaked at wide angles.

\item The visible energy satisfy
$\Evis < \frac{1}{2} \sqrt{s} - m_{\LSP,\mbox{\tiny min}}$
in the forward direction.
As above, we take $m_{\LSP,\mbox{\tiny min}} = 46$ GeV~\cite{Yao:2006px}.
However, in order to estimate the effect on the background if
this bound is increased, we perform a second analysis with
$m_{\LSP,\mbox{\tiny min}} = 100$ GeV.

\item The invariant mass of the jet-pair be larger than 2.4 GeV.
This cut is used to eliminate jets stemming from $\tau$ decays.
\end{enumerate}

Here, we examine the energy and invariant mass of the jet-pair.
We find that the same issues discussed above in the 4-jet channel
regarding potential additional cuts based on acoplanarity and
transverse momentum are also relevant in this case and thus these cuts are
not employed in our analysis. The background remaining after our cuts
are imposed is presented in Fig.~\ref{fig:2jetmubg} for the invariant
mass distribution, where
the dominant remaining SM background processes are $\gamma \gamma
\rightarrow q \bar{q},\, \gamma \gamma \rightarrow
\tau^+ \tau^-$ and $e^+e^- \rightarrow q
\bar{q}' l \bar{\nu}_l$. We see that the background distribution is roughly
the same for both beam polarizations as is to be expected for
$\gamma\gamma$ induced processes.

\begin{figure*}[hptb]
\centerline{
\epsfig{figure=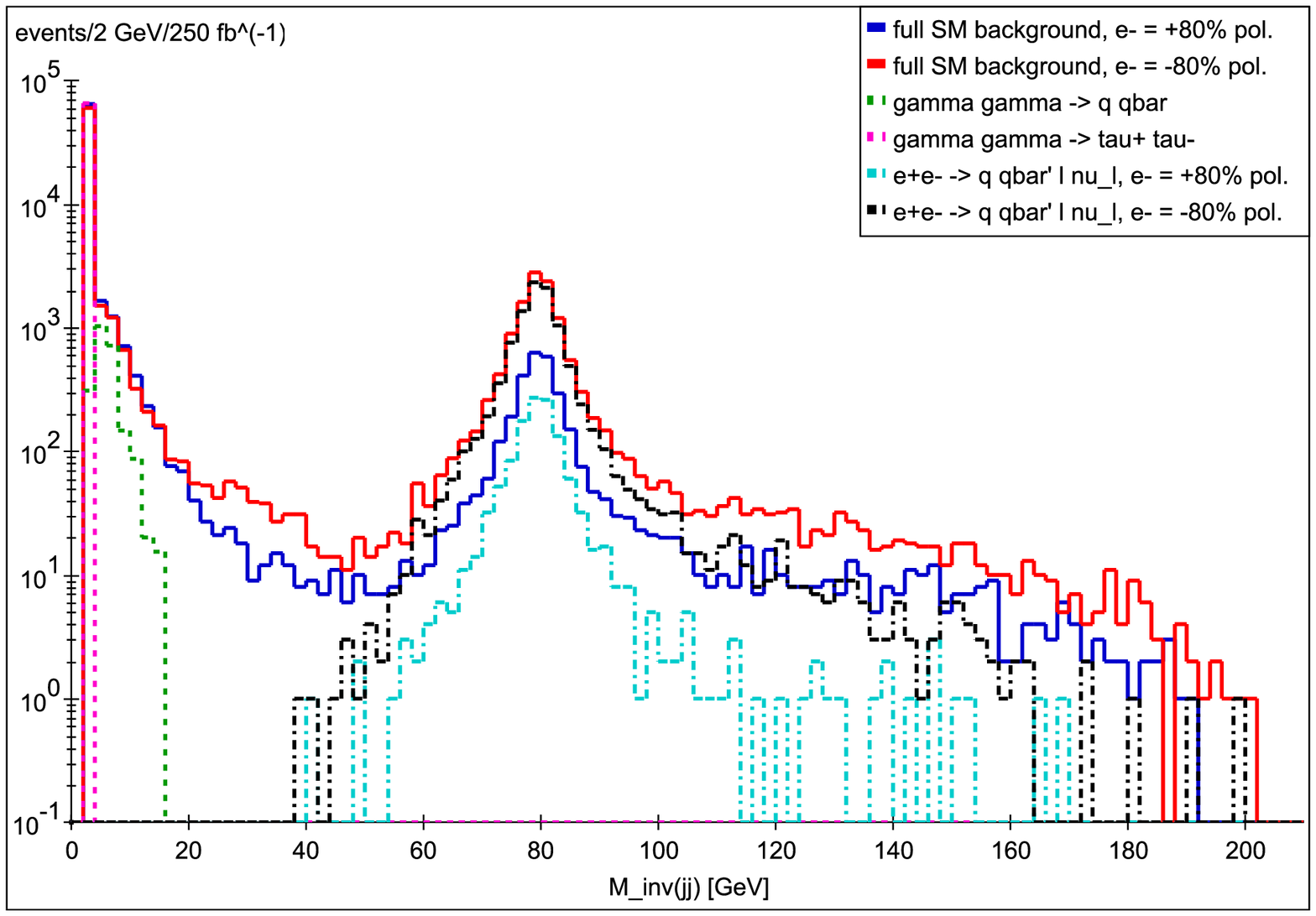,width=13cm,clip=}}
\vspace*{0.1cm}
\caption{Distribution of dijet invariant mass
from the remaining SM background after the chargino 2-jet plus muon
selection cuts listed in the text have been imposed. This is generated from
$250 \mbox{ fb}^{-1}$ of SM events with
80\% right-handed (solid blue line) and 80\% left-handed (solid red line)
electron beam polarization, and unpolarized positron beam
at $\sqrt{s} = 500$ GeV. The other dashed, and
solid lines show the main processes contributing to the
background, $\gamma \gamma \rightarrow q \bar{q}$ (dashed green line),
$\gamma \gamma \rightarrow
\tau^+ \tau^-$ (dashed pink line), which are independent of beam
polarization, and
$e^+e^- \rightarrow q
\bar{q}' l \bar{\nu}_l$,
for 80\% right-handed (solid cyan line) and 80\% left-handed (solid
black line) electron polarization.}
\label{fig:2jetmubg}
\end{figure*}

The invariant mass spectrum for the case where the charginos decay to
off-shell $W$ bosons is displayed in Fig.~\ref{mjpairmu_off} for both
beam polarizations. In both cases, the signal rises above the
background in the region of smaller ($<60$ GeV) invariant masses of the
jet pair, as is expected due to the off-shell nature of the
$W$ bosons. The model labelled as 1822 also shows a visible
signature, with a peak located at $M_{jj}\sim 80-90$ GeV. As we saw
above, this is due to $\tilde\chi_2^0$ production in this model with the
subsequent decay $\tilde\chi_2^0\to Z+\tilde\chi_1^0$ with the $Z$
decaying hadronically and is a false signal. The $M_{jj}$ distribution
for the 2 AKTW models where the charginos decay to on-shell $W$ bosons is
shown in Fig.~\ref{mjpairmu_on}. Here, we would expect to see a peak
above the SM background
in the distribution around $M_W$, and indeed, that is the case.
In summary,
we find that 23(35) of the AKTW models with kinematically
accessible charginos lead to signals in this observable with a visibility
significance ${\cal S}>5$ for RH(LH)
electron beam polarization at these integrated luminosities.
We note that none of the AKTW models are visible over the background
in the case where the minimum value of the LSP mass is increased to 100 GeV
as described in our cuts.

The second observable we examine in this analysis is the energy of the
jet pair which is displayed in Fig.~\ref{ejpairmu_off} for several AKTW
models where the chargino decays to an off-shell $W$ boson. Again, we
see that the signal rises above the background for lower values
of $E_{jj}$ ($\lsim M_W$), except for the case of model 1822 which is
a fake as described above. For comparison, the results for the two
models which have decays to on-shell $W$ bosons are shown in
Fig.~\ref{ejpairmu_on}, where we see that the $M_{jj}$ spectrum is
peaked at larger values in this case. For this observable, we
find that 26(35) of the AKTW models meet our visibility criterion.
We note that two more models are visible with RH polarization
in this observable compared to the $M_{jj}$ distribution discussed
above. Again, none of the models are visible when the minimum
value of the LSP mass is increased in the analysis.

Except for model 1822, we note that there
are no fake signals from the production of other SUSY particles
for either observable in this channel.

\begin{figure*}[hptb]
\centerline{
\includegraphics[width=13.0cm,height=10.0cm,angle=0]{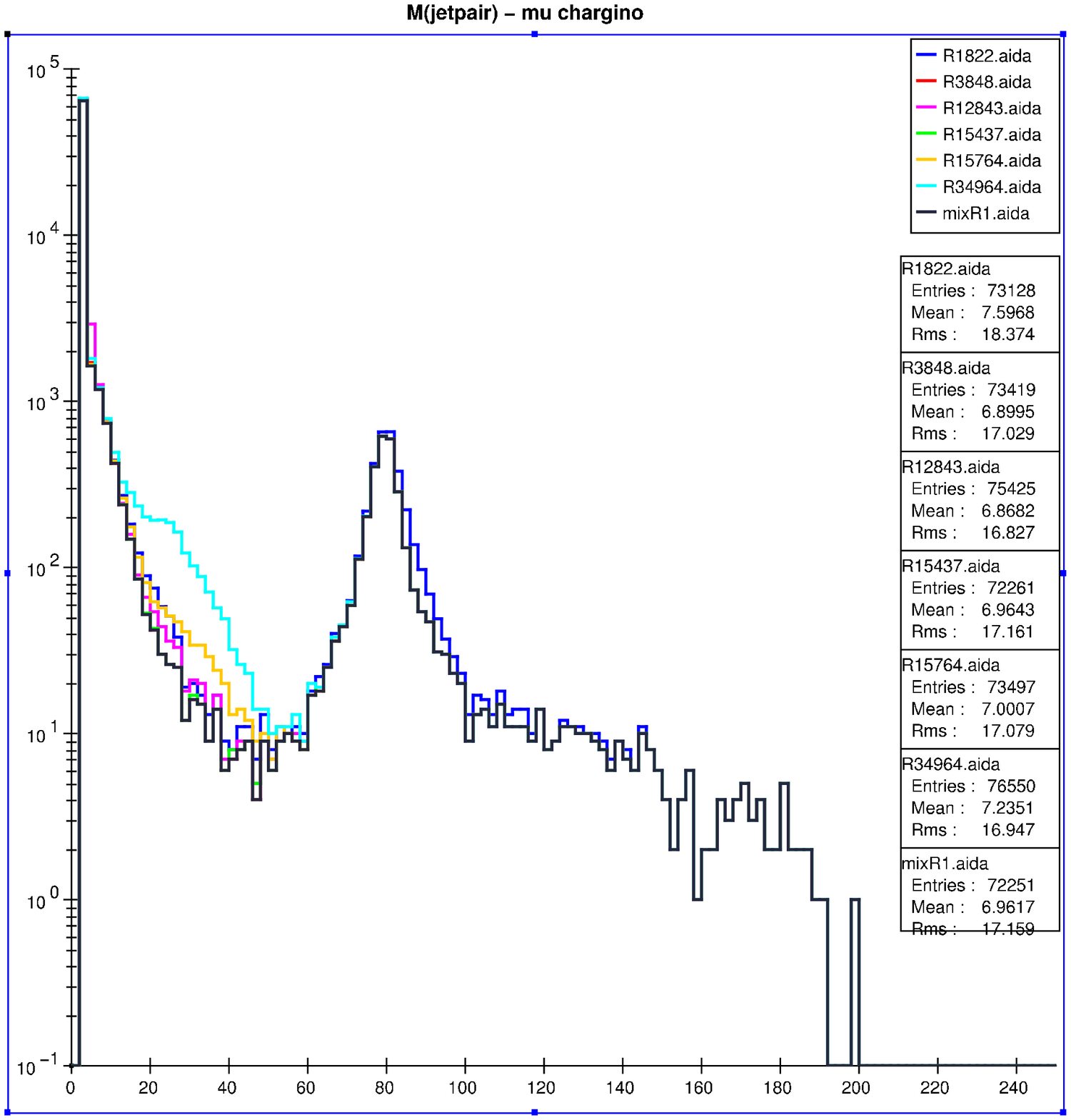}}
\vspace*{0.1cm}
\centerline{
\includegraphics[width=13.0cm,height=10.0cm,angle=0]{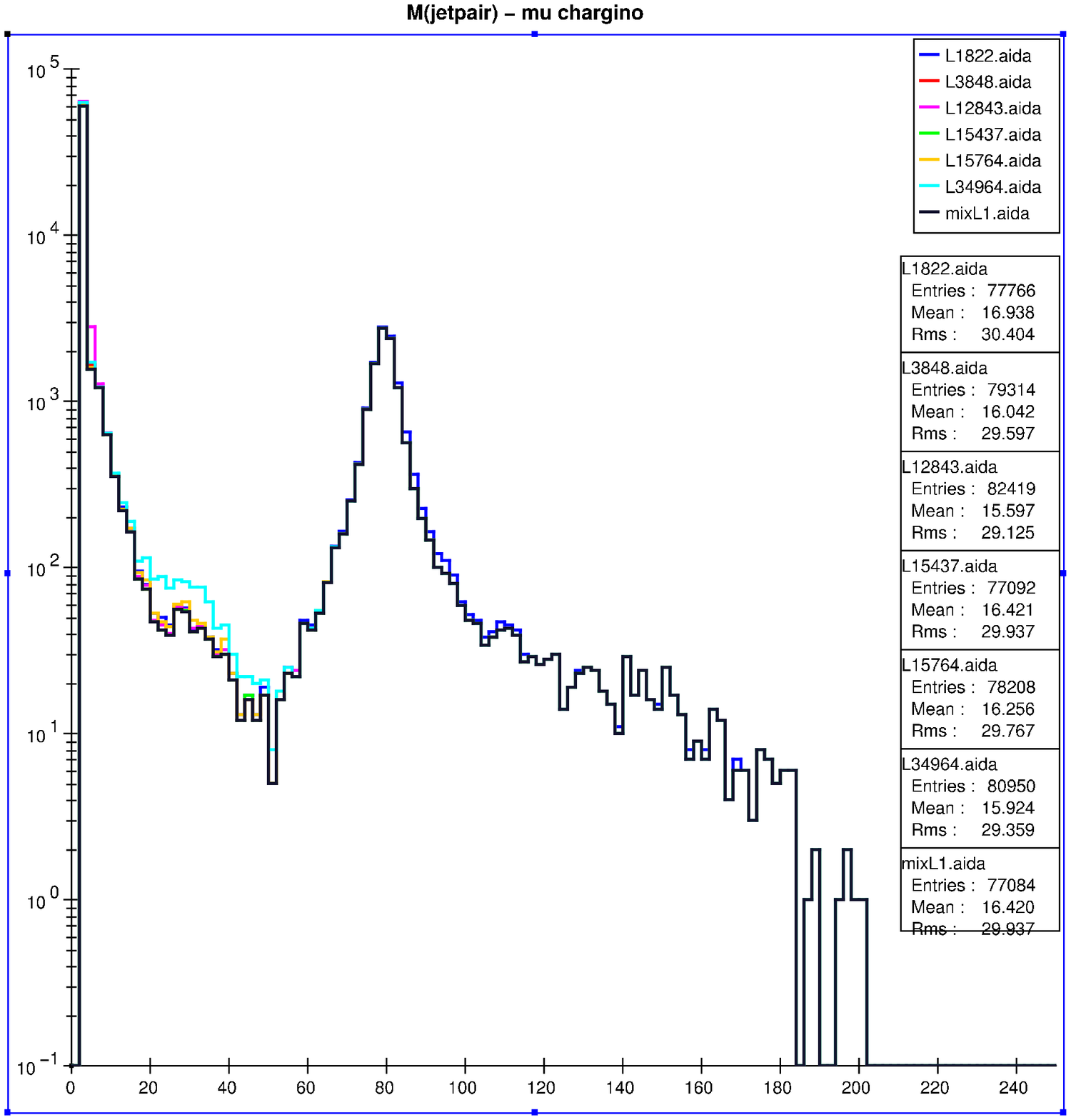}}
\vspace*{-0.1cm}
\caption{Jet-pair invariant mass distribution:
the number of events/2 GeV bin
after imposing the full set of cuts discussed in the text for the
2-jet $+\mu$ channel with $\Delta m_{\tilde\chi}<M_W$ for representative
models which are visible in this channel.
RH(LH) beam polarization is employed in the top(bottom)
panel, assuming
an integrated luminosity of 250 fb$^{-1}$
for either polarization. The SM background is shown as the black histogram.}
\label{mjpairmu_off}
\end{figure*}

\begin{figure*}[hptb]
\centerline{
\includegraphics[width=13.0cm,height=10.0cm,angle=0]{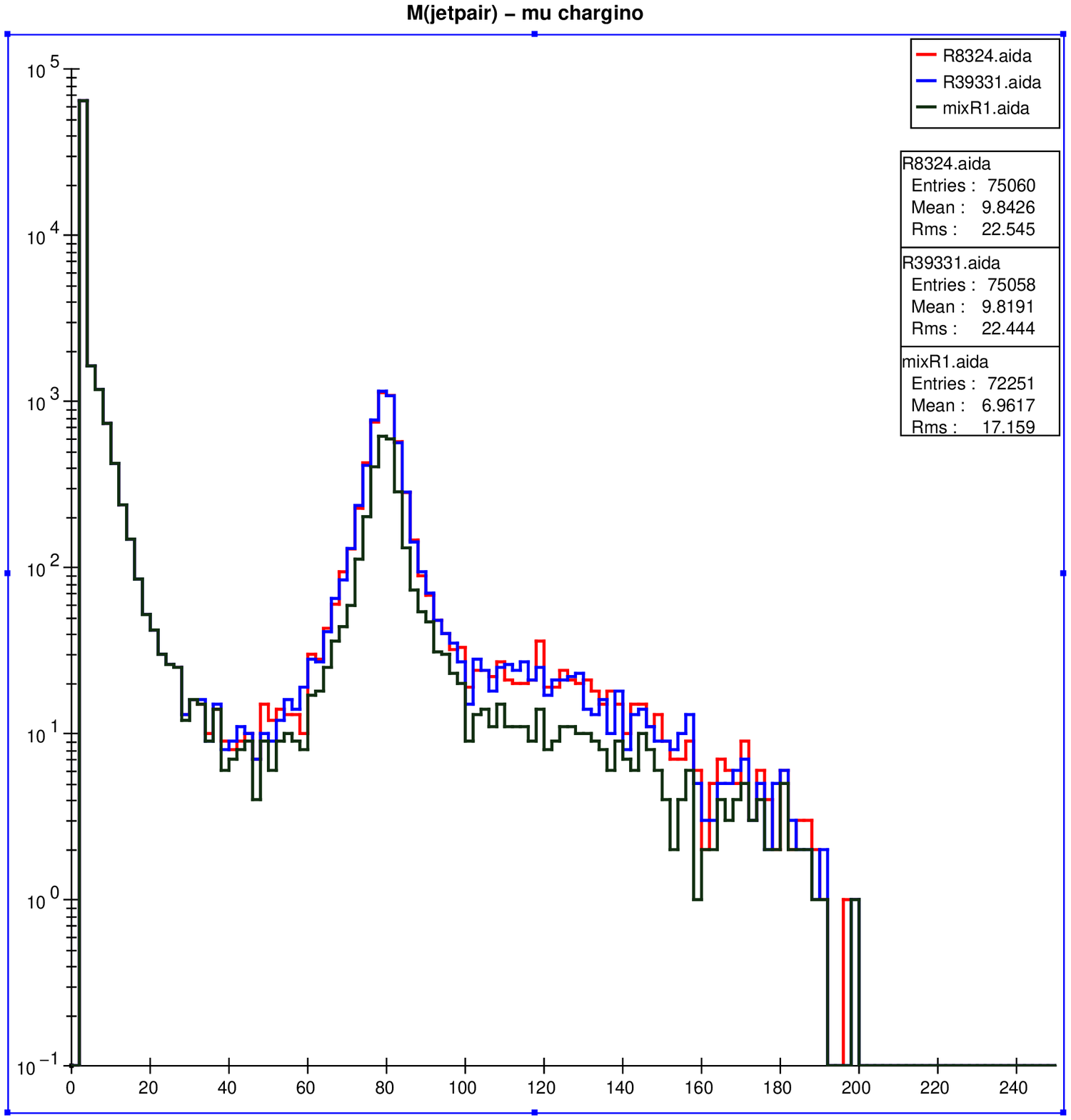}}
\vspace*{0.1cm}
\centerline{
\includegraphics[width=13.0cm,height=10.0cm,angle=0]{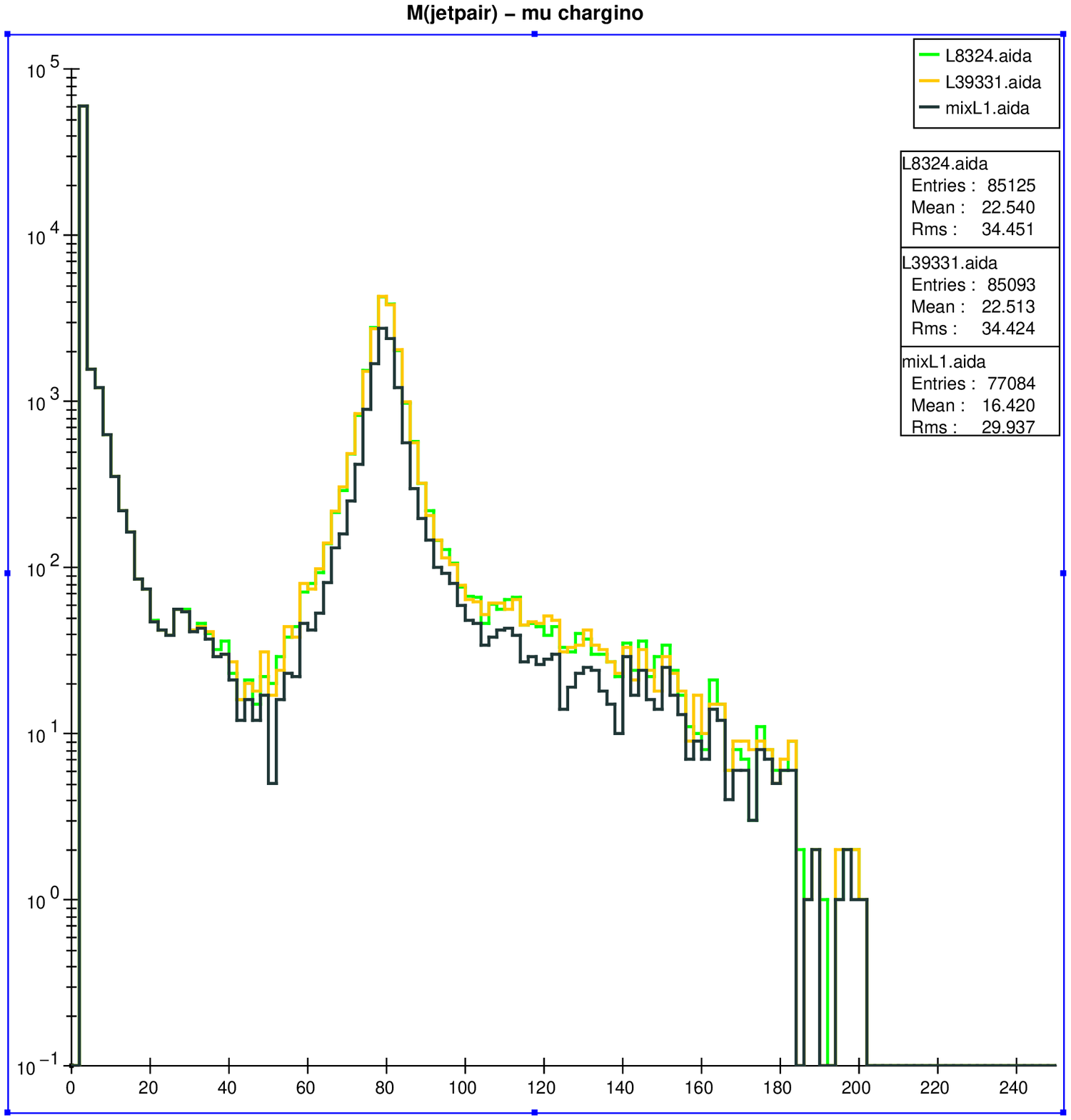}}
\vspace*{-0.1cm}
\caption{Jet-pair invariant mass distribution: the number of events/2 GeV bin
after imposing the full set of cuts discussed in the text for the
2-jet $+\mu$ channel with $\Delta m_{\tilde\chi}>M_W$ for representative
models which are visible in this channel.
RH(LH) beam polarization is employed in the top(bottom)
panel, assuming
an integrated luminosity of 250 fb$^{-1}$
for either polarization. The SM background is shown as the black histogram.}
\label{mjpairmu_on}
\end{figure*}

\begin{figure*}[hptb]
\centerline{
\includegraphics[width=13.0cm,height=10.0cm,angle=0]{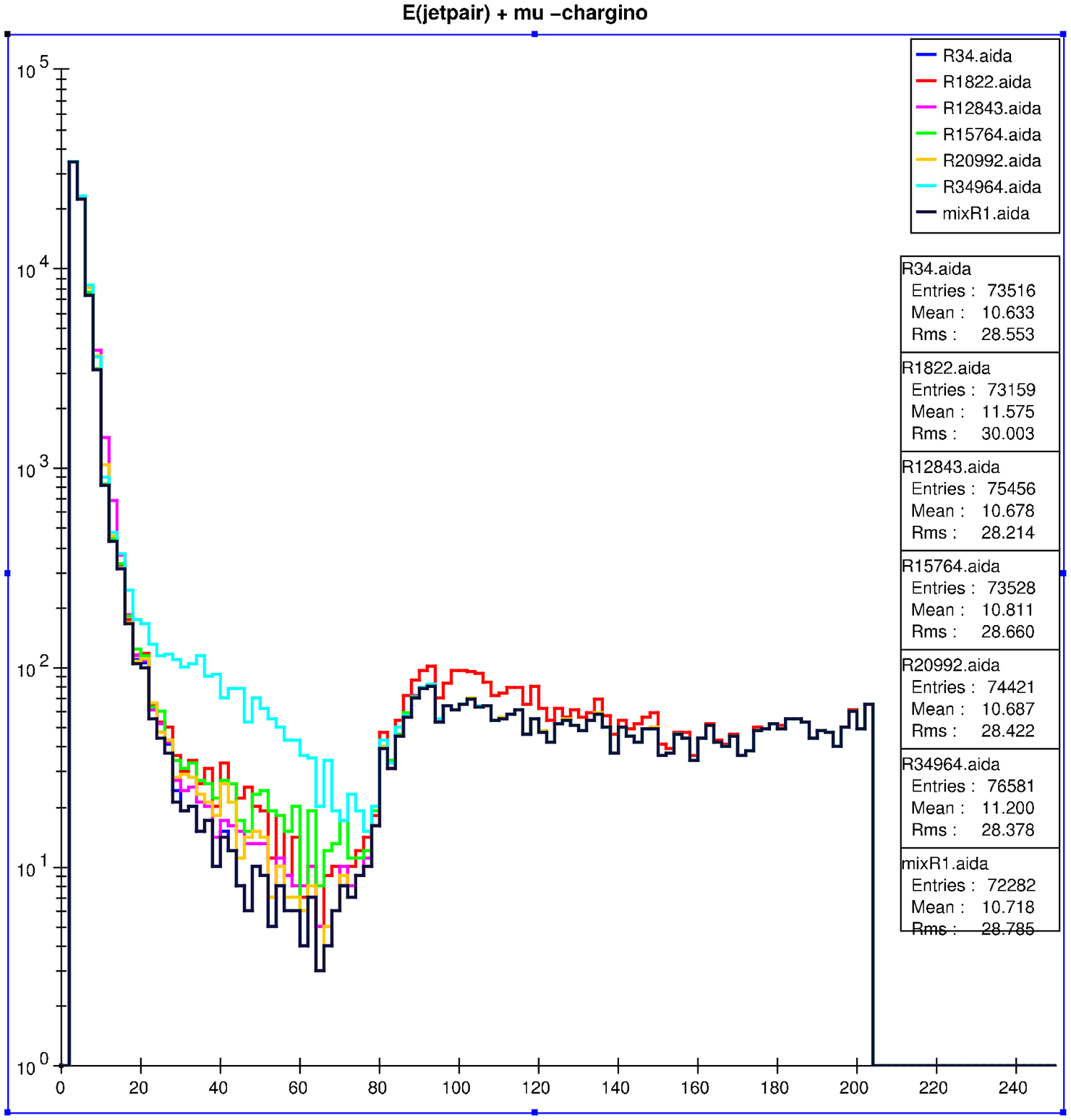}}
\vspace*{0.1cm}
\centerline{
\includegraphics[width=13.0cm,height=10.0cm,angle=0]{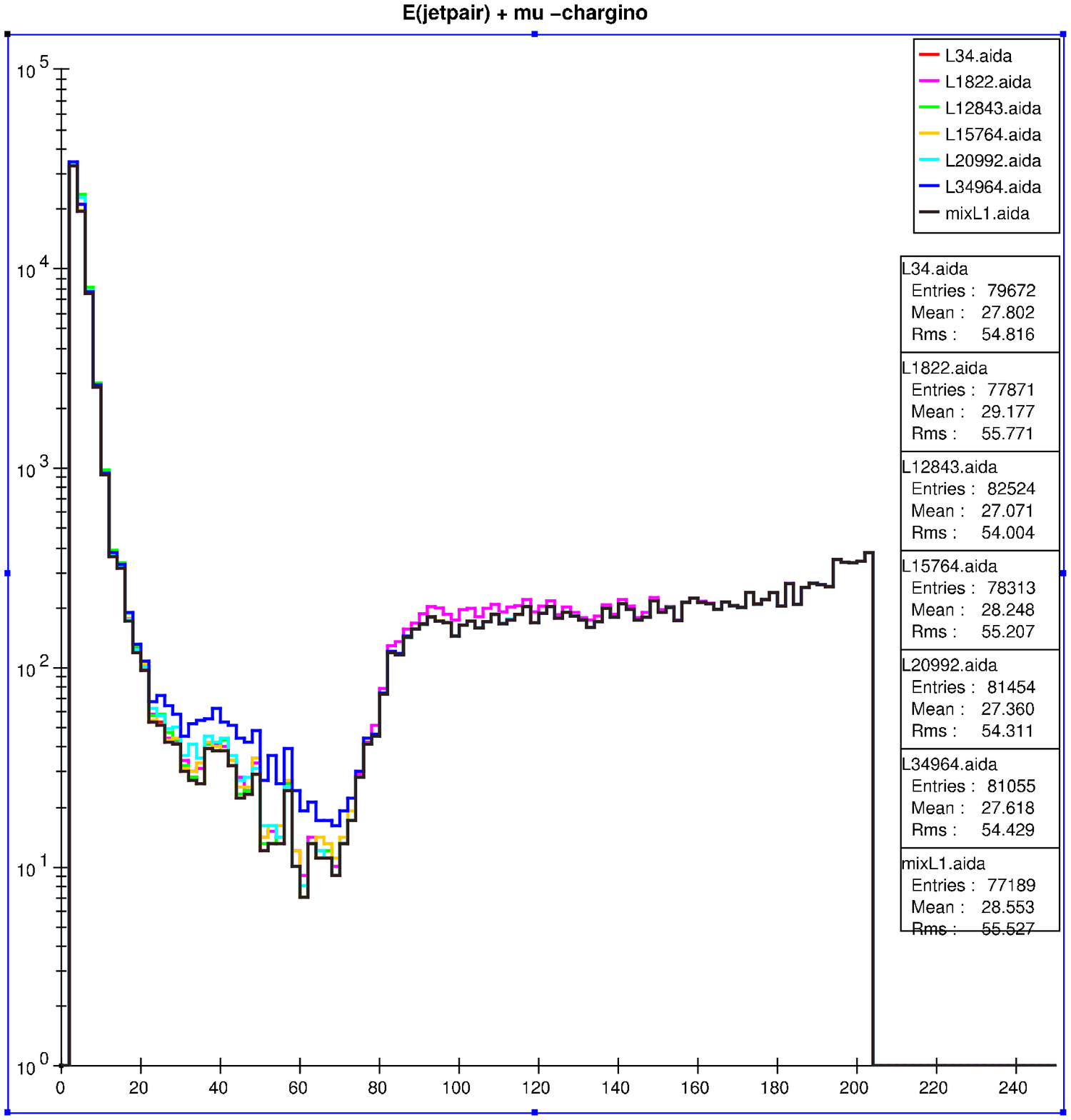}}
\vspace*{-0.1cm}
\caption{Jet-pair energy distribution: the number of events/2 GeV bin
after imposing the full set of cuts discussed in the text for the
2-jet $+\mu$ channel with $\Delta m_{\tilde\chi}<M_W$ for representative
models which are visible in this channel.
RH(LH) beam polarization is employed in the top(bottom)
panel, assuming
an integrated luminosity of 250 fb$^{-1}$
for either polarization. The SM background is shown as the black histogram.}
\label{ejpairmu_off}
\end{figure*}

\begin{figure*}[hptb]
\centerline{
\includegraphics[width=13.0cm,height=10.0cm,angle=0]{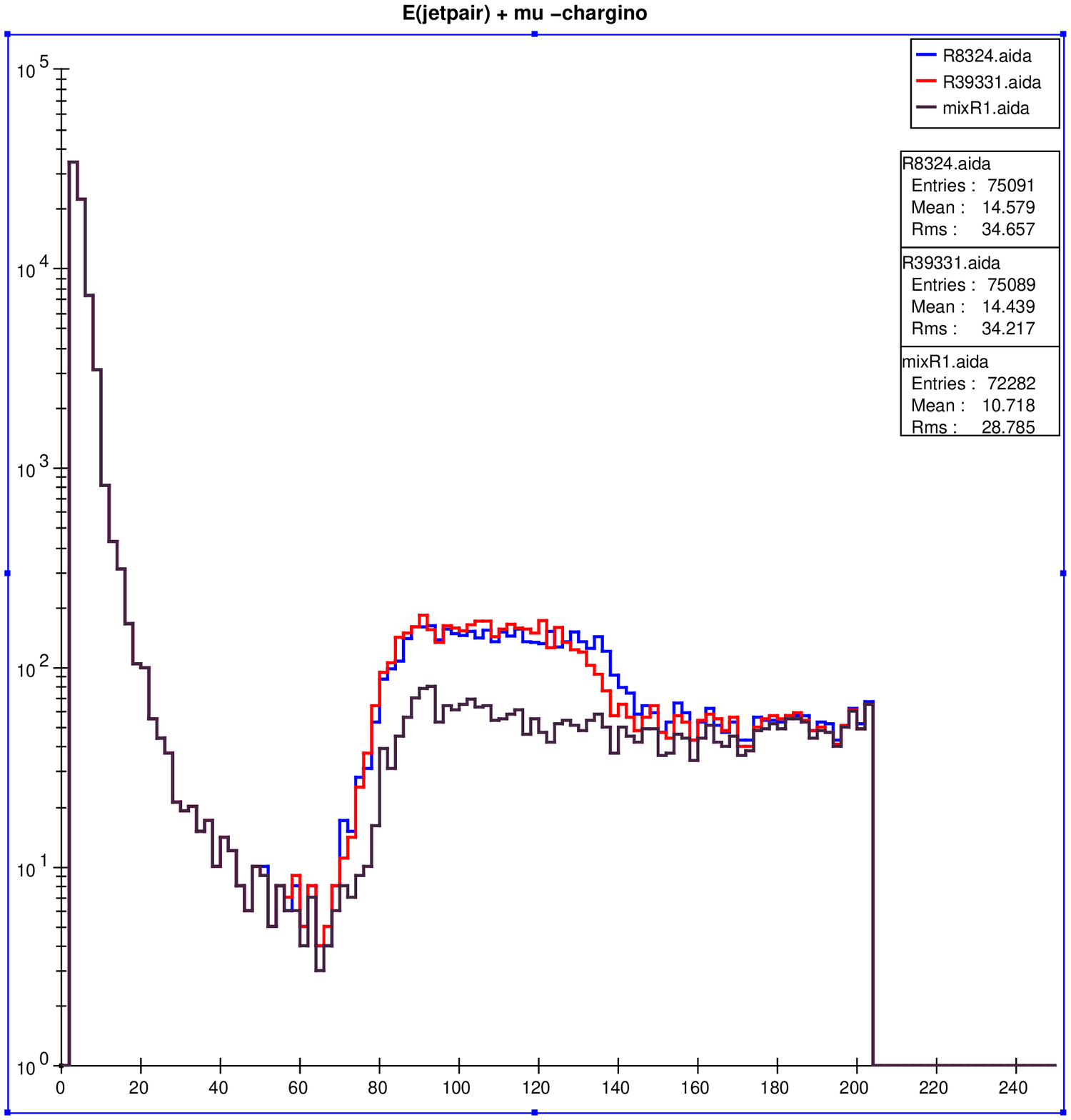}}
\vspace*{0.1cm}
\centerline{
\includegraphics[width=13.0cm,height=10.0cm,angle=0]{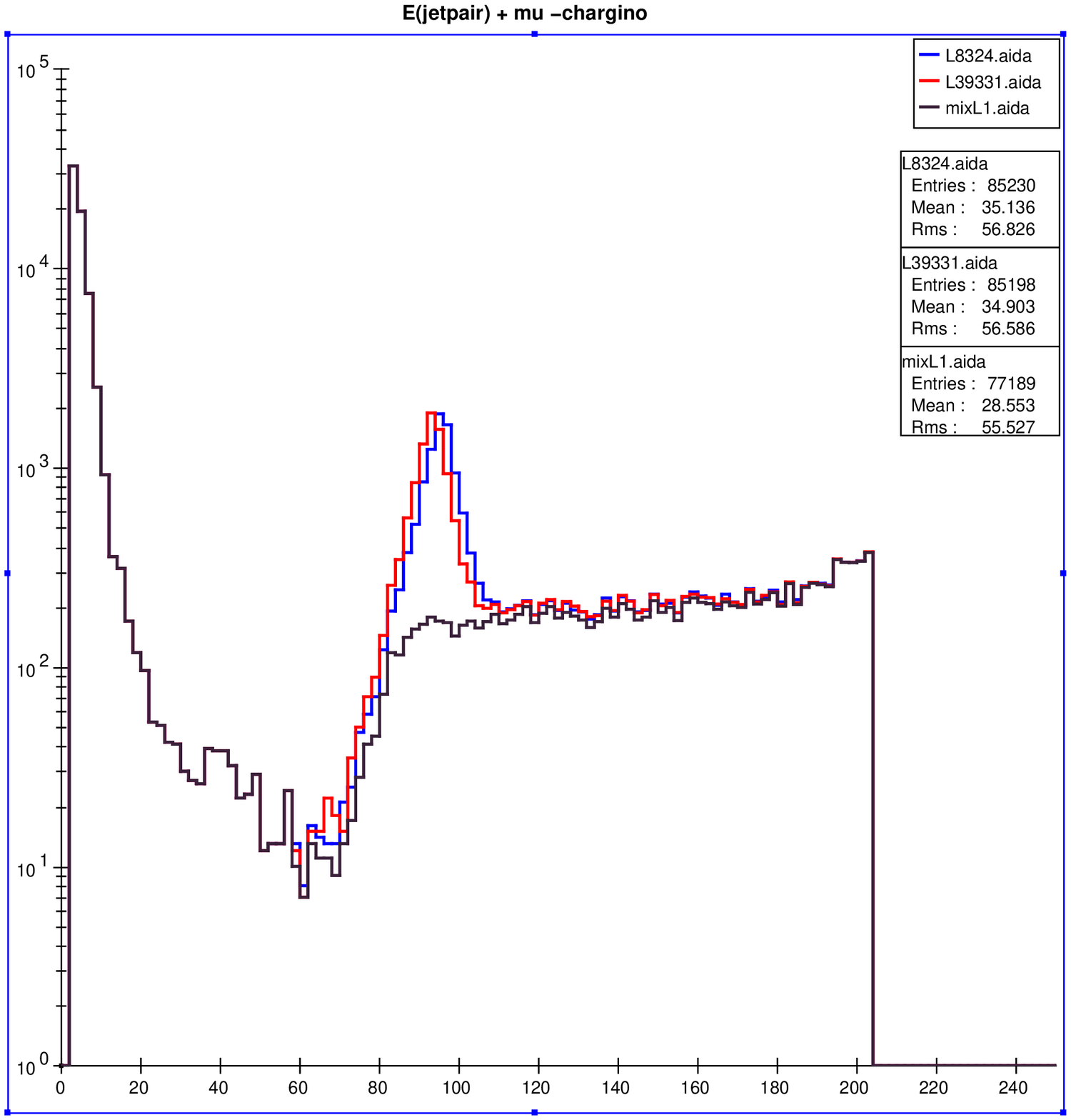}}
\vspace*{-0.1cm}
\caption{Jet-pair energy distribution: the number of events/2 GeV bin
after imposing the full set of cuts discussed in the text for the
2-jet $+\mu$ channel with $\Delta m_{\tilde\chi}<M_W$ for representative
models which are visible in this channel.
RH(LH) beam polarization is employed in the top(bottom)
panel, assuming
an integrated luminosity of 250 fb$^{-1}$
for either polarization. The SM background is shown as the black histogram.}
\label{ejpairmu_on}
\end{figure*}

We now compare these results to those for the case of the well-studied
benchmark point SPS1a'. Figures~\ref{mjpairmu_sps1a} and \ref{ejpairmu_sps1a}
display the jet pair invariant mass spectrum and energy distribution,
respectively, for both polarization choices. The chargino in this model
decays to an on-shell $W$ boson and has a large production cross section;
both of these features are observable in the Figures. The signal for
this model is clearly visible above the SM background and there is a peak
at $M_{jj}\simeq M_W$ in the invariant mass distribution.

\begin{figure*}[hptb]
\centerline{
\includegraphics[width=13.0cm,height=10.0cm,angle=0]{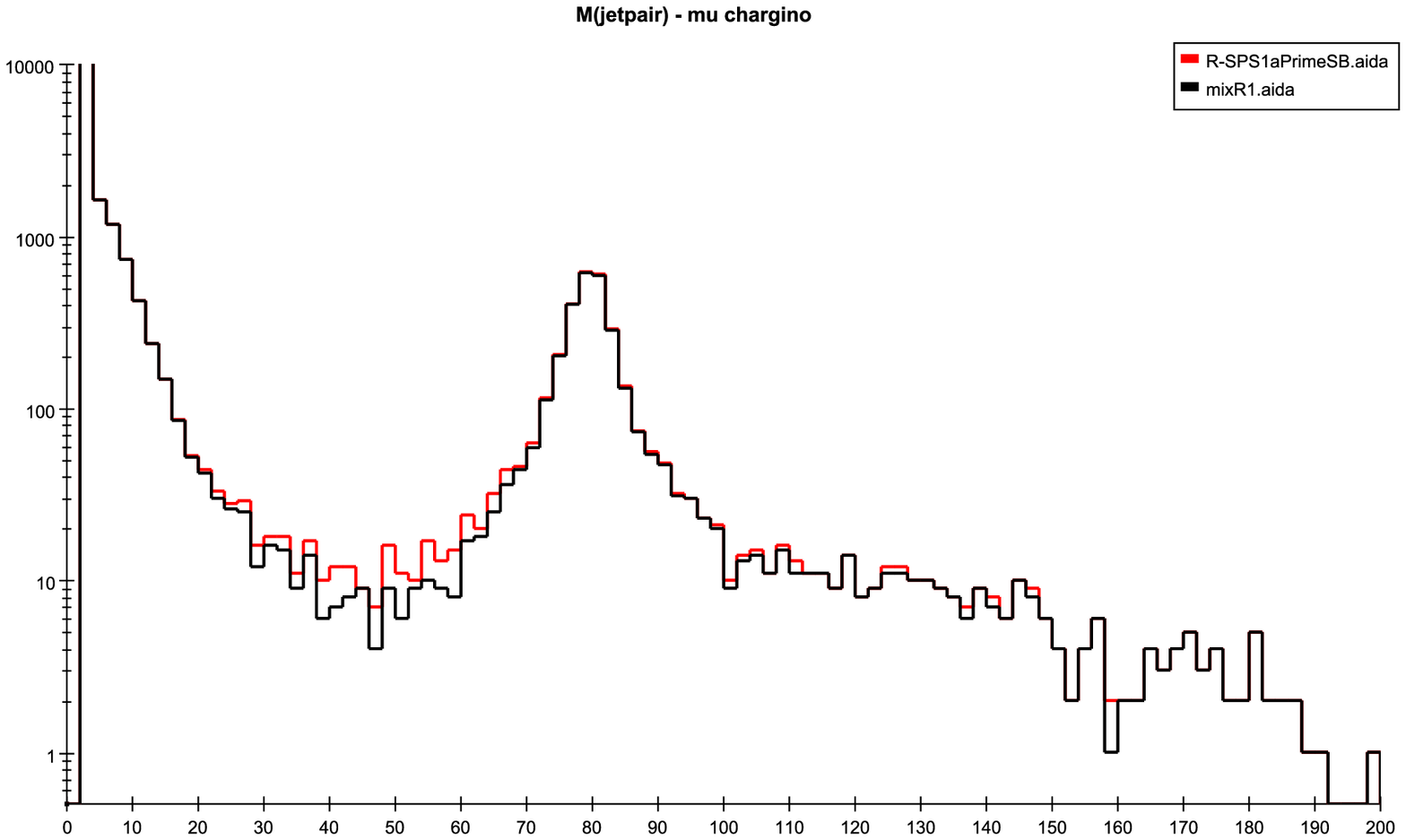}}
\vspace*{0.1cm}
\centerline{
\includegraphics[width=13.0cm,height=10.0cm,angle=0]{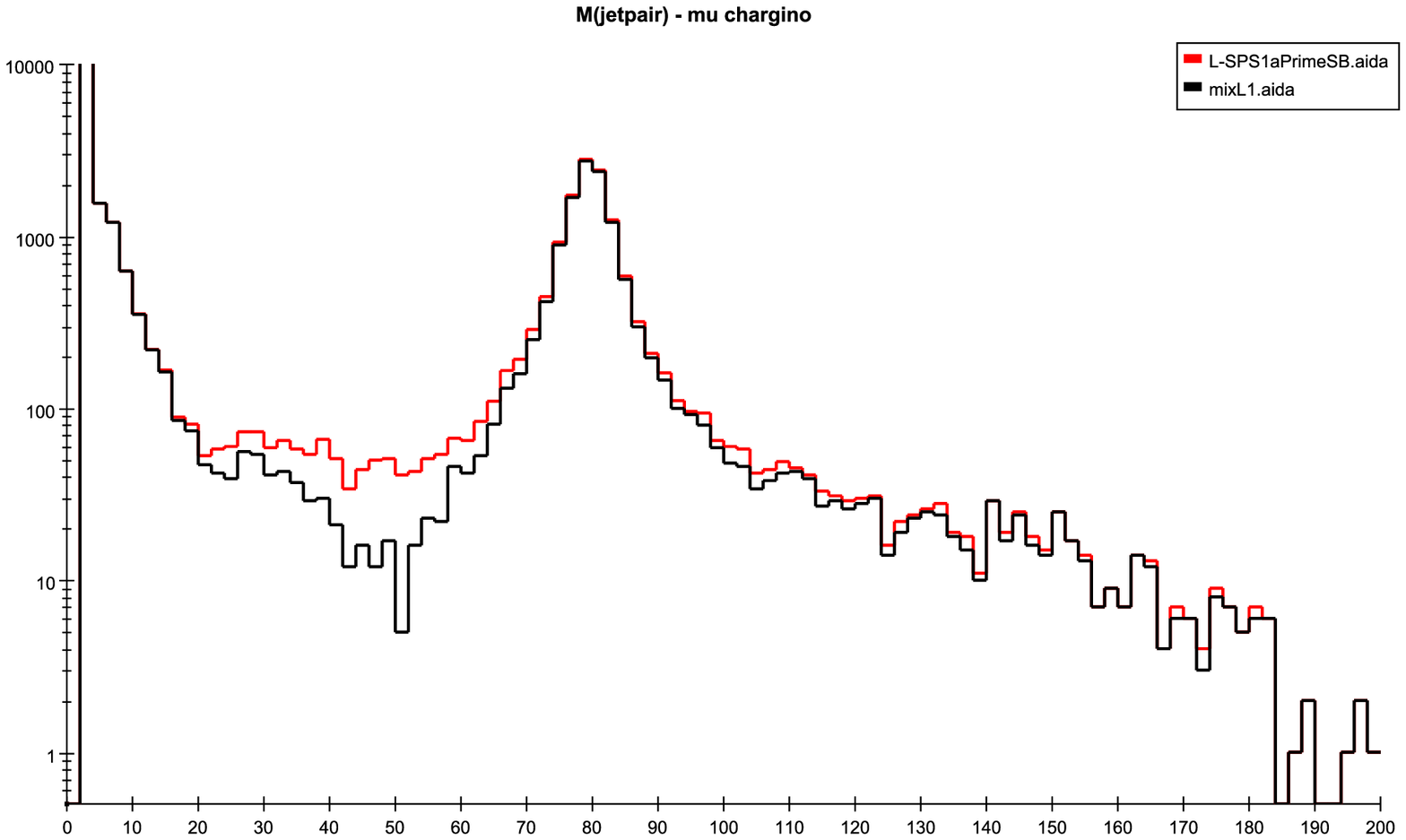}}
\vspace*{-0.1cm}
\caption{Jet-pair invariant mass distribution: the number of events/2 GeV bin
after imposing the full set of cuts discussed in the text for the
benchmark model SPS1a'.
RH(LH) beam polarization is employed in the top(bottom)
panel, assuming
an integrated luminosity of 250 fb$^{-1}$
for either polarization. The SM background is shown as the black histogram.}
\label{mjpairmu_sps1a}
\end{figure*}

\begin{figure*}[hptb]
\centerline{
\includegraphics[width=13.0cm,height=10.0cm,angle=0]{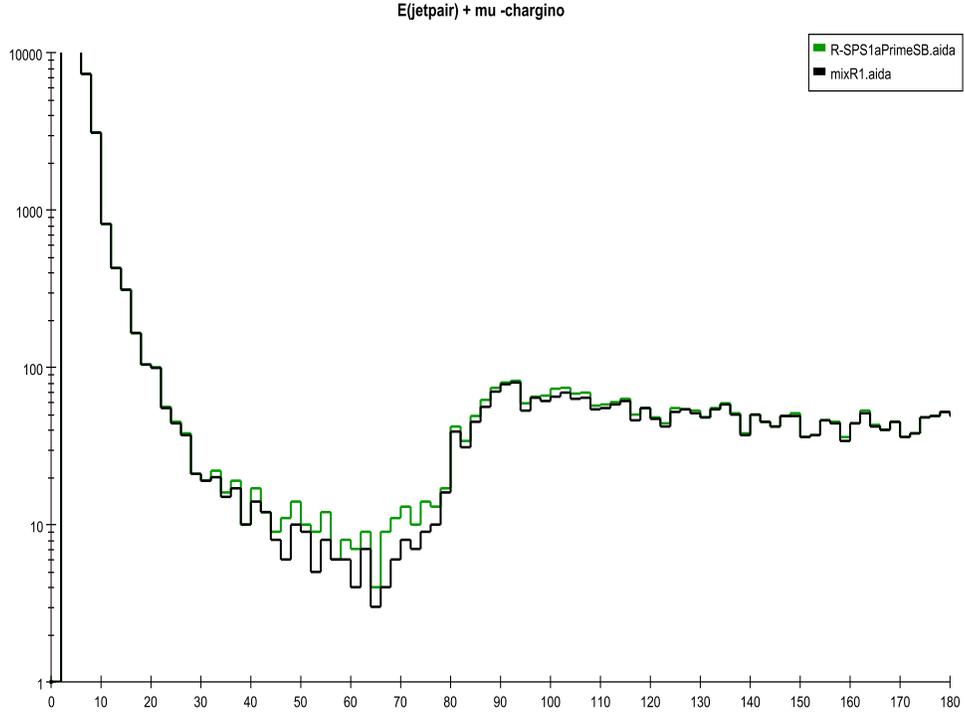}}
\vspace*{0.1cm}
\centerline{
\includegraphics[width=13.0cm,height=10.0cm,angle=0]{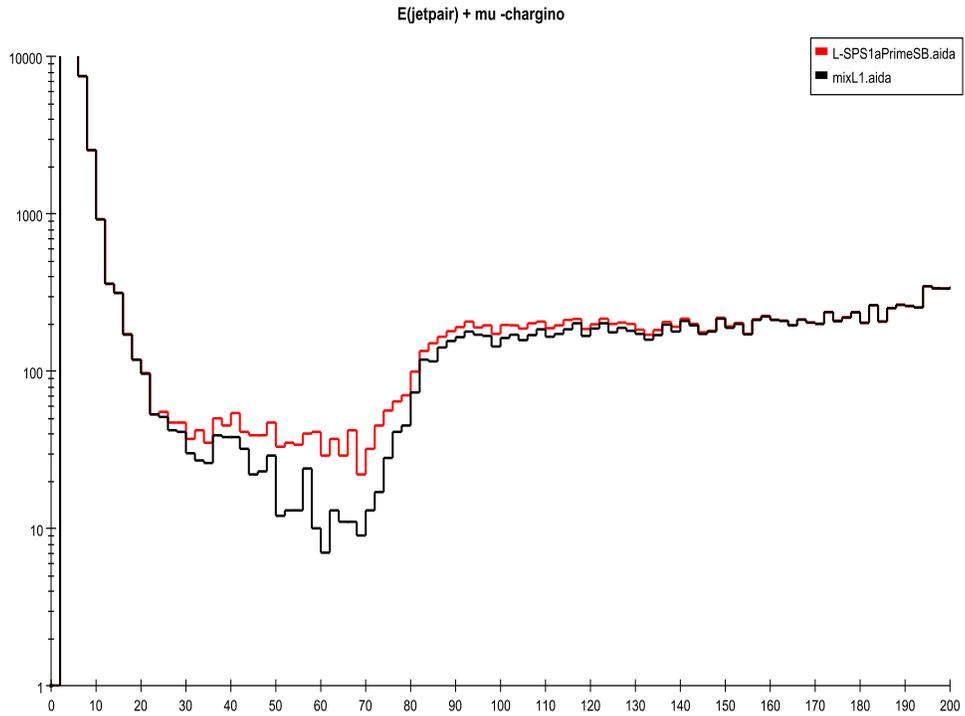}}
\vspace*{-0.1cm}
\caption{Jet-pair energy distribution: the number of events/2 GeV bin
after imposing the full set of cuts discussed in the text for the
benchmark model SPS1a'.
RH(LH) beam polarization is employed in the top(bottom)
panel, assuming
an integrated luminosity of 250 fb$^{-1}$
for either polarization. The SM background is shown as the black histogram.}
\label{ejpairmu_sps1a}
\end{figure*}

\clearpage

\subsection{Radiative Chargino Production}

As was pointed out by Gunion and
collaborators~\cite{Gunion:2001fu,Chen:1995yu,Chen:1999yf} (see also
Riles \etal {\cite {Riles:1989hd}}), in the
case where $\Delta m_{\tilde{\chi}}$ is in the approximate range
0.1 GeV $\lsim \Delta m\lsim 2$ GeV, the dominant
decay mode of the charginos is into soft pions plus the LSP (which
appears as missing
energy), as shown in Fig.~\ref{fig:charginoBF}. The dominant SM background
to this final state is from $\gamma\gamma$ interactions and has an
enormous event rate. If
$\Delta m_{\tilde{\chi}} < m_\pi$, however, the dominant chargino decay is
leptonic, $\tilde{\chi}_1^\pm \rightarrow e^\pm \nu_e \LSP$, and
does not pose a problem for detection as this essentially results in a
charged stable-particle search. In this subsection, we discuss the
radiative chargino search where the
hard photon emitted in the process $e^+e^- \to
\tilde\chi^+_1 \tilde \chi^-_1\gamma$ is tagged.
Unlike the other chargino signatures we
consider, the strength of this signal is not very dependent on the mass
splitting between the lightest chargino and the LSP neutralino.

\subsubsection{Event Generation}

An immediate issue in performing the search for radiative chargino
production is that we find PYTHIA
underestimates the rate and energy distribution of hard photon
emission from the final state charginos. Thus we use
CompHEP~\cite{Boos:2004kh} to generate the
$e^+e^- \to \tilde\chi^+_1 \tilde\chi^-_1 \gamma$
(as well as $e^+e^- \to \tilde\chi^+_1\tilde\chi^-_1$) events from the
explicit (tree-level) matrix elements.

In particular, we find that PYTHIA with default ISR and FSR options
yields a lower
cross section for chargino production with an associated photon than
does CompHEP. This is illustrated in Fig.~\ref{pychepcompare}
where we display
the cross section for $e^+e^-\to\tilde{\chi}^+\tilde{\chi}^-\gamma$ in
one of the AKTW models (labelled as model 13348)
as a function of photon transverse momentum.
For this model (where the $\tilde{\chi}^+_1$ has a mass of $\simeq 124$ GeV)
the cross section computed by PYTHIA is about $20$\% smaller than the
CompHEP cross section for all values of $p_T$. In examining the MSSM
parameter space further, we find points where
the PYTHIA generated cross section can be as low as $50$\% of
that calculated via CompHEP; for the models considered in this paper,
the PYTHIA cross section is generally $80-90$\% of that from CompHEP.
As CompHEP uses an exact (tree-level) matrix element
calculation, it is presumably more accurate. Therefore we use
CompHEP when calculating the cross sections for radiative chargino
production in each of our models and to generate the events for this
process.

CompHEP does not allow one to set an arbitrary beam polarization.
However, one may set the electron beam polarization to be,
for example, purely left-handed, by
effectively inserting the relevant projection operator into the
expression for the matrix element. Thus we can calculate the
desired cross sections and generate events for each of the two
initial helicity states that we consider here. For
each pure initial helicity state we generate two large event files and
find the relevant cross sections. We then choose the correct number
of events for each of our two partial electron
beam polarizations (80\% left-handed and 80\% right-handed)
and pipe this
through our analysis as described in Section~\ref{Sec:analysis}.

CompHEP includes the option of using a beamstrahlung spectrum
calculated from the beam dimensions and the number of particles
per bunch. However, for consistency with the rest of our signal
and background, we must use the same beam spectrum as described
above. To implement this spectrum in CompHEP, we read in the
beamspectrum as generated by GuineaPig~\cite{Guinea}\footnote{
A slight complication arises from the need to deconvolute this from
one part of the beamspectrum
that is already present in one of the precompiled CompHEP libraries
that cannot be changed by the user.}. We checked that the normalization
was correct by comparing the cross sections for $e^+e^- \to t\bar t$
as generated by our modified CompHEP code and by PYTHIA.

In calculating the cross section and generating events for the
$\tilde\chi^+_1 \tilde\chi^-_1 \gamma$ final state, we demand that the
transverse momentum of the photon be greater than 5 GeV. This cut
is much softer than that we apply in our analysis with the
detector simulation; we do
not wish to eliminate the possibility of low $p_T$ signal events
passing the final $p_T$ cut due to mismeasurement. We do, however,
need to apply a cut at this stage for the purpose of regularization.

\begin{figure*}[hptb]
\centerline{
\psfig{figure=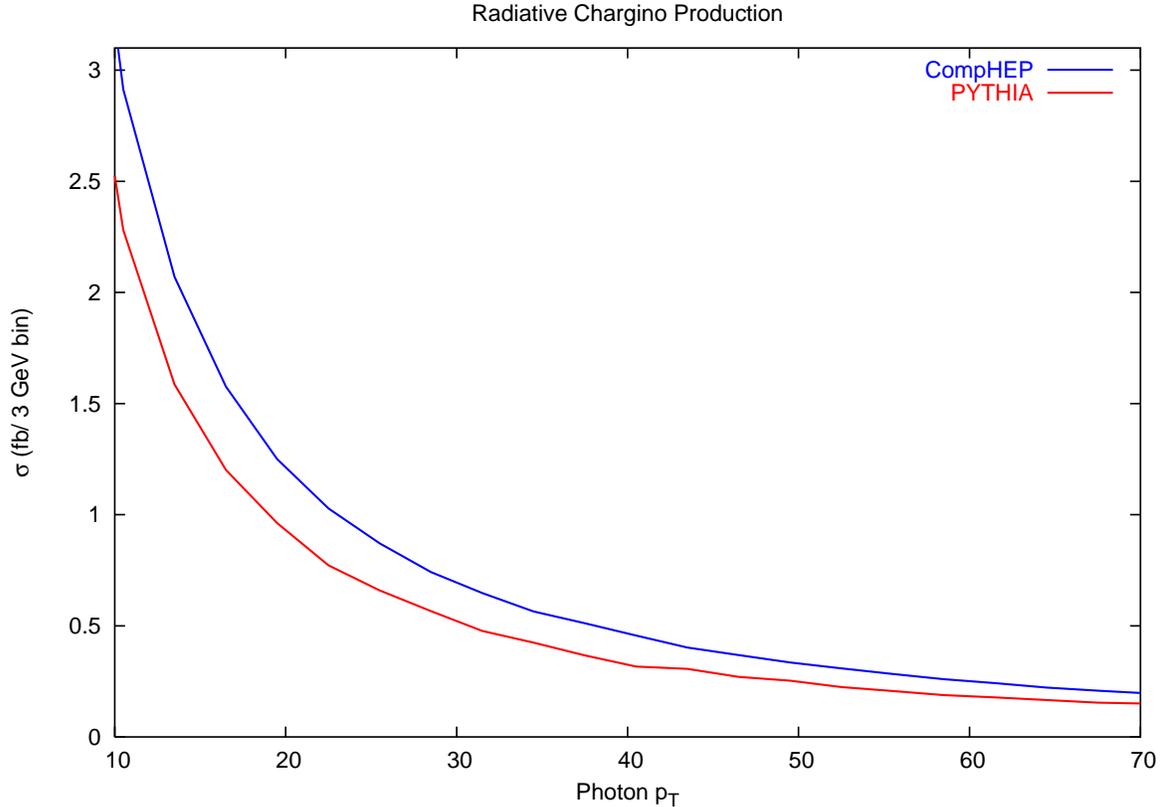,width=11cm,angle=-90,clip=}}
\vspace*{0.1cm}
\caption{Comparison of the cross section for the pair production of
the lightest chargino with
an associated photon as a function of the photon transverse momentum
as calculated by PYTHIA (red, bottom curve) and by CompHEP (blue, top curve)
in one of the AKTW models where $m_{\tilde{\chi^+}}=124$ GeV.}
\label{pychepcompare}
\end{figure*}

\subsubsection{Analysis}

In our analysis, we tag on a high-$p_T$ photon, produced by the
signal either off the initial state electron-positron pair, or
radiated off of one of the charginos. We apply the following
kinematical cuts as suggested in
\cite{Gunion:2001fu,Abbiendi:2002vz}:

\begin{enumerate}
\item There be exactly one photon in the event
with $p_T > 0.035 \sqrt{s}$ and no
other charged tracks within 25 degrees.
This isolation cut removes most of the two-$\gamma$ background.

\item There be no identified (\ie, above 142 mrad) electrons or muons
in the event. Although this cut slightly reduces the signal, we find it
dramatically decreases the background from
$\gamma \gamma$ and $e^\pm \gamma$ events. For the signal, we see
from Fig.~\ref{fig:charginoBF} that the branching fraction of
charginos to electrons or muons is less than $30-40\%$ in the
relevant $\Delta m_{\tilde{\chi}}$ range.

\item We demand that the number of charged tracks be in the range 1 to 11.
Note that below 142
mrad the detector only observes clusters of energy, however, we
nonetheless treat clusters on the same
footing as tracks. This cut removes high-multiplicity events.
In particular, the removal of high-multiplicity events
restricts this analysis to the range of $\Delta m_{\tilde{\chi}}$ that
we are targeting in this analysis.
Models with larger values of $\Delta m_{\tilde{\chi}}$ generate
harder partons in the chargino decay that radiate more gluons and hence
result in more tracks.

\item We demand that the photon energy and the energy of the
remaining visible particles satisfy
$E_{\mbox{\tiny vis, other particles}} - E_\gamma < 0.35 \sqrt{s}$.
This cut further reduces the two-$\gamma$
background. It also serves to restrict this analysis to the
relevant range of $\Delta m_{\tilde{\chi}}$, as the amount of visible
energy increases with $\Delta m_{\tilde{\chi}}$.

\item We demand that the ratio of total visible transverse momentum to
transverse energy satisfy
$\frac{p_{T\, \mbox{\tiny vis}}}{E_{T \,\mbox{\tiny vis}}} > 0.4$
and that the ratio of total visible transverse momentum to total momentum
be $\frac{p_{T \,\mbox{\tiny vis}}}{p_{\tiny tot\, \mbox{\tiny
vis}}} > 0.2$.
This removes most of the hadronic two-$\gamma$ and \epem\
initiated processes.

\item We require that the recoil mass be $M_{\mbox{\tiny recoil}} = \sqrt{s}
\sqrt{\left(1 - 2 E_\gamma/\sqrt{s}\right)} > 160$ GeV.
This is the recoil mass of the tagged photon, which should be
at least twice the current lower bound on the
chargino mass, which we take to be 160 GeV from the approximate 80 GeV
lower limit on the chargino mass from LEP II {\cite{Yao:2006px}}.
\end{enumerate}

After applying these cuts we examine the recoil mass of the
tagged photon,
\be
M_{\mbox{\tiny recoil}} = \sqrt{s}
\sqrt{\left(1 - 2 E_\gamma/\sqrt{s}\right)} \, .
\ee
The dominant remaining SM background again arises from the reaction
$e^+e^- \rightarrow
l^- \bar{\nu}_l \nu_{l'} {l'}^+$ as illustrated in Fig.~\ref{fig:Mrecoilbg}.

\begin{figure*}[hptb]
\centerline{
\epsfig{figure=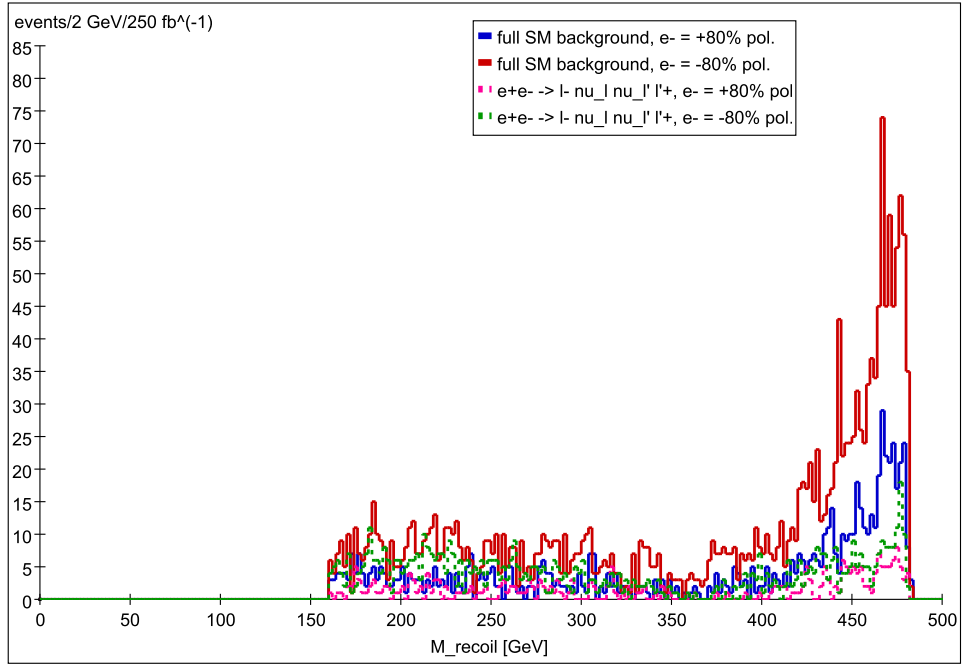,width=13cm,clip=}}
\vspace*{0.1cm}
\caption{Recoil mass of the tagged photon for the
SM background to radiative chargino production after the
cuts listed in the text have been applied. This is generated
from $250 \mbox{ fb}^{-1}$ of SM events with
80\% right-handed (solid blue line) and 80\% left-handed (solid red line)
electron beam polarization and unpolarized positron beam
at $\sqrt{s} = 500$ GeV. The dotted
lines show the main processes contributing to the
background, $e^+e^- \rightarrow
l^- \bar{\nu}_l \nu_{l'} {l'}^+$,
for 80\% right-handed (dotted pink line) and
80\% left-handed electron polarization
(dotted green line).}
\label{fig:Mrecoilbg}
\end{figure*}

\begin{figure*}[htbp]
\centerline{
\includegraphics[width=10.0cm,height=13.0cm,angle=90]{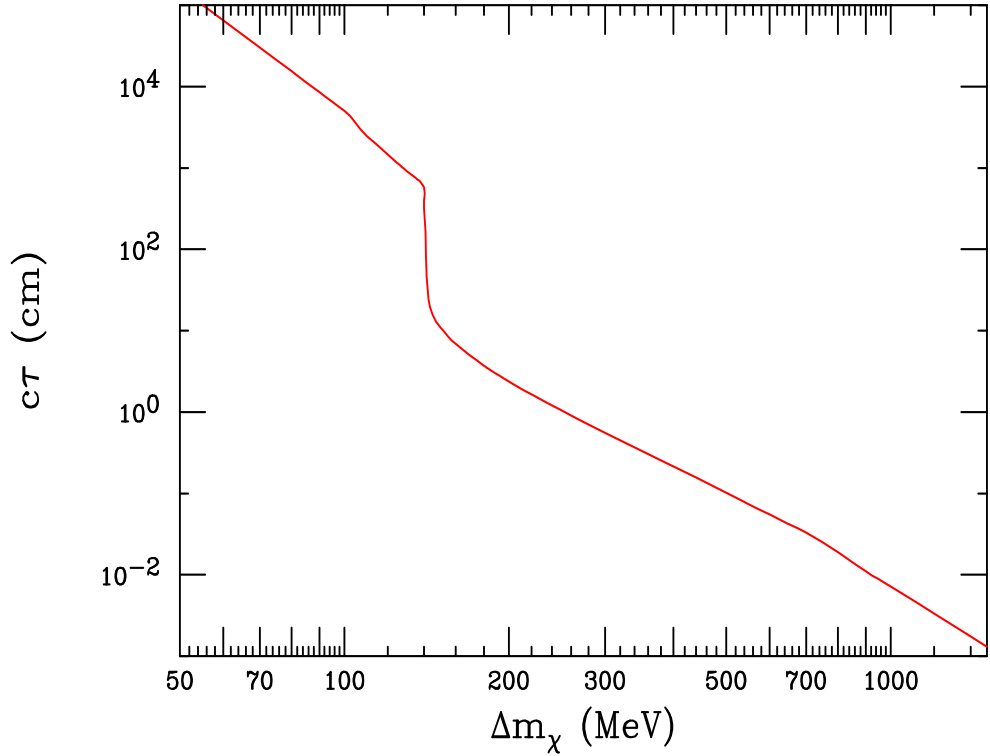}}
\vspace*{0.1cm}
\caption{Chargino lifetime as a function of the chargino-neutralino
mass splitting,
$\Delta m$.}
\label{lifetime}
\end{figure*}

This analysis is designed to only catch charginos in a relatively
narrow mass and $\Delta m_{\tilde \chi}$ range.
If $\Delta m_{\tilde \chi}$ is, \eg, greater than $\sim 3-4$ GeV, then
the kinematical properties of the average chargino decay may have difficulty
satisfying our energy, momentum and multiplicity cuts.
If the mass splitting is too small (less than
$\sim 0.15$ GeV), then the chargino will have a long
lifetime, as shown in Fig.~\ref{lifetime}, and will decay to at most 2 charged
tracks and will
not pass the above cuts. Furthermore, if the mass of the chargino is
too close to the beam energy ($\gsim 225$ GeV), then not only will the
cross section be phase space suppressed, but it
will be almost impossible for the signal
to pass the required photon $p_T$ cut.
Depending on exactly how these kinematic boundaries are chosen, we find
that only $\sim 26$ of the 53 AKTW models with
kinematically accessible charginos at $\sqrt s=500$ GeV have these necessary
properties. From this analysis we find that
there are only 14 models which are observable in the radiative channel
(for either beam polarization) over the SM
background with a significance ${\cal S} >5$. Note that although the
backgrounds are larger in the LH sample than with
RH beams, as is usual, the chargino signal in this case is {\it far}
larger (by approximately a factor of 9)
in the LH sample. This is because for small
$\Delta m_{\tilde \chi}$ the charginos in these models
are mostly wino in content.
This is illustrated in Fig.~\ref{radchargino1} which
shows our analysis results for a number of sample representative
models with either beam polarization. Model 38239, which is shown
in this Figure, provides a nice example of a case that is missed by this
analysis; this model has $\Delta m_{\tilde \chi}=0.45$
GeV while $m_{\tilde \chi_1^+}=239.75$ GeV and thus has little remaining
phase space to allow for the emission of a hard photon
with $p_T \geq 17.5$ GeV.

\begin{figure*}[hptb]
\centerline{
\includegraphics[width=13.0cm,height=10.0cm,angle=0]{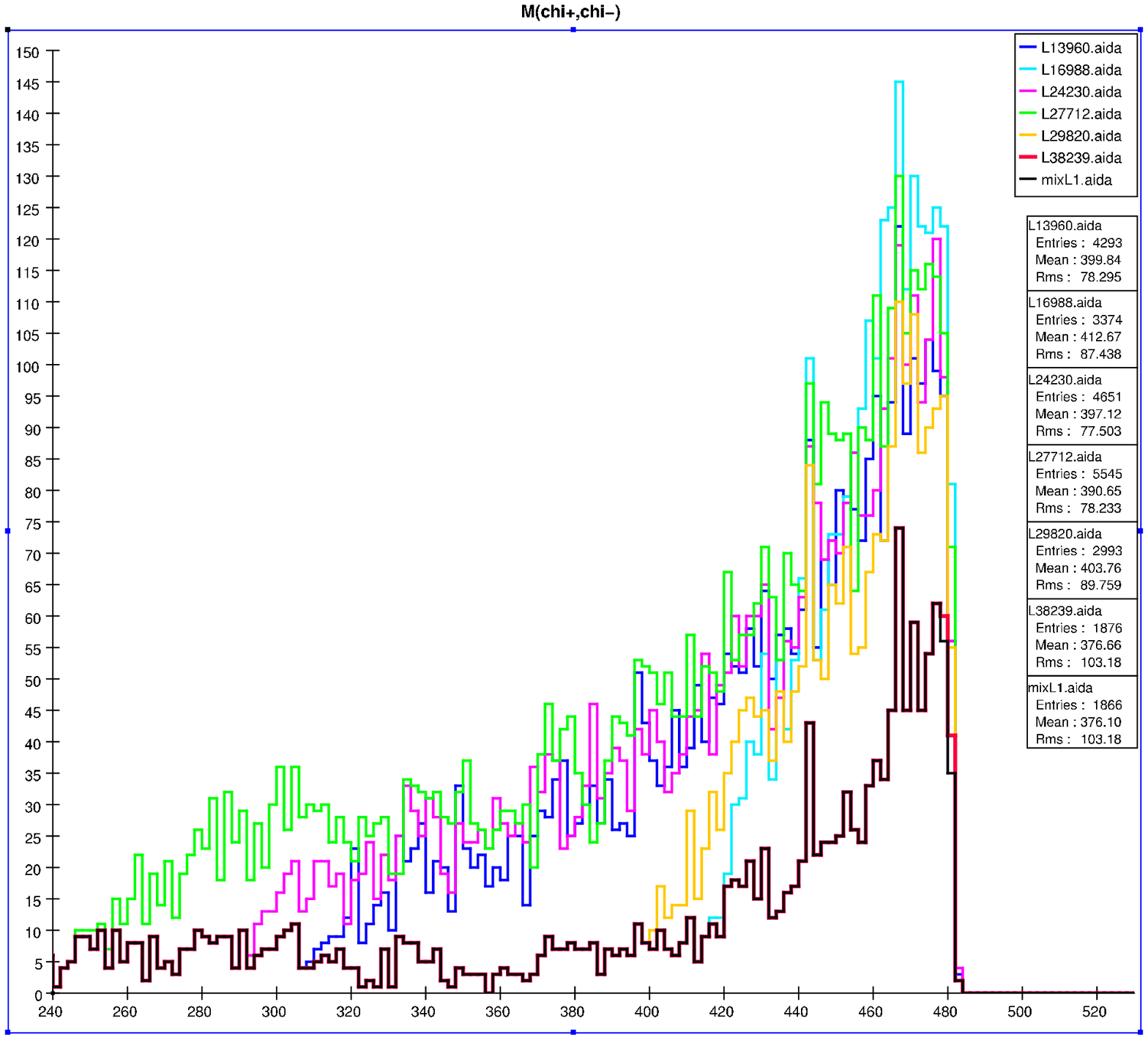}}
\vspace*{0.1cm}
\centerline{
\includegraphics[width=13.0cm,height=10.0cm,angle=0]{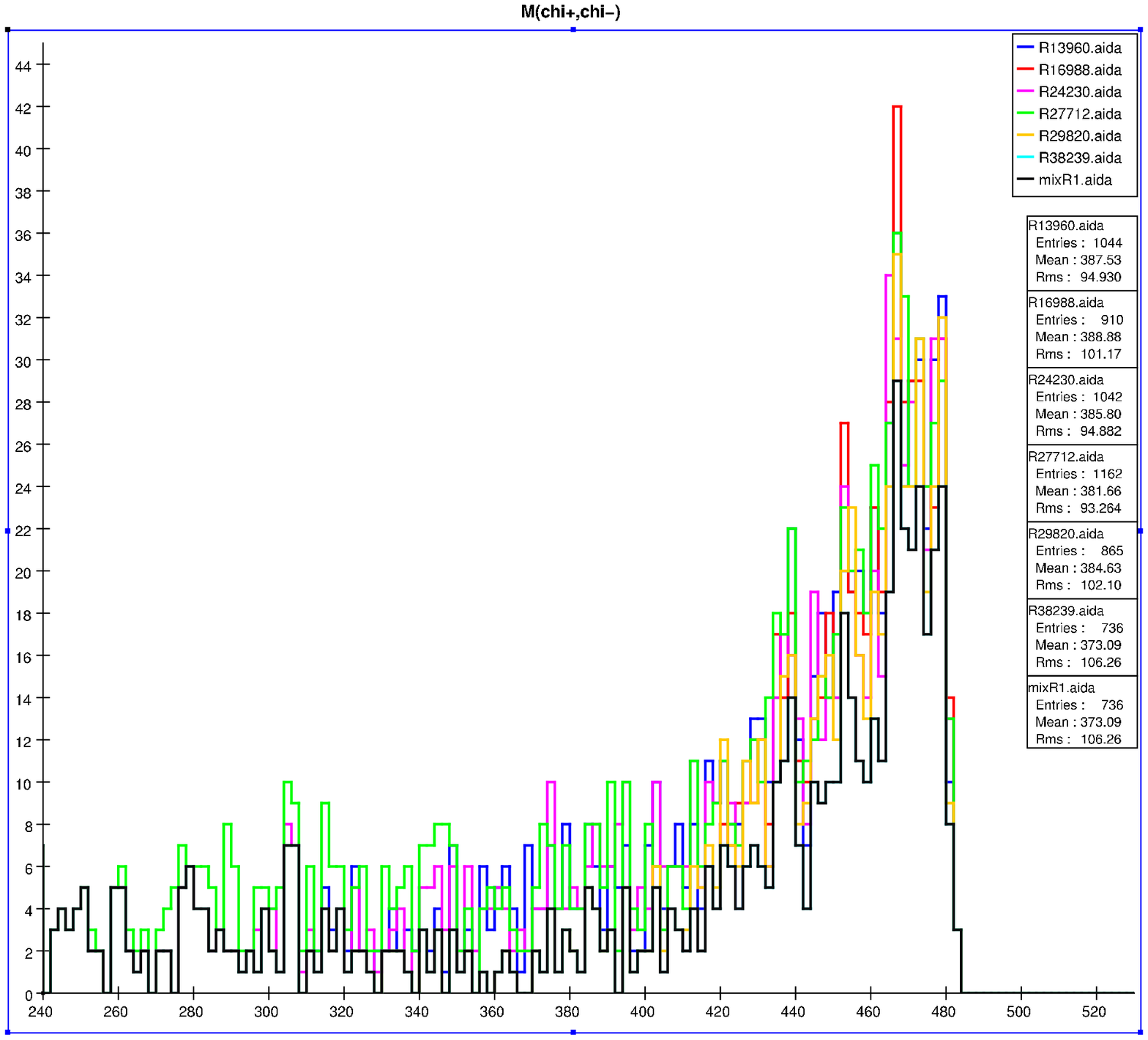}}
\vspace*{-0.1cm}
\caption{Recoil mass distribution for chargino pair production in
the AKTW models from the photon
tag analysis: number of events/2 GeV bin assuming
an integrated luminosity of 250 fb$^{-1}$ for both LH (top panel) and RH
(bottom panel) electron beam polarization.}
\label{radchargino1}
\end{figure*}

The benchmark model SPS1a' also leads to a reasonable signal excess in
this channel as shown in Fig.~\ref{radiative_sps1a}.
However, in this case, this is not the result of a small
$\tilde\chi_1^\pm-\tilde\chi_1^0$ mass difference, which is
86 GeV in this model, but rather the signal in
this channel for is a `fake' induced by $\tilde\chi_2^0$ production.
As discussed above for some
of the AKTW models, the observed signal in this case
is actually a feed down from the production of other, perhaps
more massive, states
in the SUSY spectrum as well as from
radiative associated $\tilde\chi_1^0 \tilde\chi_2^0$ production.
However, the $\tilde\chi_1^\pm$ in SPS1a' is already clearly observable
in the other channels discussed above.
We generally find that the fake contamination in the radiative channel
is less than $30\%$ and is
quite a bit smaller in many cases.

\begin{figure*}[htpb]
\centerline{
\includegraphics[height=10cm,width=13cm,angle=0]{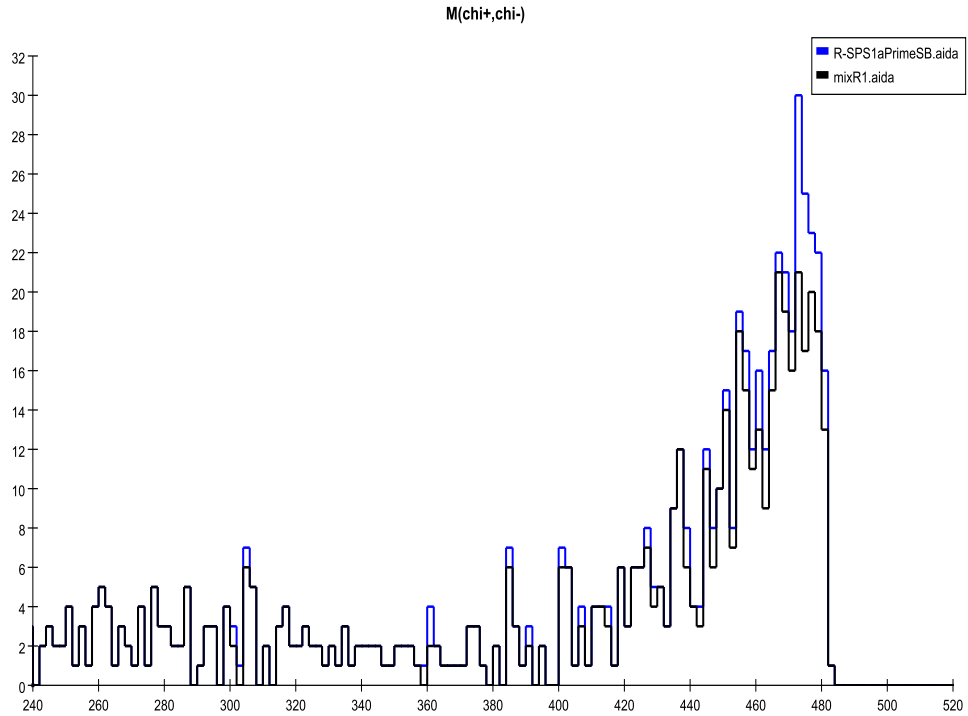}}
\vspace*{0.1cm}
\centerline{
\includegraphics[height=10cm,width=13cm,angle=0]{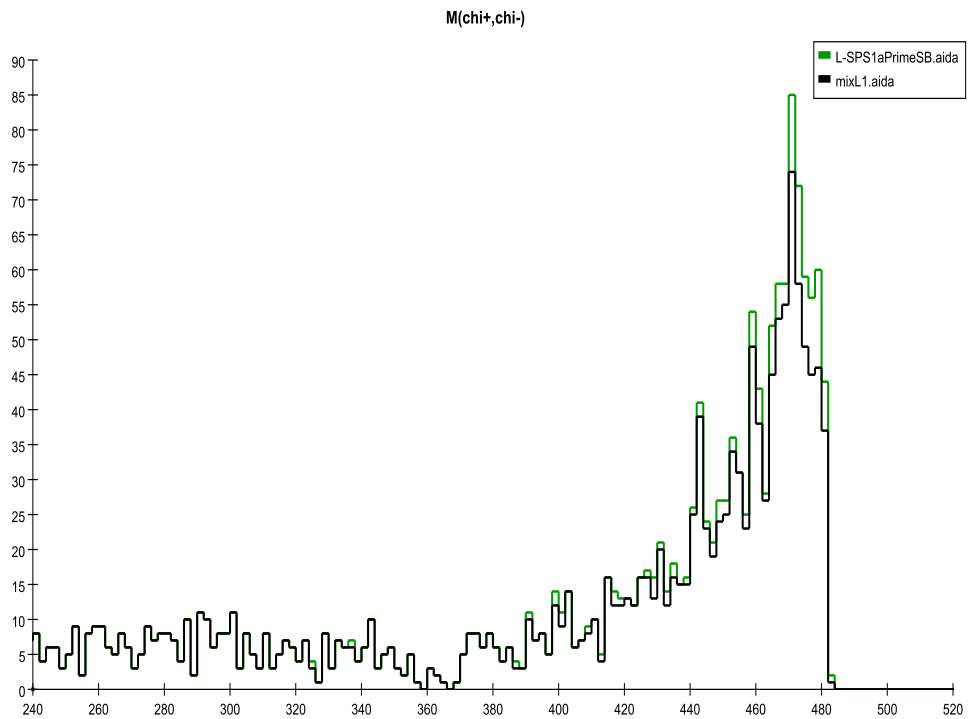}}
\vspace*{0.1cm}
\caption{Recoil mass distribution for chargino pair production in
SPS1a' from the photon
tag analysis: number of events/2 GeV bin assuming
an integrated luminosity of 250 fb$^{-1}$ for both LH (top panel) and RH
(bottom panel) electron beam polarization.}
\label{radiative_sps1a}
\end{figure*}

\clearpage

\subsection{Very Close Mass Case}
\label{Sec:closemass}

As mentioned above, when $\Delta m_{\tilde{\chi}} < m_\pi$
the decay length of the chargino is long compared with the detector size,
as is shown in Fig.~\ref{lifetime}. In this case, chargino
production may be detected by searching for two massive, essentially
stable, charged particles that traverse the full
detector. Seven of the 53 AKTW models with kinematically accessible
charginos at the 500 GeV ILC fall into this category.
To perform this search we demand:
\begin{enumerate}
\item There be only 2 massive charged tracks in the event.

\item There be no tracks, or energy clusters, within 100 mrad of the beam.

\item $\beta=\frac{p}{E} < 0.93$ for both charged tracks.
(This value is based on constraints from LEPII \cite {Yao:2006px}.)

\item The energy of the two tracks satisfy
$\sum\limits_{i = 1}^2 E_i > 0.75 \sqrt{s}$.

\end{enumerate}
The last two cuts remove most of the background from the production of
muons. After these cuts are imposed, the remaining background should
be small, aside
from detector fakes and possible tails from muon production due to
detector smearing.
We then study the $\beta=\frac{p}{E}$ spectrum for both charged tracks
and look for excesses in the region of low $\beta$.

In this analysis we search for stable charged tracks in the
final state whose energy can be reconstructed via a $dE/dx$
measurement. $\beta=p/E$ should be significantly less than 1,
which would allow us to
easily distinguish such tracks from those produced by Standard
Model particles. However, in the current public version of
org.lcsim~\cite{lcsim}, $dE/dx$ is not yet implemented\footnote{Non-fully
tested implementations
seem to be available in the contrib area, however, since these are not
yet part of the main code and hence are not fully tested and integrated,
we refrain from using them.}. We therefore employ a cheat algorithm,
and smear, by a random amount, the energy of all final state
tracks which we take from the PYTHIA input before detector simulation.
The width of a random Gaussian fluctuation should mimic the resolution
obtainable from a more realistic TOF (Time Of Flight)
or $dE/dx$ measurement. There is not yet full agreement among the ILC
detector experts as to the attainable precision which may be possible
in the determination of $\beta$ {\cite {private}},
so we perform two analyses, one with a 5\% and one with
a 10\% assumed resolution on $\beta$. (Note that an
energy smearing of $5(10)\%$ translates
into a resolution on $\beta$ of roughly $5(10)\%$.) We note that both
the ATLAS and CMS detectors have excellent $\beta$ resolutions of
$5(3)\%$ \cite{lhcbeta}, respectively, and we anticipate that any ILC detector should have
a comparable precision as demonstrated in Ref.~\cite {Martyn:2007mj}.

As shown in Fig.~\ref{fig:stablebg}, the background is indeed negligible
for an energy smearing of 5\%. However, some SM background
from Bhabha scattering with missing energy due to
initial state radiation and beamstrahlung where the
forward photon is not detected, leaks into
the analysis when an energy smearing of 10\% is assumed. The background
displayed in
Fig.~\ref{fig:stablebg} is almost entirely due to this process.
Nevertheless, we expect the ILC to have an energy resolution
better than 10\%, so this background source should not be a problem.

\begin{figure*}[hptb]
\centerline{
\epsfig{figure=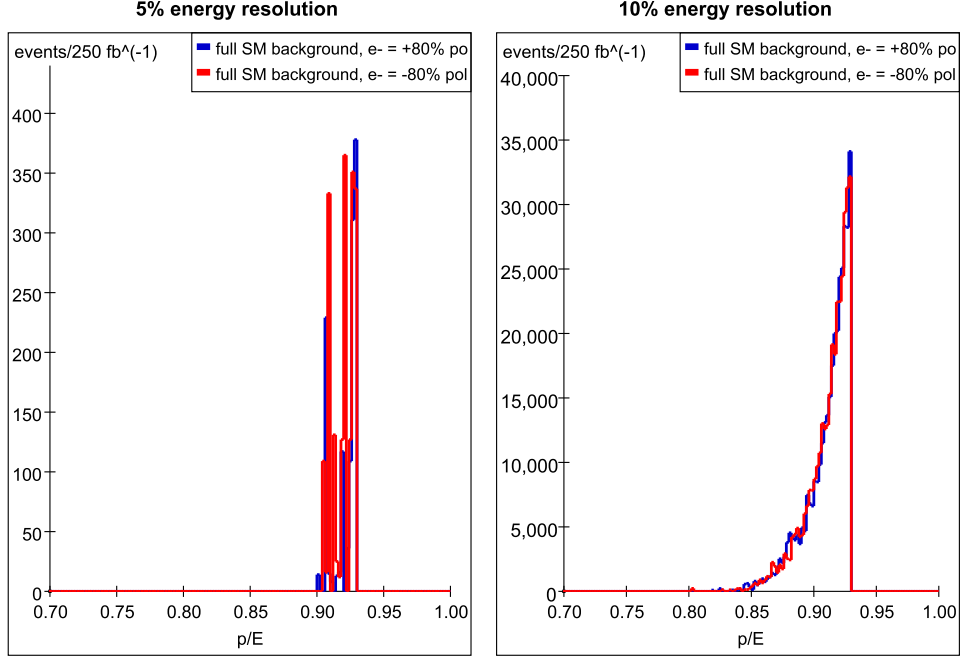,width=13cm,clip=}}
\vspace*{0.1cm}
\caption{$\beta=p/E$ distribution of the
SM background after the
kinematic cuts listed in the text have been imposed. This is generated
from $250 \mbox{ fb}^{-1}$ of SM events generated with
80\% right-handed (solid blue line) and 80\% left-handed (solid red line)
electron beam polarization and unpolarized positron beam
at $\sqrt{s} = 500$ GeV. In the Figure in the left panel, we
assume an energy
resolution of 5\%, the Figure on the right is for an energy
resolution of 10\%.}
\label{fig:stablebg}
\end{figure*}

Out of the 53 AKTW models with kinematically accessible chargino pairs,
only 7 have values of $\Delta m_{\tilde \chi}< m_\pi$ and have effectively
stable charginos as far as collider detector measurements are concerned.
As in the previous analysis for radiative chargino production,
the backgrounds are similar
for both polarizations, however the chargino production
cross sections for these models are about a factor of 9 larger in
the case of LH polarization than in the RH case. This is because
charginos with small values of $\Delta m_{\tilde \chi}$ are mostly wino
in the AKTW models, corresponding to large values of the $\mu$ parameter.
Figure~\ref{stable1}
shows these 7 models for both values of the
assumed $\beta$ resolution and we see that in either case all of these
models are clearly visible above background.

\begin{figure*}[hptb]
\centerline{
\includegraphics[width=13.0cm,height=10.0cm,angle=0]{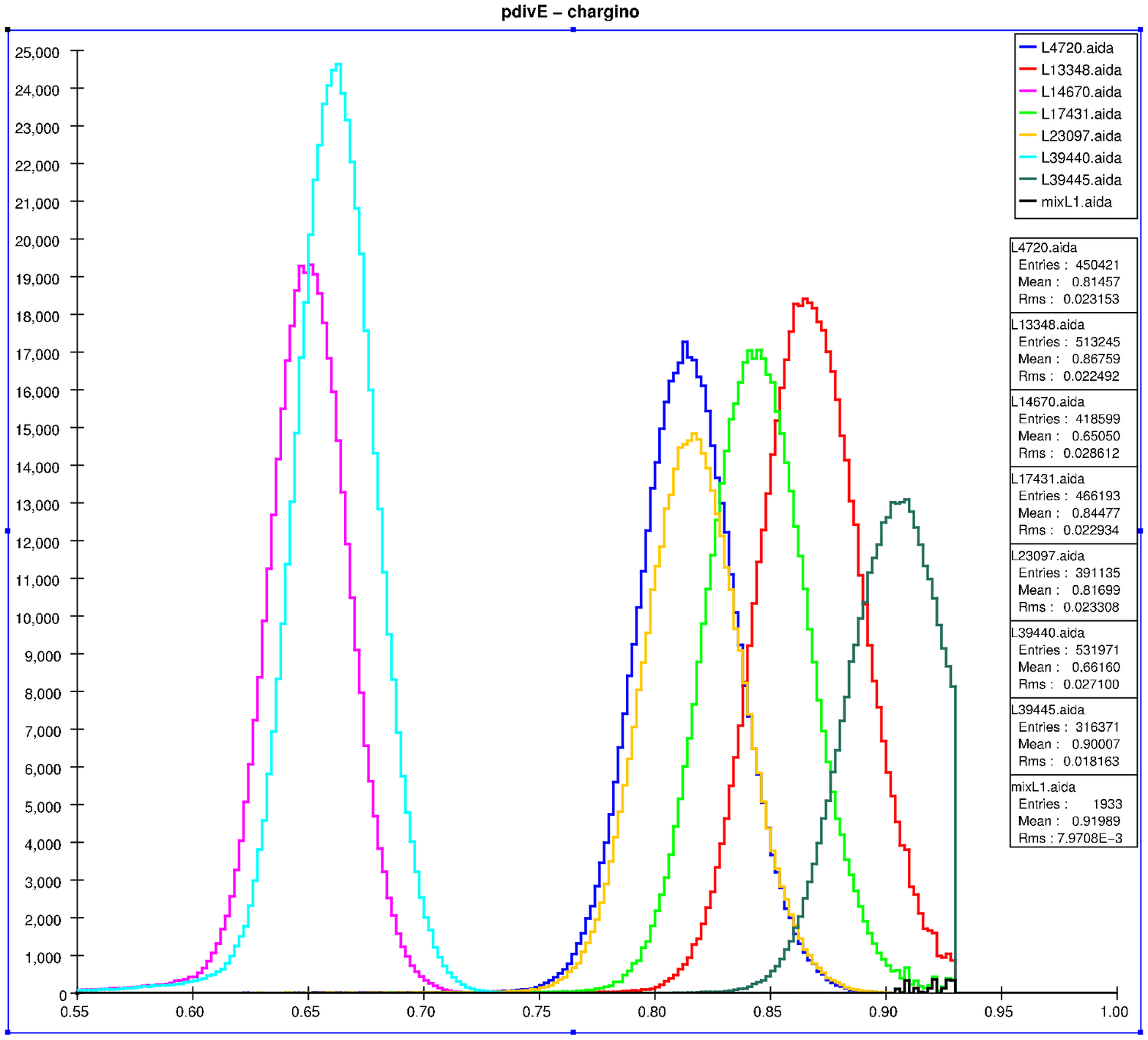}}
\vspace*{0.1cm}
\centerline{
\includegraphics[width=13.0cm,height=10.0cm,angle=0]{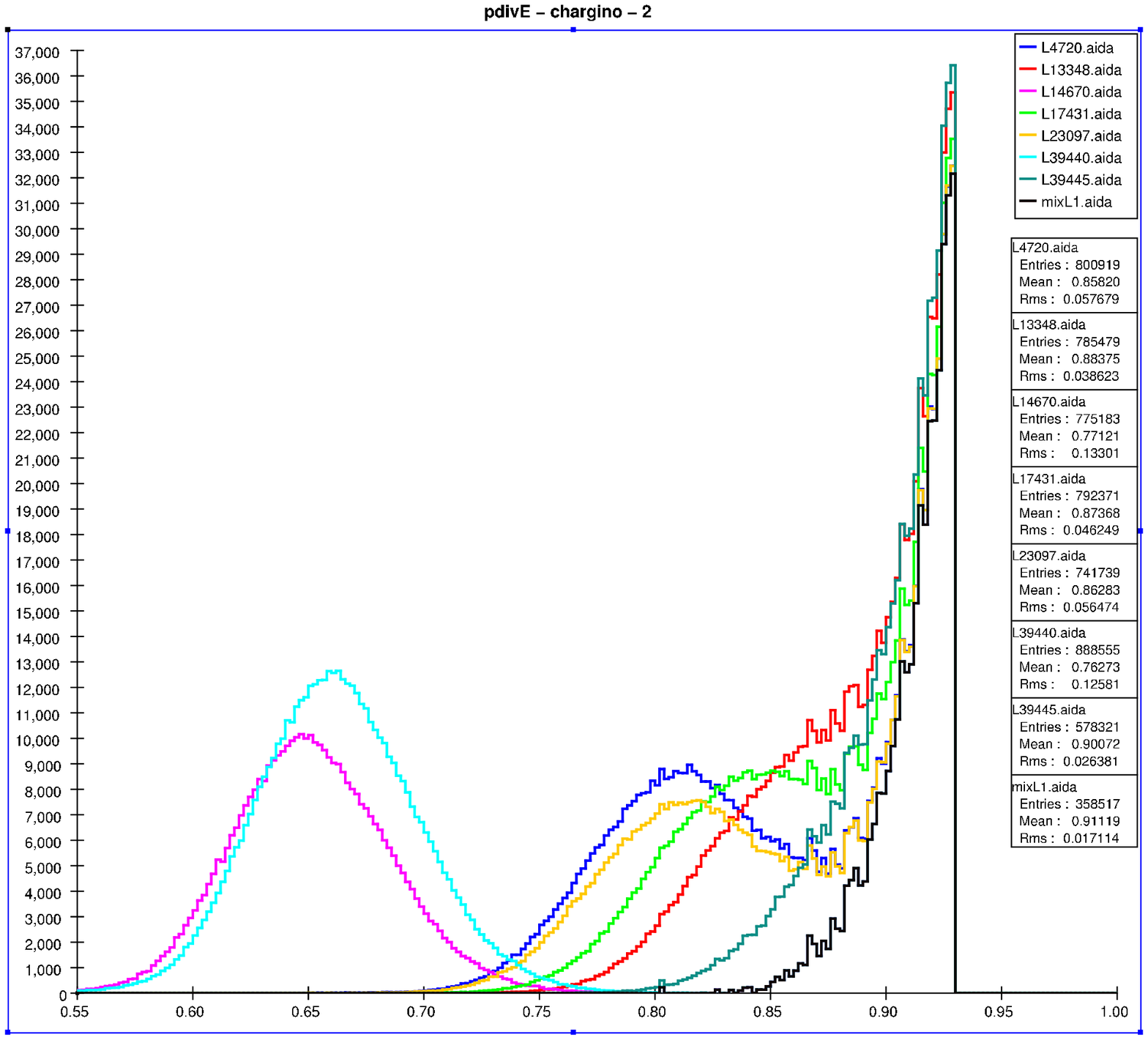}}
\vspace*{-0.1cm}
\caption{Velocity (=$\beta=p/E$) distribution for long-lived charginos
assuming an integrated luminosity of 250 fb$^{-1}$
and LH electron beam polarization. The top(bottom) panel corresponds
to a resolution of $5(10)\%$ on $\beta$.}
\label{stable1}
\end{figure*}

Of course this search will not just find charginos, but will also detect
any sufficiently long-lived charged particles in the
kinematically accessible mass range. In fact, out of the 28 AKTW models
with kinematically accessible $\tilde \tau$'s,
we find 3 models (labelled as
207, 27285 and 29334) which have staus with very long
lifetimes; in one of these cases (27285) the $\tilde\tau$ is the NLSP.
Such a situation
can occur, \eg, in gauge mediation, where the $\tilde \tau$ will decay
with a very long lifetime by emitting a gravitino {\cite {Ambrosanio:1997rv}}.
Of course in
such a model $\tilde \chi_1^0$ is not the actual LSP; thus the cosmology is
non-standard, but viable. With the above analysis,
these models should also lead to
observable signals in this channel.

These two possible candidate for stable charged particles are easily
distinguished by their angular distributions, \ie, spin-0 vs. spin-1/2,
as well as by their response
to the various electron beam polarizations as we will see below.

A similar search can be performed in the case of the three models with
long-lived staus; the results are seen in
Fig.~\ref{stable2}. Recall that in the case of stau pair production the
cross section is not only controlled by
the $\tilde \tau$ mass but also by the $\tilde \tau_{L,R}$ mixing angle
which governs the stau coupling to the $Z$ boson.
Note that the event rates shown here are
significantly lower than those of the
long-lived charginos and so the SM background is potentially more serious.
It is clear, however, that
in the case of a $5\%$ resolution on $\beta$ the staus in these 3 models
will be observable; the situation is more difficult to assess by eye
in the case of only a $10\%$ resolution. A detailed statistical study,
however,
shows that these 3 stau models will lead to
signals at the level of significance $>5$ for both choices of the
electron beam
polarization and for either assumed value of the $\beta$
resolution. We observe that stau production in these 3 special models
with both LH and RH beam polarization lead
to comparable cross sections.

We note that there are no long-lived charged particles in the case of SPS1a'.

\begin{figure*}[hptb]
\centerline{
\includegraphics[width=13.0cm,height=10.0cm,angle=0]{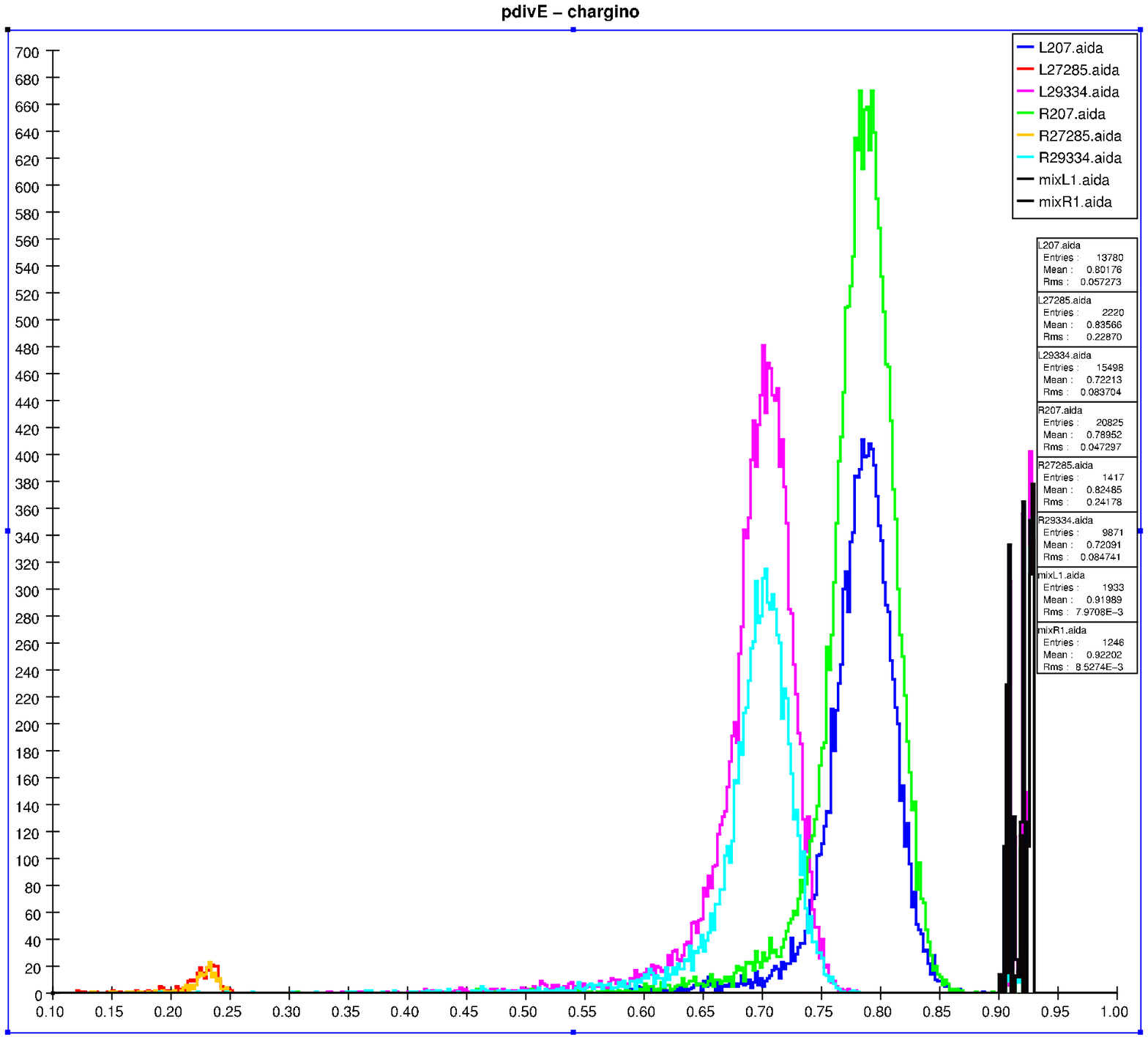}}
\vspace*{0.1cm}
\centerline{
\includegraphics[width=13.0cm,height=10.0cm,angle=0]{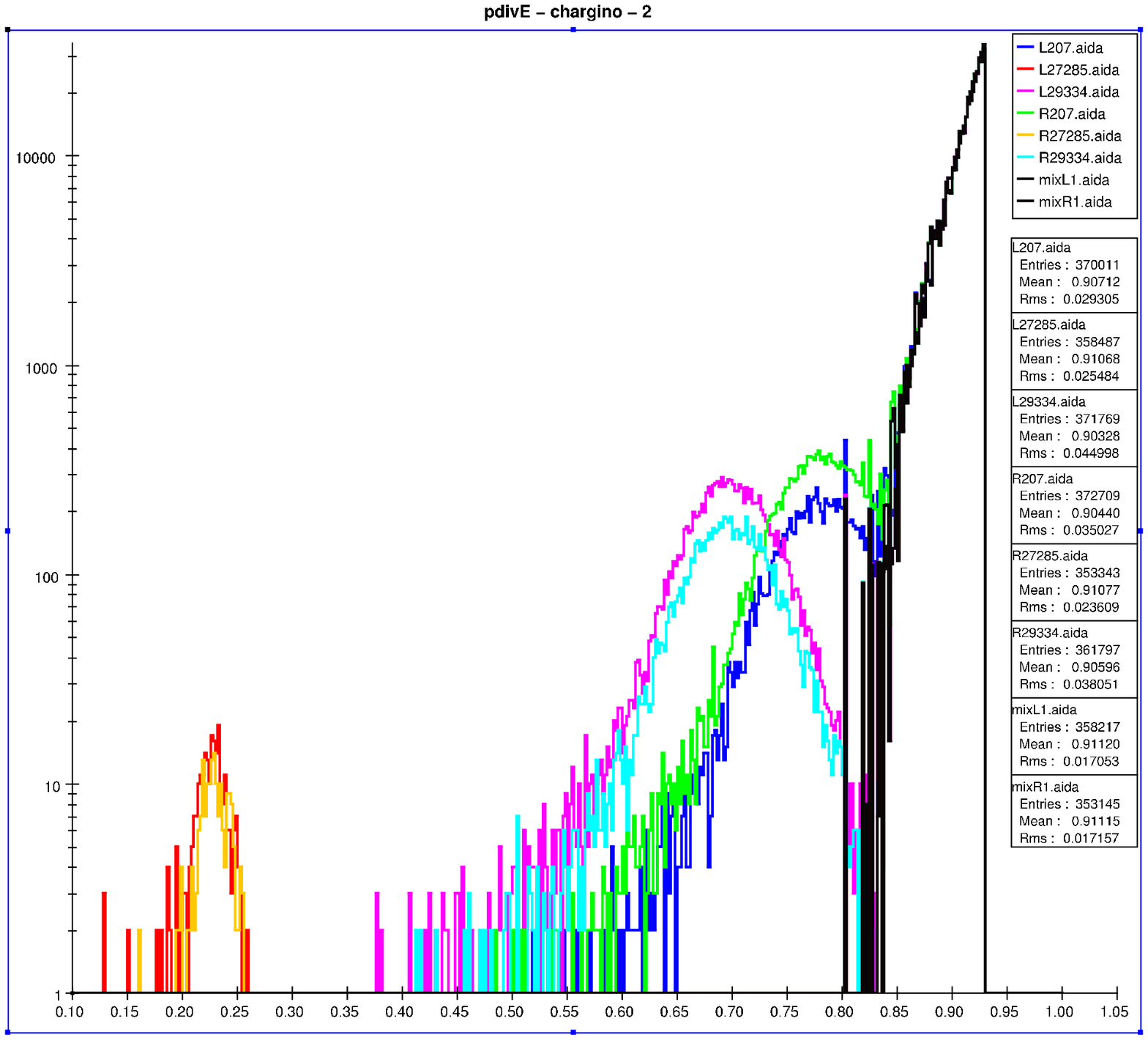}}
\vspace*{-0.1cm}
\caption{Velocity(=$\beta=p/E$) distribution for long-lived staus assuming
an integrated luminosity of 250 fb$^{-1}$
and for both electron beam polarizations as labelled.
The top(bottom) panel corresponds
to a resolution of $5(10)\%$ on $\beta$.}
\label{stable2}
\end{figure*}

\subsection{Summary of Chargino Analyses}

Here, we collect and summarize the results of the various chargino
analyses presented in this section.
We remind the reader that the set of AKTW models
contain 53 models with kinematically accessible charginos at the 500 GeV
ILC. The critical parameter that
determines the open decay channels for the chargino, and hence governs
the appropriate search analysis, is
$\Delta m_{\tilde\chi}$, the mass difference between the lightest
chargino and the $\tilde\chi_1^0$ LSP.
Of the 53 models, 7 have values of $\Delta m_{\tilde\chi}$ small enough
to render the lightest chargino essentially stable and it traverses the
full detector before it decays. These models are visible in our
stable charged particle search. An additional 7 models have
chargino-LSP mass differences in the range $\Delta m_{\tilde\chi}<1$
GeV, and result in final states with large values of missing energy plus
several soft pions.
These models are targeted by our radiative chargino analysis.
Thirty-seven of
the models have mass differences in the range $1<\Delta m_{\tilde\chi}<6$
GeV and the charginos decay into off-shell $W$ bosons. For this region,
we designed a multi-pronged search strategy using 11 observables in 3
decay channels (missing energy + $\mu\mu$, 2 jets $\mu$, 4 jets).
Lastly, 2 of the models have $\Delta m_{\tilde\chi}> M_W$ and in this
case the charginos decay into on-shell $W$ bosons. We developed only
one search analysis here, utilizing the 4-jet mode, and found this
to be a very clean channel for detecting chargino production.

A summary of how many models are visible above the SM background for
each observable in our chargino search analyses
is presented in Table~\ref{tabcharg}. Here, we employ, as always, our
visibility criteria that the signal significance ${\cal S}>5$. We see
that the channel with the mixed final state of 2-jets + muon + missing
energy and the missing energy observable in the 4-jet + missing channel
yield the highest number of observable models and thus
are the best channels for detecting chargino production in the
randomly generated AKTW models.

\begin{table*}
\centering
\begin{tabular}{|c|c|c|} \hline\hline
Observable & Visible with RH & Visible with LH \\ \hline
$E_{jj}^{\mbox\tiny on-shell}$ & 2 & 2\\
$E_{\mu\mu}$ & 12 & 10\\
$E_{\mbox jet-pair}-\mu$ & 26 & 35\\
$E_{\mbox jet-pair}-\mu$, $m_{\LSP,\mbox{\tiny min}} = 100$ GeV
&0 & 0\\
$M_{\mbox jet-pair}-\mu$ & 23 & 35\\
$M_{\mbox jet-pair}-\mu$, $m_{\LSP,\mbox{\tiny min}} = 100$ GeV
& 0 & 0\\
ME(4jets) & 9 & 30\\
ME(4jets), $m_{\LSP,\mbox{\tiny min}} = 100$ GeV& 4 & 34\\
ME(4jets), additional $p_T$ cut & 3 & 2\\
$M_{\mbox jet-pairs}$ & 2 & 4\\
$M_{\mbox jet-pairs}$, $m_{\LSP,\mbox{\tiny min}} = 100$ GeV &
2 & 2\\
$M_{\mbox jet-pairs}$, additional $p_T$ cut & 3 & 2\\
Radiative Production & 14 & 14\\ \hline\hline
\end{tabular}
\caption{Number of models that are visible above the SM background
with a significance ${\cal S}>5$ in each observable
at $\sqrt s=500$ GeV with
250 \infb\ of integrated luminosity for each electron beam polarization.}
\label{tabcharg}
\end{table*}

Figure~\ref{distribution} displays the location of each AKTW model with
$\Delta m_{\tilde\chi}<M_W$ in the $\Delta m_{\tilde\chi}-$ chargino
mass plane. The color coding of the model marker indicates whether it
is observable in any of our analyses for either beam
polarization as labelled in the Figure caption.
The location of the various model
points in this plane reveals the kinematic properties targeted by
each search technique. For example, the radiative production analysis
captures the models with low $\Delta m_{\tilde\chi}$ as it was designed
to do, and also detects charginos that are light enough to be produced
with a hard photon.
Here, we see that all 7 of the essentially stable charginos are
captured by our stable charged particle search, 3 models are only
observable via radiative chargino production, 11 models are visible
in both the radiative production channel and at least one of
the off-shell $W$ analyses, while 26 models are detectable in at
least one of the off-shell $W$ analyses but not in radiative production.
4 out of the 53 AKTW models are not observable in any of our analysis
channels. However, each of these 4 models have a lower cross section
due to phase space suppression.

\begin{figure*}[htbp]
\centerline{
\includegraphics[width=10.0cm,height=13.0cm,angle=90]{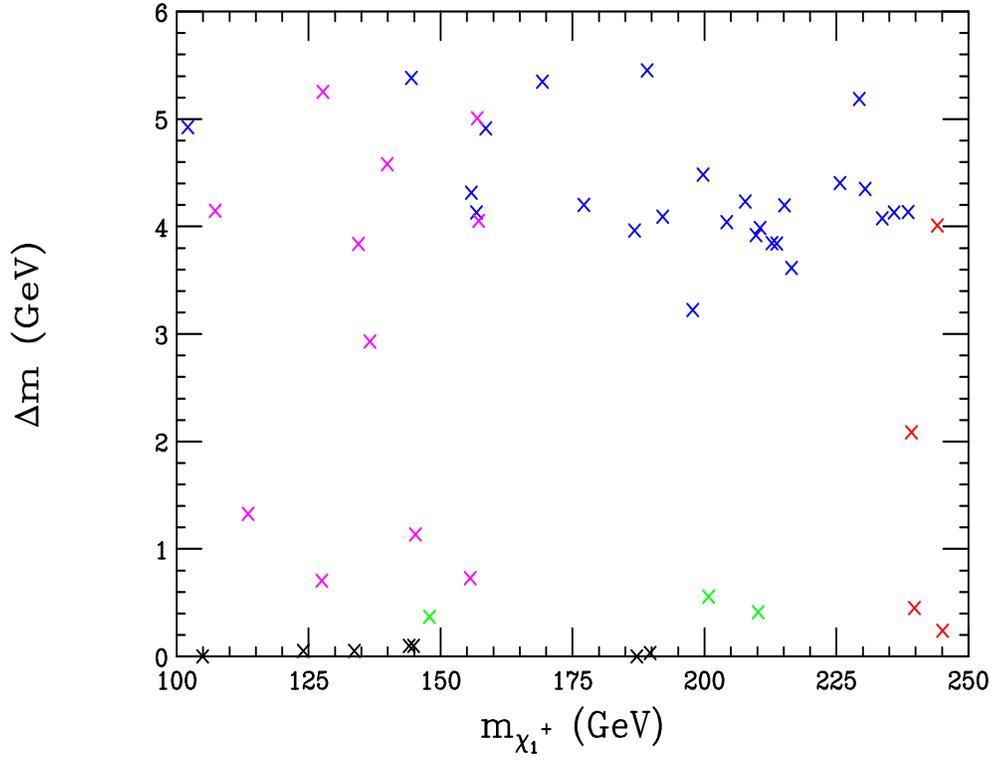}}
\vspace*{0.1cm}
\caption{Distribution of the chargino-LSP mass difference versus the chargino
mass for the AKTW models with $\Delta m_{\tilde \chi}<6$ GeV for the
$\tilde\chi_1^\pm$ states that are accessible at $\sqrt s=500$ GeV. The blue
crosses represent models that are observable in our suite of analysis channels
based on the $\tilde\chi_1^\pm$ decay via off-shell $W$ bosons. The green
crosses correspond to models that are only visible in the radiative chargino
production analysis channel, while the magenta ones represent models
that yield observable signals in both the radiative and off-shell $W$
channels. The black crosses are models that are visible in the stable
chargino analysis. The red points are the 4 models where the
$\tilde\chi_1^\pm$ state is {\it not} observable in any of our
analysis channels, essentially due to phase space restrictions.}
\label{distribution}
\end{figure*}

\clearpage

\section{Neutralino Production}
\label{Sec:neutralino}

The neutralino sector of the MSSM is the most complex as the
mass eigenstates are admixtures
between the Bino, neutral Wino, and two neutral Higgsino weak states.
Neutralinos, $\tilde \chi_i^0$, can be pair produced in
$e^+e^-$ collisions via two distinct mechanisms: $s-$channel
$Z$ boson exchange can make a neutral Wino plus a
Higgsino, while $t,u-$channel selectron exchange can produce Binos
and neutral Winos in all combinations.
These mechanisms ensure that all of the ten possible
processes $e^+e^- \to\tilde\chi_i^0 \tilde\chi_j^0$ are potentially
accessible with rates depending upon the sparticle masses and
the various mixing angle factors. At a
$\sqrt s$=500 GeV machine, it is likely that only the
first one or two of these states will be
kinematically accessible and this is indeed the case for the wide
selection of AKTW models analyzed here as
shown in Fig.~\ref{benj2} and Table ~\ref{finalstates}. If the mass
separation between $\tilde \chi_1^0$ and $\tilde \chi_2^0$ is
sufficiently large, then the decay channel $\tilde \chi_2^0 \to Z\tilde
\chi_1^0$ may lead to a clear signal if the $Z$ boson is not too far
off-shell. Unfortunately, in the models examined
here, if $\tilde \chi_2^0$ is sufficiently light to be produced it is
very close in mass to $\tilde \chi_1^0$ and we find that
such decays are almost impossible to observe. That being the case,
we only consider $\tilde \chi_1^0$ pair production with a radiated
photon, as well as
$\tilde\chi_1^0\tilde\chi_2^0$
associated production in
the discussion below. Recall that $\tilde\chi_1^0$ is the only
accessible MSSM particle in many of the AKTW models.

It is important to consider the weak eigenstate mixture of the
$\tilde \chi_1^0$ in our set of models; this is shown in
Fig.~\ref{neutcomp}, where it is interesting to observe that the
lightest neutralino is mostly a pure weak eigenstate.

\begin{figure*}[hptb]
\centerline{
\includegraphics[width=10.0cm,height=13.0cm,angle=-90]{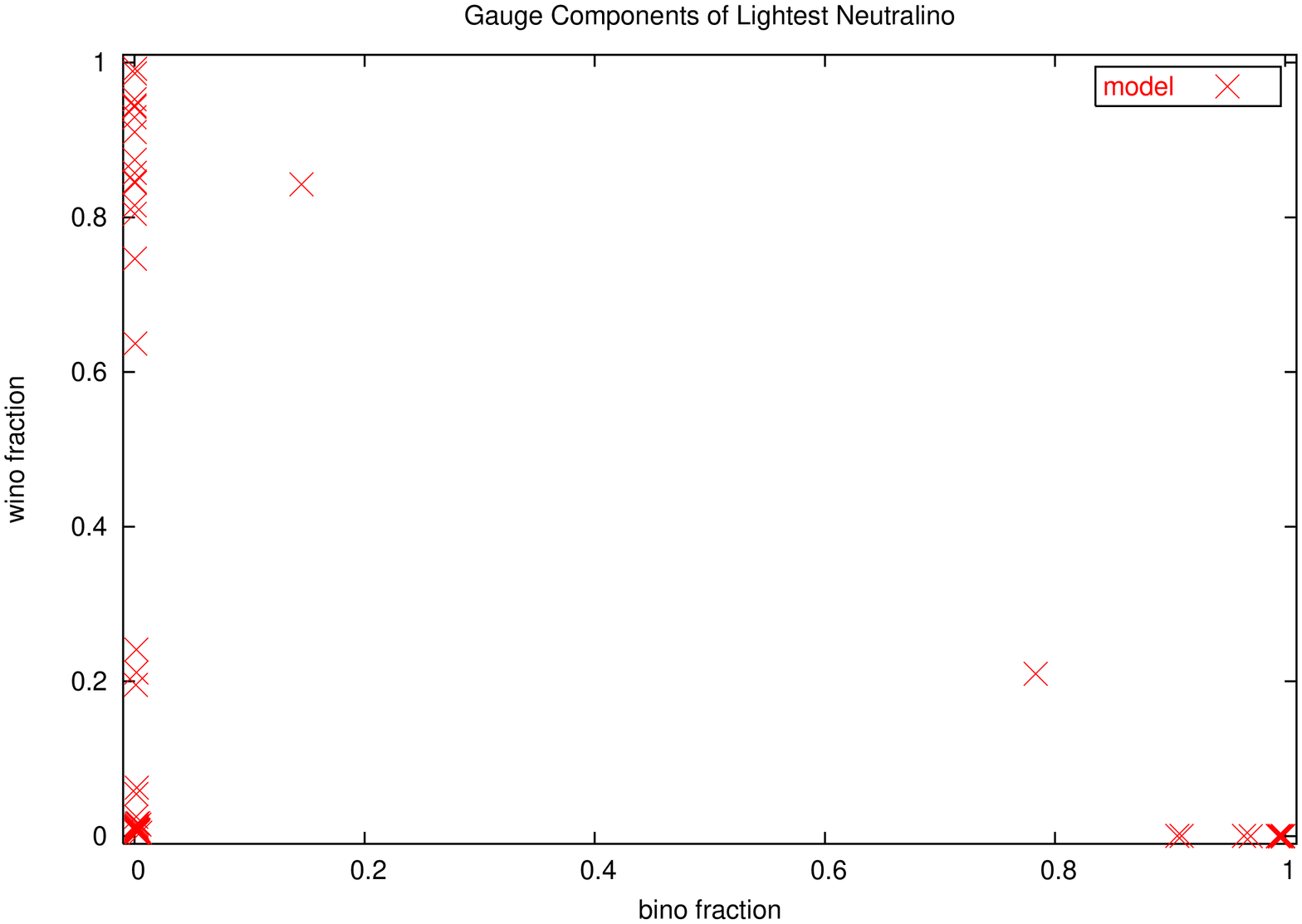}}
\vspace*{0.1cm}
\centerline{
\includegraphics[width=10.0cm,height=13.0cm,angle=-90]{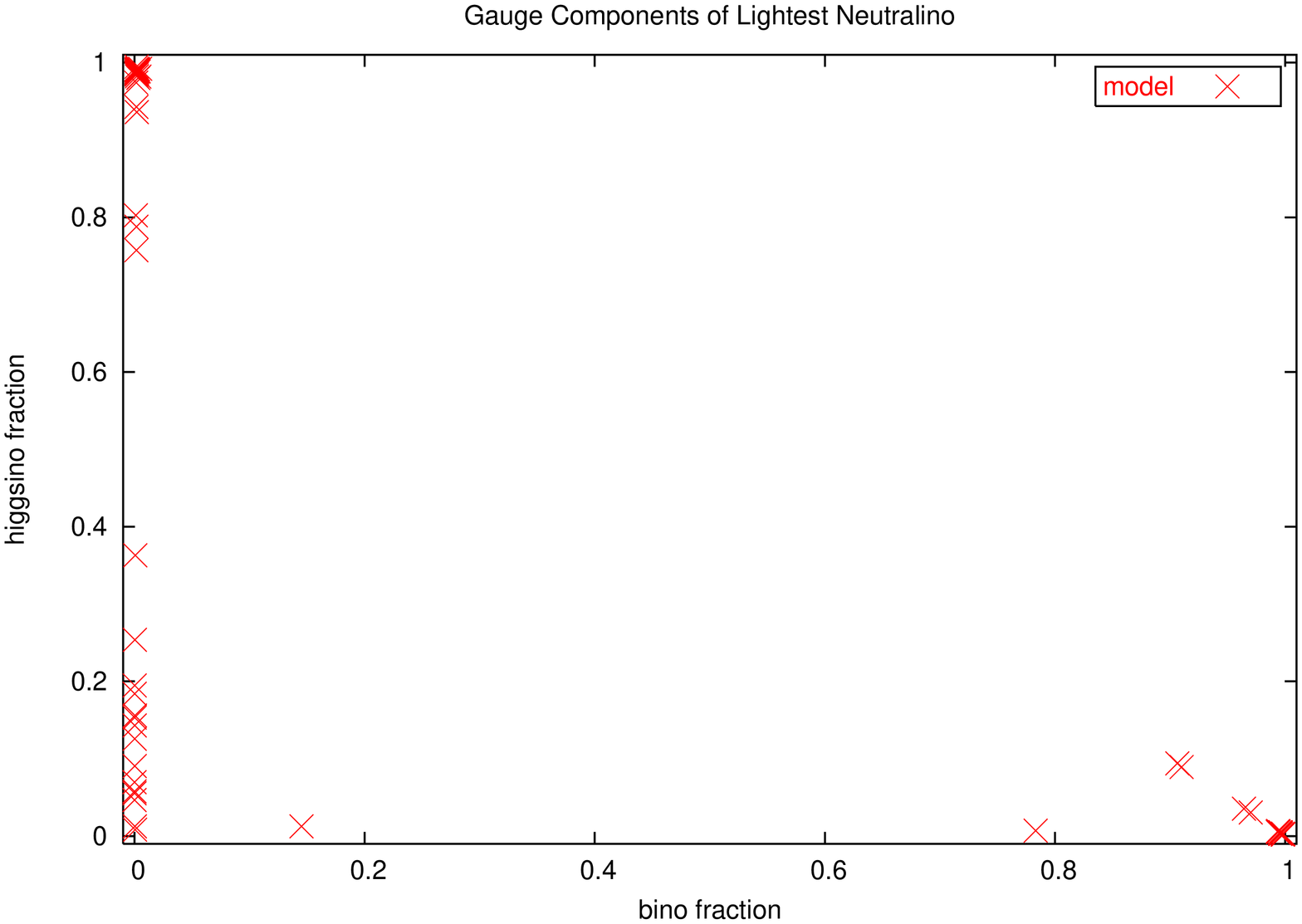}}
\vspace*{-0.1cm}
\caption{The composition of the lightest neutralino in the 242
AKTW models.}
\label{neutcomp}
\end{figure*}

\subsection{$\tilde \chi_2^0\tilde\chi_1^0$ Associated Production}
\label{Sec:chi1chi2}

In order to get a handle on the neutralino sector, it is
important to consider the associated production of
neutralinos, \ie, $e^+e^- \to \tilde \chi_2^0\tilde \chi_1^0$, which
proceeds by $Z$ boson exchange in the $s$-channel via the
Wino and Higgsino content of the $\tilde \chi_{1,2}^0$ and by
selectron exchange in the
$t,u$-channels via their corresponding Wino and Bino content.
The cross section for this process is thus sensitive to
the mixing in the neutralino sector as well as to the
masses of the exchanged $\tilde e_{L,R}$.
Note that if the selectrons are heavy, as is
the case in many of the AKTW models, then the $s$-channel transition
dominates; in this situation the associated
production process will be suppressed if either or both of the
$\tilde \chi_2^0$ or $\tilde \chi_1^0$ have a large Bino content,
which as we saw above, is a relatively common occurrence in the models
considered here.

At $\sqrt s=500$ GeV, 46/242 of the AKTW models have the final state
$\tilde\chi_2^0\tilde \chi_1^0$ kinematically accessible.
The state $\tilde \chi_2^0$ can decay in several ways depending on the
mass spectrum details in the gaugino sector. A mode which
is always present and yields a relatively clean signature is
$\tilde \chi_2^0 \to\tilde \chi_1^0 Z/H$, with the $Z/H$ being on- or
off-shell depending on the $\tilde \chi_2^0-\tilde \chi_1^0$ mass difference.
Certainly, this channel will be easier to observe in
the on-shell case since the invariant mass distribution of the
visible particles in the final state
will be peaked at the $Z/H$ mass. In either case,
we consider the decay modes
$Z,H\to jj$, with the jets not flavor tagged, as well as the leptonic
modes $Z\to \mu^+\mu^-, e^+e^-$. In order to access the viability of
this channel, we examine in
Fig.~\ref{chi2chi1} the mass splitting between the first
two neutralino states for the 46 models where this production
mechanism is kinematically accessible.
Here, we see that for most models the
mass splitting is rather small; only 8 of these models have
neutralino mass splittings larger than $M_Z$ and 13 have mass differences
larger than 20 GeV.
It is unlikely that any of the other remaining models will
produce hard enough jets and/or leptons to pass the analysis
cuts or be visible above background.
For the models with the larger mass differences, we stress again
that their signal
rates will be controlled by both the selectron masses
and the Bino content of the two neutralinos. The dominant
background we contend with arise from, \eg,
$e^+e^- \to ZZ\to jj/ \ell^+\ell^- + \nu \bar \nu$,
$\gamma e \to \nu W$ with $W\to jj$,
as well as $\gamma\gamma\to\ell^+\ell^-$.

\begin{figure*}[hptb]
\centerline{
\includegraphics[height=13cm,angle=270,width=13cm]{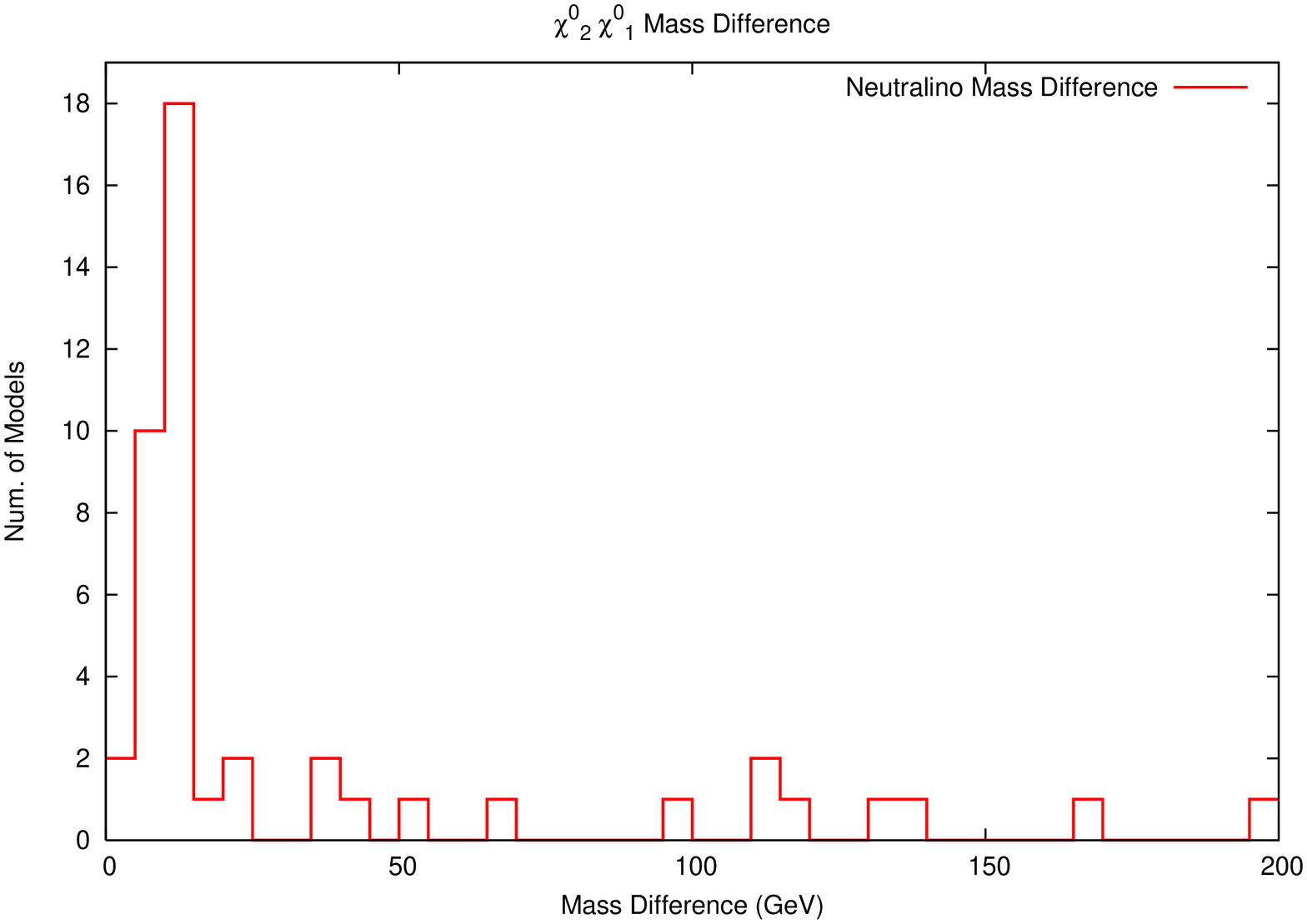}}
\vspace*{0.1cm}
\caption{$\tilde \chi_2^0-\tilde \chi_1^0$ mass difference for the
46 AKTW models which have $\tilde \chi_2^0\tilde \chi_1^0$ associated
production kinematically accessible at $\sqrt s=$ 500 GeV.}
\label{chi2chi1}
\end{figure*}

To reduce the SM background for associated neutralino production,
we demand:
\begin{enumerate}
\item There be precisely one lepton pair (electrons or muons) or one
jet-pair in the event and no other visible particles.

\item The missing energy satisfy $\Emiss > 300 $ GeV.
This removes the majority of the background arising from
$Z$ and $W$ boson production.

\item The transverse momentum for each lepton or jet satisfy
$p_T > 0.14 \sqrt{s}$.
This cut removes most of the ubiquitous $\gamma \gamma$ and
$e \gamma$ initiated backgrounds.

\item The angle between the lepton or the jet pair be $< 95$ degrees.
This further reduces the background from $W$ boson production.
\end{enumerate}
We then examine the invariant mass spectrum of the
electron-, muon-, or jet-pair. The remaining background after these
kinematic cuts are imposed is displayed for the
$\mu\mu\,, jj+$ missing energy channels in Figs.~\ref{chi21mumu}
and \ref{chi21jj}; we find that the background for the
$e^+e^-$ final state is qualitatively similar to those for muons.

\begin{figure*}[hptb]
\centerline{
\epsfig{figure=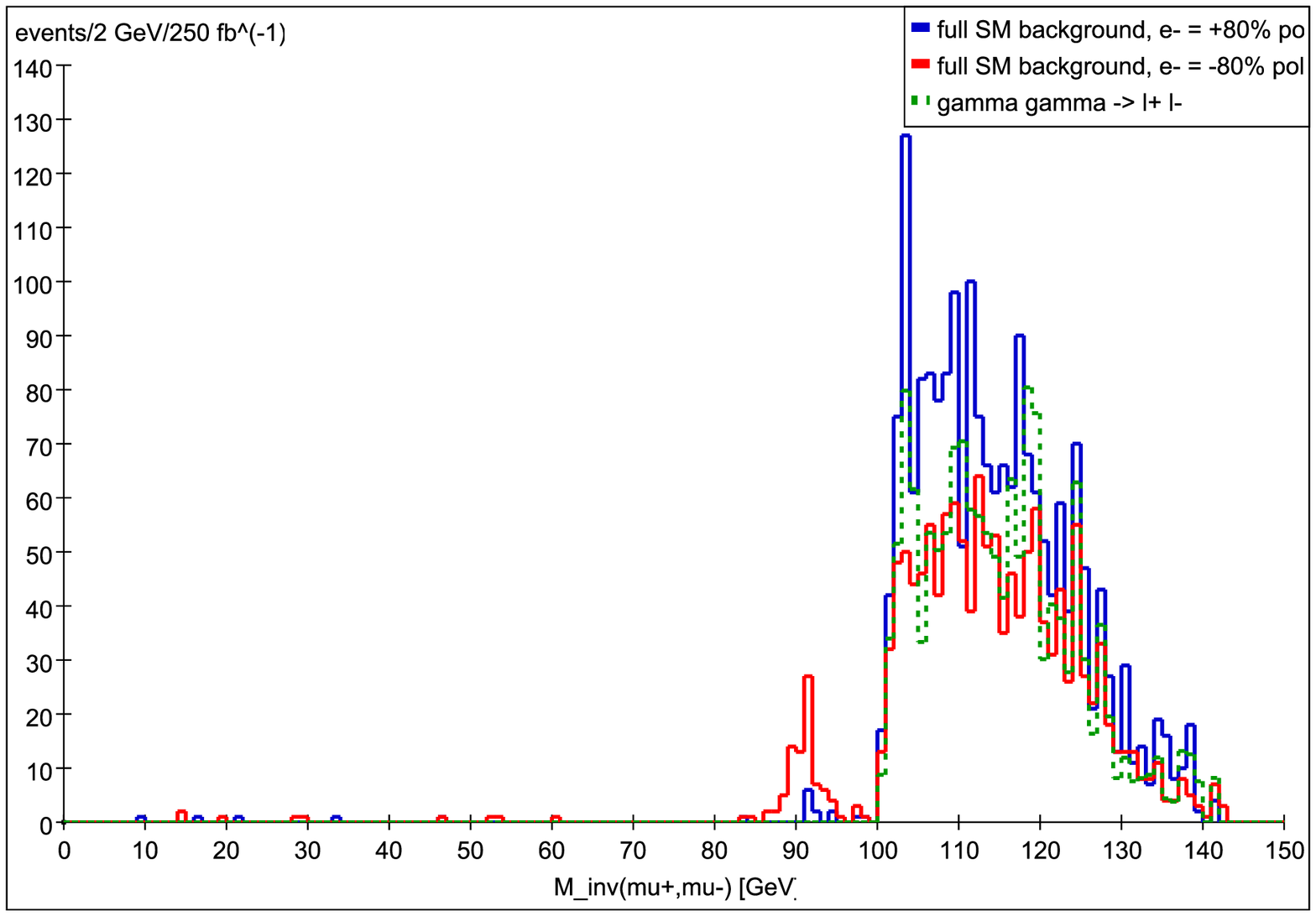,width=13cm,clip=}}
\vspace*{0.1cm}
\caption{SM background to associated neutralino production for the
$M_{inv}(\mu\mu)$ distribution. This is generated for a
250 fb$^{-1}$ sample of SM events with $80\%$ RH(solid blue) or
LH(solid red) electron beam polarization at $\sqrt s=500$ GeV.
The dotted green line represents the dominant contribution after the
cuts are imposed, $\gamma \gamma \to \ell^+\ell^-$,
which is independent of the beam polarization.}
\label{chi21mumu}
\end{figure*}

\begin{figure*}[hptb]
\centerline{
\epsfig{figure=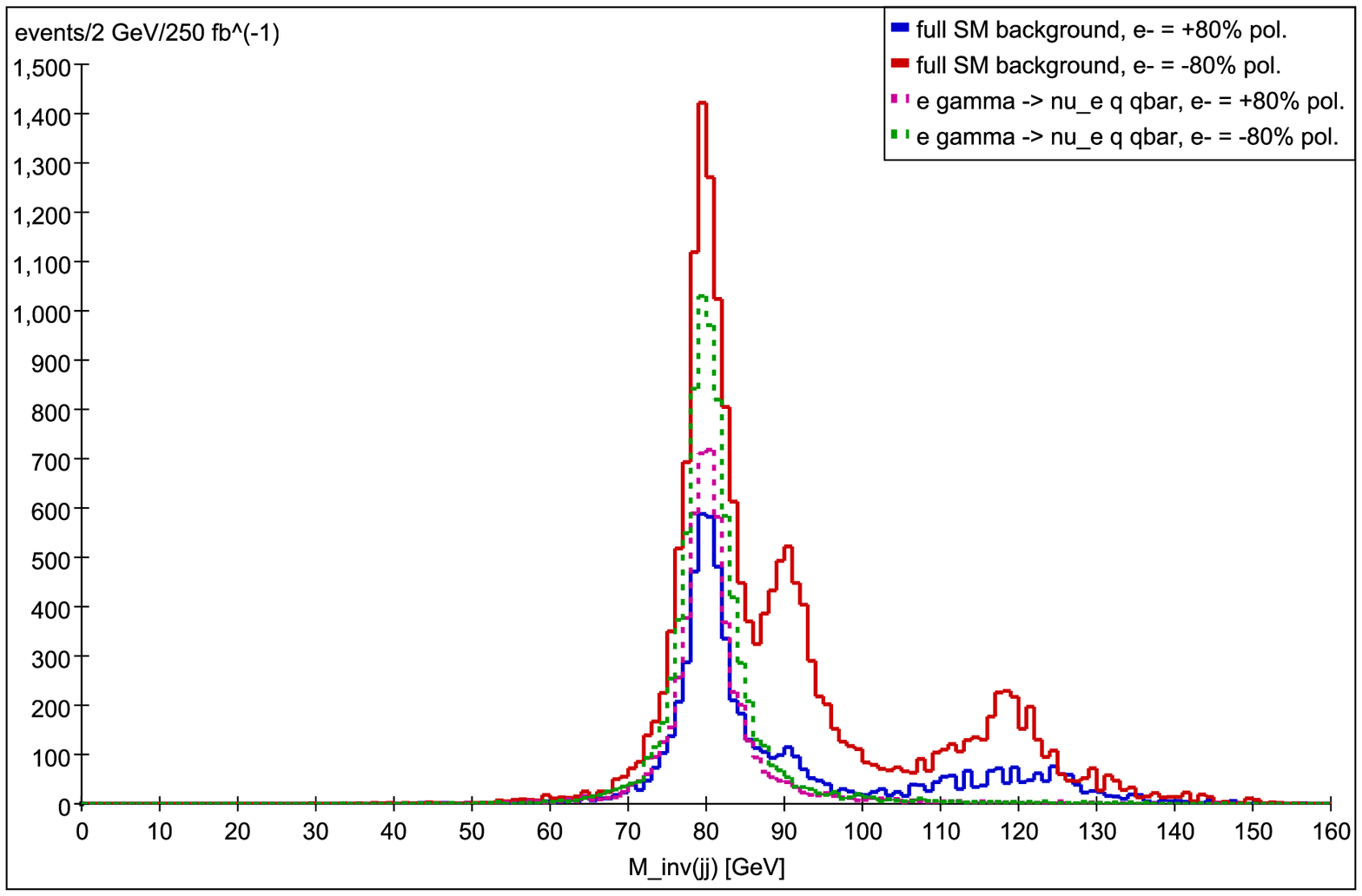,width=13cm,clip=}}
\vspace*{0.1cm}
\caption{SM background to associated neutralino production for the
$M_{inv}(jj)$ distribution. This is generated for a
250 fb$^{-1}$ sample of SM events with $80\%$ RH(solid blue) or
LH(solid red) electron beam polarization at $\sqrt s=500$ GeV.
The dominant contribution after the cuts are imposed
arise from $\gamma e \to \nu_e q\bar q$ and are shown
by the green(pink) dotted line for LH(RH) beam polarization.}
\label{chi21jj}
\end{figure*}

The signal should have a clear peak in the invariant mass spectrum that
reconstructs to the $Z$ boson although excesses may also appear
elsewhere in the distribution.
Note that jet energy resolution is crucial here as some
background sources, for example,
$e^- \gamma \rightarrow \nu_e d \bar{u}$ and $e^- \gamma \rightarrow
\nu_e s \bar{c}$, have an invariant mass peak at the $W$ boson mass.
Thus the $W$ and $Z$ boson mass peaks must be separable in the 2-jet
channel. As is common in many of our analyses, the
SM background is far lower with RH electron beam polarization as this
suppresses $W$ boson production. Note that we may also have to
deal with backgrounds arising from other SUSY production processes that can
fake the signals from associated production.

Typical results for these analyses are shown in Fig.~\ref{associated1}
for representative AKTW models.
In the case of the dijet
analysis, three peaks are observable at the masses of the $W$, $Z$ and 120
GeV Higgs boson. Some models
lead to small excesses on the $W$ peak while some have excesses at the $Z$;
others have excesses at both
locations. Five models are found to show a signal with a significance
$>5$ in the dijet channel; all these models
have small excesses at the $Z$ peak, corresponding to the
associated production channel under consideration.
Similarly, some models also show some excess at the $W$ peak arising
from a different $\tilde\chi_2^0$ decay channel:
$\tilde \chi_2^0\to W^\pm \tilde\chi_1^\mp$ with $W\to jj, \tilde\chi_1^\mp \to
\tilde \chi_1^0 +$ very
soft jets. This can happen in models with
light charginos which have a small mass splitting with the LSP.
Unfortunately, the rest of the AKTW models are
unobservable, being buried in
the dijet case by the enormous $W\to jj$ peak. There are two ways to
reduce this background: either
decrease the jet pair mass resolution to a value below $30\%/\sqrt E$
and/or employ positron
polarization to reduce the SM rate for $\gamma e \to W\nu$.

\begin{figure*}[htpb]
\centerline{
\includegraphics[width=13.0cm,height=10cm,angle=0]
{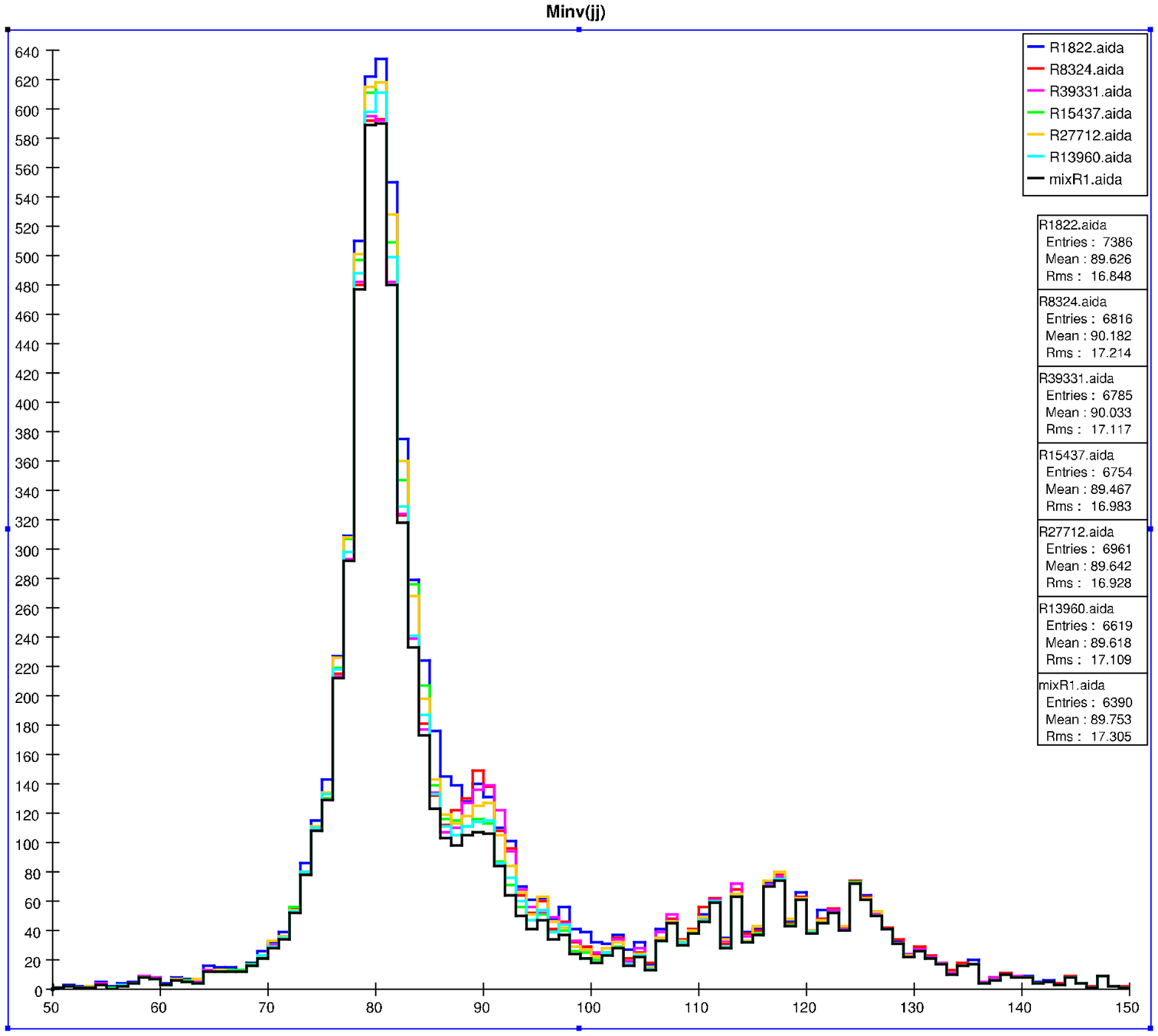}}
\vspace*{0.1cm}
\centerline{
\includegraphics[width=13.0cm,height=10cm,angle=0]
{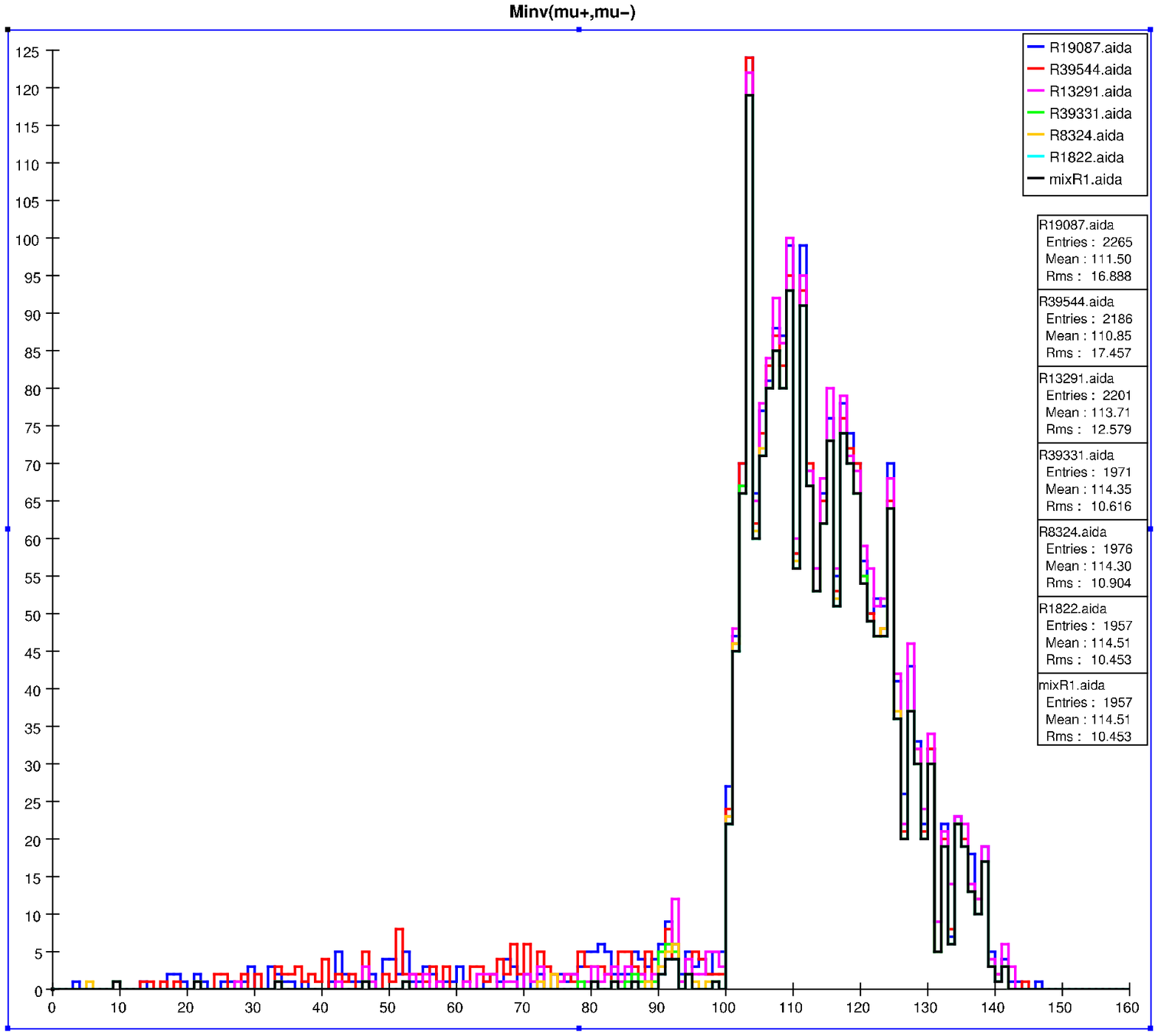}}
\vspace*{-0.1cm}
\caption{Invariant mass distribution in the dijet (dimuon) channel in the
top (bottom) panel from the analysis for associated neutralino
production for representative AKTW models:
events/2 GeV bin assuming RH polarization and 250
fb$^{-1}$ of integrated luminosity. As usual the SM background
corresponds to the black histogram.}
\label{associated1}
\end{figure*}

In the dimuon channel, the signal region is seen to have very little
background, however there are also very few signal events.
A total of ten AKTW models are found to show an excess over
background with a significance
$>5$. However, only a few
of these excesses can be seen in the $Z$ mass region.
Unfortunately, all but 2 of these models are fakes in the sense
that they do not have the $\tilde \chi_2^0 \tilde\chi_1^0$ channel
kinematically accessible; they do, however, all have visible
smuons. Some of the signal for models which populate the lower
invariant mass region
originate from additional sources, such as
$\tilde \chi_1^+\tilde \chi_1^-$, or even $\tilde\chi_3^0 \tilde\chi_1^0$,
production. It would appear from these results that perhaps
the cuts employed in this analysis
are too strong even though the signal region is essentially background
free. However, we find that relaxing the cuts, even just slightly,
overwhelms the signal
region by background. We have not found a set of cuts that
allows more of the signal to be visible over background in this channel.
A similar situation happens for LH electron beam polarization.
While some models lead to
observable signals in the dimuon channel, they are all fakes in the case
of either polarization and the apparent signal is due to feed
down from other SUSY sources. In the \epem\ channel,
7 models are observable with a significance $>5$, however only one of them
is not a fake as illustrated in Fig.~\ref{associated21}.
We note that there are fewer fake signals
in the dijet channel.

Thus at this level of statistics,
these $jj$, $\mu^+\mu^-$, and $\epem$ analyses have
captured very few of the AKTW models where
$\tilde\chi_2^0 \tilde\chi_1^0$ is kinematically
accessible.

\begin{figure*}[htpb]
\centerline{
\includegraphics[width=13.0cm,height=10cm,angle=0]
{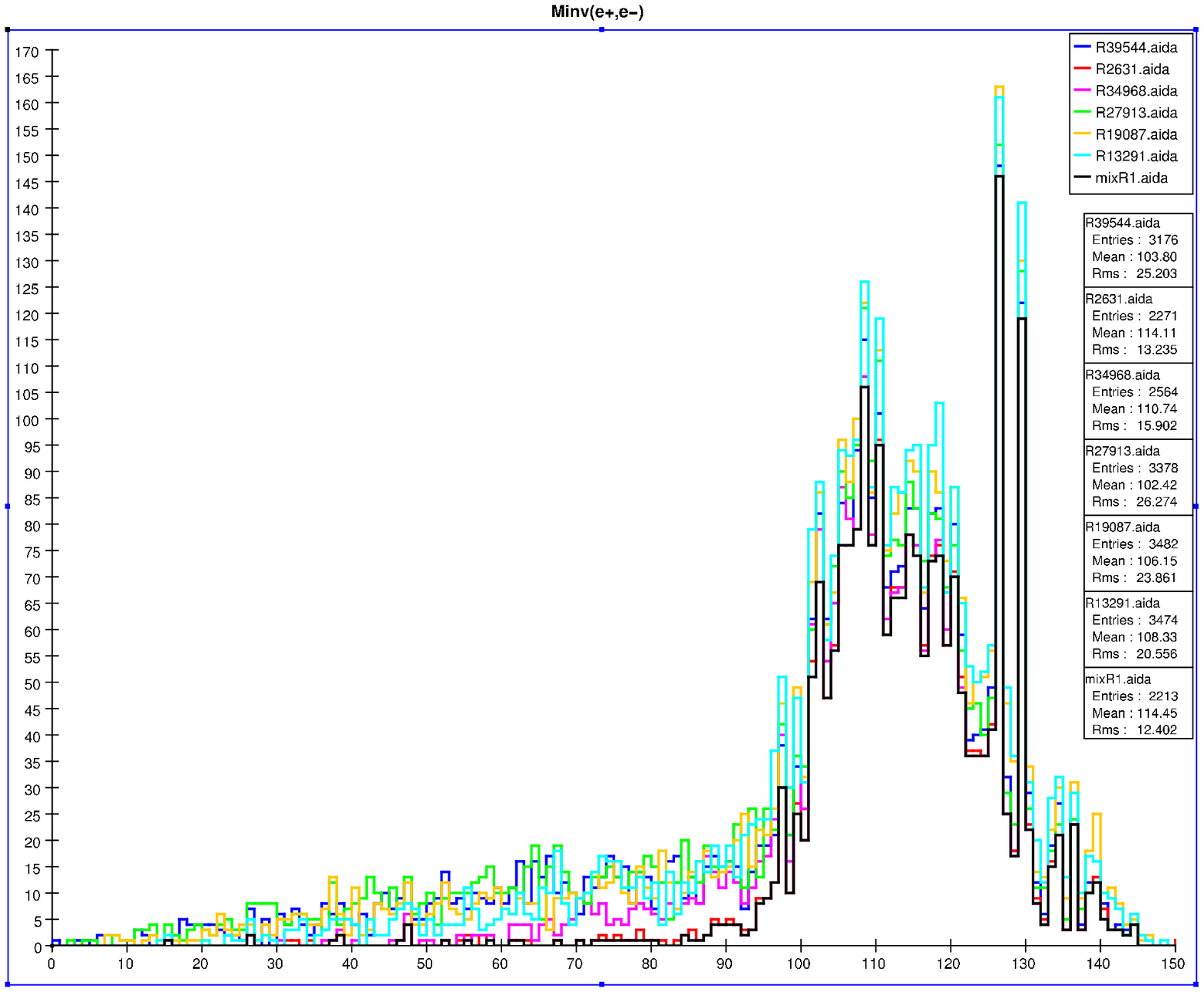}}
\vspace*{0.1cm}
\centerline{
\includegraphics[width=13.0cm,height=10cm,angle=0]
{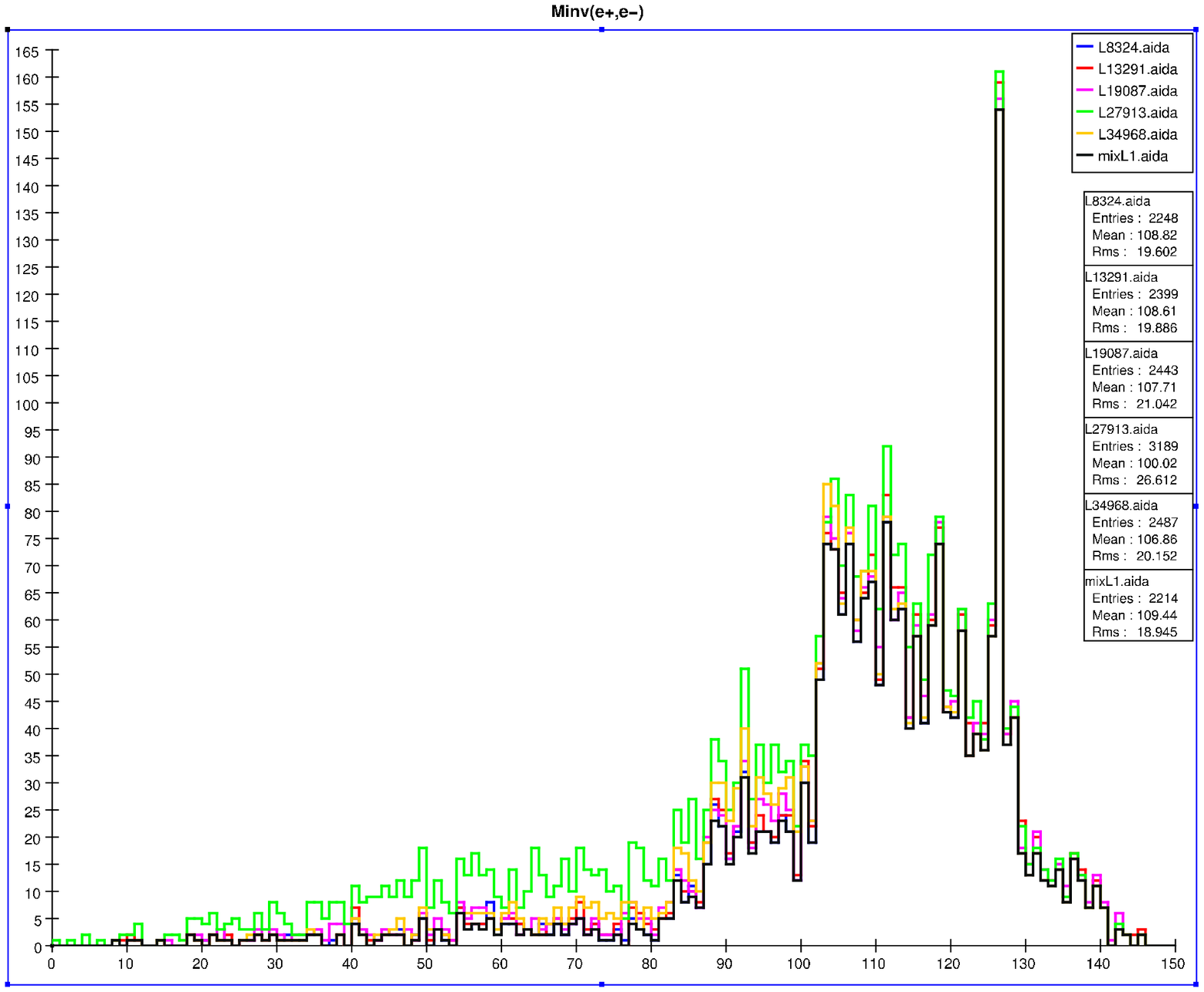}}
\vspace*{-0.1cm}
\caption{Invariant mass distribution for the dielectron channel in the
top (bottom) panel for RH (LH) polarization from the
analysis for associated neutralino production for representative AKTW
models: events/2 GeV bin assuming 250
fb$^{-1}$ of integrated luminosity. As usual the SM background
corresponds to the black histogram.}
\label{associated21}
\end{figure*}

For the conventional benchmark point SPS1a',
associated neutralino production at $\sqrt s=500$ GeV
can easily proceed as the mass
of $\tilde\chi_2^0$ is only
184 GeV. In this case, the $\tilde\chi_2^0-\tilde\chi_1^0$ mass
splitting is $\simeq 86$ GeV, \ie, $<M_Z$, and thus the signal is not
observable in the dijet channel due to the very large
SM $W$ boson background. However, a reasonable non-resonant
signal excess is observable over background in the substantially cleaner
dimuon mode. This is true for either electron beam
polarization, however the signal is more strongly observable for the
case of LH polarization, as
can be seen in
Fig.~\ref{assoc_mumu_sps1a}. Feed down to this final states from the
production of heavier chargino
and neutralino states is also present in this model.

\begin{figure*}[htpb]
\centerline{
\includegraphics[height=10cm,width=13cm,angle=0]{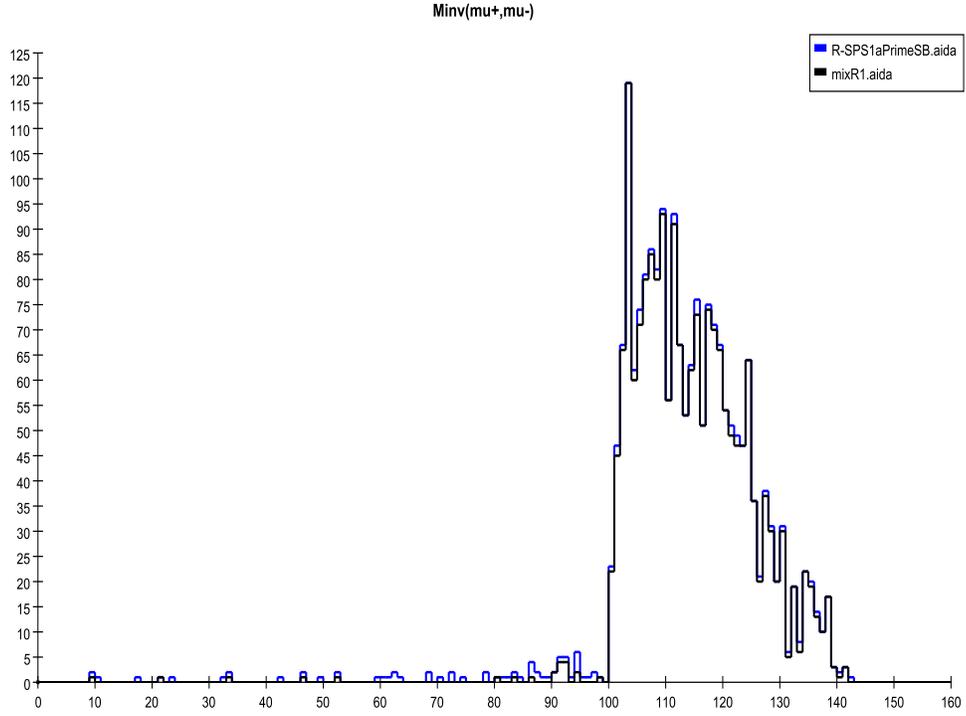}}
\vspace*{0.1cm}
\centerline{
\includegraphics[height=10cm,width=13cm,angle=0]{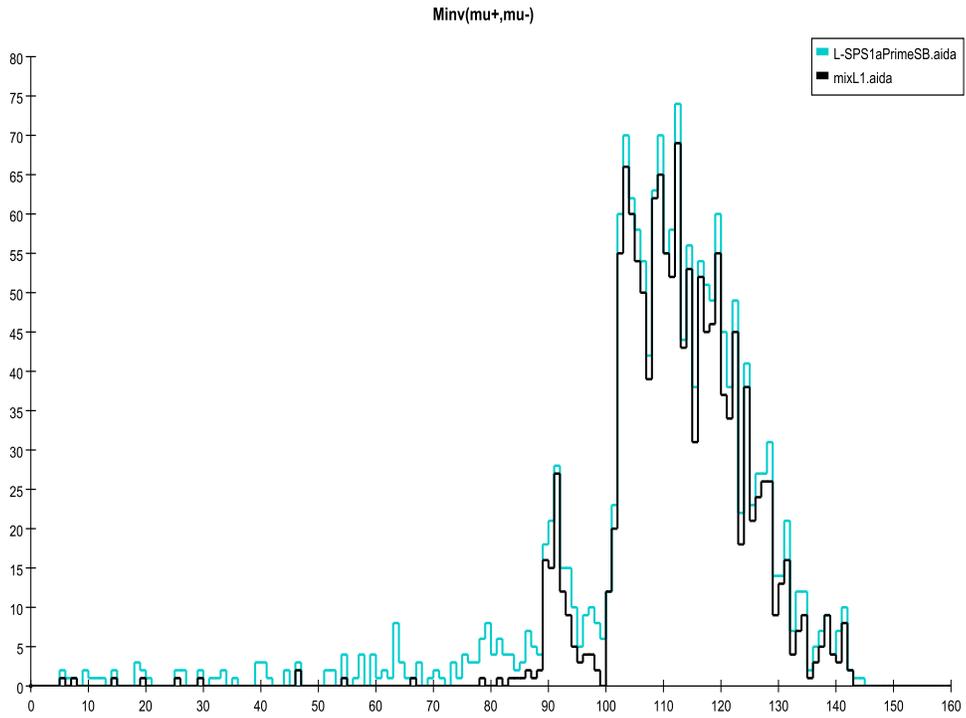}}
\vspace*{0.1cm}
\caption{Invariant mass distribution in the dijet (dimuon) channel in the
top (bottom) panel from the analysis for associated neutralino
production for model SPS1a':
events/2 GeV bin assuming RH polarization and 250
fb$^{-1}$ of integrated luminosity. As usual the SM background
corresponds to the black histogram.}
\label{assoc_mumu_sps1a}
\end{figure*}

\clearpage

\subsection{Radiative Neutralino Production}
\label{Sec:radneu}

In 91 of the 242 AKTW models the neutralino LSP $\tilde\chi_1^0$
is the only kinematically
accessible SUSY particle at $\sqrt s=$ 500 GeV.
The process $e^+e^- \to 2\tilde \chi_1^0$
is, by itself, impossible to observe as the final state particles
are stable, neutral and weakly interacting.
The only way to render $\tilde\chi_1^0$ production
observable is to tag it by the emission of
a photon off the initial state electrons
or off the intermediate $t-$channel selectron; one then looks for an
excess of events of the form
$e^+e^- \to \gamma+E_{miss}$. The SM background to such a
signature can be quite large and arises mainly from the reaction
$e^+e^- \to \nu \bar \nu \gamma$ which occurs through intermediate
$W$ and $Z$ boson exchanges. As will be discussed
below, beam polarization can play an important role in reducing this
dominant component of the SM background, as
$W$ bosons couple to electrons in a purely left-handed manner.

As noted above, we employ PYTHIA for the generation of the signal events
However, PYTHIA does not take into account photon emission from
the virtual $t$-channel selectron in neutralino pair production.
Without this contribution to the cross section for $e^+e^- \to
\tilde\chi^0_1 \tilde\chi^0_1\gamma$, the signal would always be invisible
beneath the background provided by $e^+e^- \to \nu\bar\nu\gamma$.
Thus an accurate modelling of the radiative LSP signal
at the ILC requires a more sophisticated approach.
We therefore use CompHEP to generate the full matrix element for this
process, in a manner
analogous to that described above in Section 5.2.1
for the radiative chargino analysis.
The CompHEP evaluation of the cross section for radiative LSP production
uses the complete matrix element and can be up to a factor of
2 larger than that given by PYTHIA and also generally yields
harder photons.

We tag on a high-$p_T$ photon, which is the sole visible final state
particle in the process $e^+ e^- \rightarrow \gamma \LSP
\LSP$. Clearly, right-handed
electron beam polarization should be effective in reducing the
background contributions from $W$ boson exchange in $\nu\bar\nu\gamma$
production. In fact, after the cuts described below are employed,
we find that the RH SM background event rate is about a
factor of 7-8 less than that with LH beam polarization.
(We note that in the case of $100\%$ electron beam polarization,
the LH cross section is almost 50 times larger than that for the RH case.)
In contrast, we find that the signal cross sections for the AKTW
models follow either one of two patterns:
($i$) the LH and RH polarized cross sections are
comparable in magnitude or ($ii$) the RH polarized cross section is far
larger than that of the LH case. Thus, for either of
these possibilities, RH electron beam
polarization is highly favored in order to increase
the signal and reduce the background. We will thus
limit ourselves to this polarization configuration in our analysis below.

We employ the cuts of Ref.~{\cite{AguilarSaavedra:2001rg}}, and require:
\begin{enumerate}
\item There be exactly one photon and no other visible
particle in the event.

\item The photon transverse energy satisfy $E_T^\gamma =
E_\gamma \sin \theta_\gamma > 0.03 \sqrt{s}$.
Here, $\theta_\gamma$ is the angle of the photon with the
electron beam axis.

\item The photon be present in the angular region
$\cos \theta_\gamma < 0.9$

\item The total photon energy satisfy the constraint
$E_\gamma < 0.5 \sqrt{s} - 90 $ GeV.
This removes radiative return to the $Z$-pole.
\end{enumerate}

We then examine the photon energy distribution and look
for a signal in excess of the SM background;
some typical results are presented in Fig.~\ref{lsppairs}.
Unfortunately, as can be
seen from this Figure, $S/B$ is {\it at best} $\sim 8-9\%$ for the
models shown here. This remains
true for all $180$ AKTW models that have kinematically accessible
$\tilde\chi_1^0$ states. In many cases $S/B$ is
far below the $1\%$ level. However, we find that 17 of the models
lead to a signal significance $\cal S$ greater than 5. We note that
in 4 of these 17 models, the $\tilde\chi_1^0$ is the only kinematically
accessible SUSY particle at $\sqrt s=500$ GeV.
Of course, {\it a priori}, one cannot be certain that the
neutralino LSP has been produced and discovered, as this final state
may receive sizable signal contributions from other SUSY sources
such as $e^+e^- \to \tilde \nu \tilde \nu^* \gamma$, with
$\tilde\nu\to\nu\tilde\chi_1^0$. In fact, model 36022, shown in
the Figure, is an
example of one such case. This renders it difficult to uniquely
identify the signal as
arising from only the lightest neutralino without further analysis.

\begin{figure*}[htpb]
\centerline{
\includegraphics[width=13.0cm,height=10.0cm,angle=0]{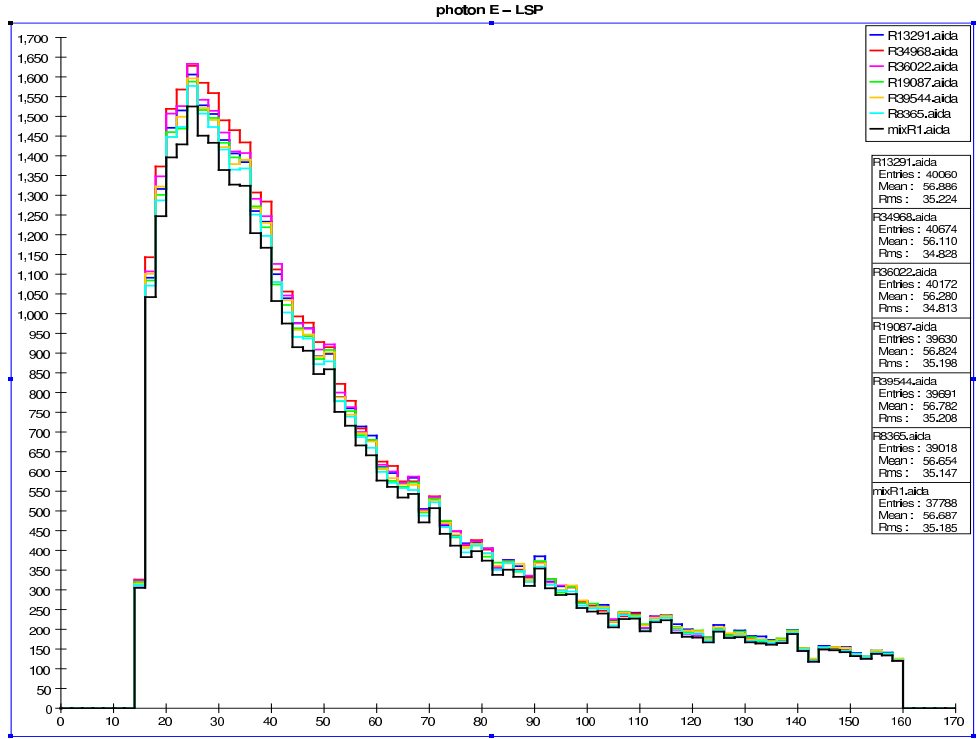}}
\vspace*{0.1cm}
\centerline{
\includegraphics[width=13.0cm,height=10.0cm,angle=0]{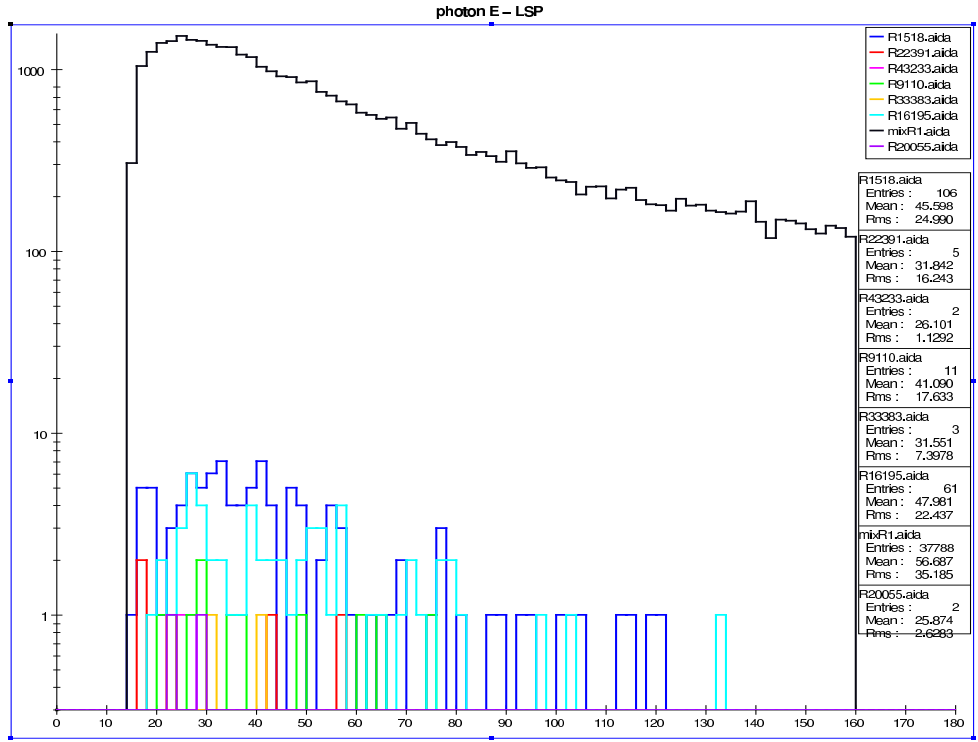}}
\vspace*{-0.1cm}
\caption{Photon energy spectra for several representative AKTW models from
the photon-tagged $\tilde\chi_1^0$ pair
production process. Shown is the event rate/2 GeV bin assuming RH
electron beam polarization and an integrated luminosity
of 250 fb$^{-1}$. The black histogram is the SM background. The top
panel shows signal plus background for models with larger event rates whereas
the bottom panel displays more typical cases with signal and the
background now being shown separately.}
\label{lsppairs}
\end{figure*}

Of course, increased luminosity or an adjustment of the cuts may make
the signal in this channel slightly more visible, but
what would be more useful, independently of the choice of cuts, would
be to include positron polarization {\cite {MoortgatPick:2005cw}}.
Having such polarization at the $30(45,60)\%$ level would reduce the
background by roughly $\simeq 44(60,73)\%$
in comparison to that with the canonical $80\%$ electron beam polarization
assumed in our analysis. The corresponding
increase in the signal in the most conservative AKTW model would be
$24(36,48)\%$ and thus significant boosts
in $S/B$ would result; these increases can be somewhat larger
depending upon the parameter values in a particular model.

The observation of radiative neutralino LSP pair production is rather
straightforward in the case of the familiar benchmark model
SPS1a', where the LSP is rather light with a mass of only 97.7 GeV.
Figure~\ref{LSPpairs_sps1a} shows that the signal is much larger in
this case than in
any of the AKTW models. In fact, an excess
in the number of events over background
can be observed for almost the entire range of the photon energy
spectrum. However, some of this
excess may be attributed to radiative sneutrino pair production
which is reasonably
significant in this model {\cite {Dreiner:2006sb}}.

\begin{figure*}[htpb]
\centerline{
\includegraphics[height=10cm,width=13cm,angle=0]{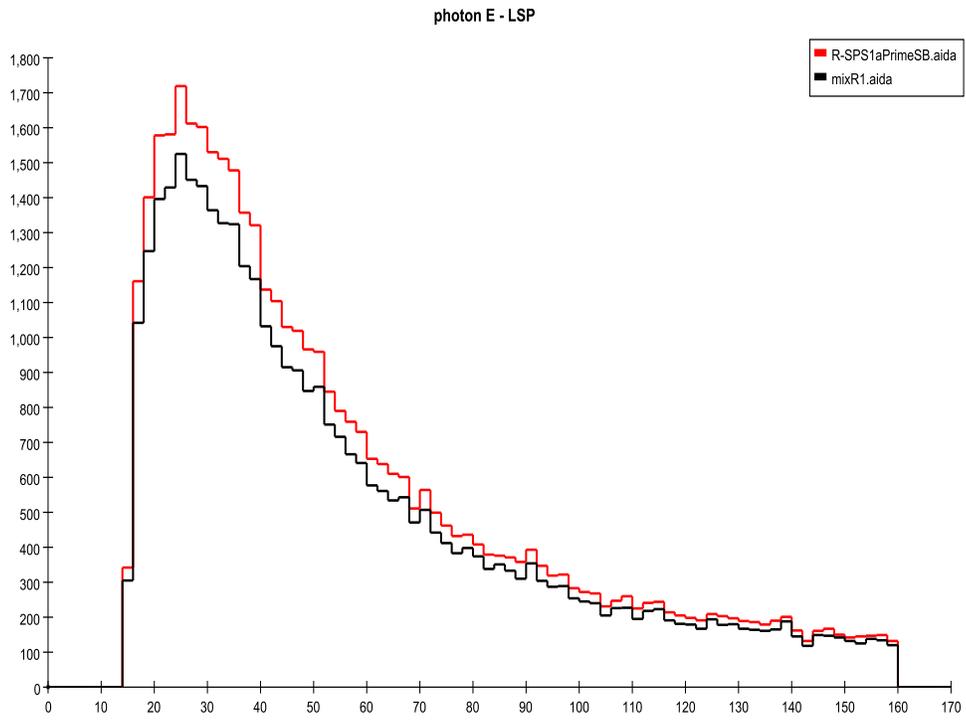}}
\vspace*{0.1cm}
\caption{Photon energy spectra for the benchmark model SPS1a' from
the photon-tagged $\tilde\chi_1^0$ pair
production process. Shown is the event rate/2 GeV bin assuming RH
electron beam polarization and an integrated luminosity
of 250 fb$^{-1}$. The black histogram is the SM background.}
\label{LSPpairs_sps1a}
\end{figure*}

\clearpage

\section{Model Visibility and Comparisons}

As discussed in the Introduction, before addressing the issue of
model differentiation
we first need to determine which SUSY particles are visible above
the SM background in each of the models under study.
In particular we would like to know how
many models contain a given SUSY
particle that is clearly observable
with a significance, $\cal S$, greater than 5 from our Likelihood
analysis discussed in Section 3.
This information can be obtained
by combining the individual results found in Sections 4 through 6;
the summary of
these analyses in terms of SUSY particle discovery is displayed in
Table~\ref{visibleparticles}.
This Table shows the number
of models with a given SUSY particle that we found to yield a visible
signal above background in our analysis relevant for that particular
SUSY state, as
compared to the number of models where
the same particle is kinematically accessible. Thus, \eg, the
$\tilde e_R$ is observable in 12 of the 15 models in which it is
kinematically accessible. We declare a particle to be visible for a given
model if it is kinematically accessible and a signal with ${\cal S}>5$
is observed in the relevant search channel.
We note that it is possible that some of these
observable signatures may be due to fakes, \ie, the production of other
SUSY states; this is certainly true in, \eg, the case of radiative
$\tilde\chi_1^0$ production.
From this Table we see
that for the set of AKTW models the ILC does an excellent job at detecting
selectrons and smuons as
well as charginos, however, staus are somewhat more problematic,
and the neutralino sector appears to be difficult.

\begin{table}
\centering
\begin{tabular}{|c|c|} \hline\hline
Particle & Number Visible \\ \hline
$\tilde e_L$ & 8/9 \\
$\tilde e_R$ & 12/15 \\
$\tilde \mu_L$ & 9/9 \\
$\tilde \mu_R$ & 12/15 \\
$\tilde \tau_{1,2}$ & 21/28 \\
$\tilde \nu_{e,\mu}$ & 0/11 \\
$\tilde \nu_\tau$ & 0/18 \\
$\tilde \chi_1^\pm$ & 49/53 \\
$\tilde \chi_1^0$ & 17/180 \\
$\tilde \chi_2^0$ & 5/46 \\ \hline\hline
\end{tabular}
\caption{Number of models, at $\sqrt s=500$ GeV, which have a
given final state
particle visible above the SM background with a significance ${\cal S}>5$
as defined in the
text, divided by the number of models where the same particle
is kinematically accessible.}
\label{visibleparticles}
\end{table}

We can now combine the results represented
in the Table and ask for the {\it total} number of models which
contain visible sparticles with a signal
significance greater than 5. Out of the 85 models which have at least
one charged SUSY partner kinematically accessible, we find that 78 have
visible sparticle signatures at the ILC. The SUSY particles in the
other 7 models are not detectable mainly due to
phase space suppression of the SUSY cross sections as discussed
in the previous Sections.
Of the 96 models which have only stable neutral SUSY partners accessible
($\tilde\chi_1^0$ or $\tilde\nu$),
4 of them are observable via the photon tag recoil analysis. Thus,
out of all the
models with at least one
accessible SUSY partner we find that 82/181 lead to detectable signals
at the ILC. This corresponds to 82 visible models out of the full set
of 242 AKTW models
(recall 61 of the models have no SUSY partners accessible at 500 GeV).
Surveying all of the models, there is a total of 129 charged
sparticles which are kinematically accessible and we find that
111 are visible in our analyses; several more may appear as `fakes'.

Using our ILC analyses,
we now pairwise compare the models that were found
to be indistinguishable at
the LHC by AKTW. We recall that out of the
original 283 model pairs, 121 were removed from our sample
due to the PYTHIA feature which shifted the LSP mass,
leaving us with 162 pairs of models to examine. Interestingly,
out of these 162 model pairs,
90 have only neutral sparticles kinematically
accessible in both models.

In order to compare signals that originate from different
models for the observables described in the previous Sections,
we perform a $\chi^2$ analysis of the generated
histogram distributions for the model pairs. To begin, recall
that we have generated two complete and statistically independent
background sets, $B1$ and $B2$, for all
of the individual analyses. Taking the set of pure signal
distributions for the two models we wish to
compare, $M1$ and $M2$, we add each signal distribution to one of the
corresponding background distributions. This forms
the combinations $R=M1+B1$ and $S=M2+B2$ for each observable.
We then perform a $\chi^2$ analysis of the two distributions
for each model pair,
accounting for the fact that the number of events in each sample
can be different:

\be
\chi^2 = \sum_i \frac{ \left( \sqrt{\frac{S}{R}}
R_i - \sqrt{\frac{R}{S}} S_i \right)^2}{
R_i + S_i},
\ee
with
\be
R = \sum_i R_i \qquad S = \sum_i S_i \, .
\ee
$R_i$ and $S_i$ denote the number of events in bin $i$ produced by
the two models (plus background) in each observable
that we compare. Note that such a $\chi^2$ test is somewhat
sensitive to the binning
of the data, especially since we compare two sets of generated
``data'' instead of
comparing a signal to a theoretical
prediction. Note further that the above $\chi^2$
prescription relatively normalizes the two
distributions so that we only
compare shapes at this point. We then add an additional term to the
$\chi^2$ which accounts for the
total number of events in both histograms and allows for
an $1\%$ systematic error in the relative
normalizations due to luminosity and cross section
normalization uncertainties.

We then compute the $\chi^2$ distributions for each of the model pairs,
for each the following
observables, taken one at a time, which were obtained after
applying the analysis cuts described in the Sections above:

\begin{itemize}

\item Selectron analysis: $E_{e^+ \mbox{\tiny{ or }}e^-}$ and
$p_{T \mbox{vis}}$.

\item Smuon analysis: $E_{\mu^+ \mbox{\tiny{ or }}\mu^-}$ and
$p_{T \mbox{vis}}$ .

\item Stau analysis: $E_{\tau}$
and $p_{T \mbox{vis}}$. We employ the $\tau$ identification procedure
described in Section~\ref{Sec:stau}, with and without the inclusion of
electrons in the final state in order to remove the
background from beam remnants.

\item Sneutrino analysis: missing energy for two channels,
4-jet plus lepton-pair and 6-jets. Each channel is analyzed
with two different assumed minimum values of the LSP mass.

\item Chargino non-close mass case: For the case of on-shell
$W$ boson production in chargino decays, we examine the
$E_{\mbox{\tiny{ jetpair}}}$ spectrum. For the case where the
charginos decay into off-shell $W$ bosons, we examined three
decay channels. Our observable for
the fully leptonic channel is $E_{\mu^+ \mbox{\tiny{ or }}\mu^-}$.
In the fully hadronic channel we analyze
the missing energy distribution for the 4-jet final state and
the invariant mass spectrum of the two jet-pairs. For these two
distributions we perform analyses with two different assumed values of the
$\tilde\chi_1^0$ mass and also with an additional cut on
$p_{T \mbox{vis}}$. In the semi-leptonic channel with the
jet-pair $+\mu +$ missing energy final state, we examine
$E_{jj}$ as well as the invariant mass of the jet-pair. In this case,
we again employ analyses with two different assumed values of the
LSP mass.

\item Chargino radiative production: the recoil mass
$M_{\mbox{\tiny {recoil}}}$ of the tagged hard photon.

\item Chargino very close mass case: $\beta=\frac{p}{E}$ of
the two massive tracks in the event, assuming an energy smearing/$\beta$
resolution of 5\% and 10\%.

\item $\tilde\chi_2^0\tilde\chi_1^0$ associated production analysis:
the invariant mass of electron, muon and jet-pairs.

\item Radiative $\tilde\chi_1^0$ analysis: $E_\gamma$.

\end{itemize}

Before examining the results of our $\chi^2$ model comparison,
we first check our procedure by comparing
the two pure background samples to verify that, though independent,
they are not statistically
distinguishable. We do indeed find this to be the case for every
observable in each analysis. Next, we examine each
observable listed above and determine
whether the comparison probability
(as given by the value of the $\chi^2$ and
number of degrees of freedom) shows a difference at the
5(or 3)$\sigma$ level for that specific distribution.
Note that we perform this comparison separately for each electron beam
polarization since we have distributions for
each polarization configuration. If there is a 5(or 3)$\sigma$ level
difference in {\it at least one}
distribution then we claim
that the two models are distinguishable at that level of confidence.
In fact, a number of observables are distinguishable at this level
in many models.
We can improve this procedure by summing over the various observables
in the $\chi^2$ computation, taking
only one distribution from the different analyses to ensure that we do
not introduce any effects from correlations.
Note that in the case where we employ only a single observable
in this comparison there are no issues of statistical
independence in contrast to when several distributions are combined.

When performing these comparisons, we find that many model pairs are
not distinguishable. This happens, \eg, in all of the 90 cases
where we compare models which have only neutral particles
kinematically accessible. The few models where we observe excess
photons in radiative $\tilde\chi_1^0$ production
are already differentiated by other analyses involving charged sparticle
production, and
thus these cases do not help with distinguishing
pairs of models containing only kinematically accessible
neutral sparticles.
This implies that we should concentrate on the
72 potentially distinguishable model pairs where at least one
member of the pair has at least one kinematically accessible
charged sparticle. In this case,
we find that a
large number of model pairs are distinguishable at the $5\sigma$
level in several different analyses.

Based on the criteria above, using our results from the {\it single}
observable comparison procedure described above, we find that 55(63)/72
model pairs where at least one model has kinematically accessible
charged sparticles are
distinguishable at the $5(3)\sigma$ level.
These results are based solely on single final state comparisons between
models. Making use of the {\bf combined} observable comparison procedure
described above, we find instead that 57(63)/72 pair of models
are distinguishable, which is only a slight improvement.

The model pairs that are found to be indistinguishable fall
into two broad classes: ($i$) those
where one model in the pair has only a kinematically accessible neutral
sparticle, \eg, the LSP,
which is not visible above background and the other model contains a sole
accessible charged sparticle which is also difficult to observe.
There are 7 model pairs in this category that cannot be
differentiated at the $5\sigma$
level. Examples are models with a heavy selectron and smuon which have
kinematically suppressed cross sections and models which only
contain $\tilde \tau$ states that are also difficult to observe
due to small production cross sections. ($ii$) The second class
consists of 8 model pairs where each model in the pair contains a single
kinematically accessible charged sparticle. In 7 of the 8 model
pairs, it is the lightest chargino state that is produced and found
to lead to an indistinguishable signature.

\section{Conclusions}

In this paper, we have performed a systematic and detailed analysis
of the capabilities of the
500 GeV ILC (with an integrated luminosity of 500 fb$^{-1}$ with $80\%$
electron beam polarization) to explore the nature of a large
number, 242, of scenarios within the MSSM. The goal of this project
was to study general model-independent MSSM signatures
and then to determine whether
the ILC could differentiate between 162 pairs of these models,
\ie, MSSM parameter space points, which were found
to be impossible to distinguish at the LHC. To do this we first had to
address the issues of
kinematic accessibility of the SUSY states, as well as the experimental
observability of the corresponding sparticle production over the
SM background. Given the present controversy about the ILC in the world-wide
community, we feel that such a detailed study of the MSSM at the ILC is critical
at the present time.

In order to accomplish this task, we employed a complete set of full matrix
element SM backgrounds for all $2\to 2$, 4 and 6 parton final states
initiated by $e^+e^-,\gamma e^\pm$ and $\gamma\gamma$ as
calculated by T. Barklow using WHIZARD/O'MEGA.
We made use of both PYTHIA and CompHEP for generating
the SUSY model signal events and employed a realistic beamspectrum
generated with WHIZARD/GuineaPig. Additionally,
we included the effects of the SiD detector by implementing a version of
the org.lcsim fast simulation.
In this analysis we assumed that the integrated luminosity was equally
split between two distinct
samples with $80\%$ LH or RH electron beam polarization. Analyses were
performed on many different SUSY
channels simultaneously in order to probe the charged slepton, sneutrino,
lightest chargino, LSP and $\tilde \chi_2^0-\tilde\chi_1^0$ sectors.
A universal set of cuts for all models was developed. This is the first realistic,
model-independent analysis of its kind of SUSY signatures at the ILC

Out of the original 242 models only 85 led to the existence of a
kinematically accessible charged SUSY
partner at 500 GeV. The remaining models either had no SUSY particles
kinematically accessible (61) or only the lightest neutralino and/or
sneutrino accessible (96).
Using log likelihood techniques, we found that 78/85
models with a charged SUSY
partner as well as 4 additional models which only had neutral particle
states accessible were visible above SM background in our analyses.
Thus, a total of 82/161 models with accessible particles were
found to be observable at the 500 GeV ILC. In performing our analysis,
beam polarization was essential in reducing the SM background and
allowed for distinguishing sparticle states in many cases. Some models
contained charged states that were found not to be visible generally
as a result of suppressed cross sections due
to phase space availability. Of the 72 pairs of models where at least
one member of the pair contains one
or more accessible charged SUSY partner, our analysis found that
57(63) of the pairs could be distinguished at the level of $5(3)\sigma$.
These results are generally quite surprising.
Perhaps this is because the "standard lore" about SUSY at the ILC derives
largely from studies of a handful of very special points in MSSM parameter
space, whose signatures are particularly easy to see.  We, on the other hand,
have studied a wide variety of MSSM parameter points.

From this analysis it is clear that the ILC with the SiD detector
does a reasonably good job at observing charged sparticles which are
kinematically accessible and distinguishing models containing
such particles. The major weakness, beyond the restricted kinematic reach,
is in the neutral sparticle sector. This problem may be resolved by
employing positron beam polarization as
well as even more sophisticated analyses.

For the future we plan to extend this analysis to the case of 1 TeV
center-of-mass energy, and include a study of the influence
of positron polarization as well as a number of detector issues.

\section{Acknowledgments}

We are indebted to Tim Barklow for extraordinary help with the Standard Model
background and for many invaluable discussions. We thank Burt Richter
for asking the right question.
We also thank Alexander Belyaev, John Conway, J. Feng,
Norman Graf, John Jaros,
Uli Martyn, Jeremy McCormick, William Morse, Steve Mrenna, Uriel Nauenberg,
Nan Phinney, Keith Riles, Bruce Schumm,
Peter Skands, Tim Tait, Jesse Thaler, Graham Wilson and Mike Woods
for useful conversations. JLH thanks the Fermilab theory group for
their hospitality while part of this work was carried out.
We would also like to thank Lupe Makasyuk for help in the preparation of this manuscript.
Furthermore, we are especially grateful to Ben Lillie of Auburn, Alabama, for
providing the ultimate feature and for his good humor and patience.
BL's research was supported in part by the US Department of Energy under contract
 DE--AC02--06CH11357.


\begin{thebibliography}{99}
\bibitem{TimB}
The SM backgrounds we employ here have been generated by
T.~Barklow.
\end{thebibliography}
\end{document}